\documentclass[%
	a4paper,
	twoside,
	numbers=noenddot,
	headsepline,
	footsepline,
	headings=small,
]{scrreprt}

\usepackage{scrhack} 

\usepackage[sectionbib]{natbib}

\usepackage[utf8]{inputenc}
\usepackage{kbordermatrix}
\usepackage{algorithm}
\usepackage{algpseudocode}

\newcommand{\MAFOAutor}{Marcel Müller}
\newcommand{\MAFOakadTitel}{M.\,Sc.}

\newcommand{\MAFOTitel}{Multi-Agent Reinforcement Learning for Deadlock Handling among Autonomous Mobile Robots}

\usepackage{lmodern}

\usepackage{microtype}
\usepackage{booktabs}
\usepackage{array}
\usepackage{tabularx}
\usepackage{multirow}
\usepackage[table]{xcolor} 
\newcolumntype{C}{>{\centering\arraybackslash}X}
\newcolumntype{L}{>{\raggedright\arraybackslash}X}
\newcolumntype{R}{>{\raggedleft\arraybackslash}X}

\usepackage{pdfpages}

\usepackage{ifpdf}

\usepackage{graphicx}
\graphicspath{{bilder/}}
\usepackage{longtable}
\ifpdf
	\DeclareGraphicsExtensions{.pdf,.jpg}
\else
	\DeclareGraphicsExtensions{.png,.jpg}
\fi

\usepackage{grffile}

\usepackage{tikz}
\usetikzlibrary{positioning}
\usetikzlibrary{arrows}
\usetikzlibrary{patterns}

\tikzset{>=latex}

\usepackage{tikz-3dplot}

\usetikzlibrary{backgrounds}
\usepackage{tikz-layers}

\usepackage{pgfplots}
\pgfplotsset{compat=newest}
\pgfplotsset{
	scriptsize/.style={
		width=4.5cm,
		height=,
		legend style={font=\scriptsize},
		tick label style={font=\scriptsize},
		label style={font=\footnotesize},
		title style={font=\footnotesize},
		every axis title shift=0pt,
		max space between ticks=25,
		every mark/.append style={mark size=7},
		major tick length=0.1cm,
		minor tick length=0.066cm,
	}
}
\pgfplotsset{
	small/.style={
		width=6.5cm,
		height=,
		tick label style={font=\footnotesize},
		label style={font=\small},
		legend style={font=\footnotesize},
		max space between ticks=30,
	}
}

\pgfplotsset{legend cell align=left}
\pgfplotsset{xmajorgrids}
\pgfplotsset{ymajorgrids}
\pgfplotsset{scale only axis}
\definecolor{matlab1}{rgb}{0,0,1}
\definecolor{matlab2}{rgb}{0,0.5,0}
\definecolor{matlab3}{rgb}{1,0,0}
\definecolor{matlab4}{rgb}{0,0.75,0.75}
\definecolor{matlab5}{rgb}{0.75,0,0.75}
\definecolor{matlab6}{rgb}{0.75,0.75,0}
\definecolor{matlab7}{rgb}{0.25,0.25,0.25}
\definecolor{applegreen}{rgb}{0.55, 0.71, 0.0}
\definecolor{darkslateblue}{rgb}{0.28, 0.24, 0.55}
\definecolor{amber}{rgb}{1.0, 0.75, 0.0}
\definecolor{electricyellow}{rgb}{1.0, 1.0, 0.0}
\pgfplotscreateplotcyclelist{matlab}{
	{matlab1,solid},
	{matlab2,dashed},
	{matlab3,dashdotted},
	{matlab4,dotted},
	{matlab5,densely dashed},
	{matlab6,densely dashdotted},
	{matlab7,densely dotted}
}
\pgfplotsset{cycle list name=matlab}
\pgfplotsset{every axis plot/.append style={line width=1pt}}
\pgfplotsset{/pgf/number format/.cd,1000 sep={\,}}

\ifpdf
	\usepgfplotslibrary{external}
	\tikzsetexternalprefix{external_figs/}
\fi

\usepgfplotslibrary{statistics}
\pgfplotsset{
   boxplot/every whisker/.style={solid},
   boxplot/every boxplot/.style={solid},
   boxplot/every box/.style={solid},
   boxplot/every average/.style={solid},
   boxplot/every median/.style={solid},
}
\pgfplotsset{every boxplot/.style={mark=*, every mark/.append style={mark size=1.5pt}}} 

\usepackage[edges]{forest}
\usetikzlibrary{shadows}

\usepackage{subfig}

\setcapindent{1em}

\usepackage{amsmath}
\usepackage{amssymb}
\usepackage{amsthm}
\theoremstyle{definition}
\newtheorem{definition}{Definition}

\usepackage{tensor}

\usepackage{setspace}
\addtokomafont{pageheadfoot}{\linespread{1}\selectfont}

\usepackage{rotating}

\usepackage{color}

\usepackage{numprint}

\usepackage[margin=2.5cm]{geometry}

\usepackage{xfrac}

\usepackage{csquotes}

\usepackage{xspace}

\usepackage{booktabs}

\RedeclareSectionCommand[%
	beforeskip=-20pt plus -2pt minus -2pt,%
	afterskip=10pt plus 2pt minus 2pt%
]{paragraph}

\RedeclareSectionCommand[%
	beforeskip=-6pt plus -2pt minus -2pt,%
	afterskip=6pt plus 2pt minus 2pt%
]{subparagraph}

\makeatletter
\renewcommand{\fps@figure}{htbp}
\renewcommand{\fps@table}{htbp}
\renewcommand{\fps@algorithm}{htbp}
\makeatother

\makeatletter
\newcommand\itemizebullet{%
{\@listi\hspace*{\dimexpr\leftmargin-\labelsep}}\llap{\labelitemi}%
\hspace{\labelsep}}
\makeatother

\usepackage{blindtext}

\usepackage[
	pdftitle={Dissertation},
	pdfauthor={\MAFOAutor},
	pdfcreator={MiKTeX, LaTeX mit hyperref und KOMA-Script},
	pdfsubject={\MAFOTitel},
	pdfkeywords={Stichwort, Stichwort, Stichwort},
	pdfstartview={Fit},
	plainpages=false, 
	pdfpagelabels	
]{hyperref}

\usepackage{graphbox}

\cleardoublepage

\phantomsection\label{listofeq}
\usepackage{hyperref}
\hypersetup{
    colorlinks=false, 
    linkcolor=black,   
    citecolor=black,   
    filecolor=black,   
    urlcolor=blue,     
    pdfborder={0 0 0}  
}

\definecolor{blueMM}{rgb}{0.0, 0.5, 1.0}
\definecolor{redMM}{rgb}{1.0, 0.13, 0.32}
\definecolor{greenMM}{rgb}{0.2, 0.8, 0.2}

\begin{document}


\pagenumbering{alph}
\pagestyle{empty}
%
%

\begin{titlepage}
	\mbox{}\vspace{1.5cm} 
	\begin{center}
		\Large
		\MAFOTitel 
	\end{center}
	\vspace{1.5cm} 
	\normalsize
	\textbf{Dissertation}
	
	\noindent zur Erlangung des akademischen Grades
	\vspace{1.5cm} 
	\begin{center}
		\Large
		Doktoringenieur
		
		(Dr.-Ing.)
	\end{center}
	\vspace{1.5cm} 
	\normalsize
	von \textbf{\MAFOAutor, \MAFOakadTitel} \\
	
	\noindent geb. am 30.10.1990 in Magdeburg \\
		
	\noindent genehmigt durch die Fakultät für Maschinenbau \\
	der Otto-von-Guericke-Universität Magdeburg
	\vfill 
	\noindent Gutachter:
	\begin{labeling}{Gutachter:}
		\item[] Prof. Dr.-Ing. Hartmut Zadek
		\item[] Prof. Dr.-Ing. Ernesto William De Luca
	\end{labeling}
	\vspace{1.5cm} 
	Promotionskolloquium am 27. Oktober 2025
\end{titlepage}



\cleardoublepage

\pagenumbering{roman}
\pagestyle{plain}
%
%

\begin{flushright}
    \vspace*{5cm}
    \emph{\enquote{Wenn ein durstiger Mann zwei Becher Wasser vor sich stehen hat, die für seinen Zweck in jeder Hinsicht gleich sind, müsste er verdursten, sofern nicht einer von beiden ihm schöner, leichter oder seiner rechten Hand näher erschiene.}}
    
    \vspace{0.8em} \footnotesize
    — Abū Hāmid al-Ghazālī ($\approx$ 1095), \textit{Tahāfut al-falāsifa (Die Inkohärenz der Philosophen)}\\*
    Problem I, S.\,25 f., dt. Übers. d. Verf. nach S.\,A.\ Kamali (1963), \textit{The Incoherence of the Philosophers}
\end{flushright}

\begin{flushright}
    \vspace*{5cm}
    \emph{\enquote{If a thirsty man has before him two glasses of water, which are equal in all respects as far as his purpose is concerned,
    he cannot take either of the two, unless he thinks that one of the two is prettier, or ligther, or nearer to his right hand [...]}}
    
    \vspace{0.8em} \footnotesize
    — Abū Hāmid al-Ghazālī ($\approx$ 1095), \textit{Tahāfut al-Falāsifa (The Incoherence of the Philosophers)}\\*
Problem I, p.\,25 f. (transl.\ S.\,A.\ Kamali 1963)
\end{flushright}


\cleardoublepage
%

\chapter*{Acknowledgements}

First and foremost, my deepest thanks go to my wife, Lorena. Her unwavering support has shaped both my academic journey and my life outside of it. Her encouragement, patience, and our countless conversations have meant everything to me during these challenging years.

I am profoundly grateful to my parents, Birgit and Matthias, for their lifelong support and the values they instilled in me. Their belief in the importance of education and their steady encouragement have laid the foundation for everything I’ve achieved.

I would also like to remember my grandmother, Gerda, who passed away while I was working on this thesis. One of her final words was her pride in her grandchildren. I hope she would be proud of this work too.

Special thanks are due to Prof. Hartmut Zadek and Prof. Ernesto William De Luca for entrusting me with this topic and giving me the freedom to pursue it in my own way. Their guidance and respectful feedback made this work so much richer.

I am especially thankful to all the students who tackled the tricky subject of deadlocks in their own theses. In particular, I want to mention Hendrik for his dedication and the insights he shared along the way.

I am also very grateful to Elke for her valuable feedback and the ideas she contributed from a logistics perspective. Her input helped shape and improve this work in meaningful ways. Thanks also to my colleagues and everyone in the academic community for the many conversations, big and small, that sparked ideas, kept me motivated, or simply made this journey a bit more enjoyable.

Finally, a special thank you to Mathias Magdowski for providing the thesis template, which made formatting and presenting this work so much easier.

\cleardoublepage
%

\section*{Kurzfassung}
Diese Dissertation untersucht den Einsatz von Multi-Agent-Reinforcement-Learning (MARL) zur Handhabung von Deadlocks in Intralogistiksystemen, die auf autonome mobile Roboter (AMR) setzen. AMRs erhöhen die betriebliche Flexibilität, steigern jedoch auch das Risiko von Deadlocks, die den Systemdurchsatz und die Zuverlässigkeit mindern. Bestehende Ansätze vernachlässigen oft die Behandlung von Deadlocks in der Planungsphase und stützen sich auf starre Steuerungsregeln, die sich nicht an dynamische Betriebsbedingungen anpassen.

Diese Arbeit entwickelt eine strukturierte Methodik zur Integration von MARL in die Logistikplanung und Betriebssteuerung. Dazu werden Referenzmodelle eingeführt, die deadlock-fähige Multi-Agent-Pathfinding-Probleme (MAPF) explizit berücksichtigen und eine systematische Bewertung von MARL-Strategien ermöglichen. Mit gitterbasierten Umgebungen und externer Simulationssoftware werden traditionelle Deadlock-Strategien mit MARL-basierten Ansätzen verglichen, wobei der Fokus auf den Algorithmen PPO und IMPALA in verschiedenen Trainings- und Ausführungsmodi liegt.

Die Ergebnisse zeigen, dass MARL-Strategien, insbesondere in Kombination mit zentralisiertem Training und dezentraler Ausführung (CTDE), in komplexen, überlasteten Umgebungen regelbasierte Verfahren übertreffen. In einfacheren Umgebungen oder bei ausreichendem Bewegungsspielraum bleiben regelbasierte Ansätze aufgrund ihres geringeren Rechenaufwands konkurrenzfähig. Diese Resultate verdeutlichen, dass MARL eine flexible und skalierbare Lösung zum Umgang mit Deadlocks in dynamischen Intralogistikszenarien bietet, jedoch eine sorgfältige Anpassung an den jeweiligen Anwendungsfall erfordert.

\section*{Abstract}
This dissertation explores the application of multi-agent reinforcement learning (MARL) for handling deadlocks in intralogistics systems that rely on autonomous mobile robots (AMRs). AMRs enhance operational flexibility but also increase the risk of deadlocks, which degrade system throughput and reliability. Existing approaches often neglect deadlock handling in the planning phase and rely on rigid control rules that cannot adapt to dynamic operational conditions.

To address these shortcomings, this work develops a structured methodology for integrating MARL into logistics planning and operational control. It introduces reference models that explicitly consider deadlock-capable multi-agent pathfinding (MAPF) problems, enabling systematic evaluation of MARL strategies. Using grid-based environments and an external simulation software, the study compares traditional deadlock handling strategies with MARL-based solutions, focusing on PPO and IMPALA algorithms under different training and execution modes.

Findings reveal that MARL-based strategies, particularly when combined with centralized training and decentralized execution (CTDE), outperform rule-based methods in complex, congested environments. In simpler environments or those with ample spatial freedom, rule-based methods remain competitive due to their lower computational demands. These results highlight that MARL provides a flexible and scalable solution for deadlock handling in dynamic intralogistics scenarios, but requires careful tailoring to the operational context.




\cleardoublepage

\pagenumbering{arabic}
\pagestyle{headings}
\tableofcontents
\listoffigures
\listoftables
\listofalgorithms
\phantomsection
\thispagestyle{empty}
\chapter*{List of Abbreviations}

\markboth{Abbreviations}{Abbreviations}

\begin{longtable}{ll}
\toprule

A2C	& Advantage Actor-Critic\\
A3C	& Asynchronous Advantage Actor-Critic\\
AGV	& Automated Guided Vehicle\\
AMR	& Autonomous Mobile Robot\\
ANN	& Artificial Neural Network\\
APPO &	Asynchronous Proximal Policy Optimization\\
AS/RS &	Automated Storage and Retrieval System\\
BC & Behavioral Cloning\\
CBS	& Conflict-based Search\\
CNN	& Convolutional Neural Network\\
COM	& Component Object Model\\
CPU	& Central Processing Unit\\
CTDE & Centralized Training with Decentralized Execution\\
CTE	& Centralized Training and Execution\\
DDPG &	Deep Deterministic Policy Gradient\\
DQN	& Deep Q-Network\\
DTE	& Decentralized Training and Execution\\
FNN	& Feedforward Neural Network\\
GAE	& Generalized Advantage Estimation\\
GNN	& Graph Neural Network\\
GPU	& Graphics Processing Unit\\
GRU	& Gated Recurrent Unit\\
IMPALA	& Importance Weighted Actor-Learner Architectures\\
KL & Kullback-Leibler\\
KPI	& Key Performance Indicator\\
LSTM &	Long Short-Term Memory\\
MA-A* & Multi-Agent A* Variant\\
MAPF &	Multi-Agent Pathfinding\\
MARL &	Multi-Agent Reinforcement Learning\\
MAS	& Multi-Agent System\\
MDP	& Markov Decision Process\\
MTTR &	Mean Time To Repair\\
PG	& Policy Gradients\\
POMDP &	Partially Observable Markov Decision Process\\
PPO	& Proximal Policy Optimization\\
RL	& Reinforcement Learning\\
RNN	& Recurrent Neural Network\\
RQ	& Research Question\\
SAC	& Soft Actor-Critic\\
SARSA &	State-Action-Reward-State-Action\\
TD3	& Twin Delayed Deep Deterministic Policy Gradient\\
TRPO &	Trust Region Policy Optimization\\
UAV	& Unmanned Aerial Vehicle\\
UCB	& Upper Confidence Bound\\
WoS	& Web of Science\\

\bottomrule
\end{longtable}
\phantomsection
\thispagestyle{empty}
\chapter*{Notation Used}

\markboth{Notation}{Notation}

\begin{longtable}{ll}
\toprule
\multicolumn{2}{l}{\textbf{Markov decision process and reinforcement learning}} \\

$S$ & State space \\
$A$ & Action space \\
$P(s'|s,a)$ & State transition probability function\\
$R(s,a,s')$ & Reward function \\
$\mu$ & Initial state distribution \\
$\gamma$ & Discount factor \\
$\pi$ & Policy \\
$v_\pi(s)$ & State-value function under policy $\pi$ \\
$q_\pi(s,a)$ & Action-value function under policy $\pi$ \\
$G_t$ & Return at timestep $t$, i.e., $\sum_{k=0}^{\infty} \gamma^k r_{t+k}$ \\
\midrule
\multicolumn{2}{l}{\textbf{Multi-agent reinforcement learning}} \\

$I$ & Set of agents $i \in \{1,\ldots,n\}$ \\
$s(t)$ & Global state at time $t$ \\
$s_i(t)$ & State of agent $i$ at time $t$ \\
$o_i(t)$ & Observation of agent $i$ at time $t$ \\
$a_i(t)$ & Action of agent $i$ at time $t$ \\
$R_i$ & Total reward of agent $i$ over all timesteps $t$ \\
$\pi_i$ & Policy of agent $i$ \\
$A_i$ & Action space of agent $i$ \\
$A(t)$ & Joint action vector at time $t$, $A(t) = (a_1(t),\ldots,a_n(t))$ \\
$\mathcal{A}$ & Joint action space $\prod_{i=1}^n A_i$ \\
$f$ & Transition function $s(t+1) = f(s(t), A(t))$ \\
$T$ & Set of discrete timesteps $\{0, 1, \dots, T_{\text{max}}\}$ \\
$T_{\text{max}}$ & Maximum number of timesteps per episode \\
\midrule
\multicolumn{2}{l}{\textbf{Grid-based environment}} \\
$\text{Pos}$ & Set of valid grid positions \\
$\text{pos}_i(t)$ & Position of agent $i$ at time $t$ \\
$G_i=(x_i,y_i)$ & Goal position for each agent $i \in I$ \\
$F_i(t)$ & Goal-reached flag of agent $i$\\
$ogrid_i(t)$ & Local grid view centered on agent $i$ at $t$ \\
$opos_i(t)$ & Encoded own position of agent $i$ \\
$ogoal_i(t)$ & Encoded goal location of agent $i$ \\
$L_i(t)$ & Indicator whether agent $i$ is at its goal at time $t$ \\
$J_i(t)$ & Indicator whether agent $i$ reaches its goal for the first time at $t$ \\
$K(t)$ & Indicator whether all agents are at their goal at time $t$ \\
$C_i(t)$ & Indicator whether agent $i$ is involved in a collision at time $t$ \\
$\alpha, \beta, \gamma$ & Weighting coefficients for different reward components \\
\midrule
\multicolumn{2}{l}{\textbf{Continuous space environment}} \\
$v_i(t)$ & Current speed of agent $i$ \\
$G_i(t)$ & Goal position of agent $i$ \\
$C_i(t)$ & Carrying status of agent $i$\\
$L_i(t)$ & Remaining route length to goal of agent $i$\\
$S = \{\text{A}, \text{B}\}$ & Set of stations\\
$status_s(t)$ & Status of station $s$ at time $t$ \\
$buffer_s(t)$ & Buffer fill level of station $s$ at time $t$ \\
$D_i(t)$ & Delivery success indicator for agent $i$ at time $t$ \\
$P^{\text{pickup}}_i(t)$ & Pickup operation indicator for agent $i$ at time $t$ \\
$P^{\text{place}}_i(t)$ & Place operation indicator for agent $i$ at time $t$ \\
$W_s(t)$ & Workload contribution of station $s$ at time $t$ \\
$Col_i(t)$ & Collision penalty indicator for agent $i$ at time $t$ \\
$\alpha, \beta, \gamma, \delta, \epsilon$ & Weighting coefficients for different reward components \\

\bottomrule
\end{longtable}



\chapter{Introduction}
\label{ch:introduction}

\section{Motivation and Problem Statement}

Recent disruptions in global supply chains, caused by events such as the COVID-19 pandemic, geopolitical tensions \citep{Blum.2020}, and wars, have exposed significant vulnerabilities in logistics operations. These events have led to irregular inbound flows, sudden demand shifts, and supply-side bottlenecks, for example, the shortage of semiconductors and supply constraints in the retail food industry \citep{ifoInstitut.2023}. Such disruptions no longer affect only global transportation but increasingly propagate into the internal material flows of warehouses and production environments. As a result, intralogistics systems are required to absorb volatility in volumes, dynamically reroute tasks, and quickly adapt to changing priorities. This transformation from stable to fluctuating operating conditions highlights the need for higher structural flexibility, both in terms of layout and control logic. 

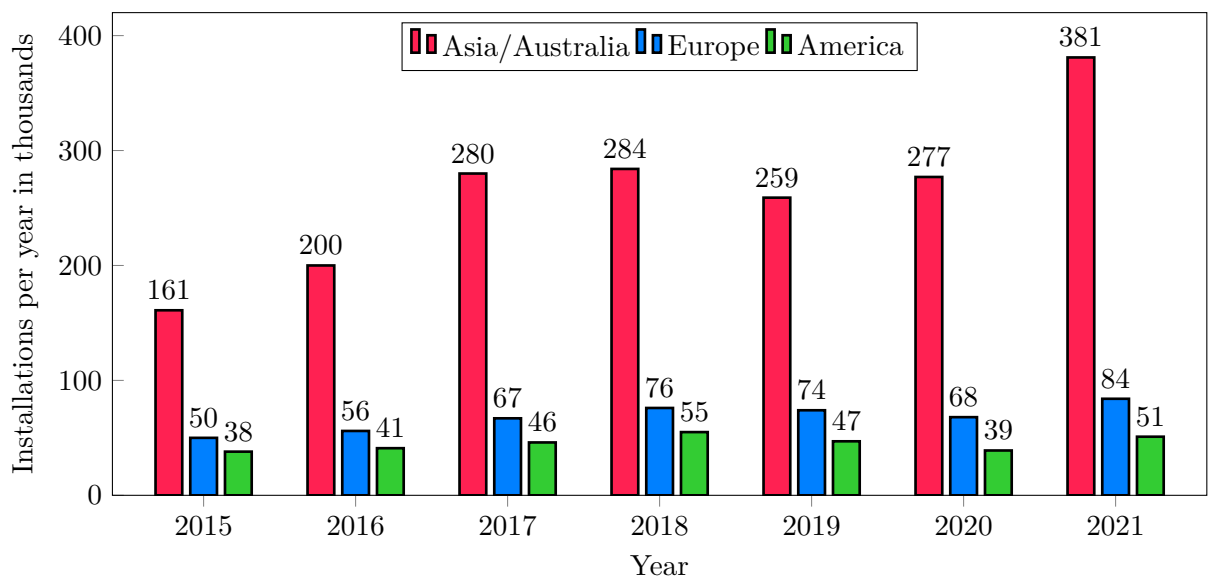
\begin{figure}[h]
\centering
\begin{tikzpicture}
    \begin{axis}[
        width=0.9\textwidth,
        height=0.4\textwidth,
        ybar=3,
        ymin=0,
        ymax=420,
        enlargelimits=0.1,
        enlarge y limits = false,
        legend style={at={(0.5,0.98), anchor=north west},
        anchor=north,legend columns=-1},
        xlabel={Year},
        ylabel={Installations per year in thousands},
        xtick={2015, 2016, 2017, 2018, 2019, 2020, 2021},
        xticklabels={2015, 2016, 2017, 2018, 2019, 2020, 2021},
        tick pos=left,
        nodes near coords,
        nodes near coords align={vertical},
        grid=none
        ]
        \addplot[fill=redMM] coordinates {(2015, 161) (2016, 200) (2017,280) (2018,284) (2019,259) (2020,277) (2021,381)};
        \addplot[fill=blueMM] coordinates {(2015, 50) (2016, 56) (2017,67) (2018,76) (2019,74) (2020,68) (2021,84)};
        \addplot[fill=greenMM] coordinates {(2015, 38) (2016, 41) (2017,46) (2018,55) (2019,47) (2020,39) (2021,51)};
        \legend{Asia/Australia, Europe, America}
    \end{axis}
\end{tikzpicture}
\caption{Number of annual installations of industrial robots worldwide \citep{MullerC.2022}.}
\label{fig:robot-installations}.
\end{figure}

At the same time, a growing shortage of skilled labor in industrialized countries is accelerating the adoption of automation in intralogistics. Unlike in the past, where fluctuations in workload could be buffered by hiring temporary workers, labor scarcity has made it necessary to rely on systems that are both automated and reconfigurable. A prime indicator of this trend is the rapid increase in the use of industrial robots. According to \cite{MullerC.2022}, the global stock of industrial robots in 2021 reached nearly 3.5 million units, an increase of 15 \% over the previous year. Since 2016, this stock has grown by an average of 14 \% per year. Figure~\ref{fig:robot-installations} illustrates that the number of annual robot installations is also rising in the European and Asia-Pacific regions, signaling a broader shift toward scalable, software-driven automation.

The use of automation in intralogistics systems has increased their complexity while improving their operational efficiency and capacity. Autonomous mobile robots (AMRs) and automated guided vehicles (AGVs) are leading this progression. Especially, the trend towards the increased usage of AMRs is leading to more decentralized decisions. AMRs are used to transport materials within facilities. The continuous operation of AMRs without human intervention significantly reduces downtime and labor costs, making them essential for achieving elevated throughput and reliability in warehouses and distribution centers. The autonomy of AMRs introduces new challenges, particularly in managing deadlocks. Deadlocks occur when two or more processes become mutually blocked, each waiting for the release of a resource by another, potentially causing significant operational disruptions or, in extreme cases, leading to complete system shutdowns. This situation is characterized by mutual exclusion, hold and wait, no preemption, and circular wait conditions, as formalized by \cite{Coffman.1971}. Developing effective deadlock handling methods is crucial, particularly in environments that heavily use AMRs.

While traditional strategies for managing deadlocks in intralogistics systems have often relied on rule-based or heuristic approaches, the unique dynamics introduced by AMRs challenge these conventional methods. Rule-based strategies require a comprehensive understanding of all possible states and events within the system, a task rendered impractical by the complexity introduced by AMRs and their autonomous decision-making capabilities \citep{Lienert.2017}. Although heuristic methods can provide efficient solutions in certain contexts \citep{Mayer.2010}, their performance is variable and often lacks the adaptability needed to cope with the ever-changing conditions and the operational intricacies of AMRs. The environmental parameters, such as the operational number of AMRs and the occurrence of general disturbances, play a crucial role in the efficacy of deadlock management strategies \citep{Muller.2020, Muller.2021}. These considerations underscore the urgency for innovative approaches that not only address the complexity brought on by AMRs but also harness their full potential to improve intralogistics systems' efficiency and resilience.

In this context, machine learning methods, and especially reinforcement learning (RL), offer a promising approach to address these challenges. Machine learning methods can recognize patterns and relationships in large data sets and derive predictions and decisions from them. RL, a form of machine learning, goes a step further, allowing systems to learn through interaction with their environment and continuously improve their performance. The ability of RL to continuously improve and adaptively agile in dynamic and complex learning environments may prove to be a decisive factor in efficiently handling deadlocks in intralogistics systems.

Although the potentials and advantages of machine learning, and particularly RL, are well documented in theory and practice, such as by \cite{Silver.2016} and \cite{Sutton.2018}, the specific application of these technologies to address deadlocks in intralogistics systems remains a relatively unexplored area. The existing literature and research focus mainly on general applications of RL, and there is a lack of detailed investigation of their specific application in intralogistics contexts. In particular, the development and validation of models and algorithms for RL that might be suitable for the unique challenges and dynamics of deadlocks in intralogistics systems have not been sufficiently explored. Chapter~\ref{ch:background} provides a detailed account of the relevant state of the art.

A notable research gap exists in the integration of RL-based methods into operational intralogistics environments. There is limited understanding of how these methods can be embedded in existing processes or aligned with human decision-making to achieve tangible improvements in system performance. Furthermore, comprehensive empirical studies evaluating the practical effectiveness of machine learning and RL in real-world intralogistics settings are lacking. Such studies are necessary to demonstrate the applicability of these methods and to benchmark their performance against traditional deadlock handling strategies.

These research gaps constrain the broader adoption and acceptance of RL-based solutions in intralogistics, but they also present promising opportunities for further research and technological development. The specific gaps addressed by this thesis, along with the underlying research hypotheses, are presented in Section~\ref{sec:methodology-researchgap}.

\section{Objective}

This thesis investigates how RL can be used to handle deadlocks in intralogistics systems that operate with AMRs. The aim is to develop learning-based methods that support adaptive and robust transport coordination, focusing on environments where static rule sets are insufficient due to structural complexity or dynamic disturbances. 
The work centers on three focal aspects. First, it explores how the ability to handle deadlocks contributes to the resilience of intralogistics systems and how this can be incorporated into planning objectives. Second, it formalizes deadlock-prone transport scenarios as multi-agent pathfinding (MAPF) problems, emphasizing the role of agent interactions and resource contention. Third, it examines how RL can be applied and trained in such settings to improve coordination outcomes compared to conventional strategies.

Rather than proposing a single fixed solution, the thesis aims to clarify under which conditions RL-based approaches are suitable for deadlock handling and how they interact with structural factors such as layout, agent density, or disruption patterns. The broader goal is to contribute to a planning methodology in which deadlocks are treated not only as implementation concerns but as design-relevant phenomena that can be addressed through adaptive behavior. To accomplish this objective, this thesis addresses the following research questions (RQ) in the context of deadlock-capable intralogistics systems with AMRs:

\begin{itemize}
    \item RQ 1.1: Is deadlock handling significant?
    \item RQ 1.2: How to integrate RL into the planning process?
    \item RQ 2.1: What do reference scenarios for deadlock-capable MAPF problems look like?
    \item RQ 2.2: When and in which configurations is multi-agent RL effective for MAPF with deadlocks?
    \item RQ 3.1: Does the deadlock handling strategy matter?
    \item RQ 3.2: Is centralized training and decentralized execution the best approach in deadlock-capable MAPF problems?
\end{itemize}

The selection of the RQs, their addressed research gap and the expected outcome of answering the RQs is described in detail in Section \ref{sec:methodology-researchgap}.

\section{Structure}

This thesis follows a systematic structure that progressively deepens the research field. The introduction outlines the motivation and key research questions, establishing a foundation for the following theoretical and methodological sections. The subsequent chapters on theoretical background and empirical investigations build upon this foundation, setting the stage for the presentation of results and their discussion in relation to existing research.

The structure adheres to a classical research process, moving from problem identification to theoretical engagement and finally to implementation and evaluation. The work is largely linear, although Chapters \ref{ch:background} and \ref{ch:evaluation} are modular in design and may be considered independently.

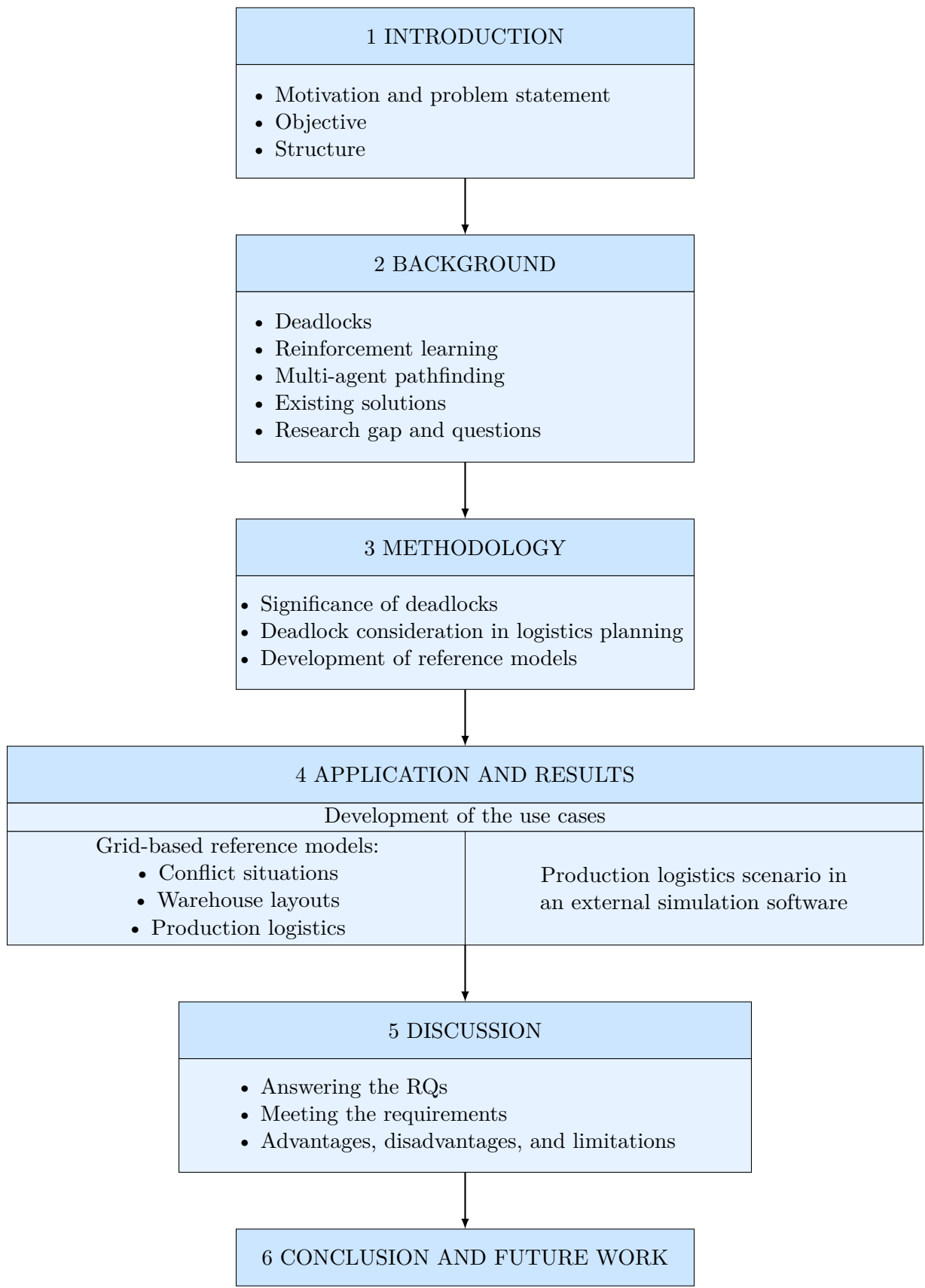
\begin{figure}[ht]
\centering
\begin{tikzpicture}    
    \draw[fill=blueMM!20] (0,11) rectangle (8,12) node[midway] {1 INTRODUCTION};
    \draw[fill=blueMM!10] (0,9) rectangle (8,11) node[midway, text width=7.5cm, align=left] 
    {
        \textbullet\ Motivation and problem statement\\ 
        \textbullet\ Objective\\ 
        \textbullet\ Structure
    };
    
    \draw[->, thick] (4,9) -- (4,8);
    
    \draw[fill=blueMM!20] (0,7) rectangle (8,8) node[midway] {2 BACKGROUND};
    \draw[fill=blueMM!10] (0,4) rectangle (8,7) node[midway, text width=7.5cm, align=left] 
    {
        \textbullet\ Deadlocks\\ 
        \textbullet\ Reinforcement learning\\ 
        \textbullet\ Multi-agent pathfinding\\ 
        \textbullet\ Existing solutions\\ 
        \textbullet\ Research gap and questions\\
    };
    
    \draw[->, thick] (4,4) -- (4,3);
    
    \draw[fill=blueMM!20] (0,2) rectangle (8,3) node[midway] {3 METHODOLOGY};
    \draw[fill=blueMM!10] (0,0) rectangle (8,2) node[midway, text width=8cm, align=left] 
    { 
        \textbullet\ Significance of deadlocks\\ 
        \textbullet\ Deadlock consideration in logistics planning\\ 
        \textbullet\ Development of reference models
    };
    
    \draw[->, thick] (4,0) -- (4,-1);
    
    \draw[fill=blueMM!20] (-4,-2) rectangle (12,-1) node[midway] {4 APPLICATION AND RESULTS};
    \draw[fill=blueMM!10] (-4,-2.5) rectangle (12,-2) node[midway] {Development of the use cases};
    \draw[fill=blueMM!10] (-4,-4.5) rectangle (4,-2.5) node[midway, text width=7.5cm, align=center] {
    Grid-based reference models:\\
        \textbullet\ Conflict situations\\ 
        \textbullet\ Warehouse layouts\\ 
        \textbullet\ Production logistics
    };
    \draw[fill=blueMM!10] (4,-4.5) rectangle (12,-2.5) node[midway, text width=7.5cm, align=center] {Production logistics scenario in an external simulation software};
    
    \draw[->, thick] (4,-4.5) -- (4,-5.5);
    
    \draw[fill=blueMM!20] (-1,-6.5) rectangle (9,-5.5) node[midway] {5 DISCUSSION};
    \draw[fill=blueMM!10] (-1,-8.5) rectangle (9,-6.5) node[midway, text width=8cm, align=left] {
        \textbullet\ Answering the RQs\\ 
        \textbullet\ Meeting the requirements\\ 
        \textbullet\ Advantages, disadvantages, and limitations\\
    };
    
    \draw[->, thick] (4,-8.5) -- (4,-9.5);
    
    \draw[fill=blueMM!20] (0,-10.5) rectangle (8,-9.5) node[midway] {6 CONCLUSION AND FUTURE WORK};
\end{tikzpicture}
\caption{Structure of this thesis.}
\label{fig:DissertationStructure}
\end{figure}

Figure \ref{fig:DissertationStructure} visualizes the overall structure of this thesis and supports the reader’s orientation. To ensure coherence, each main chapter introduces its specific subtopics at the outset. Concluding summaries in Chapters \ref{ch:background} to \ref{ch:discussion} consolidate the main points, facilitating an integrated understanding of the work.

%

\chapter{Background and Research Gaps in Deadlock Handling for Multi-Agent Systems}
\label{ch:background}
This chapter provides the essential background information required to understand this work and outlines the current state of scientific knowledge.

\begin{itemize}
    \item Section \ref{sec:background-deadlocks} defines the term \enquote{deadlock} and describes the types of deadlocks considered in this work. The section discusses the causes and forms of deadlocks and presents the most relevant strategies and solution approaches found in the literature.
    \item Section \ref{sec:background-reinforcementlearning} introduces the advancements in the field of machine learning, with a particular focus on the fundamentals and characteristics of RL. Various concepts like Markov decision processes or multi-agent systems are introduced.
    \item Section \ref{sec:background-MAPF} explains the basics of multi-agent pathfinding and reviews how path finding problems have been addressed in the literature. 
    \item Section \ref{sec:background-existing-solutions} identifies the gap in existing planning methods with regard to deadlocks. The section then proceeds with a structured literature review that assesses previous solution approaches to RL in intralogistics and the handling of deadlocks, as discussed in scholarly articles.
    \item Section \ref{sec:methodology-researchgap} derives the research gaps from the literature discussed in the sections \ref{sec:background-deadlocks} and \ref{sec:background-existing-solutions} and formulates RQs to be investigated in this thesis.
\end{itemize}

\section{Deadlocks} \label{sec:background-deadlocks}
\subsection{Definition and Distinction}
The term \textbf{\enquote{deadlock}} generally characterizes a situation in which an entity or a process is endlessly blocked. Historically, deadlocks first appeared in the literature in the realm of philosophy in the 11th century as an insoluble decision-making problem related to human free will \citep[p. 25 f.]{AlGhazzali.1963}. Although al-Ghazālī did not directly use the term \enquote{deadlock} in his writings, he describes situations that can be interpreted as deadlocks from today's perspective. In the 1970s, the term became prominent in computer science, especially in the context of structured programming and the allocation of computer resources \citep{Coffman.1971}.

\begin{figure}
    \centering
    \includegraphics[width=1\linewidth]{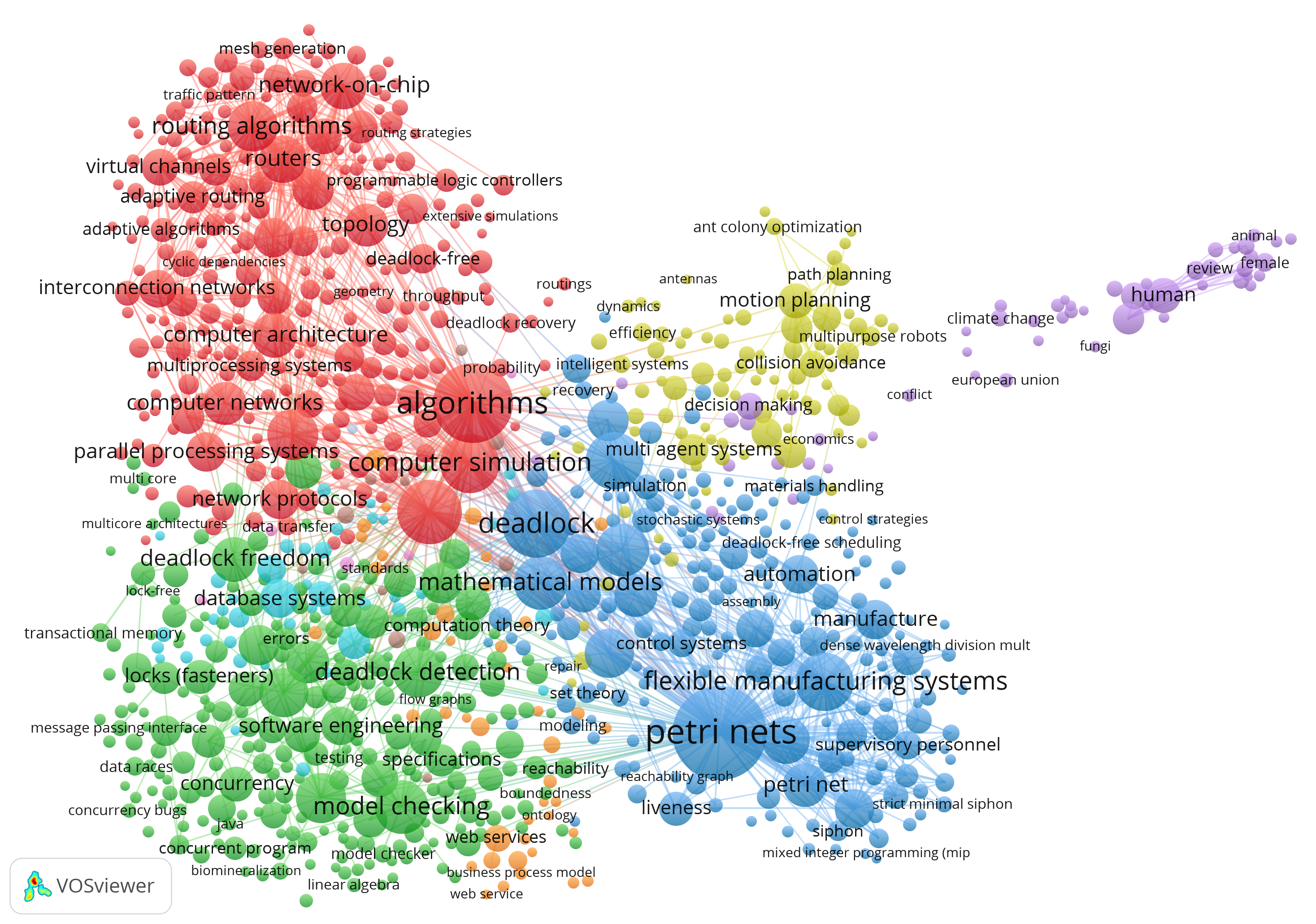}
    \caption{Bibliometric network of the term \enquote{deadlock} in VOSviewer \((n = 921)\).}
    \label{fig:bibliographic-map-deadlock}
\end{figure}

To outline the current state of scientific literature regarding deadlocks, a Scopus query was conducted on August 11, 2023. This query included publications that contained the term \enquote{deadlock} in the title, abstract, or as a keyword. The search resulted in 13,257 publications with a total of 45,337 keywords. Figure \ref{fig:bibliographic-map-deadlock} visualizes the dominant keywords, incorporating only those that occurred at least 20 times \((n = 921)\).

The bibliometric network in Figure \ref{fig:bibliographic-map-deadlock} categorizes the literature on deadlocks into different research areas: structured programming (red cluster), communication networks (green cluster), and the use of Petri nets in flexible manufacturing systems (blue cluster). The focus of this thesis is primarily on the yellow cluster, which addresses transport processes and multi-agent systems.

\begin{figure}[h]
    \centering
    \includegraphics[width=0.4\linewidth]{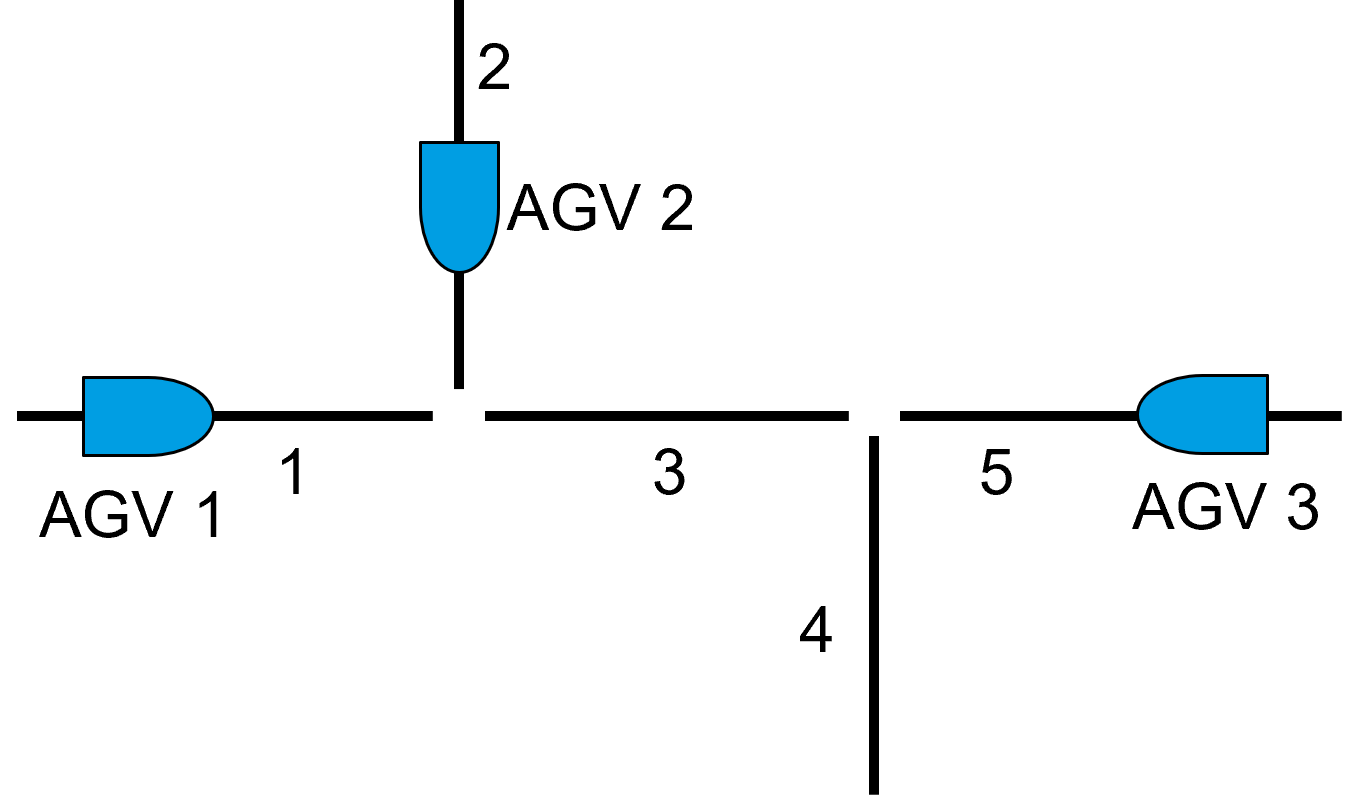}
    \caption{Sketch of an impending deadlock in an automated guided vehicle system.}
    \label{fig:Deadlock-simple-sketch}
\end{figure}

Different variants of the term \enquote{deadlock} appear in the literature. One of the most common forms is the resource deadlock \citep[p. 436 ff.]{Tanenbaum.2015}. In a resource deadlock, processes claim and request resources in such a way that they lock each other in an endless blockade. A practical example from logistics would be if AGVs on a track-bound network wait for other resources (in this case, the free path) in such a way that none can proceed. Figure \ref{fig:Deadlock-simple-sketch} illustrates this with a simple example involving a situation of three AGVs on five path sections. If AGV 3 moves from path section 5 to path section 3, a resource deadlock occurs.

Communication deadlocks, on the other hand, arise from interruptions or failures in the flow of information that cause processes to be halted \citep[p. 459 ff.]{Tanenbaum.2015}. In a logistics context, this could occur when a warehouse worker waits for information to continue their task but receives no clear instructions due to a communication error, thus preventing the process from proceeding.

A concept closely related to deadlock is \textit{livelock} \citep[p. 461 ff.]{Tanenbaum.2015}. In a livelock, agents remain active but fail to make effective progress toward their goals. Unlike in a deadlock, where all entities are blocked and unable to act, agents in a livelock continuously react to one another without reaching a stable or productive state. For example, consider two storage and retrieval machines attempting to access the same shelf location. Upon detecting a potential collision, each retreats to yield. If this mutual avoidance repeats indefinitely, both machines stay in motion, yet the task remains unfulfilled.

A distinct but related phenomenon is \textit{oscillation}. Oscillation refers to cyclic or alternating decision patterns in either agent behavior or coordination logic. In multi-robot systems, one form of oscillation occurs at the control level: if the resolution of a deadlock by modifying coordination dependencies (e.g., changing the direction of a wait-for edge) causes another deadlock that leads back to the original configuration, the system enters a resolution loop. \citet[p. 1216 f.]{Jager.2001} describe this effect in decentralized coordination systems, where repeated redirection of coordination links induces oscillation in the deadlock resolution graph.

While all such oscillations impede progress and thus fall under the broader class of livelocks, not all livelocks involve oscillatory behavior. For example, agents may cycle through disjoint avoidance maneuvers without ever revisiting previous states. In contrast, oscillation presupposes recurrence, either spatial, temporal, or structural. In this thesis, we distinguish explicitly between deadlock, livelock, and oscillation. Unless noted otherwise, deadlock refers to a resource-based stalemate, livelock denotes non-progress under continuous activity, and oscillation refers to recurring patterns, either in motion or resolution logic, that inhibit progress.

\cite{Coffman.1971} defined four conditions that must be met for a deadlock:

\begin{enumerate}
\item Tasks claim exclusive control of the resources they require (\textit{mutual exclusion} condition).
\item Tasks hold on to resources already assigned to them while waiting for additional resources (\textit{hold and wait} condition).
\item Resources needed for a task cannot be forcibly taken away until the task is completed (\textit{no preemption} condition).
\item There exists a circular chain of tasks such that each task holds one or more resources that the next task in the chain requests (\textit{circular wait} condition).
\end{enumerate}

The four necessary conditions for a deadlock, as formulated by \citet{Coffman.1971}, can be transferred to intralogistics contexts. Table~\ref{tab:deadlock-conditions-intralogistics} illustrates how each of these conditions manifests in two representative use cases: automated storage systems with storage and retrieval units, and material flow systems based on conveyor technology. The examples clarify how resource contention, process coordination, and system design can interact in ways that fulfill all conditions required for a deadlock.

\begin{table}[htbp]
    \caption{Examples of deadlock conditions in intralogistics.}
    \label{tab:deadlock-conditions-intralogistics}
    \centering
    \begin{tabular}{>{\raggedright\arraybackslash}p{0.2\linewidth}>{\raggedright\arraybackslash}p{0.35\linewidth}>{\raggedright\arraybackslash}p{0.35\linewidth}}
    \toprule
    \textbf{Deadlock condition} & \textbf{Automated storage with storage retrieval units} & \textbf{Material flow system with conveyors} \\
    \midrule
    Mutual exclusion & A specific storage retrieval unit has exclusive access to a storage area, preventing other units from accessing that area simultaneously. & A conveyor belt has exclusive access to a transfer point, so other conveyor belts cannot use this point at the same time. \\
    \midrule
    Hold and wait & The storage retrieval unit transports a product while waiting for the release of a target storage bin, currently blocked by another operation. & A package on a conveyor belt is stopped at a transfer point because the target conveyor belt is full and cannot accept more packages. \\
    \midrule
    No preemption & A product being transported by the storage retrieval unit cannot be interrupted or released until the entire movement to the target storage bin is successfully completed. & A segment of the conveyor belt remains reserved for a specific package until it is completely transferred to the next section of the conveyor system. \\
    \midrule
    Circular wait & Multiple storage retrieval units form a waiting loop, with each unit waiting for the release of a storage bin that is blocked by another unit in the sequence. & Multiple sections of the conveyor belt are interconnected in such a way that each belt waits for the release of the subsequent belt to forward packages, leading to circular wait. \\
    \bottomrule
    \end{tabular}
\end{table}

The first column of the table lists the canonical deadlock conditions. The second and third columns describe concrete intralogistics scenarios in which these conditions may be satisfied. While the specific technical implementations differ, e.g., mobile storage and retrieval units versus fixed conveyor segments, the underlying structural dependencies remain comparable. In both cases, exclusive resource allocation (mutual exclusion), resource retention while requesting additional resources (hold and wait), lack of enforced resource release (no preemption), and resource dependency cycles (circular wait) emerge as potential contributors to deadlock formation. These examples serve to bridge the abstract deadlock theory with practical system behavior in automated logistics environments.

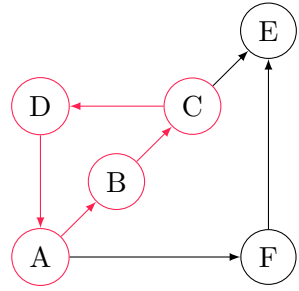
\begin{figure}[h]
    \centering
    \begin{tikzpicture}
        \begin{scope}[every node/.style={circle,thin,draw}]
            \node [draw=redMM] (A) at (0, 0) {A};
            \node [draw=redMM] (B) at (1, 1) {B};
            \node [draw=redMM] (C) at (2, 2) {C};
        	\node [draw=redMM] (D) at (0, 2) {D};
        	\node (E) at (3, 3) {E};
            \node (F) at (3, 0) {F};
        \end{scope}
    
    	\begin{scope}[every path/.style={->}]
     		\path [color=redMM] (A) edge node {} (B);
                \path [color=redMM] (B) edge node {} (C);
                \path [color=redMM] (C) edge node {} (D);
                \path (C) edge node {} (E);
                \path [color=redMM] (D) edge node {} (A);
                \path (A) edge node {} (F);
                \path (F) edge node {} (E);
        \end{scope}
    \end{tikzpicture}
    \caption{Directed graphs with a circular reference.}
    \label{fig:CircularReference}
\end{figure}

In connection with the fourth condition, \enquote{circular wait}, the term \enquote{circular reference} arises. A circular reference involves a series of references where the first object, which refers to another object, is eventually referred to by a subsequent object. Figure \ref{fig:CircularReference} presents a simple example of directed graphs with a circular reference. The nodes A, B, C, and D form a circular reference, which is depicted in red. A circular reference is a necessary but not sufficient condition for a deadlock.

In the context of deadlocks, the term \enquote{disturbance} often emerges, which is variably described in the literature. \citet[p. 27]{Brackel.2009} differentiates between deterministic and stochastic disturbances. According to \cite{Brackel.2009}, deterministic disturbances are events that, while interrupting the production process, are planned. In contrast, stochastic disturbances are identified as unforeseen or unintended interruptions, known only by their probability of occurrence \cite[p. 27]{Brackel.2009}. \citet[p. 12]{Ivanov.2023} defines a disturbance within supply chain contexts as an unexpected event that interrupts the normal flow of goods and materials in a supply chain network, resulting in significant negative impacts on the operation and performance of the supply chain.

Summarizing, a disturbance can be described as an event that interrupts operational workflow. Following the distinction made by \cite{Brackel.2009}, deadlocks are classified under the category of stochastic disturbances. Due to their specific conditions of occurrence, deadlocks are considered a special case of stochastic disturbance.

For the sake of clarity and distinction from the terms mentioned above, a definition of \enquote{deadlock} applicable to all subsequent mentions is provided here:

\begin{definition}[Deadlock]
A \textbf{deadlock} is the state of a system in which a set of processes endlessly waits for additional resources that are exclusively occupied by other processes in the same state.
\end{definition}

Deadlocks can manifest in various forms within intralogistics. The morphological box (Table \ref{tab:MorphologicalBox-Deadlock}) displays exemplary characteristics and their manifestations for deadlocks in intralogistics. The relevant characteristics for this thesis are highlighted in blue.

\begin{table}[h]
    \centering
    \caption{Morphological box for deadlocks in intralogistics.}
    \label{tab:MorphologicalBox-Deadlock}
    \small
    \setlength\tabcolsep{0.2pt} 
    \begin{tabularx}{\textwidth}{|>{\columncolor{gray!25}}c!{\vrule width 1pt}C|} 
         \hline
        \rowcolor{gray!25} 
        \textbf{Characteristic} & \textbf{Characteristic manifestation} \\
         \hline
        Cause & {\begin{tabularx}{\linewidth}{*{4}{>{\columncolor{blueMM!25}}C|}C}\vspace{5pt} Planning error & \vspace{1pt} Shortage of resources  & \vspace{5pt} Resource conflict & \vspace{1pt} Process interdependencies\end{tabularx}} \\
        \hline
        Involved resources & {\begin{tabularx}{\linewidth}{>{\columncolor{blueMM!25}}C|C|>{\columncolor{blueMM!25}}C|C}\vspace{1pt} Transport vehicles & \vspace{1pt} Storage spaces & \vspace{1pt} Paths & \vspace{1pt} Production resources \end{tabularx}} \\
        \hline
        Process behavior & {\begin{tabularx}{\linewidth}{C|>{\columncolor{blueMM!25}}C}Deterministic & Stochastic\end{tabularx}} \\
        \hline
        Prevention method & {\begin{tabularx}{\linewidth}{C|C|C|>{\columncolor{blueMM!25}}C}\vspace{1pt} Meticulous layout planning & \vspace{1pt} Generous capacity planning & \vspace{1pt} Advanced control systems & \vspace{1pt} Use of predictive analytics and machine learning \end{tabularx}} \\
        \hline
        Detection time & {\begin{tabularx}{\linewidth}{>{\columncolor{blueMM!25}}C|>{\columncolor{blueMM!25}}C} Early (preventative) & After occurrence (reactive recovery)\end{tabularx}} \\
        \hline
        Recovery mechanism & {\begin{tabularx}{\linewidth}{>{\columncolor{blueMM!25}}C|>{\columncolor{blueMM!25}}C} Resource reallocation & Temporary suspension of processes\end{tabularx}} \\
        \hline
        Risk level & {\begin{tabularx}{\linewidth}{>{\columncolor{blueMM!25}}C|C}Critical & Tolerable\end{tabularx}} \\
        \hline
        Evaluation criteria & {\begin{tabularx}{\linewidth}{>{\columncolor{blueMM!25}}C|>{\columncolor{blueMM!25}}C|C|C}\vspace{1pt} Downtime & \vspace{1pt} Detours & \vspace{1pt} Cost-benefit analysis & Implementation effort and scalability\end{tabularx}} \\
         \hline
    \end{tabularx}
\end{table}

\textbf{Logistics} today, as an interdisciplinary field, is a fundamental part of the modern economy. Historically, logistics was initially shaped by military operations. It was not until after World War II that logistics methods were first applied to the economy by the U.S. armed forces in the 1950s \citep{Morgenstern.1955}. Within the economic context, there are various definitions of logistics. Originally, logistics was primarily understood in the context of transport processes as the provision of goods. \citet[p. 3]{Gudehus.2010} describes the basic task of operational logistics as efficiently providing the required quantities of needed objects in the correct composition, at the right time, and in the right place. For the rational execution of this task, so-called analytical logistics also develops and organizes optimal processes, structures, and systems, always considering the needs of clients, consumers, and service recipients \citep{Gudehus.2010}.

Other modern definitions fundamentally describe logistics as a cross-functional activity that combines various scientific disciplines and adopts a process or system perspective. \citet[p. 2]{Fleischmann2018} defines logistics as the design of logistic systems and the control of processes within them, expanded by three core features: 1) The importance of information and communication systems for process control, 2) the holistic view of all processes in a system, and 3) the reference to physical systems and processes, thereby highlighting the interdisciplinary nature of logistics, which encompasses both technical and economic tasks and is situated in the fields of economics, engineering, and computer science.

\citet[p. 1]{Baumgarten.2013} already define logistics as a distinct scientific discipline responsible for the holistic planning, control, execution, and monitoring of all internal and cross-company flows of goods and information. In this thesis, logistics is understood according to the definition by \cite{Baumgarten.2013}.

\begin{figure}[h]
    \centering
    \begin{tikzpicture}
        \node (Performance) at (0, 4) {\textbf{Performance}};
        \node (Quality) at (-4, 0) {\textbf{Quality}};
        \node (Cost) at (4, 0) {\textbf{Costs}};

        \draw[thick] (Performance) -- (Quality);
        \draw[thick] (Quality) -- (Cost);
        \draw[thick] (Cost) -- (Performance);

        \node at (0, 1.5) {\textbf{Logistics objectives}};

        \node at (0, 5.7) {Orders};
        \node at (0, 5.3) {Throughput};
        \node at (0, 4.9) {Storage};
        \node at (0, 4.5) {Schedules};

        \node[align=right] at (-4, -0.5) {Delivery capability};
        \node[align=right] at (-4, -0.9) {Schedule adherence};
        \node[align=right] at (-4, -1.3) {Delivery quality};
        \node[align=right] at (-4, -1.7) {Flexibility};

        \node[align=left] at (4, -0.5) {Personnel};
        \node[align=left] at (4, -0.9) {Equipment};
        \node[align=left] at (4, -1.3) {Transport means};
        \node[align=left] at (4, -1.7) {Inventories};
    \end{tikzpicture}
    \caption{Logistics objectives according to \citet[p. 74]{Gudehus.2010}.}
    \label{fig:logistics_objectives}
\end{figure}
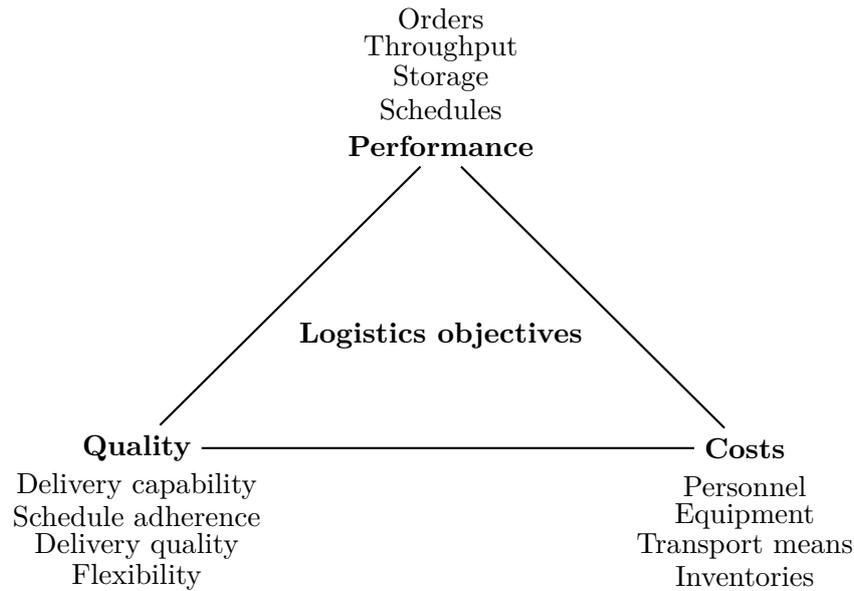

Figure \ref{fig:logistics_objectives} illustrates the core objectives of logistics in companies according to \citet[p. 74]{Gudehus.2010}: Performance, quality, and costs. Performance focuses on fulfilling orders, managing throughput, and maintaining efficient storage and scheduling. Quality emphasizes delivery capability, adherence to schedules, delivery quality, and flexibility. Costs include controlling expenses related to personnel, equipment, transport means, and inventories. The figure emphasizes the necessity of balancing these three objectives to create a resilient and efficient logistics system. Performance, quality, and cost are interdependent; an improvement in one area often impacts the others. For instance, enhancing performance through faster throughput may increase costs or require compromises in delivery quality. Similarly, rigorous cost management may affect the system's flexibility and performance.

\textbf{Intralogistics}, as a sub-discipline of logistics, pertains to the internal logistics processes and systems within a company. According to \citet[p. 1]{Arnold.2007.Intralogistik}, the term can be precisely traced back to June 30, 2003, when intralogistics was first defined during a press conference announcing the \enquote{Centrum für Materialflusstechnik} trade fair by the German Engineering Federation (VDMA). Accordingly, intralogistics encompasses the organization, control, execution, and optimization of internal material flow, information streams, and goods handling in industry, commerce, and public institutions \cite[p. 1]{Arnold.2007.Intralogistik}.

This thesis focuses on applying RL methods to resolve deadlock situations in intralogistics, with the above definition of intralogistics serving as the starting point for investigating and modeling the relevant processes and systems. The decision to narrow the research scope to intralogistics rather than the broader concept of logistics is based on two main considerations:

First, intralogistics is characterized by a significant density of processes within a confined internal space. Unlike general logistics, which deals, among other things, with supply chains over long distances and the international distribution of goods, intralogistics focuses on material and information flows within defined operational or storage facilities. It is hypothesized that this spatial concentration might lead to an increased risk of deadlocks, due to the intensive coordination of numerous processes and resources in a limited space. It is important to emphasize that this assumption is a plausible hypothesis not substantiated in the existing literature. However, the assumption serves as a plausible starting point to narrow down the topic to relevant logistics situations.

Second, focusing on intralogistics allows for a deeper scientific examination of deadlock situations. The specific conditions of intralogistics provide comparable case examples and the derivation of insights that would be difficult to achieve in a broader logistics context. This approach keeps the research effort within a practical framework and promotes profound insights tailored to the area.

Intralogistics, like logistics in the narrower sense, can be broken down into four basic tasks at the operational level: transportation, transloading, storage and picking \cite[p. 6]{Gudehus.2010}. This thesis focuses on the basic task of transportation, as it forms the backbone of intralogistics by connecting all other logistical processes. The efficiency and reliability of the transportation system directly affect the performance of the overall intralogistics system.

\begin{figure}[h]
    \centering
\tikzset{
        my node/.style={
            draw=gray,
            inner color=blueMM!25,
            outer color=blueMM!25,
            thick,
            minimum width=4cm,
            text height=1.5ex,
            text depth=0ex,
            font=\sffamily,
            drop shadow,
        }
    }
    \begin{forest}
        for tree={%
            my node,
            l sep+=5pt,
            grow'=east,
            edge={gray, thick},
            parent anchor=east,
            child anchor=west,
            if n children=0{tier=last}{},
            edge path={
                \noexpand\path [draw, \forestoption{edge}] (!u.parent anchor) -- +(10pt,0) |- (.child anchor)\forestoption{edge label};
            },
            if={isodd(n_children())}{
                for children={
                    if={equal(n,(n_children("!u")+1)/2)}{calign with current}{}
                }
            }{}
        }
        [Transportation systems
        [Continuous systems
        [Pipeline systems]
        [Belt conveyor systems]
        ]
        [Discontinuous systems
        [Conveyor systems]
        [Vehicle systems]
        ]
        ]
    \end{forest}
    \caption{Classification of transportation systems according to \citet[p. 772]{Gudehus.2010}.}
    \label{fig:Classification_transportation_systems}
\end{figure}
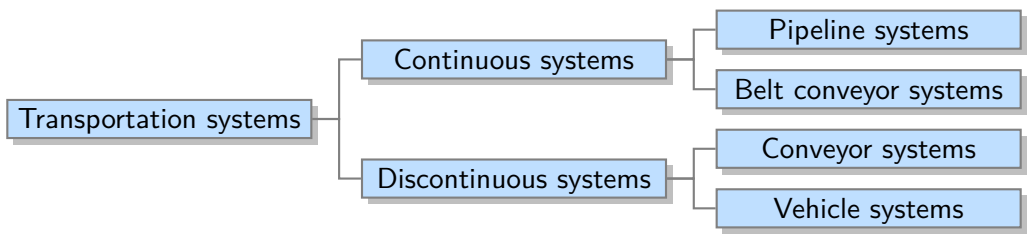

\textbf{Transportation systems}, according to \citet[p. 772]{Gudehus.2010}, can be classified into continuous and discontinuous systems. Continuous systems, such as conveyor belts and pipelines, provide an unbroken flow of materials, ideally suited for bulk goods. Discontinuous systems are further divided into conveyor and vehicle systems. The primary distinction lies in the drive mechanism: conveyor systems operate over a driven transport network, while vehicle systems, such as AGVs, AMRs, and automated storage and retrieval systems (AS/RS), possess individual drive units that move over a non-driven transport infrastructure. An overview of this classification is provided in Figure \ref{fig:Classification_transportation_systems}.

Building on this classification, \citet{fottner2021autonomous} introduce a two-dimensional framework for autonomous intralogistics systems, which accounts not only for task-specific functions (e.g., transport, storage, order picking) but also for the level of automation. Within this framework, transport systems involving AGVs and AMRs are positioned along a spectrum ranging from low to high autonomy. AMRs, in particular, represent systems that increasingly assume decentralized control functions, capable of autonomous navigation, interaction, and decision-making in dynamic environments. Their integration into logistics systems signifies a shift from rigid, centrally coordinated architectures toward flexible, adaptive networks of autonomous entities. This shift introduces new challenges in coordination, robustness, and conflict handling, most notably, deadlocks that arise due to local decision-making and shared space usage.

The choice to focus on AGVs and AMRs in exploring the application of RL for deadlock handling in this thesis is informed by their distinct characteristics and the evolving needs of modern intralogistics systems.

AGVs, which have been integral to material handling since their introduction in the early 1950s by Barrett-Cravens of Northbrook, Illinois \cite[p. 2]{Ullrich.2015.AGVS}, navigate using fixed paths and are suited for repetitive material movement in structured environments. They play a crucial role in automating logistics in monolithic and static settings but are limited by their need for predefined paths and their inflexibility to adapt to dynamic changes within the facility without substantial reconfiguration efforts.

AMRs, on the other hand, represent a significant advancement in mobile robot technology, offering superior adaptability and efficiency in dynamic environments \citep{Siegwart.2004.AMRS}. Unlike AGVs, AMRs navigate autonomously without relying on predefined paths, using sensors, cameras, and sophisticated software to perceive their surroundings and make real-time navigational decisions. This capability allows AMRs to quickly adapt to changes in the operational environment, making them ideal for dynamic settings with frequently changing layouts or processes. The versatility of AMRs in performing a wide range of tasks, from picking items off shelves to sorting packages, further underscores their suitability for various industries, including logistics, manufacturing, and healthcare.

While AGVs and AS/RS have their merits, the latter being efficient for high-volume storage and retrieval tasks within a fixed structure, they lack the flexibility and adaptability provided by AMRs. AS/RS systems operate along fixed rails or tracks and are constrained by their storage structure, making them difficult and costly to relocate or adapt to changes in business operations. 

AGVs are more flexible than AS/RS and should be used as a starting point for the learning approaches due to their smaller action space than AMRs. AMRs' ability to autonomously navigate and adapt to the environment, coupled with their ease of integration into existing operations without the need for physical infrastructure modifications, presents a fertile ground for leveraging RL. This focus is not only aimed at resolving deadlocks more effectively but also at contributing to the broader goal of advancing autonomous intralogistics solutions that can dynamically respond to changing operational demands.

The exploration of RL in the context of AMRs rather than AGVs or AS/RS is motivated by the potential of AMRs to transform intralogistics through their advanced navigation capabilities, collaborative nature, and the flexibility they offer in operational deployment. This aligns with the broader trends in industry 4.0, where adaptability, efficiency, and the ability to respond to volatile business environments are crucial.

\subsection{Causes and Strategic Approaches}
The four necessary conditions for the occurrence of deadlocks, mutual exclusion, hold and wait, no preemption, and circular wait, identified by \citet{Coffman.1971} provide a formal basis for understanding the phenomenon. However, in the context of intralogistics systems, these conditions manifest through more specific structural and procedural constellations. While much of the literature addresses deadlocks through algorithmic strategies or system-specific countermeasures, few works offer a structured analysis of their underlying causes in real-world intralogistics operations.

To address this gap, this section proposes a classification of deadlock causes derived from common patterns in warehouse automation, transport planning, and multi-agent control scenarios. The classification distinguishes between system-related and process-related causes. This analytical separation is motivated by two dominant perspectives in logistics research: the static structure of the system and the dynamic behavior of the processes it executes. The categorization draws on an inductive synthesis of failure scenarios observed in practice, reported in case studies, and encountered during the simulation-based development of this thesis.

Table~\ref{tab:System-related Causes} lists system-related causes that originate from static features of the logistics infrastructure, resource layout, and system design. These causes often emerge in the planning phase and can predispose a system to frequent deadlocks.

\begin{table}[h]
    \caption{System-related causes for deadlocks in intralogistics.}
    \label{tab:System-related Causes}
    \centering
    \begin{tabular}{p{5cm}p{10cm}}
        \toprule
        \textbf{System-related cause} & \textbf{Description} \\
        \hline
        Fixed resource allocation & Limited number of resources that are exclusively occupied, leading to waiting times and potential deadlocks. \\
        Spatial arrangement & Narrow or complex layouts in operational or storage facilities can lead to deadlocks due to movement restrictions. \\
        Resource interdependencies & Dependencies between different resources, whose simultaneous use leads to queues, conflicts, and deadlocks. \\
        Rigid system architecture & Lack of flexibility in system structure and process design, which does not allow dynamic adjustment to changed conditions. \\
        Limited access points & Restricted entry and exit points for resources and goods that can lead to bottlenecks and deadlocks. \\
        \bottomrule
    \end{tabular}
\end{table}

Table \ref{tab:Process-related Causes} presents process-related causes, including coordination deficiencies and priority conflicts, targeting the dynamics and interaction of logistic operations. This highlights that not only the structural composition but also the operational execution and control of logistics processes significantly contribute to the emergence of deadlocks.

\begin{table}[h]
    \caption{Process-related causes for deadlocks in intralogistics.}
    \label{tab:Process-related Causes}
    \centering
    \begin{tabular}{p{5cm}p{10cm}}
        \toprule
        \textbf{Process-related cause} & \textbf{Description} \\
        \hline
        Coordination deficiencies & Insufficient or incorrect coordination between processes, leading to inefficient resource use and conflicts. \\
        Priority conflicts & Contradictory or unclear priority setting in resource allocation or task execution, leading to waiting times and deadlocks. \\
        Unforeseen events & Disturbances or changes not considered in logistics system planning can suddenly lead to deadlocks. \\
        Lacking real-time feedback & Delayed or insufficient information feedback, preventing quick response to bottlenecks and deadlocks. \\
        \bottomrule
    \end{tabular}
\end{table}

Based on the strategies for handling deadlocks proposed by \citet[pp. 72 - 76]{Coffman.1971}, three fundamental strategic approaches can be identified:
\begin{itemize}
    \item Prevention
    \item Avoidance
    \item Detection and recovery
\end{itemize}

Although different strategy classifications exist in the literature, for instance, \citet[p. 181]{Holt.1972} differentiates between \enquote{Prevention}, \enquote{Detection}, and \enquote{Crash}, this thesis follows the subdivision proposed by \cite{Coffman.1971}, which is established in the scientific community.

\begin{figure}[h]
    \begin{minipage}[t]{.33\textwidth}
        \centering
        \includegraphics[width=0.95\linewidth]{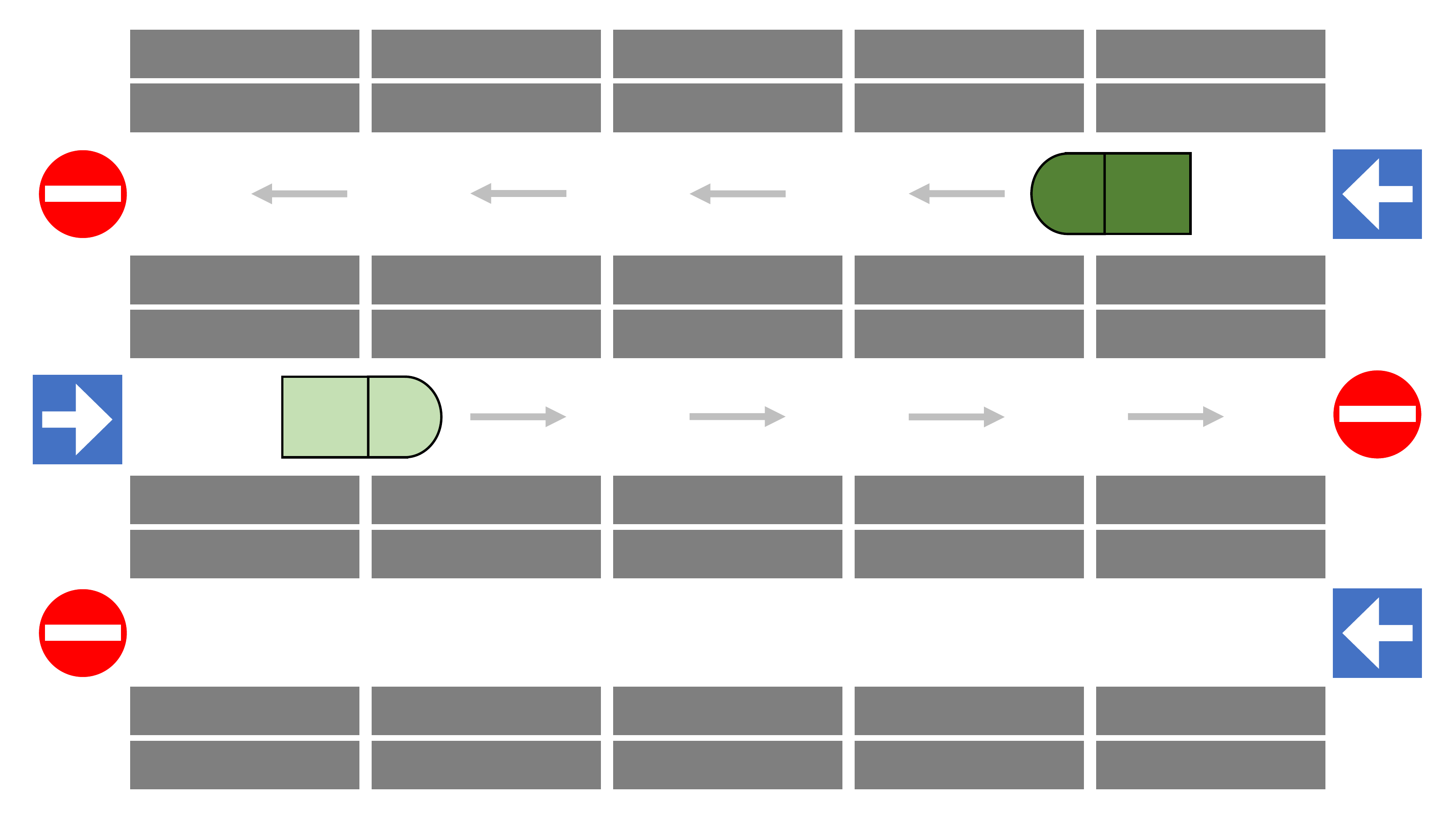}
        \caption*{Prevention}
    \end{minipage}
    \begin{minipage}[t]{.33\textwidth}
        \centering
        \includegraphics[width=0.95\linewidth]{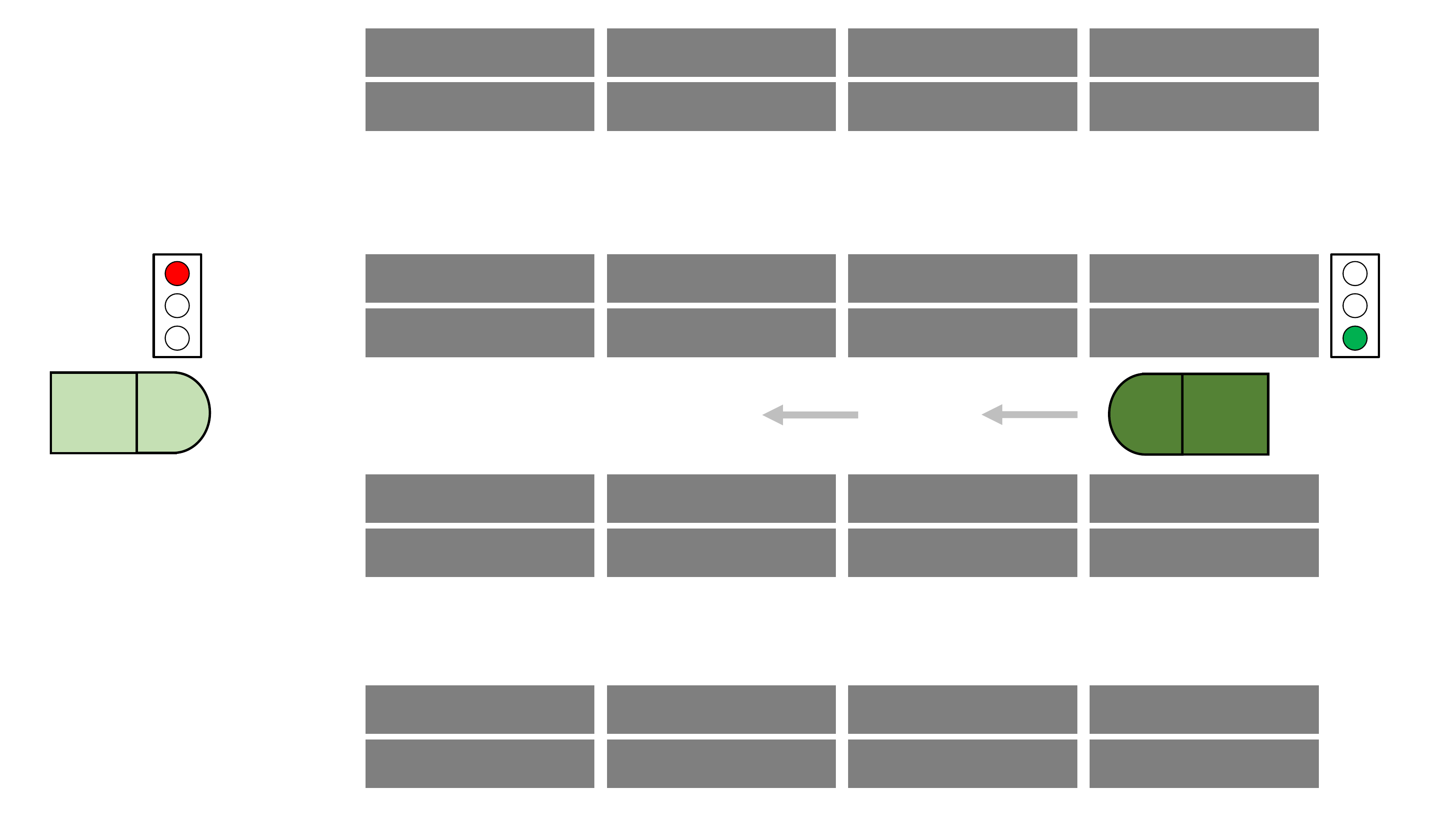}
        \caption*{Avoidance}
    \end{minipage}
    \begin{minipage}[t]{.33\textwidth}
        \centering
        \includegraphics[width=0.95\linewidth]{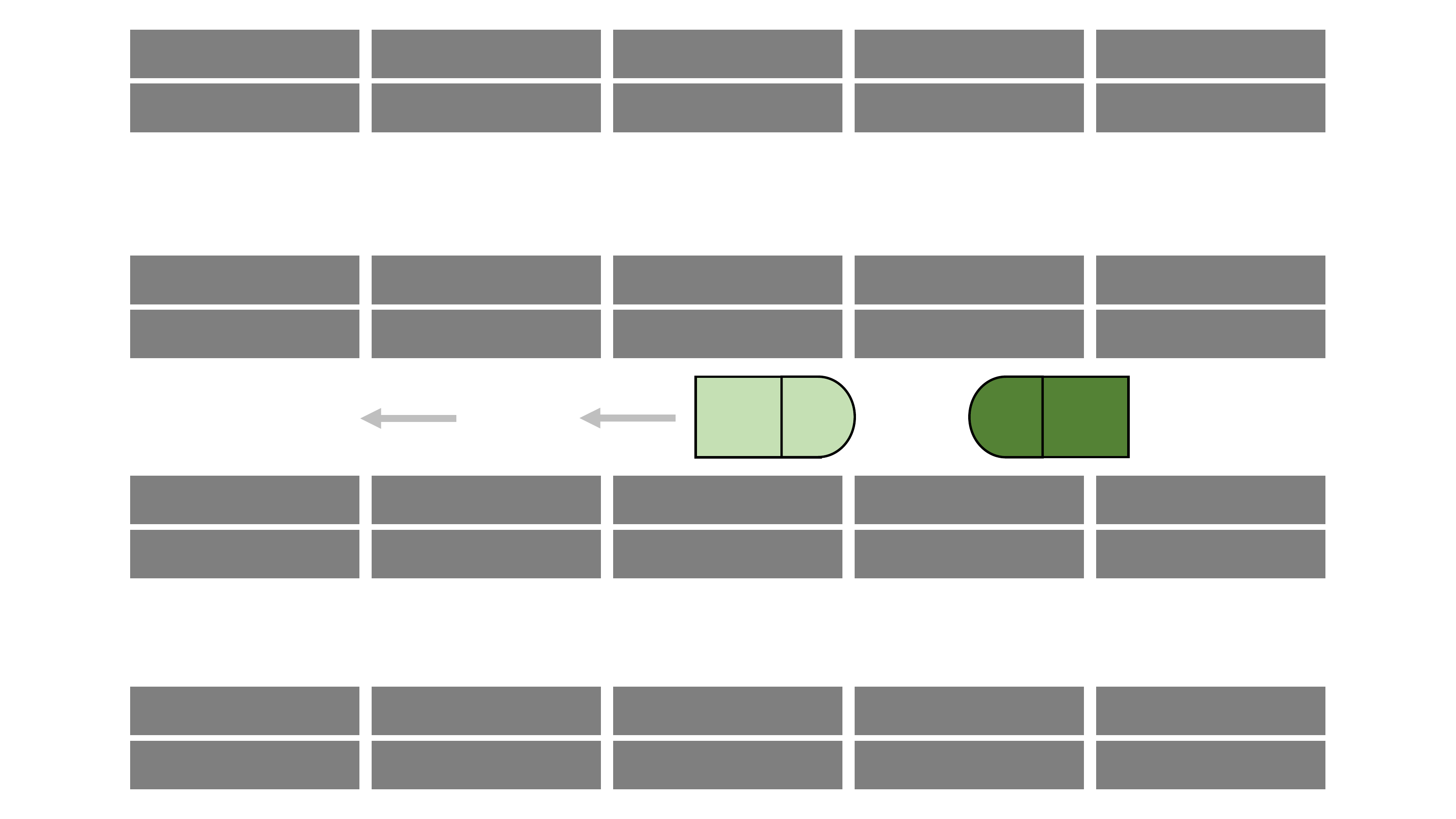}
        \caption*{Detection and recovery}
    \end{minipage}

    \caption{Illustration of the three deadlock strategy approaches prevention (left), avoidance (center), and detection \& recovery (right) inspired by \citet[p. 104]{Lehmann.2006}.}
    \label{fig:deadlock_strategies}
\end{figure}

Figure~\ref{fig:deadlock_strategies} illustrates the three classical strategies for handling deadlocks, following the taxonomy introduced by \citet{Coffman.1971}. The example depicts three narrow warehouse aisles, each permitting only unidirectional traffic. Two AMRs approach the aisles from opposite directions. If both agents enter the same aisle simultaneously, a deadlock may occur, even if physical collision is avoided.

In the prevention strategy (left), each aisle is permanently assigned a direction, ensuring that AMRs cannot face each other head-on. This eliminates deadlocks by construction but reduces flexibility, as only one aisle is accessible from each side. The avoidance strategy (center) permits bidirectional use of aisles, but employs dynamic coordination, such as a traffic light system, to ensure that access is granted to only one AMR at a time. Finally, the detection and recovery strategy (right) allows deadlocks to occur. Once a deadlock is detected, it is resolved through corrective actions such as reversing or rerouting one of the involved AMRs.

A prevention strategy in dealing with deadlocks generally aims to prevent the occurrence of deadlocks from the outset by adjusting the system and process design so that at least one of the four conditions by \cite{Coffman.1971} cannot be met. A common example in intralogistics for this is the use of unidirectional storage aisles. An avoidance strategy, on the other hand, works with control rules that avoid deadlocks through intelligent resource allocation, even though the basic structure and process design would allow deadlocks. The recovery strategy is employed when a deadlock has already occurred. This requires the deadlock to be first identified. Subsequently, measures to resolve the deadlock must be implemented. This could involve resetting or aborting a process to return the system to a deadlock-free state. In transport processes, this often means for at least one vehicle to abort the current transport process and initiate a maneuver.

In the literature, both early considerations by \cite{Coffman.1971} and later views, for example, by \cite{Tanenbaum.2015} from the perspective of computer science, saw the strategic approach of prevention primarily as changes in the way resources are utilized, referring to proposals by \cite{Havender.1968}. This view was not always followed in production and logistics. Either prevention was not clearly distinguished from avoidance to the point of synonymous use \citep{Kim.2006, Mayer.2010}, or, as \cite{Lehmann.2006} understood under deadlock prevention, the design of a logistical system, process, and infrastructure in such a way that no deadlock could occur. In designing the system, reference was often made only to the rules for resource utilization \cite[p. 2]{Kim.1997,Lienert.2017}, instead of seeing the system structure and sizing of system capacities as an integral part of deadlock prevention. The condition of mutual exclusion according to \cite{Coffman.1971} could be addressed through clever infrastructure planning, which has not been detailed in the literature so far.

\subsection{Modeling of Deadlocks}

Resource allocation graphs are a classic tool for modeling deadlocks. A resource allocation graph is a directed graph consisting of a set of nodes \( V \) divided into two disjoint subsets: \( P = \{P_1, P_2, \ldots, P_n\} \), representing the active processes in the system, and \( R = \{R_1, R_2, \ldots, R_m\} \), standing for the resource types in the system.

In this graph, a directed edge from a process \( P_i \) to a resource type \( R_j \), depicted as \( P_i \rightarrow R_j \), represents the request of an instance of the resource type by the process. Conversely, a directed edge from \( R_j \) to \( P_i \), noted as \( R_j \rightarrow P_i \), indicates that an instance of the resource type \( R_j \) has been allocated to the process \( P_i \). Here, \( P_i \rightarrow R_j \) is referred to as a request edge, and \( R_j \rightarrow P_i \) as an allocation edge.

The graphical representation of processes and resources is not uniform in the literature. While \citet[p. 192]{Holt.1972} describes resources as circles and processes as squares, \citet[p. 440]{Tanenbaum.2015}, citing \citet{Holt.1972}, define the representation conversely, with processes as circles and resources as squares. This notation is also applied by \citet[p. 319]{Silberschatz.2012}, where the resource nodes are depicted as rectangles. Despite the apparent confusion at \cite{Tanenbaum.2015}, this form of representation has become established and is therefore used in this work.

In this thesis, processes are visualized as circles and resource types as squares. If instances of the resource types are to be considered, they are represented as filled circles within the square of the resource type. Since a resource type \( R_j \) can encompass multiple instances, a request edge connects the process \( P_i \) only to the corresponding square \( R_j \), while an allocation edge starts from a specific point within the square that symbolizes the allocated resource instance.

When a process \( P_i \) requests an instance of the resource type \( R_j \), a request edge is added to the resource allocation graph. Upon fulfillment of this request, the request edge immediately transforms into an allocation edge. When the process completes its activity and no longer needs the resource, it is released, leading to the deletion of the allocation edge and updating the graph.

The resource allocation graphs in Figure \ref{fig:resource-allocation-graph-general} illustrate the following situation. The sets of processes \( P \), resource types \( R \), and edges \( E_l \) and \( E_r \) are considered:
    \begin{itemize}
        \item \( P = \{P_1, P_2, P_3\} \)
        \item \( R = \{R_1, R_2, R_3\} \)
        \item \( E_l = \{P_1 \rightarrow R_2, P_2 \rightarrow R_1, P_3 \rightarrow R_1, R_1 \rightarrow P_1, R_2 \rightarrow P_2, R_2 \rightarrow P_3, R_3 \rightarrow P_3\} \)
        \item \( E_r = \{P_1 \rightarrow R_2, P_2 \rightarrow R_1, P_3 \rightarrow R_1, R_1 \rightarrow P_1, R_2 \rightarrow P_2, R_3 \rightarrow P_3\} \)
    \end{itemize}
In the left graph of Figure~\ref{fig:resource-allocation-graph-general}, the system defines one instance of resource type \(R_1\), two instances of \(R_2\), and three instances of \(R_3\). In contrast, the right graph contains only one instance per resource type. This difference in available instances explains the variation in the number of edges: while multiple instances of \(R_2\) allow multiple simultaneous allocations in the left graph, the right graph permits only a single assignment per resource type. The red-colored edges and nodes highlight the circular wait conditions that form a deadlock. All involved processes and resources are marked in red to emphasize their participation in the deadlock cycle. Nodes and edges not involved in the deadlock remain black.

\begin{figure}
    \begin{minipage}[t]{.49\textwidth}
        \centering
        \begin{tikzpicture}[node distance=2cm,
            process/.style={circle, draw, minimum size=1cm},
            resource/.style={rectangle, draw, minimum size=1cm},
            instance/.style={circle, fill, inner sep=1.5pt},]
    
            \node[process][color=redMM] (p1) {\(P_1\)};
            \node[process, right of=p1] [color=redMM] (p2) {\(P_2\)};
            \node[process, right of=p2] (p3) {\(P_3\)};
        
            \node[resource, below of=p1, label=below:\(R_1\)]  (r1) {};
            \node[resource, below of=p2, label=below:\(R_2\)] (r2) {};
            \node[resource, below of=p3, label=below:\(R_3\)] (r3) {};
    
            \node[instance] [color=redMM] (i1) at (r1.center) {};
            \node[instance] [color=redMM] (i2) at ([xshift=-5pt]r2.center) {};
            \node[instance] (i3) at ([xshift=5pt]r2.center) {}; 
            \node[instance] (i4) at ([yshift=5pt]r3.center) {};
            \node[instance] at ([xshift=-5pt, yshift=-5pt]r3.center) {};
            \node[instance] at ([xshift=5pt, yshift=-5pt]r3.center) {};  
            
            \begin{scope}[every path/.style={->}]
                \draw [color=redMM] (i1) -- (p1);
                \draw [color=redMM] (i2) -- (p2);
                \draw (i3) -- (p3);
                \draw (i4) -- (p3);
                
                \draw [color=redMM] (p1) -- (r2);
                \draw [color=redMM] (p2) -- (r1);
                \draw (p3) -- (r1);
            \end{scope}
        \end{tikzpicture}
        \caption*{Notation with multiple resource instances.}
    \end{minipage}
    \hfill
    \begin{minipage}[t]{.49\textwidth}
        \centering
        \begin{tikzpicture}[node distance=2cm,
            process/.style={circle, draw, minimum size=1cm},
            resource/.style={rectangle, draw, minimum size=1cm},
            instance/.style={circle, fill, inner sep=1.5pt},]
    
            \node[process] [color=redMM] (p1) {\(P_1\)};
            \node[process, right of=p1] [color=redMM] (p2) {\(P_2\)};
            \node[process, right of=p2] (p3) {\(P_3\)};
        
            \node[resource, below of=p1] [color=redMM](r1) {\(R_1\)};
            \node[resource, below of=p2, label=below:\phantom{}] [color=redMM] (r2) {\(R_2\)};
            \node[resource, below of=p3] (r3) {\(R_3\)};

            \begin{scope}[every path/.style={->}]
                \draw [color=redMM] (r1) -- (p1);
                \draw [color=redMM] (r2) -- (p2);
                \draw (r3) -- (p3);
                
                \draw [color=redMM] (p1) -- (r2);
                \draw [color=redMM] (p2) -- (r1);
                \draw (p3) -- (r1);
            \end{scope}
        \end{tikzpicture}
        \caption*{Simplified notation.}
    \end{minipage}
    \caption{Resource allocation graphs for three processes and three resource types.}
    \label{fig:resource-allocation-graph-general}
\end{figure}

The states of the processes in the left graph of Figure \ref{fig:resource-allocation-graph-general} are characterized as follows:
\begin{itemize}
    \item[] Process \(P_1\) holds an instance of resource type \(R_1\) and requests an instance of resource type \(R_2\). Process \(P_2\) holds an instance of \(R_2\) and requests \(R_1\), while \(P_3\) holds instances of \(R_2\) and \(R_3\) and also requests \(R_1\). In the right graph, the states are similar, however, without the assignment of \(R_2\) to \(P_3\), leading to a deadlock between \(P_1\) and \(P_2\) since all instances of \(R_1\) and \(R_2\) are occupied. \(P_3\) is also blocked due to the deadlock. 
\end{itemize}

The states of the processes can also be represented by two adjacency matrices: the allocation matrix (\(C\), for \textit{\textbf{c}urrently allocated}) and the request matrix (\(\mathcal{R}\), for \textit{\textbf{r}equired}). To distinguish between the two scenarios depicted in Figure~\ref{fig:resource-allocation-graph-general}, subscripts \(l\) and \(r\) denote the matrices corresponding to the left and right graphs, respectively. The matrices for the left graph in Figure~\ref{fig:resource-allocation-graph-general} are:

\begin{minipage}[t]{.49\textwidth}
    \centering
   Allocation matrix \(C_l\):
    
    $C_l = \kbordermatrix{
        & R_1 & R_2 & R_3 \\
    P_1 & 1 & 0 & 0 \\
    P_2 & 0 & 1 & 0 \\
    P_3 & 0 & 1 & 1
    }$
\end{minipage}
\begin{minipage}[t]{.49\textwidth}
    \centering
    Request matrix \(\mathcal{R}_l\):    
    
    $\mathcal{R}_l = \kbordermatrix{
        & R_1 & R_2 & R_3 \\
    P_1 & 0 & 1 & 0 \\
    P_2 & 1 & 0 & 0 \\
    P_3 & 1 & 0 & 0
    }$
    \vspace{5mm}
\end{minipage}

The matrices for the right graph in Figure \ref{fig:resource-allocation-graph-general} without the additional allocation of \(R_2\) to \(P_3\) are:

\begin{minipage}[t]{.49\textwidth}
    \centering
    Allocation matrix \(C_r\):
    
    $C_r = \kbordermatrix{
        & R_1 & R_2 & R_3 \\
    P_1 & 1 & 0 & 0 \\
    P_2 & 0 & 1 & 0 \\
    P_3 & 0 & 0 & 1
    }$
\end{minipage}
\begin{minipage}[t]{.49\textwidth}
    \centering    
    
    Request matrix \(\mathcal{R}_r\):
    
    $\mathcal{R}_r = \kbordermatrix{
        & R_1 & R_2 & R_3 \\
    P_1 & 0 & 1 & 0 \\
    P_2 & 1 & 0 & 0 \\
    P_3 & 1 & 0 & 0
    }$
    \vspace{5mm}
\end{minipage}

In these matrices, the rows represent the processes \(P_i\) and the columns the resource types \(R_j\). An entry of \enquote{1} in the allocation matrix indicates that a resource is allocated to a process, while in the request matrix, a \enquote{1} represents a request for a resource by a process. The request and allocation matrices serve as the basis for algorithms to detect and avoid deadlocks.

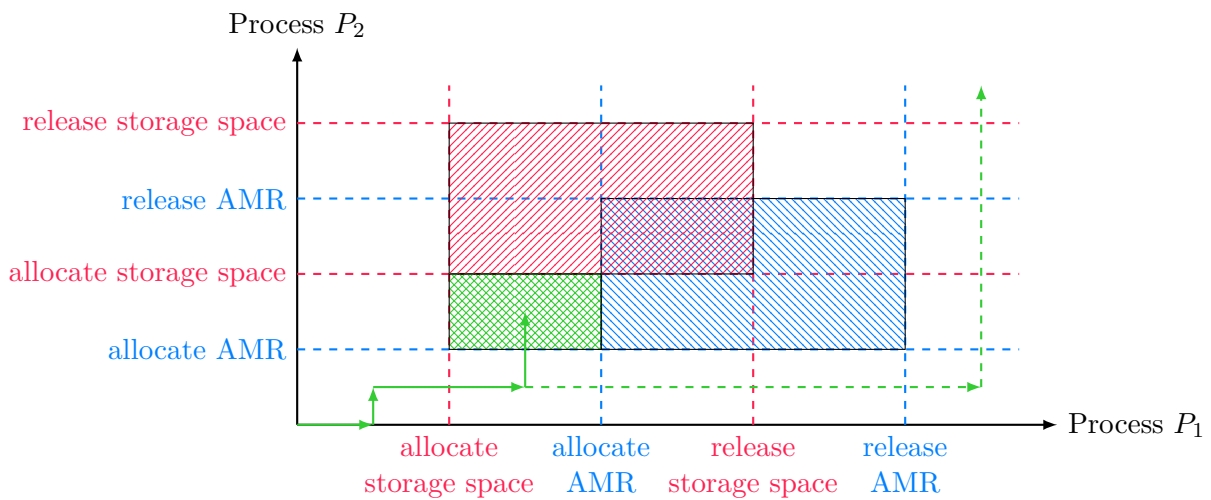
\begin{figure}[h]
    \centering
    \begin{tikzpicture}[scale=1, every node/.style={scale=1}]
      \draw[thick,->] (0,0) -- (0,5) node[above] {Process \( P_2 \)};
      \draw[thick,->] (0,0) -- (10,0) node[right] {Process \( P_1 \)};
    
      \draw[thick, dashed, color=redMM] (2,0) -- (2,4.5) node[at start, below, align=center] {allocate\\ storage space};
      \draw[thick, dashed, color=blueMM] (4,0) -- (4,4.5) node[at start, below, align=center] {allocate\\ AMR};
      \draw[thick, dashed, color=redMM] (6,0) -- (6,4.5) node[at start, below, align=center] {release\\ storage space};   
      \draw[thick, dashed, color=blueMM] (8,0) -- (8,4.5) node[at start, below, align=center] {release\\ AMR};
      
      \draw[thick, dashed, color=blueMM] (0,1) -- (9.5,1) node[at start, left] {allocate AMR};
      \draw[thick, dashed, color=redMM] (0,2) -- (9.5,2) node[at start, left] {allocate storage space};
      \draw[thick, dashed, color=blueMM] (0,3) -- (9.5,3) node[at start, left] {release AMR};
      \draw[thick, dashed, color=redMM] (0,4) -- (9.5,4) node[at start, left] {release storage space};
    
      \draw[pattern=north west lines, pattern color=blueMM] (4,1) rectangle (8,3);
      \draw[pattern=north east lines, pattern color=redMM] (2,2) rectangle (6,4);
      \draw[pattern=crosshatch, pattern color=greenMM] (2,1) rectangle (4,2);

      \draw[thick, ->, color=greenMM] (0,0) -- (1,0) {};
      \draw[thick, ->, color=greenMM] (1,0) -- (1,0.5) {};
      \draw[thick, ->, color=greenMM] (1,0.5) -- (3,0.5) {};
      \draw[thick, ->, color=greenMM] (3, 0.5) -- (3,1.5) {};
      \draw[thick, ->, dashed, color=greenMM] (3, 0.5) -- (9,0.5) {};
      \draw[thick, ->, dashed, color=greenMM] (9, 0.5) -- (9,4.5) {};

    \end{tikzpicture}
    \caption{Resource diagram based on \citet[p. 71]{Coffman.1971}.}
    \label{fig:resource-diagram-trace}
\end{figure}

Another form of modeling deadlocks is through the use of a resource diagram and a resource trace. Figure \ref{fig:resource-diagram-trace} exemplifies this with the two transport processes \(P_1\) and \(P_2\), considering two instances of resources: a storage space and an AMR. The x-axis represents the progress of process \(P_1\) and the y-axis represents the progress of process \(P_2\).

A continuous green resource trace illustrates the sequential allocation of resources by the processes along the timeline. Process \(P_1\) initiates the sequence, while \(P_2\) remains in a waiting position. Without an initial resource allocation, both processes continue their activities. Subsequently, \(P_1\) first occupies the storage space. Consecutively, the AMR is utilized by \(P_2\), leading both processes into a deadlock situation. In this constellation, further progress for both processes would require passage through the hatched areas, which would imply simultaneous allocation of the same resource by both processes – a configuration that is systemically unfeasible. The green hatched area thus visualizes an unsafe system state in which a deadlock becomes inevitable. As an alternative to the deadlock situation, \(P_2\) could wait and initially leave all resource requests to \(P_1\). This would lead to the dashed resource trace. The system would be in a safe state the entire time. If the AMR is requested by \(P_2\) after \(P_1\) has occupied the storage space, the system enters an unsafe state since there is no longer a sequence of resource requests that allows both processes to be completed. As a circular reference arises in this case, the unsafe state also leads to a deadlock. In this work, the two state types, safe and unsafe, are defined as follows:

\begin{definition}[Safe state]
A system is in a \textbf{safe state} if a sequence of resource requests by the processes exists such that all processes can be completed.
\end{definition}

\begin{definition}[Unsafe state]
A system is in an \textbf{unsafe state} if no sequence of resource requests by the processes exists that allows the execution of all processes.
\end{definition}

An unsafe state does not necessarily imply that a deadlock is imminent, but that the potential for a deadlock exists under certain conditions. The distinction between safe and unsafe states is central to deadlock avoidance algorithms, which aim to keep systems in safe states.

Another form of modeling deadlocks can be achieved using Petri nets \citep{Petri.1962}. However, Petri nets are not further used in this thesis and therefore not discussed in detail here.

\subsection{Deadlock Handling Algorithms}
Besides problem-specific algorithms for handling deadlocks, there are several general algorithms that have become established in the literature for dealing with deadlocks and are briefly explained below.

The \textbf{ostrich algorithm} represents a pragmatic approach to dealing with deadlocks and other serious system problems. Its name metaphorically refers to the behavior of the ostrich, which, in a figurative sense, sticks its head in the sand to ignore dangers. In the context of deadlocks, this means that deadlocks are deliberately ignored or left untreated, rather than implementing complicated or resource-intensive avoidance or recovery strategies.

The basic principle of the ostrich algorithm is the acceptance that not all systemic problems are efficiently solvable. Instead of allocating resources to avoid or solve rarely occurring or difficult-to-solve problems, the decision is made to ignore them. The assumption is that the costs of developing a solution would exceed the potential benefits, or that the probability of the problem occurring is so low that it is more economical to manually handle the rare cases.

In practice, the ostrich algorithm is applied in systems where the costs or complexity of problem avoidance outweigh the potential risks. An example in intralogistics would be not to implement extensive deadlock detection and  mechanisms for an AGV system, but instead to leave the responsibility to a maintenance team to detect and fix the deadlock, for example, by restarting the vehicle or the entire system.

While the ostrich algorithm can be a cost-effective solution for rare or less severe deadlocks, it carries the risk that untreated deadlocks may escalate and lead to serious system failures and performance losses. Criticism of the ostrich algorithm is mainly directed against the shortsightedness of the approach and the potential neglect of preventive measures that could ensure long-term system stability and safety.

The \textbf{Banker's algorithm}, designed by Edsger W. Dijkstra \citeyearpar{Dijkstra.1982}, is a method for avoiding deadlocks in operating systems and other computer-controlled environments that require resource management. Its fundamental analogy to a bank's risk management, where loans are only granted if the bank can ensure that all customers can repay their debts under all circumstances, illustrates the principle of avoiding unsafe states in resource management. The Banker's algorithm uses four data structures to manage its state, where \(m\) is the number of resource types and \(n\) is the number of processes \cite[p. 331]{Silberschatz.2012}:

\begin{itemize}
    \item \textbf{Available} ($Available_{R_j}$): This vector, of length \(m\), specifies the count of each type of resource that is currently available. For example: 

\[
Available_{R_j} =
\begin{cases}
k, & \text{there are \(k\) available instances of resource type } R_j, \\
0, & \text{otherwise.}
\end{cases}
\]

    \item \textbf{Maximum request} ($Max_{P_i,R_j}$): Represented as an \(n \times m\) matrix, it denotes the peak demand of each process for every type of resource. Specifically,

\[
Max_{P_i,R_j} =
\begin{cases}
k, & \parbox[t]{0.6\linewidth}{being \(k\) means that process \(P_i\) may potentially request up to \(k\) instances of resource type \(R_j\),} \\
0, & \text{otherwise.}
\end{cases}
\]

    \item \textbf{Allocation} ($Allocation_{P_i,R_j}$): This \(n \times m\) matrix records the quantity of each resource type currently allocated to each process. 

\[
Allocation_{P_i,R_j} =
\begin{cases}
k, & \parbox[t]{0.6\linewidth}{indicates that process \(P_i\) has been allocated \(k\) instances of resource type \(R_j\),} \\
0, & \text{otherwise.}
\end{cases}
\]

    \item \textbf{Need} ($Need_{P_i,R_j}$): Also structured as an \(n \times m\) matrix, it reflects how many additional instances of each resource type each process requires to complete its execution. 

    \[
    Need_{P_i,R_j} =
    \begin{cases}
    k, & \parbox[t]{0.6\linewidth}{being \(k\) implies that process \(P_i\) still requires \(k\) instances of resource type \(R_j\) to finish,} \\
    0, & \text{otherwise.}
    \end{cases}
    \]
    The computation for $Need_{P_i,R_j}$ is given by  $Max_{P_i,R_j} - Allocation_{P_i,R_j}$ for each process $P_i$ and resource $R_i$.

\end{itemize}

The Banker's algorithm checks with each request whether granting this request would transition the system into a safe state. The pseudo code for Algorithm \ref{alg:bankers-algorithm} shows the procedure of the Banker's algorithm for checking the system state based on \citet[p. 332]{Silberschatz.2012}. The checking process of the system state is achieved by simulating for each resource request whether sufficient resources would be available to execute at least one process to completion, assuming that all other processes, in the worst case, make their maximum requests.

        
        

\begin{algorithm}
\caption{Banker's algorithm}
\label{alg:bankers-algorithm}
\begin{algorithmic}[1]
    \State \textbf{Input:} \(Available_{R_j},\ Max_{P_i,R_j},\ Allocation_{P_i,R_j},\ Need_{P_i,R_j}\)
    \State \textbf{Output:} System state (safe or unsafe)
    
    \State $CurrentAvailable_{R_j} \gets Available_{R_j}$
    \State $Completed_{P_i} \gets$ [false, false, ..., false] \Comment{Length \(n\), one for each process}
    
    \While{there exists a process \(P_i\) such that \(Completed_{P_i} = \text{false}\) and \(\forall R_j,\ Need_{P_i,R_j} \leq CurrentAvailable_{R_j}\)}
        \State \textbf{/* Simulate process \(P_i\) finishing */}
        \ForAll{resource types \(R_j\)}
            \State \(CurrentAvailable_{R_j} \gets CurrentAvailable_{R_j} + Allocation_{P_i,R_j}\)
        \EndFor
        \State \(Completed_{P_i} \gets \text{true}\)
    \EndWhile

    \If{\(Completed_{P_i} = \text{true}\) for all \(i = 1, \dots, n\)}
        \State \Return \textbf{true} \Comment{System is in a safe state}
    \Else
        \State \Return \textbf{false} \Comment{System is in an unsafe state}
    \EndIf
\end{algorithmic}
\end{algorithm}

Despite its theoretical relevance and suitability as a teaching example for deadlock avoidance, the practical application of the Banker's algorithm is limited. The challenges include:

\begin{itemize}
    \item Predictability of maximum requests: In real systems, it is often not possible to predict the maximum resource requests of all processes.
    \item Dynamic process numbers: The assumption of a fixed number of processes contradicts the dynamics of many modern systems, where processes are dynamically created and terminated.
    \item Resource utilization efficiency: The need to reserve resources for potential maximum requests can lead to suboptimal use of system resources.
\end{itemize}

The Banker's algorithm remains a foundational concept in the study of deadlock avoidance. It illustrates how resource safety can be guaranteed through anticipatory planning and centralized admission control based on maximum demand assumptions. These assumptions do not transfer to MAPF problems in intralogistics. In MAPF, agents represent mobile entities with spatial goals and movement constraints rather than abstract processes with quantifiable resource demands. Resources are acquired dynamically, and agent interactions are governed by local motion conflicts rather than global allocation rules. As a result, the Banker's algorithm cannot be directly applied in this setting. Instead, classical MAPF algorithms are used as baselines to resolve spatial conflicts and evaluate the effectiveness of the learning-based methods presented in this thesis. The Banker's algorithm is included here as a conceptual benchmark, highlighting the transition from static resource safety to adaptive and decentralized coordination.

\section{Reinforcement Learning} \label{sec:background-reinforcementlearning}

\subsection{Multi-Agent Systems}
In the context of complex systems where autonomous entities interact, such as environments involving autonomous mobile robots, traditional considerations of deadlocks reach their limits. The previous focus on processes and resources provides important fundamentals but neglects some aspects of interaction that occur in real systems. The concept of multi-agent systems (MAS) offers an expanded perspective that allows for a more nuanced examination of the system. The concept of MAS brings decisions within the system to the forefront, and the interaction between actors (agents) in the system can be better described than with a selection of processes and resources.

Agents, defined as distinct hardware and/or software units with defined goals relating to the control of a technical system, enable a more realistic modeling of systems \cite[p. 4]{vdi2018agentensysteme}. It is noted by \citet[p. 5]{wooldridge2002agents} that there is no consensus on a universal definition of an agent, but autonomy is the minimal consensus regarding the definition. Agents can respond to changes in their environment, develop strategies, and coordinate their actions with other agents. These properties are crucial for understanding complex scenarios such as pathfinding and deadlock avoidance in an environment with multiple autonomous vehicles and developing effective solutions.

In addition to the fundamental property of autonomy, \citet[p. 15 ff.]{wooldridge2002agents} identifies four classes for agents, each highlighting certain aspects of agent behavior and decision-making. The following Table \ref{tab:agent_classes} provides an overview of these four classes of agents and their characteristic decision-making mechanisms.

\begin{table}[h]
    \centering
    \caption{Classes of agents according to \citet[p. 15 ff.]{wooldridge2002agents}.}
    \label{tab:agent_classes}
    \begin{tabular}{>{\arraybackslash}p{0.3\linewidth}>{\arraybackslash}p{0.6\linewidth}}
        \toprule
        \textbf{Class of agents} & \textbf{Description} \\
        \hline
        Logic-based agents & Decision-making through logical deduction. \\
        Reactive agents & Decision-making through direct mapping of situations to actions. \\
        Belief-desire-intention agents & Decision-making based on manipulating data structures representing the agent's beliefs, desires, and intentions. \\
        Layered architectures & Decision-making through various software layers, each reasoning about the environment at different levels of abstraction. \\
        \bottomrule
    \end{tabular}
\end{table}

In this thesis, a focus is placed on reactive agents, as they are utilized in RL approaches.

\subsection{The Role of Reinforcement Learning}
RL is a branch of machine learning. Machine Learning is a field of artificial intelligence. Machine learning is characterized by the artificial generation of knowledge and not the direct transfer of human knowledge into an algorithm. \citet[p. 2]{mitchell1997machinelearning} describes the learning of a machine, a computer program, as \enquote{A computer program is said to learn from experience E with respect to some class of tasks T and performance measure P, if its performance at tasks in T, as measured by P, improves with experience E.} Machine learning can be categorized according to various criteria. A common categorization is based on the criterion of how the training, i.e. the learning process, is monitored. A distinction is made between supervised learning, unsupervised learning and RL.

In \textbf{supervised learning}, the foundational element is a dataset that correlates input data with their corresponding output data. Through this dataset, an algorithm is trained to discern the relationship between the inputs and outputs, enabling it to accurately predict outcomes for new, unseen input data. This method depends on the availability of labeled data, where each input is associated with a correct output, facilitating the algorithm's learning process by providing clear, correct examples to learn from.

An example of using supervised learning to address deadlocks in intralogistics could involve training a predictive model to identify potential deadlock situations in warehouse operations. In this scenario, the model would analyze data from warehouse management systems, including the location and movement patterns of goods, AMRs, and human-operated machines within a warehouse. The input data could consist of timestamps, coordinates of items and vehicles, paths taken, and the sequence of operations. The output data would indicate whether a given set of conditions would lead to a deadlock situation, enabling preemptive action to avoid operational disruptions. The training dataset would need to be labeled with instances of both deadlock and non-deadlock scenarios. By learning from this dataset, the supervised learning model could predict deadlock situations based on real-time data, allowing for interventions such as rerouting vehicles, rescheduling operations, or adjusting loading and unloading sequences to maintain smooth intralogistics operations.

If we consider using a supervised learning algorithm to directly manage deadlocks, the drawbacks of this approach quickly become evident. The core issue lies in the need for labeled data that specifies the \enquote{correct} decision for each specific situation. In practice, this kind of data is rarely, if ever, available and creating it artificially poses significant challenges due to the highly context-dependent nature of each problem. Supervised learning relies heavily on having detailed, accurate examples to learn from, but the dynamic and complex environment of deadlock management means that the exact conditions and solutions are unique each time, making it hard to prepare a dataset that covers all possible scenarios effectively. Despite the potential applications for deadlock detection, this thesis does not consider supervised learning due to the disadvantages mentioned.

\textbf{Unsupervised learning} algorithms analyze and cluster data based on inherent patterns and relationships, seeking to understand the structure of the data without explicit instructions on what to predict or how to act. In the context of deadlocks, an unsupervised learning model could be used to analyze transactional data from warehouse operations, identifying patterns that frequently lead up to deadlock situations. For example, it might uncover that certain combinations of machine movements, loading patterns, or sequences of operations tend to precede deadlocks. This insight could then be used to adjust operations to avoid these high-risk patterns, potentially preventing deadlocks before they occur.

While unsupervised learning can highlight problematic patterns or conditions leading to deadlocks, it does not inherently provide solutions or preventive measures. It excels at identifying the conditions under which deadlocks are likely to occur but does not specify what actions to take to avoid them. Implementing effective solutions based on these insights would still require human intervention and decision-making. Intralogistics environments are dynamic, with constant changes in layout, operations, and workflows. Patterns identified by unsupervised learning at one point in time may not remain relevant as the operational context evolves, requiring continuous analysis and adjustment that may not be feasible in real-time operations. Due to mentioned disadvantages, the unsupervised learning approach is not pursued further in this thesis.

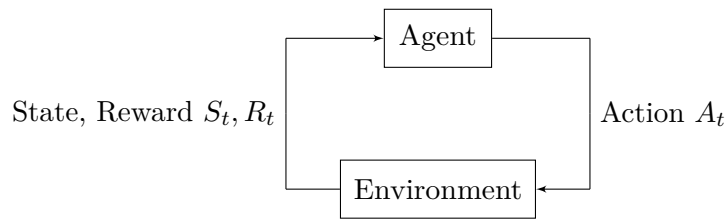
\begin{figure}[H]
\centering
\begin{tikzpicture}[node distance=2cm, auto,>=latex']
    \node [rectangle, draw, text width=3em, text centered, minimum height=2em] (agent) {Agent};
    \node [rectangle, draw, below of=agent, text width=6em, text centered, minimum height=2em] (environment) {Environment};
    
    \draw (agent) -- (2,0);
    \draw (2,0) -- (2,-1) node[anchor=west] {Action \( A_t \)};
    \draw (2,-1) -- (2,-2);
    \draw [->] (2,-2) -- (environment);

    \draw (environment) -- (-2,-2);
    \draw (-2,-2) -- (-2,-1) node[anchor=east] {State, Reward \( S_t, R_t \)};
    \draw (-2,-1) -- (-2,0);
    \draw [->] (-2,0) -- (agent);

\end{tikzpicture}
\caption{Interaction model in RL.}
\label{fig:rl-interaction}
\end{figure}

According to \citet[p. 20]{albrecht2024marl}, \textbf{RL algorithms} \enquote{[...] learn solutions for sequential decision processes via repeated interaction with an environment.} Unlike its counterparts, supervised learning and unsupervised learning, RL distinguishes itself by focusing on the development of optimal strategies through interactions with an environment. Consequently, there is a clearer focus on learning a certain policy for decision-making over time compared to the other machine learning branches. In these interactions between agent and environment, an agent makes decisions, performs actions, and receives feedback in the form of rewards or penalties. This feedback mechanism enables the agent to iteratively refine its actions based on the outcomes of past decisions, thereby learning to achieve specific goals within uncertain and complex environments. The dynamic between agent and environment is illustrated in Figure \ref{fig:rl-interaction}.

At each timestep \(t\), the agent observes a state \(s_t\) from the environment and then selects an action \(a_t\) based on its strategy. The environment responds to the agent's action by transitioning to a new state and providing the agent with a reward \(r_t\). This reward serves as feedback for the agent to evaluate and adjust its strategy.

Unlike supervised and unsupervised learning, RL does not just identify potential deadlocks. The RL algorithm actively learns and proposes a policy as a solution. By exploring different strategies to maximize its reward, an RL agent can discover innovative ways to manage and mitigate deadlocks, potentially uncovering strategies that human operators might not have considered. RL is adept at making decisions in dynamic environments, where conditions change rapidly, and actions must be adjusted accordingly. This makes it ideal for intralogistics, where variables like machine location, task priorities, and operational sequences are constantly in flux. For these reasons, this thesis focus on and pursue an RL approach.

\subsection{Markov Decision Processes}
The modeling of the decision making process in RL is based on the Markov decision process (MDP). An MDP provides a mathematical formulation for modeling decision making where outcomes are partly under the control of a decision maker and partly random \citep{bellman1957markovian}. An MDP is formally defined as a 4-tuple $(S, A, P, R)$ or 5-tuple $(S, A, P, R, \mu)$ as follows:

\begin{itemize}
    \item \textbf{State space ($S$):} a finite or infinite set of states of the environment.
    \item \textbf{Action space ($A$):} a finite or infinite set of actions.
    \item \textbf{State transition probability function ($P$):} $S \times A \times S \rightarrow [0, 1]$. $P(s'|s, a)$ denotes the probability of transitioning from state $s$ to state $s'$ upon taking action $a$.
    \item \textbf{Reward function ($R(s,a,s')$):} $S \times A \rightarrow \mathbb{R}$ provides the expected immediate reward received after transitioning from state $s$ to state $s'$ via action $a$. Alternatively, $R: S \times A \times S \rightarrow \mathbb{R}$ if the reward also depends on the resulting state $s'$.
    \item  \textbf{Initial state distribution ($\mu$):} $S \rightarrow [0, 1]$. The initial state $S_0$ is sampled from $\mu$.
\end{itemize}

The MDP is described in literature in certain variations\cite[p. 8]{schulman2016PhDThesis}. Definitions as a 4-tuple \cite[p. 14]{lang2023bestärkendesLernen, foerster2018DeepMARL} or 5-tuple \cite[p. 22]{albrecht2024marl} are common. There are also definition specified on a finite MDP, which is defined by a finite state space and a finite action space \cite[p. 22]{Sutton.2018,albrecht2024marl}. 

\begin{figure}
    \centering
    \begin{tikzpicture}
        \tikzstyle{state}=[circle,draw=black,fill=none,minimum size=65pt,inner sep=0pt]
        \tikzstyle{action}=[circle,draw=black,fill=none,minimum size=17pt,inner sep=0pt]

        \node[state] (state_0) at (-1,3) {$S_0: start$};
        \node[action] (state_0_action_0) at (-1,1) {$a_0$};
        \node[action] (state_0_action_1) at (1,4) {$a_1$};

        \node[state, fill=gray!20] (state_1) at (3,6) {$S_1: collision$};

        \node[state] (state_2) at (7,3) {$S_2: moving$};
        \node[action] (state_2_action_0) at (6,4.5) {$a_0$};
        \node[action] (state_2_action_1) at (7,1) {$a_1$};
        
        \node[state] (state_3) at (3,0) {$S_3: delivery$};
        \node[action] (state_3_action_0) at (1,0) {$a_0$};
        \node[action] (state_3_action_1) at (3,2) {$a_1$};

        \draw[->] (state_0) -- (state_0_action_0);
        \draw[->] (state_0_action_0.west) .. controls +(left:1cm) and +(left:1cm) .. node[midway, below left, align=left] {$p=1$\\$r=0$} (state_0.west);
        \draw[->] (state_0) -- (state_0_action_1);
        \draw[->] (state_0_action_1) -- node[midway, above left, align=left] {$p=0.05$\\$r=-1$} (state_1);
        \draw[->] (state_0_action_1) -- node[midway, above right, align=left] {$p=0.95$\\$r=0$} (state_2);

        \draw[->] (state_2) -- (state_2_action_0);
        \draw[->] (state_2_action_0.east) .. controls +(right:0.5cm) .. node[midway, above right, align=left] {$p=1$\\$r=-0.1$} (state_2.north);
        \draw[->] (state_2) -- (state_2_action_1);
        \draw[->] (state_2_action_1) -- node[midway, below right, align=left] {$p=0.9$\\$r=1$} (state_3);
        \draw[->] (state_2_action_1) .. controls (10,1) and (12,7)  .. node[midway, below right, align=left] {$p=0.1$\\$r=-1$} (state_1);

        \draw[->] (state_3) -- (state_3_action_0);
        \draw[->] (state_3_action_0.south) .. controls +(down:0.5cm) .. node[midway, below left, align=left] {$p=1$\\$r=0$} (state_3);
        \draw[->] (state_3) -- (state_3_action_1);
        \draw[->] (state_3_action_1) -- node[midway, below left, align=left] {$p=0.5$\\$r=0.1$} (state_0);
        \draw[->] (state_3_action_1) -- node[midway, below, align=left] {$p=0.5$\\$r=0$} (state_2);
        
    \end{tikzpicture}
    \caption{Exemplary MDP for an AMR with two actions $a_0$ (stop) and $a_1$ (move).}
    \label{fig:mdp_amr}
\end{figure}
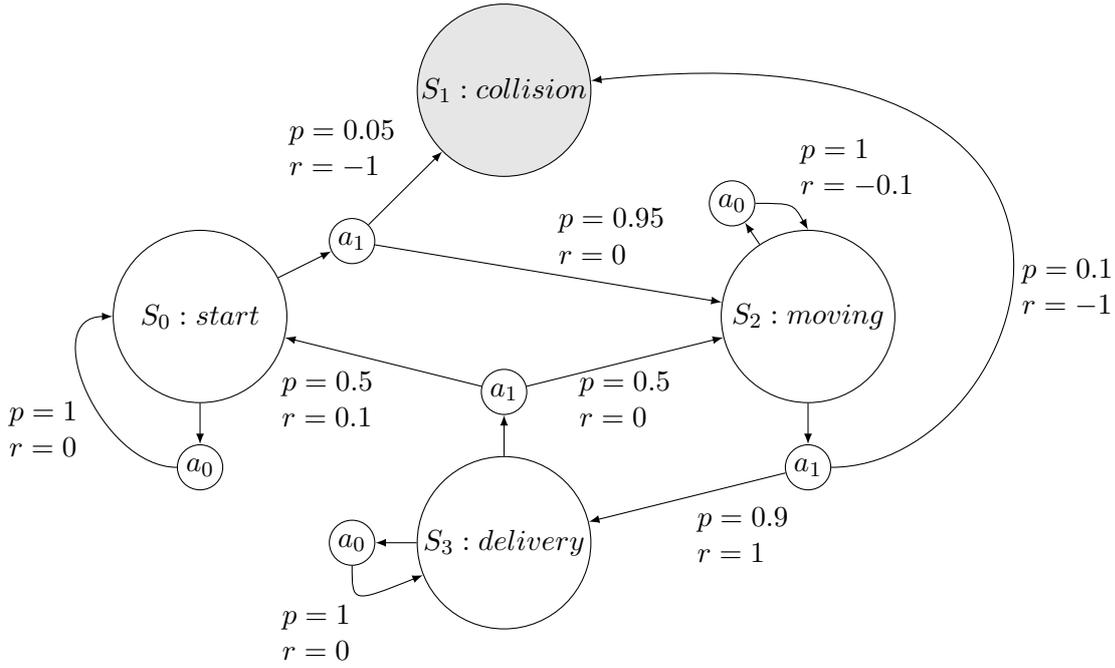

Figure \ref{fig:mdp_amr} presents an exemplary state diagram for a MDP involving an AMR. The states are represented by circles, where \( S_0 \) is the starting state, \( S_1 \) indicates a collision, \( S_2 \) represents the robot in motion, and \( S_3 \) signifies a successful delivery. Arrows between states denote transitions with the possible actions \( a_0 \) to stop and \( a_1 \) to move. Each transition with an action is labeled with two parameters: the probability of the transition \( p \) and the immediate reward \( r \). For instance, from state \( S_0 \), taking action \( a_1 \) leads to state \( S_1 \) with a probability of 0.05 and a negative reward of -1, indicating a collision, or to state \( S_2 \) with a probability of 0.95 and a reward of 0. The gray node corresponding to state \( S_1 \) signifies that \( S_1 \) is a terminal state. This indicates that once the system enters state \( S_1 \) no subsequent transitions to other states are possible. The process effectively terminates upon reaching state \( S_1 \).

The Markov property is a key assumption in MDPs and RL, stating that the future state of a process depends only on the current state and action, not on the sequence of events that preceded it. Formally, for a state $S_t$ and action $A_t$ at time $t$, the Markov property is given by:

\begin{equation}
P(S_{t+1} | S_t, A_t, S_{t-1}, A_{t-1}, \ldots, S_0, A_0) = P(S_{t+1} | S_t, A_t)
\end{equation}

The Markov property simplifies the analysis and prediction of future states in complex systems by reducing the amount of historical data that must be considered. In the context of MDPs, it allows for the modeling of decision-making processes and the development of optimization strategies without needing to track and analyze the entire history of states and actions, focusing instead on the current state and the decisions made in that state.

The objective in an MDP is to find a policy $\pi$ that maps states to actions to maximize the cumulative reward. This is done by maximizing the expected return $G_t$, which is the sum of discounted rewards from timestep $t$. The value of a state $s$ under a policy $\pi$, denoted as $v_\pi(s)$, is the expected return when starting in $s$ and following $\pi$ thereafter. The \textit{return} \( G_t \) at timestep \( t \) is the total discounted reward from that timestep onwards, which is calculated as \cite[p. 55]{Sutton.2018}:

\begin{equation}
G_t = R_{t+1} + \gamma R_{t+2} + \gamma^2 R_{t+3} + \dots = \sum_{k=0}^{\infty} \gamma^k R_{t+k+1}
\end{equation}

where:
\begin{itemize}
    \item \textbf{Index variable (\( k \)):} represents the number of timesteps from the current timestep \( t \). 
    \item \( R_{t+k+1} \) is the reward received after \( k+1 \) timesteps from \( t \).
    \item \textbf{Discount factor (\( \gamma \)):} a number between 0 and 1 (\( \gamma \in [0,1] \)) that represents the present value of future rewards, effectively weighting the importance of sooner rewards higher than those received later.
\end{itemize}

The \textit{value function} \(v_\pi(s) \) under a policy \( \pi \) for a state \( s \) is the expected return when starting in \( s \) and following policy \( \pi \) \cite[p. 58]{Sutton.2018}. It is defined as:
\begin{equation}
v_\pi(s) \doteq \mathbb{E}[G_t | S_t = s] = \mathbb{E}\left[\sum_{k=0}^{\infty} \gamma^k R_{t+k+1} \mid S_t = s\right]
\end{equation}

where:
\begin{itemize}
    \item \( S_t \) denotes the state at time \( t \).
    \item \( \mathbb{E}[\cdot | S_t = s] \) denotes the expected value given that the agent is in state \( s \) at time \( t \) and then follows policy \( \pi \), with the expectation accounting for the stochastic nature of the policy, state transitions, and reward function.
    \item The \textbf{policy \( \pi \)} is a strategy that assigns a probability distribution over actions for each state, dictating the behavior of the agent.
\end{itemize}

\(v_\pi(s) \) is called \textit{state-value function} for policy \( \pi \). In addition to the state-value function, the \textit{action-value function} \( q_\pi(s, a) \) under a policy \( \pi \) for a state \( s \) and action \( a \) is the expected return after taking an action \( a \) in state \( s \) and then following policy \( \pi \). It is defined as \cite[p. 58]{Sutton.2018}:
\begin{equation}
q_\pi(s, a) \doteq \mathbb{E}[G_t | S_t = s, A_t = a] = \mathbb{E}\left[\sum_{k=0}^{\infty} \gamma^k R_{t+k+1} \mid S_t = s, A_t = a\right]
\end{equation}

where:
\begin{itemize}
    \item \( A_t \) denotes the action taken at time \( t \).
    \item \( \mathbb{E}[\cdot | S_t = s, A_t = a] \) denotes the expected value given that the agent is in state \( s \) and takes action \( a \) at time \( t \), and then follows policy \( \pi \), with the expectation accounting for the stochastic nature of subsequent actions, state transitions, and rewards.
\end{itemize}

Agents do not always have access to the full state of the environment. For example, an AMR may perceive only its local surroundings through onboard sensors. In such cases, the environment is more accurately modeled as a \textit{partially observable Markov decision process} (POMDP). Unlike an MDP, a POMDP accounts for the fact that the agent observes only partial information, which may be noisy or ambiguous. As a result, decisions must be made under uncertainty about the true system state.

Formally, a POMDP extends the MDP by adding:
\begin{itemize}
    \item An observation space $O$,
    \item An observation function $P_O(o \mid s', a)$ that defines the probability of receiving observation $o$ after taking action $a$ and arriving in state $s'$.
\end{itemize}

The agent receives $o \in O$ instead of the true state $s \in S$ and must infer useful internal representations, for instance, belief distributions over $S$. This increases the complexity of policy learning, especially in multi-agent settings where partial observability is common. POMDPs are relevant in real-world scenarios where direct access to the full state is impractical or impossible.

\subsection{Artificial Neural Networks}
In real-world multi-agent scenarios involving AMRs, the state space can be large and complex. An exact calculation of state values or policies is computationally infeasible in such environments, especially if for the agents the environment is only partially observable. To address this challenge, nonlinear function approximation methods are employed, among which artificial neural networks (ANNs) are prominent in the context of RL \cite[p. 223]{Sutton.2018}. ANNs are advantageous due to their capability to approximate complex functions that may not be explicitly defined or easily derivable through traditional mathematical means.

\begin{figure}
    \centering
        \begin{tikzpicture}[>=stealth,shorten >=1pt,node distance=2 cm, on grid, auto]
            \tikzstyle{neuron}=[circle,draw=black,fill=none,minimum size=17pt,inner sep=0pt]
            \tikzstyle{input neuron}=[neuron];
            \tikzstyle{output neuron}=[neuron];
            \tikzstyle{hidden neuron}=[neuron];
            \tikzstyle{annot} = [text width=4em, text centered]
            \tikzstyle{arrow} = [->,>=latex]

            \foreach \name / \y in {1,...,4}
                \node[input neuron] (I-\name) at (0,-\y-0.5) {};
            
            \foreach \name / \y in {1,...,5}
                \node[hidden neuron] (H-\name) at (2,-\y cm) {};

            \foreach \name / \y in {1,...,5}
                \node[hidden neuron] (H2-\name) at (4,-\y cm) {};
                    
            \foreach \name / \y in {1,...,3}
                \node[output neuron] (O-\name) at (6,-\y-1) {};
            
            \foreach \source in {1,...,4}
                \foreach \dest in {1,...,5}
                    \draw[arrow] (I-\source) -- (H-\dest);
            
            \foreach \source in {1,...,5}
                \foreach \dest in {1,...,5}
                    \draw[arrow] (H-\source) -- (H2-\dest);

            \foreach \source in {1,...,5}
                \foreach \dest in {1,...,3}
                    \draw[arrow] (H2-\source) -- (O-\dest);
            
            \node[annot,above of=H-1, node distance=1cm] {First hidden layer};
            \node[annot,above of=H2-1, node distance=1cm] {Second hidden layer};
            \node[annot,above of=I-1, node distance=1cm] {Input layer};
            \node[annot,above of=O-1, node distance=1cm] {Output layer};
            \node[annot,left of=H-3, node distance=4.5cm] {Observations};
            \node[annot,right of=O-2, node distance=2cm] {Predictions};

            \foreach \y in {1,...,4}
                \draw[arrow] (-1,-\y-0.5) -- (I-\y);
                
            \foreach \y in {1,...,3}
                \draw[arrow] (O-\y) -- (7,-\y-1);

        \end{tikzpicture}
    \caption{Schematic representation of a feedforward neural network with two hidden layers.}
    \label{fig:neural_network}
\end{figure}
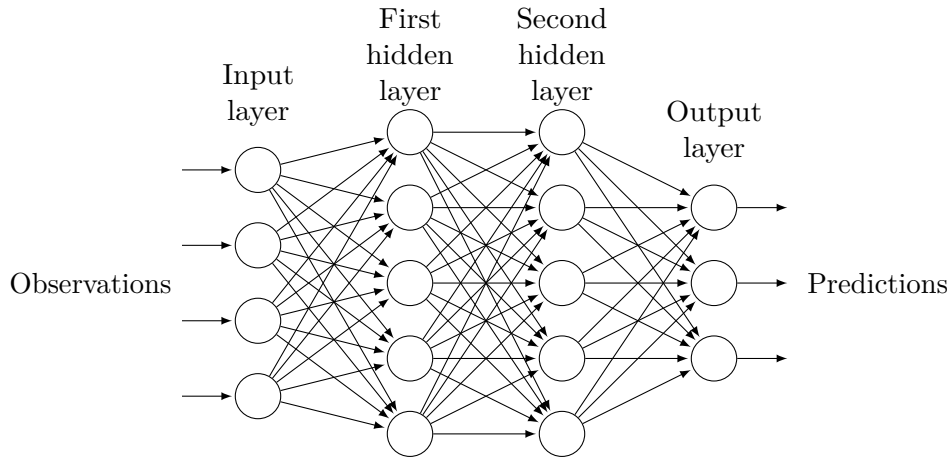

ANNs consist of interconnected nodes called neurons, organized in layers, that process inputs through weighted connections to produce outputs. The strength of these connections, called weights, is adjusted during training to minimize the error in predicting the desired output. This learning capability enables ANNs to capture and represent high-dimensional, nonlinear relationships inherent in the state and action spaces. Figure \ref{fig:neural_network} shows a schematic representation of a general feedforward neural network with two hidden layers. The input layer receives the observations, and the output layer provides the predictions based on the learned features.

There are diverse types of ANNs, each tailored for specific tasks and problem domains. Among the most common are feedforward neural networks (FNNs). As illustrated in Figure~\ref{fig:neural_network}, FNNs typically consist of multiple layers, including input, hidden, and output layers \citep{rosenblatt1958perceptron}. The data flows in one direction, from input to output, making these networks well-suited for tasks like regression and classification.

Convolutional neural networks (CNNs) are a type of FNN specifically designed to process data with a grid-like topology, such as images. CNNs employ a mathematical operation called convolution, which allows them to capture spatial hierarchies and patterns in the data. By using this technique, CNNs have significantly advanced the field of computer vision, outperforming other models in tasks such as image and video recognition, image classification, and medical image analysis \citep{ciregan2012imageclassification}.

Recurrent neural networks (RNNs) form another common category of ANNs and have a bi-directional data flow. Long short-term memory (LSTM) networks are a type of RNN and possess the unique feature of maintaining a memory of previous inputs in their internal state, which is used to influence the current and future outputs \citep{hochreiter1997lstm}. LSTMs are designed to overcome the limitations of traditional RNNs, such as the vanishing gradient problem, by incorporating memory cells that can maintain information in memory for long periods of time. This makes LSTMs exceptionally good at tasks involving sequences with long-range temporal dependencies, like sequence prediction that requires understanding context over extended time frames. Gated Recurrent Units (GRUs) are another variant of RNNs, introduced to address some of the same issues as LSTMs while being computationally more efficient \citep{cho2014learning}. GRUs combine the input and forget gates into a single update gate and merge the cell state and hidden state, making them simpler than LSTMs. Despite their simplicity, GRUs have shown comparable performance to LSTMs in many sequence modeling tasks \citep{chung2014empirical}.

Transformers are a type of neural network architecture that extensively utilizes attention mechanisms, particularly self-attention, to process sequential data. Attention mechanisms are components of neural networks that dynamically focus on the most relevant parts of the input data when making predictions. Introduced by \cite{vaswani2017attention}, transformers consist of an encoder-decoder structure, with both the encoder and decoder composed of multiple layers of self-attention and FNN. Positional encoding is added to input embeddings to retain positional information since transformers do not inherently capture sequence order. Transformers offer significant advantages over traditional RNNs, including parallel processing and effective handling of long-range dependencies. Initially designed for natural language processing tasks such as language translation and text generation, transformers have since been adapted for use in various other domains, including computer vision \citep{dosovitskiy2021} and RL \citep{chen2022MAPF-transformer}.

Graph neural networks (GNNs) are a distinct class of ANNs tailored for data structured as graphs. GNNs excel in capturing dependencies and relationships in graph-structured data, making them powerful for tasks such as node classification, link prediction, and graph classification. The foundational work by \citet{scarselli2008graph} introduced the GNN model, which leverages message passing between nodes to aggregate information from a node's neighbors, effectively learning representations that capture the graph's structure. This approach has been further refined, leading to models like the graph convolutional network proposed by \citet{kipf2017semi}, which is widely used for semi-supervised learning on graph-structured data. Comprehensive reviews, such as the one by \citet{zhou2020graph}, summarize the advancements and applications of GNNs across various domains, including social network analysis, recommendation systems, and bioinformatics.

\subsection{RL Algorithms}

RL algorithms can be broadly categorized based on the components they use, such as the policy, value function, and model. Figure \ref{fig:rl-taxonomy} presents a Venn diagram of these categories and the corresponding approaches in RL:

\begin{figure}
    \centering
    \begin{tikzpicture}
      \begin{scope}[blend group = soft light]
        \fill[redMM!30!white]   ( 90:2) circle (3);
        \fill[greenMM!30!white] (210:2) circle (3);
        \fill[blueMM!30!white]  (330:2) circle (3);
      \end{scope}
      \node at ( 90:3)    {\textit{Policy}};
      \node at ( 210:3.5)   {\textit{Value function}};
      \node at ( 330:3.5)   {\textit{Model}};
      \node at (250:2.5)  {Value-based};
      \node at (270:0.5) {Model-based};
      \node at ( 50:2.5)  {Policy-based};
      \node at (165:1) {Actor-critic};
      \node at (150:2)  {Model-free};
\end{tikzpicture}
    \caption{Venn diagram of RL algoritms based on \cite{silver2015introRL}.}
    \label{fig:rl-taxonomy}
\end{figure}

Model-based approaches, as visualized in the overlapping segment of the \enquote{Model} domain within the Venn diagram in Figure \ref{fig:rl-taxonomy}, empower agents with a model of the environment. In this context model means a function which predicts the state transitions and rewards and gives the agents a possible foresight during the learning process. Popular examples of a model-based approach are AlphaZero \citep{silver2018alphazero} and DreamerV3 \citep{hafner2024dreamerv3}. The gained foresight of the agents can be distilled into policies, enhancing the agent's sample efficiency dramatically when contrasted with model-free counterparts. One major issue with model-based RL is the lack of a precise, real-world environmental model for the agent to access. To address this, an agent must develop its model through direct interactions with the environment, which is fraught with difficulties. A key problem here is that inaccuracies or biases in the learned model might lead the agent to make decisions that seem optimal in the context of the model but are far from ideal in the actual environment. Building a reliable model is inherently challenging and resource-intensive, and despite significant investment in time and computational power, success is not guaranteed.

Model-free methods, which avoid using a model of the environment, are often chosen for their straightforwardness in setup and adjustment. They may not be as efficient in learning from limited data as model-based strategies, but their simplicity makes them more user-friendly. These algorithms are particularly prominent and well-refined, thanks to their proven success in various practical applications \citep{mnih2015human-levelDRL}.

Another distinguishing feature between the algorithms in RL is on-policy and off-policy learning, which applies across both policy-based and value-based algorithms. On-policy algorithms learn about and improve the same policy that is used to make decisions and interact with the environment. This alignment means that the learning directly reflects the decisions made by the policy. Off-policy algorithms learn a potentially different policy from the one used to gather data. This approach allows these algorithms to benefit from a broader range of experiences, including those from past policies or exploratory actions, enhancing learning efficiency and effectiveness.

Value-based algorithms, focus on learning the value function, which estimates how good it is for the agent to be in a given state, or how good it is to perform a certain action in a state. The main idea is to maximize the value function, which in turn maximizes the long-term reward. Value-based algorithms are typically easier to understand and implement for problems with discrete and finite action spaces. Typical algorithms are based on deep Q-networks (DQN), which were introduced by \cite{mnih2013dqn} and improved with various techniques \citep{andrychowicz2017hindsight,hessel2018rainbow}. DQN is an example for off-policy learning. Another example of a value-based algorithm is state-action-reward-state-action (SARSA), which stands as a counterpoint to DQN by being an on-policy algorithm \citep{zhao2016sarsa}. In SARSA, the agent learns from the actions it actually takes, as opposed to using data from a different policy. This means the policy used to select actions is the same policy that is improved based on the learning process. This direct approach to learning from the same policy that is being evaluated helps ensure that the policy and value function are closely aligned, potentially offering more stable learning in certain environments \citep{zhao2016sarsa}.

Policy-based algorithms in RL focus directly on optimizing the policy function, which defines the agent's actions in various states. These algorithms are particularly effective in environments with high-dimensional action spaces or where actions are continuous, as they can optimize policies that handle complex, probabilistic action spaces. This adaptability makes them ideal for scenarios where actions are not merely discrete choices but involve degrees of variability and decision-making under uncertainty. Typical examples of on-policy policy-based algorithms include REINFORCE \citep{williams1992simple} and policy gradients (PG) as described by \cite{sutton1999policy}. These algorithms update the policy by using gradients to maximize the expected reward, relying solely on the data generated by the policy itself, ensuring that the learning is directly aligned with the policy's performance.

For off-policy policy-based learning, an example is behavioral cloning (BC), which learns policies directly from a dataset of recorded actions without requiring interaction with the environment during training. BC is particularly useful in scenarios where collecting real-time data is expensive or dangerous. BC from observation, detailed by \cite{torabi2018behavioral}, focuses on learning policies from state observations only, eliminating the need for action information \citep{torabi2018behavioral}. Exponentially weighted imitation learning, introduced by \cite{wang2018exponentially}, addresses the challenge of learning from batched historical trajectories where the learner lacks access to a simulator or environment oracle. The method employs a monotonic advantage reweighted imitation learning strategy, making it applicable to problems with complex nonlinear function approximation and hybrid (discrete and continuous) action spaces. It does not rely on the knowledge of the behavior policy, allowing it to learn from data generated by unknown policies \citep{wang2018exponentially}.

Both policy-based and value-based algorithms can overlap with model-based approaches, but they do not necessarily have to. They can be applied in a model-free context, where the agent learns purely from interaction with the environment without any understanding of the environmental dynamics. The overlap with the model circle would occur when these algorithms are used in conjunction with a model of the environment. For instance, a policy-based algorithm might use a model to simulate future states and make better-informed decisions about the policy. Similarly, value-based algorithms can use a model to predict future rewards and adjust the value function accordingly. However, this integration is not always necessary. Policy-based algorithms can learn effective strategies without the need for a model by exploring and exploiting the environment through interaction. Value-based algorithms can accurately estimate the value function using observed rewards and transitions, again without needing to predict or understand the underlying environment model. 

The previously mentioned value-based and policy-based algorithms as well as the following actor-critic algorithms are assigned to the model-free methods for the sake of clarity, as they usually function without a model.

Actor-critic algorithms in RL represent a further development and synthesis of policy-based and value-based approaches, aiming to leverage the strengths of both. These algorithms maintain both a policy (actor) that specifies how to act and a value function (critic) that evaluates how good the chosen actions are. This dual approach allows for more robust updates during training, as the actor adjusts its policy based on feedback from the critic. In actor-critic algorithms, the policy is updated in a direction suggested by the critic's evaluations, which are based on the temporal-difference error, a measure of the difference between the predicted rewards and the rewards actually received. This setup helps in reducing the high variance of policy-based algorithms while retaining the ability to deal with high-dimensional action spaces and continuous action domains. Typical actor-critic algorithms are:

\begin{itemize}
    \item Asynchronous advantage actor-critic (A3C) utilizes multiple agents (actors) running in parallel on multiple instances of the environment. This allows each actor to learn independently from its own sequence of experiences without waiting for other actors, thus reducing the correlation in the update data. A3C updates a global policy and value network asynchronously, speeding up learning and improving the robustness of the policy. This method combines the advantages of on-policy gradient descent with the efficiency of asynchronous updates. \citep{mniha2016a3c}
    \item Advantage actor-critic (A2C) is a synchronous variant of A3C that simplifies implementation while maintaining performance. In A2C, multiple agents interact with their environments in parallel and share their experience with a centralized policy and value function, which updates synchronously. This approach retains the benefits of parallelism without the complexity of asynchronous updates. \citep{mniha2016a3c}
    \item Deep deterministic policy gradient (DDPG) is an off-policy algorithm that combines the deterministic policy gradient algorithm with Q-learning, making it effective for continuous action spaces. DDPG uses a separate target network to stabilize training and employs a replay buffer to store and sample experiences, which helps in decorrelating the training data. \citep{lillicrap2019ddpg}
    \item Twin delayed deep deterministic policy gradient (TD3) is an off-policy algorithm designed to address the function approximation errors observed in DDPG by using three critical tricks: clipped double Q-learning, delayed policy updates, and target policy smoothing. These improvements enhance the stability and performance of the actor-critic framework, especially in environments with continuous action spaces. \citep{fujimoto2018td3}
    \item Trust Region Policy Optimization (TRPO) is designed to ensure large updates in policy do not lead to performance collapse. TRPO achieves this by enforcing a constraint on the size of policy updates using a trust region, which ensures each update improves the policy performance within a certain bound. This leads to more stable and reliable learning. \citep{schulman2015trpo}
    \item Proximal policy optimization (PPO) is designed for environments with either continuous or discrete action spaces and aims to simplify and improve the stability and reliability of policy gradient approaches. PPO builds upon TRPO by using a clipped objective function to limit the updates, thus avoiding large policy updates that could destabilize training. Additionally, PPO employs Kullback-Leibler (KL) divergence to measure the difference between the old and new policy. This divergence acts as a regularization mechanism to prevent the policy from deviating too drastically after each update. To further stabilize learning, PPO commonly also incorporates generalized advantage estimation (GAE), a method to compute more accurate and stable advantage estimates by balancing the trade-off between bias and variance \citep{schulman2018GAE}. GAE improves the reliability of the critic's feedback, making the updates more robust. These improvements make PPO more efficient in practice compared to TRPO. \citep{schulman2017ppo}
    \item Asynchronous proximal policy optimization (APPO) adapts the synchronous nature of PPO into an asynchronous framework. APPO leverages parallelism similar to A3C, allowing for more frequent and diverse updates, which can speed up learning and enhance performance in complex environments. \citep{schulman2017ppo}    
    \item Soft actor-critic (SAC) is an off-policy algorithm that optimizes a stochastic policy, making it effective for tasks with continuous action spaces. SAC introduces maximum entropy RL to the actor-critic framework, which encourages exploratory behavior during learning, enhancing the robustness and sample efficiency of the policy learned. \citep{haarnoja2018SAC}
    \item Importance weighted actor-learner architectures (IMPALA) is designed to handle very large-scale distributed training scenarios. IMPALA efficiently deals with the off-policy learning problem by using a V-trace off-policy correction method, allowing IMPALA to scale with increased numbers of parallel actors without a significant drop in learning performance. \citep{espeholt2018impala}
\end{itemize}

The distinction between value-based, policy-based and action-critic methods is not always clear-cut. For example, PPO can certainly be understood as a policy-based algorithm that uses the actor-critic framework. Table \ref{tab:rl_algorithms} therefore provides a brief overview of how the listed model-free algorithms are assigned in this thesis. This is not an exhaustive list of all algorithms in the relevant area, but rather a selection.

\begin{table}[h]
\centering
\caption{Classification of model-free RL algorithms by learning approach and type.}
\label{tab:rl_algorithms}
\begin{tabular}{lll}
    \toprule
    \textbf{Algorithm type} & \textbf{On-policy} & \textbf{Off-policy} \\ 
    \hline
    \textbf{Policy-based} & REINFORCE, PG & BC \\
    \textbf{Value-based} & SARSA & DQN \\
    \textbf{Actor-critic} & A3C, A2C, TRPO, PPO, APPO & SAC, DDPG, TD3, IMPALA \\
    \bottomrule
\end{tabular}
\end{table}

\subsection{Multi-Agent Reinforcement Learning}
Multi-agent reinforcement learning (MARL) extends the foundational principles of single-agent RL to complex environments involving multiple decision-making entities. MARL addresses scenarios where multiple agents interact within a shared space, each pursuing potentially divergent objectives, and where their actions can significantly affect the outcomes for other agents. Usually, not all agents have access to all information, which is why MARL often is represented as a POMDP and only observations are available to the agents instead of the complete states. Figure \ref{fig:marl-interaction} illustrates the interaction model for MARL based on \citet[p. 6]{albrecht2024marl} and shows several agents (Agent 1, Agent 2, ..., Agent $n$) interacting within a shared environment. Each agent \( i \), where \( i \) ranges from 1 to \( n \), represents a decision-making entity in the system. The figure includes \textit{Agent 1} and \textit{Agent 2} explicitly and uses \textit{Agent n} to denote the last agent in a sequence, illustrating the setup for an arbitrary number of agents. Each agent takes an action based on its local observation, contributing to a joint action \( A_{\text{joint}} \). This joint action influences the state of the shared environment, which in turn generates new observations \( O_i \) and rewards \( R_i \) for each agent \( i \) according to the state changes produced by the joint action. This cycle captures the essence of MARL, where agents must cooperate or compete under partial observability.

\begin{figure}[h]
\centering
\begin{tikzpicture}[node distance=2cm, auto,>=latex']
    \node [rectangle, draw, text centered, minimum height=2em] (agent1) {Agent 1};
    \node [rectangle, draw, below of=agent1, text centered, minimum height=2em] (agent2) {Agent 2};
    \node [rectangle, draw, below of=agent2, text centered, minimum height=2em] (agentn) {Agent $n$};
    \node [rectangle, draw, below of=agentn, text centered, minimum height=2em] (environment) {Environment};
    \node [draw, circle] (joint_action) at (3,-4) {};

    \draw (joint_action.south east) -- (joint_action.north west);
    \draw (joint_action.south west) -- (joint_action.north east);

    \draw (agent1) -- (3,0) node[midway, below] {Action \( A_1 \)};
    \draw (agent2) -- (3,-2) node[midway, below] {Action \( A_2 \)};
    \draw [->] (agentn) -- (joint_action.west) node[midway, below] {Action \( A_i \)};
    \draw [->] (3,0) -- (joint_action.north) ;
    \draw (joint_action.south) -- (3, -6) node[midway, right, align=left] {Joint action $A$};
    \draw [->] (3,-6) -- (environment.east);

    \draw (environment.west) -- (-6,-6);
    \draw (-6,-6) -- (-6,0) node[near end, above right, align=left] {Observation $O_1$\\Reward $R_1$};
    \draw (-5,-6) -- (-5,-2) node[near end, right, align=left] {Observation $O_2$\\Reward $R_2$};
    \draw (-2,-6) -- (-2,-4) node[midway, align=right] {Observation $O_i$\\Reward $R_i$};
    \draw [->] (-6,0) -- (agent1.west);
    \draw [->] (-5,-2) -- (agent2.west);
    \draw [->] (-2,-4) -- (agentn.west);

    \draw [loosely dotted](0, -2.5) -- (0, -3.5);

\end{tikzpicture}
\caption{Interaction model for MARL based on \citet[p. 6]{albrecht2024marl}.}
\label{fig:marl-interaction}
\end{figure}

The conceptual shift to MARL is motivated by the considered real-world problems of the coordination of AGVs and AMR in industrial settings and to reduce the action space for the control of various agents. In these environments, agents must learn to make decisions that not only optimize their individual performance but also mitigate conflicts and facilitate cooperation with other agents. This is critical in the considered applications of deadlock-capable transport system, where multiple robots operate in confined spaces and must navigate without interference.

Stochastic games, also referred to as Markov games, generalize the concept of MDPs to incorporate multiple interacting agents within a single environment \citep{shapley1953stochastic}. These games, derived from the field of game theory, model the strategic interactions between agents in a dynamic setting. In the context of this thesis, the \enquote{game} from game theory is synonymous with the environment where the agents operate.

As highlighted by \citet[p. 59]{albrecht2024marl}, there is a notable alignment of terms between RL and game theory, indicating their shared foundations and interconnections. Stochastic games serve as a mathematical framework in MARL, considering not only the actions of individual agents but also how these actions influence the state of the environment over time. A stochastic game is formally defined by several key components:

\begin{itemize}
    \item A finite set of agents \( I = \{1, \ldots, n\} \).
    \item State space ($S$): a finite or infinite set of states of the environment.
    \item For each agent \( i \in I \):
    \begin{itemize}
        \item Action space ($A_i$): a finite or infinite set of actions.
        \item Reward function ($R_i$): \( S \times A \times S \rightarrow \mathbb{R} \), where \( A = A_1 \times \ldots \times A_n \) represents the joint action space.
    \end{itemize}
    \item State transition probability function ($P$): \( S \times A \times S \rightarrow [0, 1] \), satisfying the condition:
    \begin{equation}
        \sum_{s' \in S} P(s, a, s') = 1 \quad \forall s \in S, a \in A
    \end{equation}
    \item Initial state distribution ($\mu$): \( S \rightarrow [0, 1] \).
\end{itemize}

The environment begins in an initial state \( s_0 \) sampled from \( \mu \). At each timestep \( t \), every agent \( i \) observes the current state \( s_t \), selects an action \( a_{t_i} \) based on its policy, resulting in a joint action \( a_t = (a_{t_1}, \ldots, a_{t_n}) \). The system transitions to a new state \( s_{t+1} \) with probability \( P(s_t, a_t, s_{t+1}) \), and each agent \( i \) receives a reward \( r_{t_i} = R_i(s_t, a_t, s_{t+1}) \). These steps are repeated until the game reaches a terminal state or a predefined number of timesteps. Stochastic games have also the Markov property, asserting that the probability of transitioning to the next state and receiving a reward depends only on the current state and joint action, independent of past states and actions.

Stochastic games can be classified based on the relationships between the agents' rewards, which significantly influence the agents' strategies and the overall dynamics of the system. Three primary classifications are commonly discussed:

In cooperative games \cite[p. 16]{foerster2018DeepMARL}, also referred to as common-reward games \cite[p. 45]{albrecht2024marl}, all agents share the same reward, termed a team reward. This setup aligns the objectives of all agents towards a common goal, which simplifies strategic complexity but necessitates effective coordination and communication among agents. The reward function in such games is uniformly defined across all agents, expressed as:
\begin{equation}
R_i(s, a) = R(s, a), \quad \forall i \in I
\end{equation}
where \( R_i \) denotes the reward function for agent \( i \), \( s \) represents the state of the environment, \( a \) is the joint action taken by all agents, and \( I \) is the set of all agents. The uniformity of the rewards in cooperative games promotes collaborative strategies and is particularly relevant in scenarios where group performance is the critical success factor.

In zero-sum games every scenario is configured as a strictly competitive environment where the combined rewards for all agents sum to zero for any given state and joint action:
\begin{equation}
\sum_{i \in I} R_i(a) = 0, \quad \forall a \in A
\end{equation}
In zero-sum games, the gain of one agent is exactly balanced by the loss of another, fostering a highly competitive environment. Strategies in zero-sum games often require cunning tactics such as bluffing or exploiting the opponent's weaknesses, crucial for gaining an advantage under the zero-sum rule. This framework necessitates that agents develop sophisticated methods to predict and counteract the actions of their adversaries, making it suitable for scenarios where direct competition is integral to the interaction dynamics.

General-sum games represent a more complex and realistic model of multi-agent interaction, where the sum of the agents' rewards does not necessarily adhere to any fixed sum condition. In general-sum games, each agent's reward can be influenced by the collective actions of all agents, but unlike zero-sum games, these do not require that one agent's gain be exactly offset by another's loss. The outcomes of the actions can be beneficial for all, harmful for all, or beneficial for some and harmful for others, depending on the actions taken. This flexibility allows for a wide range of possible interactions, including both competitive and cooperative elements.

General-sum games are particularly relevant to scenarios where agents may have overlapping but not identical interests. For example, consider a setting in a large warehouse where multiple AMRs are deployed to manage inventory and fulfill orders. While each AMR operates independently to complete its specific tasks, they all benefit from the overall efficiency of the warehouse operations. Situations may arise where AMRs need to negotiate right-of-way or share charging stations, requiring strategies that involve cooperation to avoid collisions and minimize downtime, while also competing to complete their individual tasks promptly. This dynamic illustrates the nuanced nature of general-sum games where the actions of each agent can have varying impacts on themselves and others, reflecting the complexity of real-world interactions among autonomous systems.

In general-sum games, the concept of Nash equilibrium plays a central role in analyzing the stability of the policies adopted by the agents \citep{nash1950equilibrium}. A Nash equilibrium occurs when no agent can improve their payoff by unilaterally changing their policy, given the policies of all other agents. This state represents a stable set of policies for all agents, where each agent's policy is the best response to the policies of the others. Following the definition by \citet[p. 68]{albrecht2024marl} a Nash equilibrium is formally defined when a joint policy $\pi = (\pi_1, \ldots, \pi_n)$ with $n$ agents holds:
\begin{equation}
\forall i, \pi'_i : U_i(\pi'_i, \pi_{-i}) \leq U_i(\pi) \quad 
\end{equation}

where \( U_i(\pi'_i, \pi_{-i}) \) represents the utility that agent \( i \) would receive if they unilaterally deviated to policy \( \pi'_i \), while all other agents maintained their current policies \( \pi_{-i} \). \( U_i(\pi) \) denotes the utility for agent \( i \) when all agents, including \( i \), adhere to their current policies \( \pi \).

This equilibrium concept captures the notion of strategic interdependence among agents, where the best policy for one agent is contingent upon the policies of others. In such an equilibrium, each agent assumes that other agents will remain with their current policies, and as such, there is no incentive for any agent to deviate from their current policy, because doing so does not lead to a better outcome. The Nash equilibrium is particularly insightful in the analysis of MAS where agents interact within a shared environment. It is in these settings that the equilibrium concept becomes instrumental in designing autonomous agents that are capable of behaving optimally in a distributed and often decentralized decision-making framework.

For instance, in a system deploying AMRs for logistical operations, the Nash equilibrium can inform the development of routing algorithms that minimize the potential for deadlocks. Each AMR operates under the assumption that other robots will continue following their set routes, and therefore, each AMR's routing policy takes into account these predicted behaviors to optimize its own route. If a Nash equilibrium is reached, no AMR would benefit from deviating from its route, as doing so would either lead to a less efficient outcome or increase the risk of creating a deadlock.

The application of Nash equilibrium in MAS underscores the balance between individual agent optimization and system-wide efficiency. It ensures that each agent, while operating under its own policy, implicitly coordinates with other agents in the environment to achieve an optimal flow of operations. Computational approaches to identify Nash equilibria in these systems often employ advanced algorithms that consider both the current state of the environment and the probabilistic outcomes of agents' interactions. The challenge lies in the computational complexity of finding equilibria in large-scale and dynamic environments, which remains an active area of research in the field of MARL.

MARL algorithms can be categorized based on the availability and usage of information during the training and execution phases. This distinction provides insight into the algorithms' capabilities and limitations in various application scenarios. This thesis follows the three primary categorizations based on \citep[p. 220 - 222]{albrecht2024marl}:
\begin{itemize}
    \item Centralized training and execution (CTE)
    \item Decentralized training and execution (DTE)
    \item Centralized training with decentralized execution (CTDE)
\end{itemize}

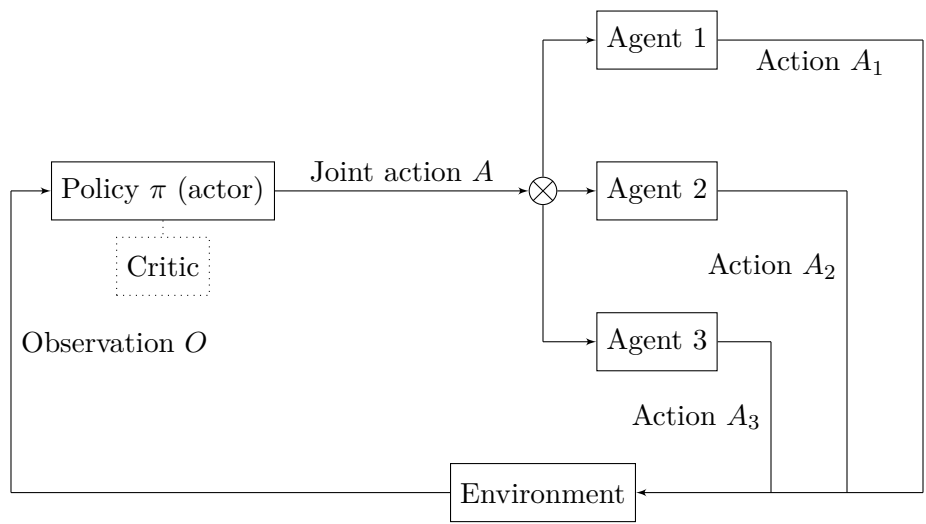
\begin{figure}[h]
\centering
\begin{tikzpicture}[node distance=2cm, auto,>=latex']
    \node [rectangle, draw, text centered, minimum height=2em] (policy_center) at (-5,-2){Policy $\pi$ (actor)};
    \node [rectangle, draw, text centered, minimum height=2em, dotted] (critic) at (-5,-3) {Critic};
    \node [rectangle, draw, text centered, minimum height=2em] (agent1) at (1.5,0){Agent 1};
    \node [rectangle, draw, below of=agent1, text centered, minimum height=2em] (agent2) {Agent 2};
    \node [rectangle, draw, below of=agent2, text centered, minimum height=2em] (agent3) {Agent 3};
    \node [rectangle, draw, below of=agent3, text centered, minimum height=2em] at (0,-4) (environment) {Environment};
    \node [draw, circle] (joint_action) at (0,-2) {};

    \draw (joint_action.south east) -- (joint_action.north west);
    \draw (joint_action.south west) -- (joint_action.north east);

    \draw (agent1.east) -- (5,0) node[midway, below] {Action \( A_1 \)};
    \draw (5,0) -- (5,-6);
    \draw (agent2.east) -- (4,-2);
    \draw (4,-2) -- (4,-6) node[near start, left] {Action \( A_2 \)};
    \draw (agent3.east) -- (3,-4);
    \draw (3,-4) -- (3,-6) node[midway, left] {Action \( A_3 \)};
    \draw (joint_action.north) -- (0,0);
    \draw [->] (0,0) -- (agent1.west);
    \draw [->] (joint_action.east) -- (agent2.west);
    \draw (joint_action.south) -- (0,-4);
    \draw [->] (0,-4) -- (agent3.west);
    \draw [->] (5,-6) -- (environment.east);

    \draw (environment.west) -- (-7,-6);
    \draw (-7,-6) -- (-7,-2) node[midway, right, align=left] {Observation $O$};
    \draw [->] (-7,-2) -- (policy_center.west);
    \draw [->] (policy_center.east) -- (joint_action.west) node[midway, above, align=left] {Joint action $A$};

    \draw [dotted](policy_center.south) -- (critic.north);

\end{tikzpicture}
\caption{CTE for an actor-critic approach of three agents.}
\label{fig:centralized-training}
\end{figure}

CTE mean that both the learning of agent policies and their subsequent execution rely on centrally shared information. This approach includes sharing agents' observation histories, value functions, and the policies themselves. Central learning is a common example of this approach, where a central unit is responsible for learning and controlling the policies of all agents. Figure \ref{fig:centralized-training} illustrates this concept using an actor-critic framework. In this setup, a single policy denoted as $\pi$ (the actor) is responsible for generating joint actions, while a critic evaluates the potential outcomes. The central unit, comprising the actor and critic, receives global observations $O$ and determines joint actions $A$ for all agents. This centralized process ensures effective coordination and adaptability in complex environments.

Each agent, Agent 1 through Agent 3, performs actions ($A_1$, $A_2$, $A_3$) within the environment based on the joint action decision. The dashed lines indicate the connection between actor and critic. In off-policy or value-based algorithms, the value functions and replay buffers can be shared in place of the critic. Despite the potential for advanced coordination, CTE faces scalability challenges and may not be feasible in real-time systems like AMRs due to the exponential growth in action spaces and communication constraints \citep{muller2022singleRL}.

DTE maintain the autonomy of agents throughout both the learning and operational phases. Agents rely solely on local information and are trained independently of one another. This modality aligns well with scenarios where agents cannot share information or are physically distributed. A notable example of DTE is independent learning \citep{tan1993independentlearning}. Here, agents train their policies using single-agent RL techniques, treating other agents as part of the environment's dynamics. While scalable and applicable in distributed settings, independent learning may suffer from non-stationarity and an inability to leverage information about other agents, potentially leading to unstable learning dynamics and suboptimal convergence.

\begin{figure}[h]
\centering
\begin{tikzpicture}[node distance=2cm, auto,>=latex']
    \node [rectangle, draw, align=left, minimum height=4em, minimum width=9em] (agent1) at (0,0){};
    \node [anchor=west](agent1_name) at (agent1.west){Agent 1};
    \node [anchor=north east, rectangle, draw, dotted, outer sep=3pt](agent1_actor) at (agent1.north east){Policy $\pi_1$};
    \node [anchor=south east, rectangle, draw, dotted, outer sep=3pt](agent1_critic) at (agent1.south east){Critic};
    
    \node [rectangle, draw, below of=agent1, text centered, minimum height=4em, minimum width=9em] (agent2) {};
    \node [anchor=west](agent2_name) at (agent2.west){Agent 2};
    \node [anchor=north east, rectangle, draw, dotted, outer sep=3pt](agent2_actor) at (agent2.north east){Policy $\pi_2$};
    \node [anchor=south east, rectangle, draw, dotted, outer sep=3pt](agent2_critic) at (agent2.south east){Critic};
    
    \node [rectangle, draw, below of=agent2, text centered,minimum height=4em, minimum width=9em] (agent3) {};
    \node [anchor=west](agent3_name) at (agent3.west){Agent 3};
    \node [anchor=north east, rectangle, draw, dotted, outer sep=3pt](agent3_actor) at (agent3.north east){Policy $\pi_3$};
    \node [anchor=south east, rectangle, draw, dotted, outer sep=3pt](agent3_critic) at (agent3.south east){Critic};
    
    \node [rectangle, draw, below of=agent3, text centered, minimum height=2em] at (0,-4) (environment) {Environment};

    \draw (agent1.east) -- (6,0) node[midway, below] {Action \( A_1 \)};
    \draw (6,0) -- (6,-6);
    \draw (agent2.east) -- (5,-2) node[midway, below] {Action \( A_2 \)};
    \draw (5,-2) -- (5,-6);
    \draw (agent3.east) -- (4,-4) node[midway, below] {Action \( A_3 \)};
    \draw (4,-4) -- (4,-6);

    \draw [->] (6,-6) -- (environment.east);

    \draw (environment.west) -- (-7,-6);
    \draw (-7,-6) -- (-7,0);
    \draw (-6,-6) -- (-6,-2);
    \draw (-5,-6) -- (-5,-4);
    \draw [->] (-7,0) -- (agent1.west) node[midway, below, align=left] {Observation $O_1$};
    \draw [->] (-6,-2) -- (agent2.west) node[midway, below, align=left] {Observation $O_2$};
    \draw [->] (-5,-4) -- (agent3.west) node[midway, below, align=left] {Observation $O_3$};

\end{tikzpicture}
\caption{Independent learning for an actor-critic approach of three agents.}
\label{fig:indpendent-learning}
\end{figure}
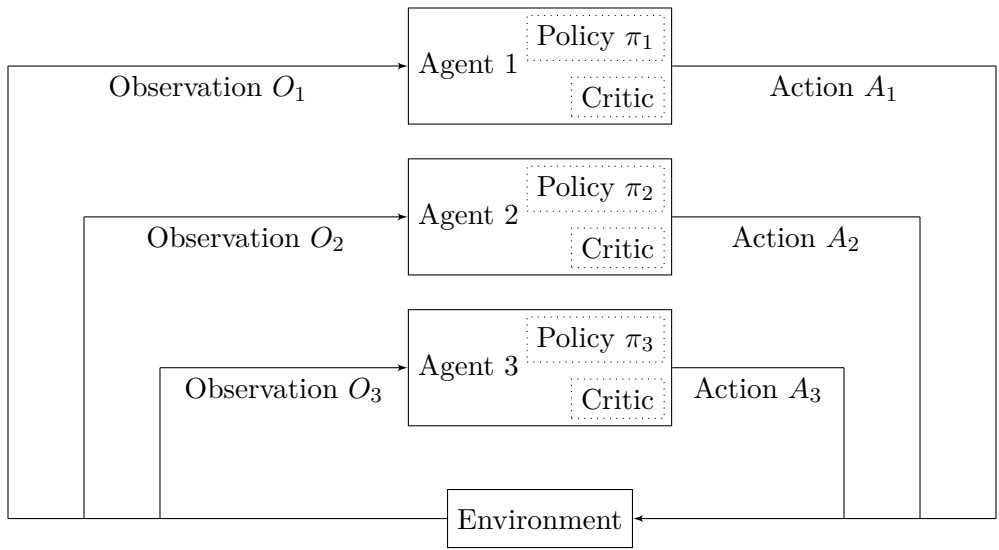

In Figure \ref{fig:indpendent-learning}, each agent (Agent 1, Agent 2, and Agent 3) is depicted with its own policy $\pi_i$ and critic. The agents independently process their unique observations $O_1$, $O_2$, and $O_3$ from the environment and take actions $A_1$, $A_2$, and $A_3$ without any central coordination. The critics provide feedback on the actions taken based on the individual experiences of the agents. Figure \ref{fig:indpendent-learning} illustrates an independent training mode of each agent within the multi-agent environment.

The hybrid approach of CTDE aims to leverage the advantages of centralized training while ensuring that agents can operate independently during execution \citep{lowe2017multi}. This paradigm allows the use of shared information during the training phase to enhance policy learning, while the execution of policies remains local to each agent. CTDE algorithms have gained prominence in MARL due to their efficacy in managing computational complexity while still enabling agents to condition their policies on privileged information during training. For instance, in a multi-agent actor-critic setup, the critic can utilize joint observations for more accurate value estimations, but during execution, only the policy, which is conditioned on individual agent observations, is required \citep{lowe2017multi}.

Parameter sharing in CTDE involves all agents using a single neural network with the same shared parameters, which facilitates efficient learning from a collective experience \citep{foerster2016learning_to_communicate}. This approach is scaled across multiple agents, enhancing both individual and collective decision-making in cooperative scenarios \citep{yang2018meanfield}. An agent-specific indicator can be added to address the limitations of homogeneous policy application across diverse agent roles and environments. This adaptation allows for parameter sharing even in scenarios with heterogeneous action and observation spaces, thereby supporting optimal policy convergence across varied multi-agent configurations \citep{terry2023revisiting_parametersharing}.

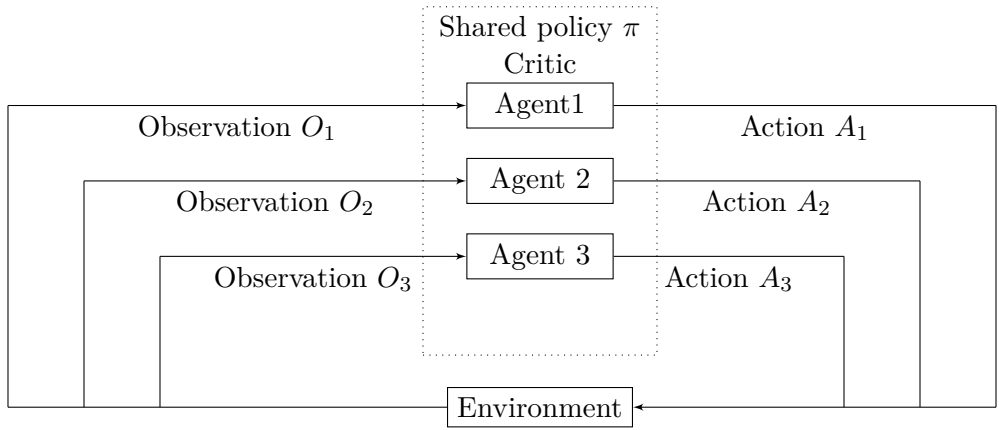
\begin{figure}[h]
\centering
\begin{tikzpicture}[node distance=1cm, auto,>=latex']
    \node [rectangle, draw, minimum width=5em] (agent1) at (0,0){Agent1};
    \node [rectangle, draw, below of=agent1, minimum width=5em] (agent2) {Agent 2};
    \node [rectangle, draw, below of=agent2, minimum width=5em] (agent3) {Agent 3};
    \node [rectangle, draw, dotted, below of=agent1, minimum width=8em, minimum height=12 em] (shared_policy) {};
    \node [anchor=north, align=center](shared_policy_name) at (shared_policy.north){Shared policy $\pi$ \\ Critic};
    \node [rectangle, draw, below of=agent3, text centered] at (0,-3) (environment) {Environment};

    \draw (agent1.east) -- (6,0) node[midway, below] {Action \( A_1 \)};
    \draw (6,0) -- (6,-4);
    \draw (agent2.east) -- (5,-1) node[midway, below] {Action \( A_2 \)};
    \draw (5,-1) -- (5,-4);
    \draw (agent3.east) -- (4,-2) node[midway, below] {Action \( A_3 \)};
    \draw (4,-2) -- (4,-4);

    \draw [->] (6,-4) -- (environment.east);

    \draw (environment.west) -- (-7,-4);
    \draw (-7,-4) -- (-7,0);
    \draw (-6,-4) -- (-6,-1);
    \draw (-5,-4) -- (-5,-2);
    \draw [->] (-7,0) -- (agent1.west) node[midway, below, align=left] {Observation $O_1$};
    \draw [->] (-6,-1) -- (agent2.west) node[midway, below, align=left] {Observation $O_2$};
    \draw [->] (-5,-2) -- (agent3.west) node[midway, below, align=left] {Observation $O_3$};

\end{tikzpicture}
\caption{Parameter sharing in CTDE for an actor-critic approach of three agents.}
\label{fig:parameter-sharing}
\end{figure}

Figure \ref{fig:parameter-sharing} illustrates the CTDE approach with parameter sharing where all agents, Agent 1, Agent 2, and Agent 3, operate under a shared policy $\pi$ and a single critic. This shared setup allows the agents to learn from a collective experience, enhancing their ability to make decisions both as individuals and as a group. Actions $A_1$, $A_2$, and $A_3$ are taken based on the shared policy after receiving individual observations $O_1$, $O_2$, and $O_3$ from the environment. 

Each of the mentioned training and execution modes presents unique advantages and challenges. The choice of a training approach and the particular MARL algorithm depends on the specific requirements of the application domain, such as the need for real-time decision-making, the degree of environmental observability, and the scalability to a large number of agents.

\section{Multi-Agent Pathfinding}\label{sec:background-MAPF}

In logistics and transportation systems using AMRs, the principal operational challenge is often multi-agent pathfinding (MAPF). \cite{li2021lifelongMAPF} describes MAPF as \enquote{[\ldots] the problem of moving a team of agents from their start locations to their goal locations while avoiding collisions.} Similarly, \cite{stern2019MAPF} defines MAPF as \enquote{[\ldots] to plan paths for multiple agents, where the key constraint is that the agents will be able to follow these paths concurrently without colliding with each other.} Implicit in these definitions is the expectation of a deadlock-free solution, which ensures not only collision avoidance but also the continuous movement of all agents within the system. This thesis introduces the following definition to explicitly incorporate the consideration of deadlocks into the study of MAPF:

\begin{definition}[Multi-agent pathfinding]
MAPF is the computational challenge of finding paths for multiple agents where each agent must move from a specified starting location to a designated goal with the constraints that the paths must allow all agents to move concurrently without any collisions or deadlocks.
\end{definition}

In this context, a collision occurs when two agents occupy the same location at the same time, while a deadlock results when agents block each other in a way that makes further movement impossible, thereby indefinitely delaying at least one agent from reaching its goal.

A MAPF problem can be represented in different ways, such as grid-based representations or general graph-based approaches. Figures \ref{fig:MAPF-grid} and \ref{fig:MAPF-graph} illustrate these two representations for the same problem:

\begin{figure}[h]
    \centering
    \begin{minipage}{0.45\textwidth}
        \centering
        \begin{tikzpicture}[scale=0.5]    
            \draw[step=1cm,gray,very thin] (0,0) grid (10,10);
            
            \fill[gray] (4,4) rectangle (5,5);
            \fill[gray] (5,5) rectangle (6,6);
            \fill[gray] (6,6) rectangle (7,7);
            \fill[gray] (7,7) rectangle (8,8);
            \fill[gray] (8,3) rectangle (9,4);
            \fill[gray] (1,1) rectangle (3,3);
            \fill[gray] (6,1) rectangle (7,4);
            
            \draw[thick, redMM, ->] (1.5,8.5) -- (8.5,8.5) -- (8.5,4.5) -- (7.5,4.5) -- (7.5,2.5) -- (8.5,2.5);        
            \draw[thick, blueMM, ->] (0.5,1.5) -- (0.5,3.5) -- (5.5,3.5);        
            \draw[thick, greenMM, ->] (3.5,0.5) -- (3.5,6.5);
            
            \node[rotate=270, fill=redMM, regular polygon, regular polygon sides=3, inner sep=2pt] at (1.5,8.5) {};
            \node[rotate=0, fill=blueMM, regular polygon, regular polygon sides=3, inner sep=2pt] at (0.5,1.5) {};
            \node[rotate=0, fill=greenMM, regular polygon, regular polygon sides=3, inner sep=2pt] at (3.5,0.5) {};
            
        \end{tikzpicture}
        \caption{Grid-based MAPF.}
        \label{fig:MAPF-grid}
    \end{minipage}
    \hfill
    \begin{minipage}{0.45\textwidth}
        \centering
        \begin{tikzpicture}[scale=0.5]
            \begin{scope}[on background layer]                
                \foreach \x in {0,...,9} {
                    \foreach \y in {0,...,9} {
                        \ifthenelse{\(\x=4 \AND \y=4\) \OR 
                                    \(\x=5 \AND \y=5\) \OR 
                                    \(\x=6 \AND \y=6\) \OR 
                                    \(\x=7 \AND \y=7\) \OR 
                                    \(\x=8 \AND \y=3\) \OR 
                                    \(\x=6 \AND \y=3\) \OR 
                                    \(\x=6 \AND \y=2\) \OR 
                                    \(\x=6 \AND \y=1\) \OR 
                                    \(\x=2 \AND \y=2\) \OR 
                                    \(\x=1 \AND \y=1\) \OR 
                                    \(\x=1 \AND \y=2\) \OR 
                                    \(\x=2 \AND \y=1\) \OR 
                                    \(\x=2 \AND \y=2\) 
                                    }{
                        }{
                            \node[draw, circle, inner sep=3pt] (node\x\y) at (\x,\y) {};
                            \ifthenelse{\(\x>0\) \AND \NOT \(
                                                            \(\x=5 \AND \y=4\) \OR 
                                                            \(\x=6 \AND \y=5\) \OR 
                                                            \(\x=7 \AND \y=6\) \OR 
                                                            \(\x=8 \AND \y=7\) \OR 
                                                            \(\x=9 \AND \y=3\) \OR 
                                                            \(\x=7 \AND \y=3\) \OR 
                                                            \(\x=7 \AND \y=2\) \OR 
                                                            \(\x=7 \AND \y=1\) \OR 
                                                            \(\x=2 \AND \y=1\) \OR 
                                                            \(\x=2 \AND \y=2\) \OR 
                                                            \(\x=3 \AND \y=1\) \OR 
                                                            \(\x=3 \AND \y=2\) 
                                                            \)
                                                            }{
                                \pgfmathtruncatemacro{\xminus}{\x-1}
                                \draw (node\xminus\y) -- (node\x\y);
    
                            }{}
                            \ifthenelse{\(\y>0\) \AND \NOT \(
                                                            \(\x=4 \AND \y=5\) \OR 
                                                            \(\x=5 \AND \y=6\) \OR 
                                                            \(\x=6 \AND \y=7\) \OR 
                                                            \(\x=7 \AND \y=8\) \OR 
                                                            \(\x=8 \AND \y=4\) \OR 
                                                            \(\x=6 \AND \y=4\) \OR 
                                                            \(\x=6 \AND \y=3\) \OR 
                                                            \(\x=6 \AND \y=2\) \OR 
                                                            \(\x=1 \AND \y=2\) \OR 
                                                            \(\x=1 \AND \y=3\) \OR 
                                                            \(\x=2 \AND \y=2\) \OR 
                                                            \(\x=2 \AND \y=3\) 
                                                            \)
                                                            }{
                                \pgfmathtruncatemacro{\yminus}{\y-1}
                                \draw (node\x\yminus) -- (node\x\y);
    
                            }{}
                        }
                    }
                }
            \end{scope}

            \begin{scope}[on above layer]  
                \draw[thick, redMM, ->] (node18) -- (node88) -- (node84) -- (node74) -- (node72) --(node82.center);        
                \draw[thick, blueMM, ->] (node01) -- (node03) -- (node53.center);        
                \draw[thick, greenMM, ->] (node30) -- (node36.center);
            \end{scope}
            \node[rotate=270, fill=redMM, regular polygon, regular polygon sides=3, inner sep=2pt] at (node18) {};
            \node[rotate=0, fill=blueMM, regular polygon, regular polygon sides=3, inner sep=2pt] at (node01) {};
            \node[rotate=0, fill=greenMM, regular polygon, regular polygon sides=3, inner sep=2pt] at (node30) {};
            
        \end{tikzpicture}
        \caption{Graph-based MAPF.}
        \label{fig:MAPF-graph}
    \end{minipage}
\end{figure}

Figure \ref{fig:MAPF-grid} demonstrates a grid-based MAPF scenario on a 10x10 grid with obstacles represented as grey boxes. Each agent, represented by a color-coded triangle, navigates to a designated endpoint while avoiding collisions and obstacles. The red agent travels horizontally from the upper left and then descends to the lower right corner. The blue agent starts from the lower left and proceeds to the center, while the green agent moves straight up from the bottom center to the middle. The paths of the blue and green agents intersect, creating a potential risk for collisions or deadlocks if both agents reach the intersection simultaneously without implementing conflict avoidance mechanisms. Figure \ref{fig:MAPF-graph} illustrates a graph-based representation of the same problem. Nodes represent locations and edges represent possible paths. Agents follow paths along the edges while avoiding collisions at nodes or along edges. This representation can accommodate more complex environments where paths are not strictly aligned to a grid. Both representations are essential for solving MAPF problems. The grid-based approach, which is more commonly used due to its simplicity and ease of visualization, simplifies conflict detection in structured environments. The graph-based approach provides flexibility for more complex scenarios where paths do not adhere to a regular grid, enabling the modeling of a wider range of environments and movement patterns. 

The classical MAPF problem can be formally described by a tuple \( \langle G, s, t \rangle \), where \( G = (V, E) \) is an undirected graph representing the environment \citep{stern2019MAPF}. Here, \( s : [1, \ldots, k] \rightarrow V \) maps an agent to a start vertex, and \( t : [1, \ldots, k] \rightarrow V \) maps an agent to a target vertex. Agents perform actions within discrete timesteps, moving to an adjacent vertex or waiting at their current vertex \citep{stern2019MAPF}.

\cite{stern2019MAPF} classify common conflicts in MAPF as demonstrated in Figure \ref{fig:MAPF-conflicts}.

\begin{figure}
    \begin{minipage}[t]{.17\textwidth}
        \centering
        \begin{tikzpicture}[node distance=1.5cm, auto,>=latex']
            \node[draw, circle](A){A};
            \node[draw, circle, right of=A](B){B};
            \node[below of=A]{}; 
            \draw[](A) -- (B){};
            \fill[redMM] ([yshift=5pt] A.north) circle(3pt);
            \fill[blueMM] ([yshift=-5pt] A.south) circle(3pt);
            \draw[->, redMM]([xshift=5pt, yshift=5pt] A.north) -- ([xshift=30pt, yshift=5pt] A.north){};
            \draw[->, blueMM]([xshift=5pt, yshift=-5pt] A.south) -- ([xshift=30pt, yshift=-5pt] A.south){};
        \end{tikzpicture}
        \caption*{(a)}
    \end{minipage}
    \hfill
    \begin{minipage}[t]{.17\textwidth}
        \centering
        \begin{tikzpicture}[node distance=1.5cm, auto,>=latex']
            \node[draw, circle](A){A};
            \node[draw, circle, below of=A](B){B};
            \node[draw, circle](C) at (1,-0.75){C};
            \draw[](A) -- (C){};
            \draw[](B) -- (C){};
            \fill[redMM] ([yshift=10pt] A.north) circle(3pt);
            \fill[blueMM] ([yshift=-10pt] B.south) circle(3pt);
            \draw[->, redMM]([xshift=5pt, yshift=6pt] A.north) -- ([xshift=30pt, yshift=-14pt] A.north){};
            \draw[->, blueMM]([xshift=5pt, yshift=-6pt] B.south) -- ([xshift=30pt, yshift=14pt] B.south){};
        \end{tikzpicture}
        \caption*{(b)}
    \end{minipage}
    \begin{minipage}[t]{.25\textwidth}
        \centering
        \begin{tikzpicture}[node distance=1.5cm, auto,>=latex']
            \node[draw, circle](A){A};
            \node[draw, circle, right of=A](B){B};
            \node[draw, circle, right of=B](C){C};
            \node[below of=A]{}; 
            \draw[](A) -- (B){};
            \draw[](B) -- (C){};
            \fill[redMM] ([yshift=5pt] A.north) circle(3pt);
            \fill[blueMM] ([yshift=-5pt] B.south) circle(3pt);
            \draw[->, redMM]([xshift=5pt, yshift=5pt] A.north) -- ([xshift=30pt, yshift=5pt] A.north){};
            \draw[->, blueMM]([xshift=5pt, yshift=-5pt] B.south) -- ([xshift=30pt, yshift=-5pt] B.south){};
        \end{tikzpicture}
        \caption*{(c)}
    \end{minipage}
    \begin{minipage}[t]{.19\textwidth}
        \centering
        \begin{tikzpicture}[node distance=1.5cm, auto,>=latex']
            \node[draw, circle](A){A};
            \node[draw, circle, right of=A](B){B};
            \node[draw, circle, below of=A](D){D};
            \node[draw, circle, below of=B](C){C};
            \draw[](A) -- (B){};
            \draw[](A) -- (D){};
            \draw[](B) -- (C){};
            \draw[](C) -- (D){};
            \fill[redMM] ([yshift=5pt] A.north) circle(3pt);
            \fill[blueMM] ([xshift=5pt] B.east) circle(3pt);
            \fill[greenMM] ([yshift=-5pt] C.south) circle(3pt);
            \fill[magenta] ([xshift=-5pt] D.west) circle(3pt);
            \draw[->, redMM]([xshift=5pt, yshift=5pt] A.north) -- ([xshift=30pt, yshift=5pt] A.north){};
            \draw[->, blueMM]([xshift=5pt, yshift=-5pt] B.east) -- ([xshift=5pt, yshift=-30pt] B.east){};
            \draw[->, greenMM]([xshift=-5pt, yshift=-5pt] C.south) -- ([xshift=-30pt, yshift=-5pt] C.south){};
            \draw[->, magenta]([xshift=-5pt, yshift=5pt] D.west) -- ([xshift=-5pt, yshift=30pt] D.west){};
        \end{tikzpicture}
        \caption*{(d)}
    \end{minipage}
    \begin{minipage}[t]{.17\textwidth}
        \centering
        \begin{tikzpicture}[node distance=1.5cm, auto,>=latex']
            \node[draw, circle](A){A};
            \node[draw, circle, right of=A](B){B};
            \node[below of=A]{}; 
            \draw[](A) -- (B){};
            \fill[redMM] ([yshift=5pt] A.north) circle(3pt);
            \fill[blueMM] ([yshift=-5pt] B.south) circle(3pt);
            \draw[->, redMM]([xshift=5pt, yshift=5pt] A.north) -- ([xshift=30pt, yshift=5pt] A.north){};
            \draw[->, blueMM]([xshift=-5pt, yshift=-5pt] B.south) -- ([xshift=-30pt, yshift=-5pt] B.south){};
        \end{tikzpicture}        
        \caption*{(e)}
    \end{minipage}
    \caption{Common conflicts in MAPF according to \cite{stern2019MAPF}.}
    \label{fig:MAPF-conflicts}
\end{figure}
                                                    
This thesis follows the descriptions of the common conflict categories for MAPF by \cite{stern2019MAPF} with small deviation regarding the notation. Let $\pi_i(t)$ denote the location of agent $i$ at discrete timestep $t$. The letter of the conflict category refers to Figure \ref{fig:MAPF-conflicts}:
\begin{description}
    \item[Edge conflict (a):] An edge conflict between $\pi_i$ and $\pi_j$ arises when both agents are scheduled to traverse the same edge in the same direction at the same timestep. Formally, an edge conflict exists if for a timestep $t$, $\pi_i(t) = \pi_j(t)$ and $\pi_i(t + 1) = \pi_j(t + 1)$.
    
    \item[Vertex conflict (b):] A vertex conflict occurs when $\pi_i$ and $\pi_j$ plan for the agents to occupy the same vertex simultaneously. Specifically, a vertex conflict exists if for any timestep $t$, $\pi_i(t) = \pi_j(t)$.
    
    \item[Following conflict (c):] This type of conflict happens if $\pi_i$ schedules an agent to occupy a vertex immediately after it was vacated by $\pi_j$. There is a following conflict if there exists a timestep $t$ such that $\pi_i(t + 1) = \pi_j(t)$.
    
    \item[Cycle conflict (d):] A cycle conflict involves a series of agents moving into vertices previously occupied by another agent in the same timestep, creating a rotational movement pattern. Formally, this occurs if for a set of plans $\pi_i, \pi_{i+1}, \dots, \pi_j$ and a timestep $t$, the transitions form a cycle $\pi_i(t + 1) = \pi_{i+1}(t), \dots, \pi_j(t + 1) = \pi_i(t)$. Cycle conflicts fulfill the condition for a deadlock. 
    
    \item[Swapping conflict (e):] A swapping conflict occurs when two agents $\pi_i$ and $\pi_j$ plan to exchange their positions in one timestep. Formally, this exists if at timestep $t$, $\pi_i(t + 1) = \pi_j(t)$ and $\pi_j(t + 1) = \pi_i(t)$. This type of conflict is often referred to as an edge conflict in contemporary MAPF literature. A swapping conflict can lead to a deadlock.
\end{description}

In MAPF, there are two typical objective functions used to evaluate the effectiveness of solutions: makespan and sum of costs. Makespan is defined as the maximum number of timesteps required for all agents $I=\{1, 2, \ldots, k\}$ to reach their goals, formally expressed as $\max_{1 \leq i \leq k} |\pi_i|$. Sum of costs, also referred to as flowtime, accumulates the timesteps each agent takes to reach its goal, calculated as $\sum_{1 \leq i \leq k} |\pi_i|$. These measures assess the total completion time and the collective effort of all agents, respectively. Adjustments may be made based on whether agents' waiting times at their goals are included in the sum of costs.

The development of efficient algorithms to solve MAPF problems has been a key focus within the field and resulted in various improved algorithms in recent years. In this context, prioritized planning represents a fundamental approach in MAPF to manage the combinatorial complexity by decomposing the problem into manageable single-agent pathfinding problems  \citep{cap2015prioritized}. Agents are assigned unique priorities, and each agent plans its path sequentially, considering the paths of higher priority agents to avoid conflicts.
Despite its computational efficiency and relative simplicity, prioritized planning is neither complete nor optimal \cite[p. 101]{stern2019MAPF}. Prioritized planning may fail to find a solution even if one exists (non-completeness) and the solutions prioritized planning finds may not be the global optimum of the given objective functions, such as the sum of costs or makespan. However, the practical effectiveness and adaptability of prioritized planning, as seen in algorithms like windowed hierarchical cooperative A* \citep{Silver.2005}, which plans over a limited horizon, have made it a popular choice among practitioners.

Is it necessary for a solution to guarantee a certain quality, the focus shifts to optimal MAPF solvers that ensure solutions are optimal with respect to specific objective functions. Optimal MAPF algorithms can be categorized into several distinct types \cite[p. 103]{stern2019MAPF}:

\begin{itemize}
    \item \textbf{Enhanced A* variants} expand the traditional A* search algorithm to accommodate multiple agents, optimizing their collective pathfinding effectively.
    \item \textbf{Increasing cost tree search} approaches the MAPF problem by initially estimating the incremental cost each agent contributes and subsequently seeking a feasible path that integrates these costs \citep{sharon2013CTSMAPF}.
    \item \textbf{Conflict-based search (CBS)} handles MAPF by addressing the individual pathfinding problems and integrating constraints progressively to ensure that the solution is collision-free, complete and optimal \citep{sharon2015conflict}.
    \item \textbf{Constraint programming} transforms the MAPF into a constraint satisfaction problem, which is then solved using general-purpose constraint solvers such as mixed integer programming solvers.
\end{itemize}

\section{Evaluation of Existing Solutions} \label{sec:background-existing-solutions}
\subsection{Gaps in Planning Methods}
Despite the potential negative impacts of deadlocks, they are not considered in the general methodologies for planning processes and systems in logistics. In the planning of logistics systems and processes, deadlocks, if considered at all, only play a role late in the operational detailed planning of control strategies. For a deadlock prevention approach, this is often too late. The classical methodologies for planning in logistics, such as:

\begin{itemize}
    \item Process chain-based planning according to \cite{kuhn1995prozessketten},
    \item Guidelines for commissioning systems VDI 3590 sheet 2 \citep{vdi2002kommissioniersysteme},
    \item Phases of planning and implementation of logistics systems according to \citet[p. 70 ff.]{Gudehus.2010},
    \item Four-step approach to the planning of transport systems according to \citet[p. 291 ff.]{vanBonn2013basisdaten},
\end{itemize}

do not consider deadlocks neither in resource sizing nor in infrastructure planning. For example, \citet[p. 869 ff.]{Gudehus.2010} explains in detail how goals such as transport optimization and minimization of space can be achieved in the design of logistics halls through gate arrangement and other measures. Yet, neither in the design of the logistics hall nor in the planning and sizing of the transport system does \citet[p. 817 ff. and p. 869 ff.]{Gudehus.2010} address potential deadlock situations that could occur in what is supposedly an optimal layout of logistics halls and transport systems. In the process chain-based planning by \cite{kuhn1995prozessketten}, the VDI guideline 3590 sheet 2 for the system finding of commissioning systems \citep{vdi2002kommissioniersysteme}, and the approach to the planning of transport systems after \citet[p. 291 ff.]{vanBonn2013basisdaten}, the issue of deadlocks is also not mentioned in any way. The effects of identified deadlock situations on the mentioned methodologies are unknown and thus represent a research gap.

In the area of factory planning, which often serves as a model for logistics planning approaches \citep{schenk2013fabrikplanung, tempelmeier2018planung}, the issue of deadlocks is also not addressed. The process according to the VDI Society Production Engineering \citep{vdi2009fabrikplanung} does not explicitly mention the prevention of deadlocks. However, it would be sensible to implement an approach to deadlock prevention both in Phase 3 \enquote{Concept Planning} during the sizing of operating resources and subsequently in Phase 4 \enquote{Detailed Planning}. In Phase 4, during the sub-process of fine planning, the planning of material flows, information flows, and communication flows in the form of process representations and process descriptions occurs \cite[p. 8]{vdi2009fabrikplanung}. This also involves the assignment of products and resources to the processes, the sequence of process steps, their organizational integration, and the work aids used \cite[p. 8]{vdi2009fabrikplanung}. Whether the process representations or the allocation of resources make deadlock situations visible, and how the concept or detailed plan may then need to be adjusted, is not explained by the VDI guideline.

While there are methodologies for logistics planning in the literature into which consideration and analysis of potential deadlock situations could fit, no detailed methodology for such analysis has yet been explained. This may lead to deadlocks remaining undetected in the planning stage and only being identified during the testing of control and decision rules, for instance, in a simulation model where control rules are tested, and certain configurations within the examined system lead to a deadlock situation. Handling deadlock issues in logistics then often occurs with additional problem-specific control rules \citep{Mayer.2010} or with preventive, deterministic measures through advance planning and reservation of resources \citep{Kim.2006, Lienert.2017}, provided that the information situation is sufficient.

Both approaches scale poorly with larger systems. With large systems, the implementation of problem-specific control rules is very complex, and there is a risk of overlooking special cases. For the second approach, such as the use of reservation systems, deterministic assumptions (e.g., vehicle X is at location Y at time Z) can render the planned solution impractical in reality due to disturbances or other stochastic events. For larger systems, prediction with deterministic assumptions becomes more difficult. Additionally, creating reservation plans can lead to high computation times when many processes and resources lead to numerous possible combinations of resource reservation. The changing availability of processes and resources due to the dynamics of the logistics scenario constantly provides new optimal solutions, necessitating frequent recalculations of the reservation plan. 

Decentralized control approaches in logistics, which promise such scalability, have also relied on problem-specific avoidance rules through optimization algorithms for deadlocks \citep{schonung2011dezentrale, seibold2016logical} and on agent-based approaches \citep{forget2009study, termors2010marp, yalcin2017marp, lu2019framework}.

\subsection{Systematic Review of Existing Approaches}
\label{subsec:SOTA-analysis}
This section reviews the current state of the art in applying MARL to deadlock handling. \cite{wesselhoft2022controlling} provide a brief literature review on the control of AMR fleets using RL. While their review offers valuable insights, it underscores the need for a more detailed exploration tailored to the complexities of deadlock handling within MARL frameworks. In response, this thesis conducts a comprehensive literature review, employing up-to-date research and focusing precisely on how MARL is used for MAPF in deadlock-capable MAPF problems. 

MARL can be used directly to handle deadlocks, or to manage the control of AMRs, thereby handling deadlocks indirectly. There are several possible search terms, but all should have a MARL approach in common. The structured analysis is divided into two parts: a preliminary analysis of relevant search terms and a comprehensive analysis of the selected search term. In the preliminary analysis, the first 20 search results for each search term are used to assess the relevance, quality, and scope of the research under these terms. The preliminary analysis helps to identify the most promising combinations of search terms for a comprehensive review aimed at identifying where MARL converge in the context of AMR deadlock handling. Table \ref{tab:search_terms_abbreviations} presents the search terms considered and the corresponding abbreviations for further use in the structured state of the art analysis in the application of MARL for deadlock handling.

\begin{table}[h]
    \centering
    \caption{Search terms and abbreviations for the preliminary analysis of the state of the art.}
    \label{tab:search_terms_abbreviations}
    \begin{tabular}{p{10cm}p{4cm}}
    \toprule
    \textbf{Search terms} & \textbf{Abbreviation} \\
    \hline
    multi AND agent AND reinforcement AND learning AND deadlock & MARL + deadlock\\
    multi AND agent AND reinforcement AND learning AND path AND finding & MARL + MAPF\\
    multi AND agent AND reinforcement AND learning AND autonomous AND mobile AND robot & MARL + AMR\\
    multi AND agent AND reinforcement AND learning AND automated AND guided AND vehicle & MARL + AGV\\
    multi AND agent AND reinforcement AND learning AND intralogistics & MARL + intralogistics\\
    \bottomrule
    \end{tabular}
\end{table}

The Scopus and Web of Science (WoS) databases are used for the preliminary analysis of the literature on MARL applications in deadlock handling. In order to include only recent research, the search results are limited to publications since 2019. The search is refined by ensuring that relevant terms appear in the titles, abstracts, or keywords (indexing for WoS) of the publications. The databases were queried on May 10, 2024. Table \ref{tab:preliminary-analysis} shows the search results for Scopus and WoS for the proposed search topics.

\begin{table}[h]
    \centering
    \caption{Number of results and relevance of the first 20 exported results for Scopus and WoS.}
    \label{tab:preliminary-analysis}
    \begin{tabular}{lcccc}
    \toprule
    \textbf{Search topic} & \textbf{Results Scopus} & \textbf{Rel. Scopus}  & \textbf{Results WoS} & \textbf{Rel. WoS}\\
    \hline
    MARL + deadlock & 21 & 12 & 9 & 7\\
    MARL + MAPF & 100 & 10 & 176 & 8\\
    MARL + AMR & 57 & 9 & 456 & 9\\
    MARL + AGV & 26 & 15 & 15 & 9\\
    MARL + intralogistics & 4 & 4 & 3 & 3\\
    \bottomrule
    \end{tabular}
\end{table}

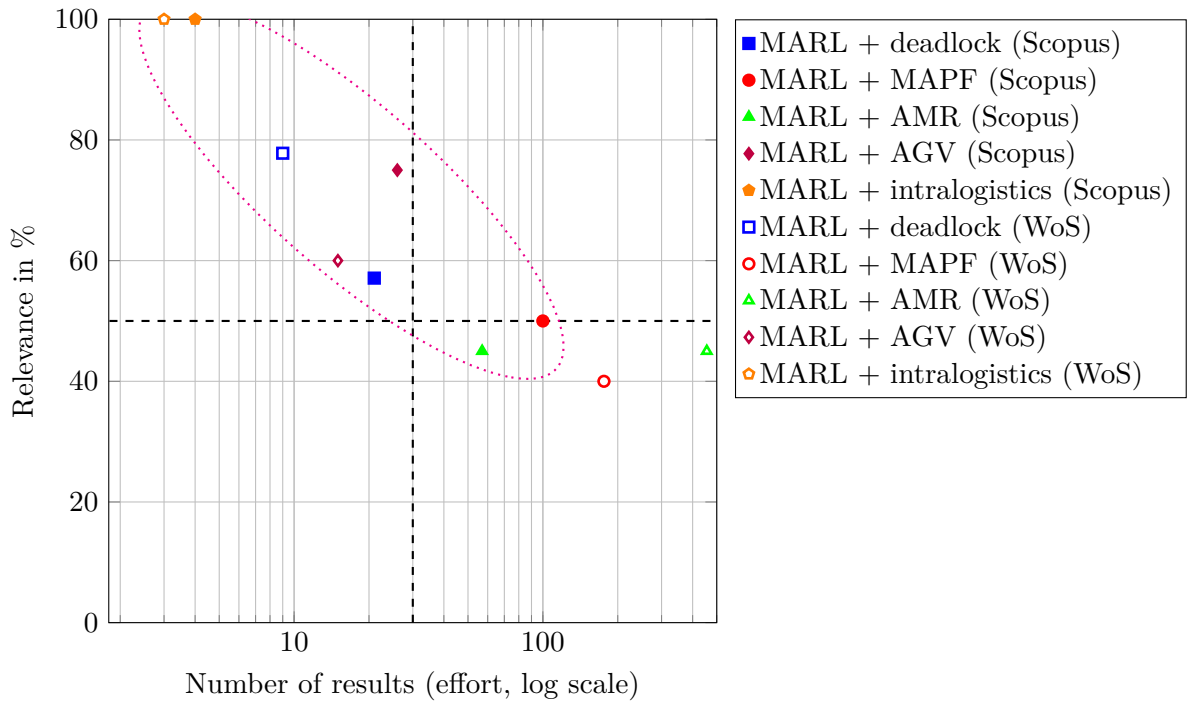
\begin{figure}[h]
    \centering
    \begin{tikzpicture}
        \begin{axis}[
            scatter/classes={
                a={mark=square*,blue}, 
                b={mark=*,red}, 
                c={mark=triangle*,green}, 
                d={mark=diamond*,purple},
                e={mark=pentagon*,orange},
                f={mark=square*,blue,fill=white},
                g={mark=*,red,fill=white},
                h={mark=triangle*,green,fill=white},
                i={mark=diamond*,purple,fill=white},
                j={mark=pentagon*,orange,fill=white}
            },
            xlabel={Number of results (effort, log scale)},
            ylabel={Relevance in \%},
            legend entries={
                MARL + deadlock (Scopus),
                MARL + MAPF (Scopus),
                MARL + AMR (Scopus),
                MARL + AGV (Scopus),
                MARL + intralogistics (Scopus),
                MARL + deadlock (WoS),
                MARL + MAPF (WoS),
                MARL + AMR (WoS),
                MARL + AGV (WoS),
                MARL + intralogistics (WoS)
            },
            legend pos=outer north east,
            width=0.5\textwidth,
            height=0.5\textwidth,
            grid=both,
            xmax=500,
            xmode=log,
            log ticks with fixed point,
            ymin=0, ymax=100,
            ymajorgrids=true,
            xmajorgrids=true
        ]
        
        \draw[thick, dashed] (30,0) -- (30,100);
        \draw[thick, dashed] (1,50) -- (500,50);
        
        \addplot[scatter,only marks,scatter src=explicit symbolic]
            coordinates {
                (21,57.1)[a]
                (100,50)[b]
                (57,45)[c]
                (26,75)[d]
                (4,100)[e]
                (9,77.8)[f]
                (176,40)[g]
                (456,45)[h]
                (15,60)[i]
                (3,100)[j]
            };

        \draw[thick, dotted, magenta,rotate around={-50:(axis cs:17,72)}] (axis cs:17,72) ellipse (10 and 15);
        \end{axis}
    \end{tikzpicture}
    \caption{Portfolio analysis comparing the relevance and effort for different search topics.}
    \label{fig:portfolio-analysis-search-terms}
\end{figure}

During the preliminary analysis, certain types of papers were not relevant: review papers, conference proceedings without content, papers focused primarily on communication, search or exploration routines rather than deadlock handling. To determine the most effective search terms for the comprehensive analysis, Figure \ref{fig:portfolio-analysis-search-terms} shows a portfolio analysis to compare the relevance of the search results with the effort required to analyze them. Relevance is measured as the percentage of relevant articles found in the initial 20 results. Effort is inferred from the total number of results returned by each search query, as a higher number of results indicates greater effort needed for a comprehensive review.

The four areas of the portfolio analysis in Figure \ref{fig:portfolio-analysis-search-terms} can be described as follows:
\begin{itemize}
    \item \textbf{High relevance, low effort (top-left)}: These terms are highly relevant and require less effort to analyze, making them ideal for comprehensive review.
    \item \textbf{High relevance, high effort (top-right)}: These terms are highly relevant but require significant effort, suggesting they are important but resource-intensive.
    \item \textbf{Low relevance, low effort (bottom-left)}: These terms are less relevant but easy to analyze, which may make them suitable for supplementary review.
    \item \textbf{Low relevance, high effort (bottom-right)}: These terms are less relevant and require high effort, making them less attractive for in-depth analysis.
\end{itemize}

Based on these insights, the search topics \enquote{MARL + deadlock}, \enquote{MARL + AGV}, and \enquote{MARL + intralogistics} from both databases should be included in the comprehensive literature analysis due to their high relevance and manageable effort. To ensure a thorough analysis and not overlook potential insights, the results for \enquote{MARL + MAPF} and \enquote{MARL + AMR} from the Scopus database should also be included in the comprehensive analysis. All selected search topics are within the ellipse shown in Figure \ref{fig:portfolio-analysis-search-terms}.
The comprehensive analysis includes a total of 235 search results and aims to identify the primary RL algorithms and training approaches, such as CTE, DTE, and CTDE, utilized in handling deadlocks. Additionally, the analysis seeks to uncover other relevant domains beyond AMRs and AGVs where MARL is effectively applied in handling deadlocks.

The initial literature search yields 235 results. Removing 38 duplicates leaves 197 unique publications. Excluding 31 placeholder entries reduces this number to 166. Of these, 82 publications fulfill the relevance criteria. A publication is relevant if it applies RL to MAPF or deadlock handling in logistics contexts. Not all relevant publications explicitly address deadlocks, but their methods generalize to deadlock-capable systems. All publications selected are in English.

The selected articles are categorized based on the type of agent they address and their consideration of deadlocks. Agent types are divided into AGVs, AMRs, cars, conveyors, generic agents, persons, trains, and unmanned aerial vehicles (UAVs). UAVs are represented as a separate category, even though they could also be classified under AMRs due to their autonomous nature. Table \ref{tab:selected-articles-agent-type} provides a detailed classification of these articles. For each agent type, the articles are further subdivided into those that are deadlock-capable, those that do not consider deadlocks, and those where the deadlock consideration is unclear. This classification helps in understanding the scope and applicability of the research within the context of deadlock handling in logistics and multi-agent systems.

\begin{table}[h]
    \centering
    \caption{Classification of selected articles based on the agent type and consideration of deadlocks.}
    \label{tab:selected-articles-agent-type}
    \begin{tabular}{p{3.5cm}p{11.6cm}}
        \toprule
         Agent type, deadlock consideration & References\\
         \hline
         \textbf{AGV} & \\
         Deadlock-capable & \cite{gong2024, hu2023, li2022mapf-communication-learning, muller2023mappo, reijnen2020, stephen2021, ye2023}\\
         No deadlock & \cite{arques2020swarm}\\
         Unclear & \cite{choi2022AGVcontrol, choi2022optimal-route-agv, durst2023agv, Gannouni2023digital-twin-transport, Hu2021bt, jiang2022, li2021multi-agv, li2023coop-control, popper2021, rhazzaf2021, tang2022, xue2022, zhang2022multi-agv}\\
         \textbf{AMR} & \\
         Deadlock-capable & \cite{chan2022MAPF-rot, damani2021primal2,gao2024,kozjek2021, ma2021distributed, sartoretti2019, skrynnik2023, wang2023, xu2021}\\
         No deadlock & \cite{malus2020}\\
         Unclear & \cite{ahn2023learning, bhattathiri2023mulation, cao2024safeRL, chen2023mitigating}; \cite{chen2022MAPF-transformer,chen2023HotSupervision,kamezaki2023, liu2019amr, liu2021amr, liu2023mappo-amr, ma2022attention}; \cite{ma2024end-to-end, mohseni2019, parooei2024map3f, said2021, setyawan2022, skrynnik2024learn, song2023, wu2022, zhang2024amr-ppo}; \cite{zhang2020robotnavigation, zhu2021}\\
         \textbf{Car} & \\
         Deadlock-capable & \cite{goto2022, hook2021, li2023coor-plt, li2023load}\\
         Unclear & \cite{chen2023parking, mohseni2019, yan2023, zhang2023cav}\\
         \textbf{Conveyor} & \\
         Deadlock-capable & \cite{sorensen2020routing} \\
         Unclear & \cite{kim2020}\\
         \textbf{Generic} & \\
         Deadlock-capable & \cite{ayalasomayajula2022design, davydov2021q-mixing, lin2023sacha, van2021time, ye2022mapf-deadlockdetection}\\
         Unclear & \cite{guan2022, ling2021integrating, ma2023learning, zhang2024iql, zhao2023}\\
         \textbf{Person} & \\
         Deadlock-capable & \cite{chen2023game}\\
         \textbf{Train} & \\
         Deadlock-capable & \cite{bretas2023addressing, cao2022train, laurent2021flatland, mohapatra2022, walter2020aiding}\\
         \textbf{UAV} & \\
         Unclear & \cite{cysne2023UAV, ding2022drone-delivery, liu2021uav, zhao2024}\\
         \bottomrule
    \end{tabular}
\end{table}

\begin{figure}[h]
    \centering
    \begin{tikzpicture}
        \begin{axis}[
            ybar stacked,
            bar width=1cm,
            width=14cm,
            height=8cm,
            ymin=0,
            xlabel={Agent type},
            ylabel={Number of articles},
            symbolic x coords={AGV, AMR, Car, Conveyor, Generic, Person, Train, UAV},
            xtick=data,
            x tick label style={rotate=45, anchor=east},
            legend style={at={(0.5,-0.2)}, anchor=north, legend columns=-1},
            nodes near coords,
            every node near coord/.append style={font=\small}
        ]
            \addplot+[ybar, draw=black, text=black, fill=darkslateblue!75] plot coordinates {(AGV,7) (AMR,9) (Car,4) (Conveyor,1) (Generic,5) (Person,1) (Train,5) (UAV,0)};
            \addplot+[ybar, draw=black, text=black, fill=electricyellow] plot coordinates {(AGV,1) (AMR,1) (Car,0) (Conveyor,0) (Generic,0) (Person,0) (Train,0) (UAV,0)};
            \addplot+[ybar, draw=black, text=black, fill=applegreen!80] plot coordinates {(AGV,13) (AMR,22) (Car,4) (Conveyor,1) (Generic,5) (Person,0) (Train,0) (UAV,4)};
            
            \legend{Deadlock-capable, No deadlock, Unclear}
        \end{axis}
    \end{tikzpicture}
    \caption{Distribution of articles based on agent type and deadlock consideration.}
    \label{fig:agent-type-deadlock-consideration}
\end{figure}
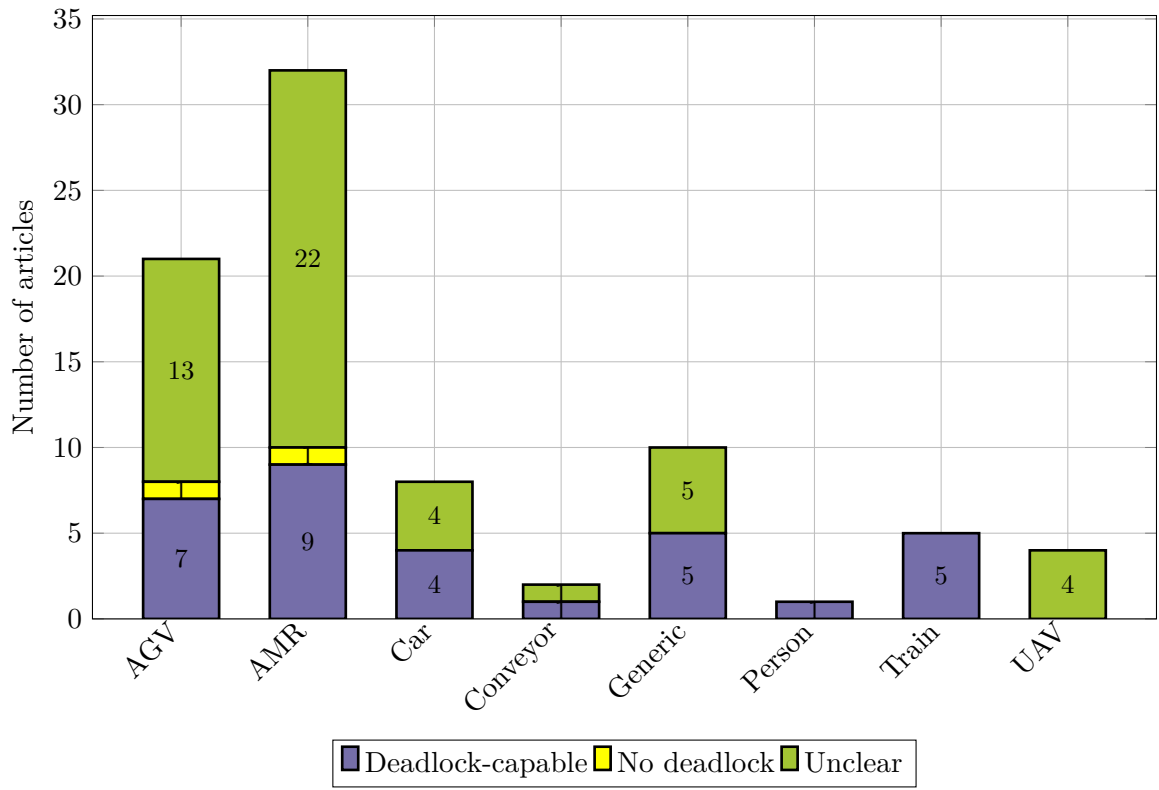

Figure \ref{fig:agent-type-deadlock-consideration} presents a stacked bar chart showing the total number of articles for various agent types, divided into the categories of Table \ref{tab:selected-articles-agent-type} based on their consideration of deadlocks: deadlock-capable, no deadlock, and unclear. This visualization allows for a clear comparison of the main application areas (via agent types) of the literature review and how frequently deadlock issues are addressed across different agent types. From Figure \ref{fig:agent-type-deadlock-consideration}, it is apparent that AMRs and AGVs are the most prevalent agent types, likely due to the search terms used in the literature review. The selected articles reveal that the types of MAPF problems and the possible movements of AGVs, AMRs, cars, and generic solution approaches are quite similar, resulting in comparable distributions in terms of deadlock consideration. These categories tend to exhibit a balanced mix of deadlock-capable and unclear articles, with a few explicitly stating no deadlock considerations.

UAVs and trains demonstrate distinct MAPF problems and possible movements. UAVs, with their freedom of movement in the air, find it easier to avoid deadlocks, resulting in fewer articles explicitly addressing deadlock issues. Trains, restricted by rail systems, face more significant challenges in deadlock handling, as indicated by a higher proportion of deadlock-capable articles.

It is noteworthy that while collision avoidance is frequently mentioned and clarified in many articles, the explicit handling of deadlocks is less common. This indicates a research gap in deadlock consideration within the application of MARL for MAPF and other intralogistic problems.

The following categorization of the selected articles is according to the used algorithm and the training and execution mode. This categorization is based on the proposed solution approach of a publication rather than direct comparisons mentioned in an article. For example, if a paper proposes a DDPG approach and compares it to a DQN benchmark, it is categorized under DDPG. Variants of Q-learning, such as QMIX \citep{rashid2020qmix} and dueling double DQN \citep{wang2016duelingdqn}, are summarized under DQN. Algorithms that are rarely mentioned, like multi-agent posthumous credit assignment \citep{cohen2021use} are summarized under custom algorithms. Different naming conventions, like independent PPO as IPPO or multi-agent PPO as MAPPO, are categorized under the core algorithm, in this case, PPO. Counterfactual multi-agent policy gradients \citep{foerster2018counterfactual} is summarized under PG. This summarize approach ensures a streamlined and coherent classification of various RL algorithms.
Table \ref{tab:selected-articles-algorithms} provides a comprehensive classification of selected articles based on the algorithm and the training and execution mode. 

\begin{table}[h]
    \centering
    \caption{Classification of selected articles based on algorithm, training and execution mode.}
    \label{tab:selected-articles-algorithms}
    \begin{tabular}{p{2.5cm}p{12.6cm}}
        \toprule
         Algorithm, training mode & References\\
         \hline
         \textbf{A2C} & \\
         CTDE & \cite{ayalasomayajula2022design}\\
         Single-agent & \cite{chen2023game} \\
         \textbf{A3C} & \\
         CTDE & \cite{chan2022MAPF-rot, damani2021primal2, laurent2021flatland, li2023load}; \cite{li2022mapf-communication-learning}\\
         DTE & \cite{sartoretti2019} \\
         \textbf{Custom} & \\         
         CTDE & \cite{goto2022, guan2022, kamezaki2023, li2021multi-agv, liu2019amr, zhang2022multi-agv}; \cite{zhang2020robotnavigation}\\
         CTE & \cite{jiang2022, ma2024end-to-end} \\
         \textbf{DDPG} & \\
         CTDE & \cite{gong2024, hu2023, li2021multi-agv, li2023coop-control, wu2022, ye2023, zhao2024}\\
         DTE & \cite{setyawan2022}\\
         \textbf{DQN} & \\         
         CTDE & \cite{cao2022train,chen2022MAPF-transformer,chen2023HotSupervision, choi2022AGVcontrol, choi2022optimal-route-agv, davydov2021q-mixing, ding2022drone-delivery,gao2024, Hu2021bt, laurent2021flatland, li2023coor-plt, ling2021integrating, liu2021amr, mohapatra2022, stephen2021, tang2022, van2021time, ye2022mapf-deadlockdetection, zhao2023} \\
         CTE & \cite{rhazzaf2021, sorensen2020routing, van2021time} \\
         DTE & \cite{hook2021, ma2021distributed, parooei2024map3f, said2021,song2023, walter2020aiding, xu2021, zhang2024iql} \\
         Unclear & \cite{bhattathiri2023mulation, kim2020, zhu2021}\\
         \textbf{PG} & \\
         CTDE & \cite{ling2021integrating, song2023}\\
         \textbf{PPO} & \\         
         CTDE & \cite{arques2020swarm,bretas2023addressing,chen2023mitigating}; \cite{chen2023parking, cysne2023UAV, laurent2021flatland, muller2023mappo, reijnen2020, skrynnik2024learn, skrynnik2023, zhang2024amr-ppo, zhao2024}\\
         CTE & \cite{durst2023agv, Gannouni2023digital-twin-transport, kozjek2021}\\
         DTE & \cite{liu2023mappo-amr, popper2021, wang2023}\\
         \textbf{REINFORCE} & \\
         CTDE & \cite{ahn2023learning}\\
         CTE & \cite{zhang2023cav}\\
         \textbf{SAC} & \\
         CTDE & \cite{lin2023sacha, xue2022}\\
         DTE & \cite{ma2023learning, ma2022attention} \\
         \textbf{SARSA} & \\
         CTE & \cite{liu2021uav}\\
        \textbf{TD3} & \\
         CTDE & \cite{cao2024safeRL}\\
         DTE & \cite{malus2020} \\
         \textbf{TRPO} & \\
         CTDE & \cite{yan2023} \\
         DTE & \cite{mohseni2019} \\
         \bottomrule
    \end{tabular}
\end{table}

Notably, the CTDE paradigm is frequently chosen, reflecting its effectiveness in balancing computational complexity and leveraging shared information during training. However, direct comparisons of different training and execution modes within the same article are rare, indicating an area for future research to explore optimal configurations and their impacts on performance.

In analyzing the selected articles, this thesis also examines the usage of various ANN structures across different RL algorithms. If an article utilized multiple network structures for a single algorithm, each structure was counted separately in the respective categories. The \enquote{MLP} category includes papers that use only MLPs without incorporating any other layer types. For instance, a network comprising a convolutional layer, two fully connected layers, and one LSTM layer would be counted under both \enquote{CNN} and \enquote{LSTM} categories. \enquote{Attention} refers to the usage of attention layers, whereas networks that employ a complete transformer architecture are categorized under \enquote{Transformer}.

\begin{figure}[h]
    \centering
    \includegraphics[width=1\linewidth]{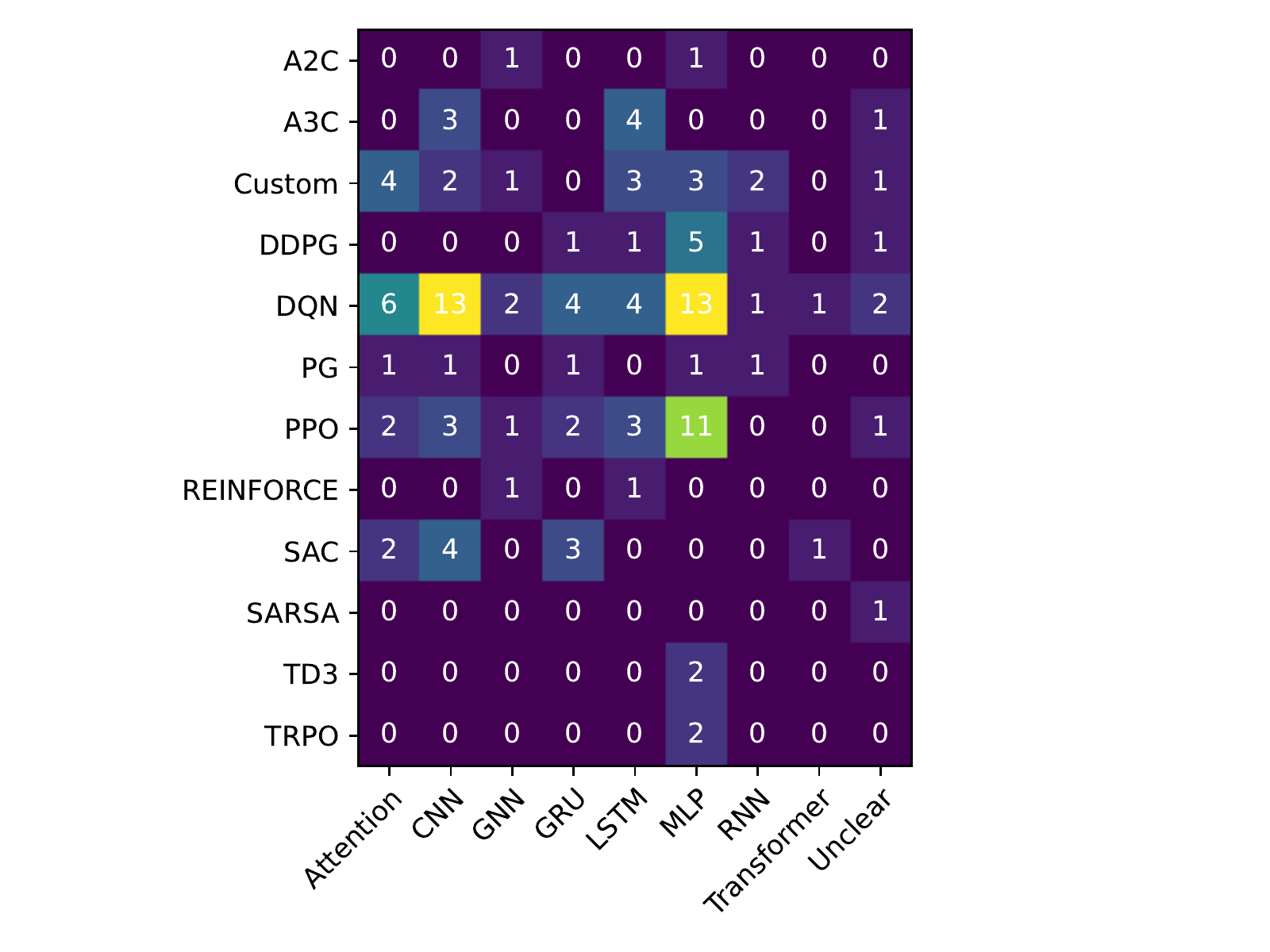}
    \caption{Heatmap of algorithms and ANN structures in selected articles.}
    \label{fig:ch2-heatmap-algorithms}
\end{figure}

The heatmap in Figure \ref{fig:ch2-heatmap-algorithms} provides a detailed visualization of the usage of various ANN structures across different RL algorithms in the selected articles. From the heatmap, it is evident that DQN and PPO are the most common algorithms, and some ANN structures are preferred for certain algorithms over others. For example, the DQN algorithm frequently employs MLP and CNN structures, with counts of 13 and 10, respectively. This suggests a strong preference for these structures within the DQN studies reviewed.

The PPO algorithm shows a notable usage of MLP structures, with a count of 11. Other structures such as CNN and LSTM are also employed, albeit less frequently. The "Custom" category indicates diverse ANN structure usage, particularly combining attention layers, CNN, and LSTM. This diversity reflects the flexibility and experimental nature of custom algorithms in exploring various ANN structures. The A3C algorithm prominently features CNN and LSTM structures, highlighting a trend towards combining these layers in the reviewed articles. In contrast, algorithms like A2C, SARSA, TD3, TRPO and REINFORCE show limited usage, indicating outdated or unpopular algorithms.

The analysis of the heatmap reveals several key trends in the usage of ANN structures across different RL algorithms. MLP structures are the most commonly employed, reflecting their fundamental role in many neural network architectures. CNN structures also see significant usage because grid environments are often represented as (stacked) arrays, which are well-suited to the spatial processing capabilities of CNNs. The frequent combination of CNN and LSTM layers in certain algorithms underscores the importance of both spatial and temporal data handling in advanced RL tasks. The varied usage of GRU and RNN structures demonstrates the exploration of different recurrent architectures to address specific challenges in RL scenarios. Attention mechanisms, while less commonly used overall, are a critical component in custom and cutting-edge algorithm designs, indicating ongoing research and innovation in this area. The presence of transformer architectures, though relatively sparse, highlights the potential for more sophisticated sequence modeling in future RL applications. 

\section{Research Gaps and Research Questions} \label{sec:methodology-researchgap}

The analysis of existing planning methods exposes a significant oversight in addressing deadlocks during the planning phase of intralogistics systems. The comprehensive literature review in Subsection \ref{subsec:SOTA-analysis} displays also a deficiency of deadlock consideration in the problem description in MAPF. To systematically derive the RQs, the research gaps are categorized into three main areas: fundamental research, transfer research, and application. Figure \ref{fig:research-questions} provides an overview of the identified research gaps, corresponding RQs, and the expected outcome for each category.

\begin{figure}[h]
    \centering
    \begin{tikzpicture}

        \def\boxwidth{4.5cm}
        \def\boxheight{2cm}
        \def\xsep{0.25cm}
        \def\ysep{0cm}

        \coordinate (A1) at (0, 0);
        \coordinate (A2) at (\boxwidth + \xsep, 0);
        \coordinate (A3) at (2*\boxwidth + 2*\xsep, 0);

        \coordinate (B1) at (0, -\boxheight - \ysep);
        \coordinate (B2) at (\boxwidth + \xsep, -\boxheight - \ysep);
        \coordinate (B3) at (2*\boxwidth + 2*\xsep, -\boxheight - \ysep);

        \coordinate (C1) at (0, -2*\boxheight -1cm);
        \coordinate (C2) at (\boxwidth + \xsep, -2*\boxheight -1cm);
        \coordinate (C3) at (2*\boxwidth + 2*\xsep, -2*\boxheight -1cm);

        \coordinate (D1) at (0, -3*\boxheight -1cm);
        \coordinate (D2) at (\boxwidth + \xsep, -3*\boxheight -1cm);
        \coordinate (D3) at (2*\boxwidth + 2*\xsep, -3*\boxheight -1cm);

        \coordinate (E1) at (0, -4*\boxheight -2cm);
        \coordinate (E2) at (\boxwidth + \xsep, -4*\boxheight -2cm);
        \coordinate (E3) at (2*\boxwidth + 2*\xsep, -4*\boxheight -2cm);

        \coordinate (F1) at (0, -5*\boxheight -2cm);
        \coordinate (F2) at (\boxwidth + \xsep, -5*\boxheight -2cm);
        \coordinate (F3) at (2*\boxwidth + 2*\xsep, -5*\boxheight -2cm);

        \node[draw, minimum width=\boxwidth, minimum height=\boxheight, align=left, text width=\boxwidth] at (A1) {Deficiency of deadlock consideration in planning and problem description};
        \node[draw, minimum width=\boxwidth, minimum height=\boxheight, align=left, text width=\boxwidth] at (A2) {RQ 1.1: Is deadlock handling significant?};
        \node[draw, minimum width=\boxwidth, minimum height=\boxheight, align=left, text width=\boxwidth] at (A3) {Simulation results about the significance of deadlock handling};

        \node[draw, minimum width=\boxwidth, minimum height=\boxheight, align=left, text width=\boxwidth] at (B1) {Deficiency of procedure model to apply RL};
        \node[draw, minimum width=\boxwidth, minimum height=\boxheight, align=left, text width=\boxwidth] at (B2) {RQ 1.2: How to integrate RL into the planning process?};
        \node[draw, minimum width=\boxwidth, minimum height=\boxheight, align=left, text width=\boxwidth] at (B3) {Procedure model for applying RL};

        \node [above of=A1, yshift=0.3cm] {Research gap};
        \node [above of=A2, yshift=0.3cm] {Research question};
        \node [above of=A3, yshift=0.3cm] {Expected outcome};
        \node [rotate=90, left of=B1, yshift=2.8cm, xshift=2cm] {Fundamental research};

        \draw (-2.35,-3) -- (4.5,-4) -- (11.85,-3);

        \node[draw, minimum width=\boxwidth, minimum height=\boxheight, align=left, text width=\boxwidth] at (C1) {Deficiency of deadlock reference scenarios for MAPF};
        \node[draw, minimum width=\boxwidth, minimum height=\boxheight, align=left, text width=\boxwidth,  inner ysep=0cm] at (C2) {RQ 2.1: What do reference scenarios for deadlock-capable MAPF problems look like?};
        \node[draw, minimum width=\boxwidth, minimum height=\boxheight, align=left, text width=\boxwidth] at (C3) {Deadlock-capable reference models for MAPF};

        \node[draw, minimum width=\boxwidth, minimum height=\boxheight, align=left, text width=\boxwidth, inner ysep=0cm] at (D1) {Deficiency of comprehensive analysis of MARL performance in deadlock-capable MAPF};
        \node[draw, minimum width=\boxwidth, minimum height=\boxheight, align=left, text width=\boxwidth] at (D2) {RQ 2.2: When and in which configurations is MARL effective for MAPF with deadlocks?};
        \node[draw, minimum width=\boxwidth, minimum height=\boxheight, align=left, text width=\boxwidth] at (D3) {Evaluation of simulation results and comparison with traditional solutions};

        \node [rotate=90, left of=D1, yshift=2.8cm, xshift=2cm] {Transfer research};

        \draw (-2.35,-8) -- (4.5,-9) -- (11.85,-8);
        
        \node[draw, minimum width=\boxwidth, minimum height=\boxheight, align=left, text width=\boxwidth,] at (E1) {Deficiency of comparison of deadlock handling strategies};
        \node[draw, minimum width=\boxwidth, minimum height=\boxheight, align=left, text width=\boxwidth] at (E2) {RQ 3.1: Does the deadlock handling strategy matter?};
        \node[draw, minimum width=\boxwidth, minimum height=\boxheight, align=left, text width=\boxwidth] at (E3) {Simulation results of different deadlock handling strategies};

        \node[draw, minimum width=\boxwidth, minimum height=\boxheight, align=left, text width=\boxwidth,] at (F1) {Deficiency of comparison of training and execution modes};
        \node[draw, minimum width=\boxwidth, minimum height=\boxheight, align=left, text width=\boxwidth] at (F2) {RQ 3.2: Is CTDE the best approach in deadlock-capable MAPF problems?};
        \node[draw, minimum width=\boxwidth, minimum height=\boxheight, align=left, text width=\boxwidth] at (F3) {Simulation results of different training and execution modes};

        \node [rotate=90, left of=F1, yshift=2.8cm, xshift=2cm] {Application research};
    \end{tikzpicture}

    \caption{Overview of research gaps, research questions and the expected outcome in this thesis.}
    \label{fig:research-questions}
\end{figure}

Fundamental research focuses on acquiring new knowledge about the basic principles and underlying mechanisms of a subject without any immediate application in mind. It aims to build a solid theoretical foundation that can be used for future applied research. In the context of this thesis, fundamental research involves understanding the core issues related to deadlock handling in intralogistics systems and developing new algorithms and models to address these issues. In the category of fundamental research, the primary gap identified is the deficiency of deadlock consideration in the planning and problem description of intralogistics systems. This foundational gap highlights the need for a deeper understanding of the significance of deadlock handling. The associated RQ 1.1 asks: \enquote{Is deadlock handling significant?} The expected outcome from this investigation includes simulation results that demonstrate the importance of incorporating deadlock handling in logistics planning. Additionally, there is a gap in procedural models for applying RL to deadlock situations. The relevant RQ 1.2 focuses on how to integrate RL into the planning process, with the anticipated outcome being a detailed procedure model for applying RL in intralogistics.

Transfer research bridges the gap between fundamental research and practical application. It involves specifying the applicability of methods to particular categories, such as relating the general problem of deadlocks to the MAPF problem, which involves movement and transport as processes and paths as resources, and focusing MARL methods on specific agent types like AMRs or AGVs. Transfer research takes theoretical concepts and adapts them for practical use, often through the development of prototypes, frameworks, or reference models. This type of research validates the feasibility and effectiveness of theoretical solutions in more realistic settings, paving the way for their practical implementation. A key gap identified in the literature review is the deficiency of deadlock reference scenarios for MAPF. This gap underscores the necessity of developing standard reference scenarios to test and validate deadlock-capable MAPF problems. The RQ 2.1 asks: \enquote{What do reference scenarios for deadlock-capable MAPF problems look like?} The expected outcome is the creation of deadlock-capable reference models for MAPF. Furthermore, there is a deficiency in the comprehensive analysis of MARL performance in deadlock-capable MAPF environments. RQ 2.2 examines the appropriate conditions for applying MARL algorithms to deadlock-capable MAPF scenarios and identifies the optimal configurations for their application. The goal is to evaluate simulation outcomes across various MARL configurations and compare these results with traditional MAPF approaches. This evaluation aims to determine under which circumstances MARL offers a viable solution.

In the application research category, the emphasis is on using methods and algorithms for specific use cases or case studies. This involves working with a specific environment, potentially modeled after real-world scenarios, and dealing with specific numbers of agents. The experiments may include variations on the numbers based on the planned experiment series but are conducted with a specific investigative goal in mind. A key gap identified in this category is the deficiency of comparative studies among various deadlock handling strategies. The RQ 3.1 explores whether the choice of deadlock handling strategy significantly impacts performance. The expected outcome is simulation results that compare different strategies, providing insights into their relative effectiveness and practical implications. RQ 3.2 investigates whether CTDE is the most effective training and execution mode for applying MARL in deadlock-capable MAPF problems, given its prevalent use in the literature. The expected outcome from this investigation includes simulation results comparing different training and execution modes, further informing the optimal strategies for practical implementation.

\section{Conclusion}

This chapter provides a comprehensive exploration of the foundational concepts and literature related to deadlocks, RL, and MAPF within the context of intralogistics. The literature review starts with a detailed examination of deadlocks, including their definition, underlying causes, and the strategic approaches to their handling such as prevention, avoidance, and detection \& recovery. This is followed by an in-depth review of RL, emphasizing its application within MAS and its potential to address deadlocks through various RL algorithms. The chapter further analyzes the role of MAPF as the main problem in logistics transport operations.

The analysis of existing planning methods shows a lack of deadlock consideration in the planning phase of intralogistics systems. The structured literature review identifies key gaps in the current state of deadlock consideration in MAPF, in particular the need for more deadlock consideration in the problem description and the application of RL algorithms specifically for deadlock-capable MAPF problems. Several RL algorithms present viable solution approaches, with the CTDE mode emerging as the predominant method for training and execution.

By formulating the RQs that drive this thesis, this chapter establishes the theoretical foundation for investigating RL-based approaches to deadlock handling. The research gaps identified in the literature motivate the development of novel methodologies that integrate deadlock considerations into MAPF and intralogistics planning. The following chapters build on this foundation by presenting the methodological framework, experimental design, and empirical evaluation of RL-based deadlock handling strategies. Having established the theoretical basis and research gaps, the next chapter focuses on the methodological approach adopted in this research. Chapter \ref{ch:methodology} introduces the deadlock consideration in logistics planning, details the integration of MARL into deadlock-aware MAPF, and outlines the experimental design for evaluating the proposed methods.

%

\chapter{A RL-based Methodology for Deadlock Handling between AMRs} \label{ch:methodology}
Chapter \ref{ch:methodology} presents the methodological framework for addressing the RQs. This chapter also includes preliminary findings that provide answers to the initial RQs.
\begin{itemize}
    \item In Section~\ref{sec:methodology-significance}, the significance of considering deadlocks in logistics planning is examined. This includes investigating which deadlock handling strategy have a statistically significant impact on logistic key figures. The results of this section are based on previously published research \cite{Muller.2020} and \cite{Muller.2021}.
    \item Section \ref{sec:methodology-logisticsplanning} presents a framework to integrate deadlock considerations into logistics planning and applying RL for deadlock handling. The framework was introduced in \citet{muller2023framework}.
    \item Section \ref{sec:methodology-reference-models} develops reference models for MAPF problems where deadlocks are explicitly considered, providing a structured basis for evaluating MAPF algorithms.
    \item Section \ref{sec:methodology-mapf-algorithms} details the selection and implementation of traditional MAPF algorithms, establishing a comparison baseline for RL-based approaches.
\end{itemize}

\section{Analysis of the Significance of Deadlock Handling} \label{sec:methodology-significance}

Addressing the RQ 1.1 \enquote{Is deadlock handling significant?} and RQ 3.1 \enquote{Does the deadlock handling strategy matter?}, this section analyzes the impact of deadlocks and the choice between different deadlock handling strategies. The analysis draws upon the insights from the two previous published paper. The first paper by \cite{Muller.2020} provides a flexible simulation model for evaluating various deadlock handling strategies in a warehouse setting. The second paper by \cite{Muller.2021} builds on the first paper and investigates the influence of disruptions on deadlock handling with AGVs.

\begin{figure}
    \centering
    \includegraphics[width=1\linewidth]{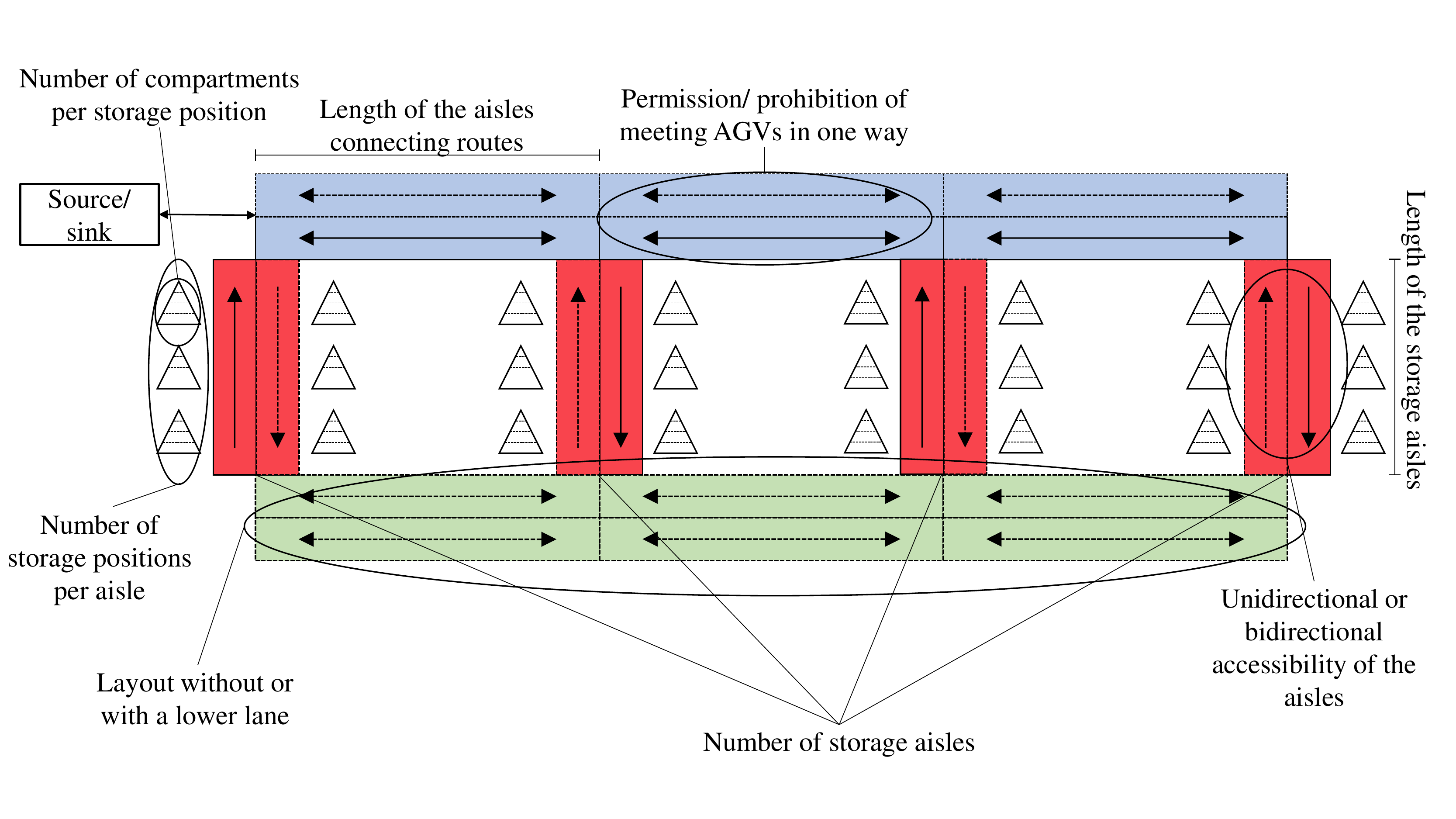}
    \caption{The considered intralogistic system as in \cite{Muller.2020}.}
    \label{fig:conceptual-model-warehouse-wsc2020}
\end{figure}

To assess the significance of deadlock handling, we consider a typical warehouse environment with various floor-bound AGVs and different layout variants. A flexible model in infrastructure design allows to investigate possible deadlocks in the warehouse for various scenarios. Figure \ref{fig:conceptual-model-warehouse-wsc2020} illustrates the conceptual model of the considered system, showcasing the adjustable parameters of the warehouse and its infrastructure.

The system boundary is located in the upper left corner, fixed and represented by a source and a sink. This is the central hub where AGVs start and end their routes. From this boundary, the routes for the AGVs extend to reach every storage space location within the rectangular layout of the warehouse. The length of each route and the number of aisles are flexible parameters. There is an option to exclude the lower lane, depicted in green, which means that the aisles can only be accessed from one side. This feature demonstrates the adaptability of the warehouse layout to different operational needs. The horizontal connecting routes, shown in blue and green, can be adjusted to switch between single and two lanes, allowing for various traffic configurations. The vertical storage aisles are fixed with one lane, but they can be either unidirectional or bidirectional, providing different levels of accessibility and complexity for AGV navigation. The unidirectional aisles help streamline the movement, reducing potential conflicts, while bidirectional aisles increase flexibility but require more sophisticated management to avoid or recover deadlocks.

Six layouts are identified as typical representations of the various route type variations discussed. Table \ref{tab:layout-variants-warehouse-wsc2020} provides a categorized overview of these layouts. Layouts 1 and 2 feature unidirectional storage aisles, which aim to prevent deadlocks within the aisles themselves. In Layouts 5 and 6, the use of solely unidirectional routes enforces a deadlock prevention policy by allowing only one AGV to enter a storage aisle at a time.

The detection and recovery strategy is not implemented for layouts 5 and 6 due to the lack of alternative routes. For the other layouts, the AGV routes are calculated based on the shortest distance. An AGV does not enter a single-lane section if another AGV is traveling in the opposite direction. In such cases, one of the vehicles is redirected to avoid a potential deadlock. The vehicle to be redirected is determined based on predefined criteria and then redirected accordingly. If additional vehicles are traveling in the same direction as the initially redirected AGV, these other AGVs must also be redirected.

\begin{table}[h]
    \caption{The different considered layout variants as in \cite{Muller.2020}.}
    \label{tab:layout-variants-warehouse-wsc2020}
    \centering
    \begin{tabular}{lccc}    
        \toprule
        & \textbf{Unidirectional} & \multicolumn{2}{c}{\textbf{Bidirectional storage aisles}} \\ 
        & \textbf{storage aisles} & \textbf{Two connection routes} & \textbf{One connection route} \\ \hline
         \multirow{4}{3cm}{Connection routes with single lane} & Layout 1 & Layout 3 & Layout 5 \\ 
        & \begin{tikzpicture}[scale=0.6]
            \draw[->] (0,0) -- (0,2); \draw[<->] (0,2) -- (1,2); \draw[->] (1,2) -- (1,0); \draw[<->] (1,0) -- (0,0);
            \draw[->] (2,0) -- (2,2); \draw[<->] (1,2) -- (2,2); \draw[->] (3,2) -- (3,0); \draw[<->] (2,0) -- (1,0); \draw[<->] (2,0) -- (3,0); \draw[<->] (2,2) -- (3,2);
        \end{tikzpicture}
        & \begin{tikzpicture}[scale=0.6]
            \draw[<->] (0,0) -- (0,2); \draw[<->] (0,2) -- (1,2); \draw[<->] (1,2) -- (1,0); \draw[<->] (1,0) -- (0,0);
            \draw[<->] (2,0) -- (2,2); \draw[<->] (1,2) -- (2,2); \draw[<->] (3,2) -- (3,0); \draw[<->] (2,0) -- (1,0); \draw[<->] (2,0) -- (3,0); \draw[<->] (2,2) -- (3,2);
        \end{tikzpicture}
        & \begin{tikzpicture}[scale=0.6]
            \draw[<->] (0,0) -- (0,2); \draw[<->] (0,2) -- (1,2); \draw[<->] (1,2) -- (1,0); 
            \draw[<->] (2,0) -- (2,2); \draw[<->] (1,2) -- (2,2); \draw[<->] (3,2) -- (3,0); \draw[<->] (2,2) -- (3,2);
        \end{tikzpicture} \\ 
        \multirow{4}{3cm}{Connection routes with two lanes} & Layout 2 & Layout 4 & Layout 6 \\ 
        & \begin{tikzpicture}[scale=0.6]
            \draw[->] (0,0) -- (0,2); \draw[<-] (0,2) -- (1,2); \draw[->] (0,2.1) -- (1,2.1); \draw[->] (1,2) -- (1,0); \draw[<-] (0,0) -- (1,0); \draw[->] (0,-0.1) -- (1,-0.1);
            \draw[->] (2,0) -- (2,2); \draw[<-] (1,2) -- (2,2); \draw[->] (1,2.1) -- (2,2.1); \draw[->] (3,2) -- (3,0); \draw[<-] (1,0) -- (2,0); \draw[->] (1,-0.1) -- (2,-0.1); \draw[<-] (2,0) -- (3,0); \draw[->] (2,-0.1) -- (3,-0.1);\draw[<-] (2,2) -- (3,2); \draw[->] (2,2.1) -- (3,2.1);
        \end{tikzpicture}
        & \begin{tikzpicture}[scale=0.6]
            \draw[<->] (0,0) -- (0,2); \draw[<-] (0,2) -- (1,2); \draw[->] (0,2.1) -- (1,2.1); \draw[<->] (1,2) -- (1,0); \draw[<-] (0,0) -- (1,0); \draw[->] (0,-0.1) -- (1,-0.1);
            \draw[<->] (2,0) -- (2,2); \draw[<-] (1,2) -- (2,2); \draw[->] (1,2.1) -- (2,2.1); \draw[<->] (3,2) -- (3,0); \draw[<-] (1,0) -- (2,0); \draw[->] (1,-0.1) -- (2,-0.1); \draw[<-] (2,0) -- (3,0); \draw[->] (2,-0.1) -- (3,-0.1);\draw[<-] (2,2) -- (3,2); \draw[->] (2,2.1) -- (3,2.1);
        \end{tikzpicture}
        & \begin{tikzpicture}[scale=0.6]
            \draw[<->] (0,0) -- (0,2); \draw[<-] (0,2) -- (1,2); \draw[->] (0,2.1) -- (1,2.1); \draw[<->] (1,2) -- (1,0); 
            \draw[<->] (2,0) -- (2,2); \draw[<-] (1,2) -- (2,2); \draw[->] (1,2.1) -- (2,2.1); \draw[<->] (3,2) -- (3,0); \draw[<-] (2,2) -- (3,2); \draw[->] (2,2.1) -- (3,2.1);
        \end{tikzpicture} \\ 
        \bottomrule
    \end{tabular}

\end{table}

Table \ref{tab:deadlock-examples-warehouse-wsc2020} shows the four identified deadlock scenarios and examines their potential occurrence within the defined layout types. The \enquote{possible solution} figures are provided as examples and are not the only methods available for resolution. The selection of the vehicle that must change its route may differ from the procedure shown in these examples.

\begin{table}[h]
    \caption{Typical examples of deadlocks and possible solution based on \cite{Muller.2020}.}
    \label{tab:deadlock-examples-warehouse-wsc2020}
    \centering
    \begin{tabular}{p{1.8cm}cccc}    
        \toprule
        & \textbf{Deadlock 1} & \textbf{Deadlock 2} & \textbf{Deadlock 3} & \textbf{Deadlock 4}  \\ 
        & Layout 1 \& 3 & Layout 1 \& 3 & Layout 3 \& 4 & Layout 3\\ \hline
         Deadlock situation & \includegraphics[align=c, scale=0.4]{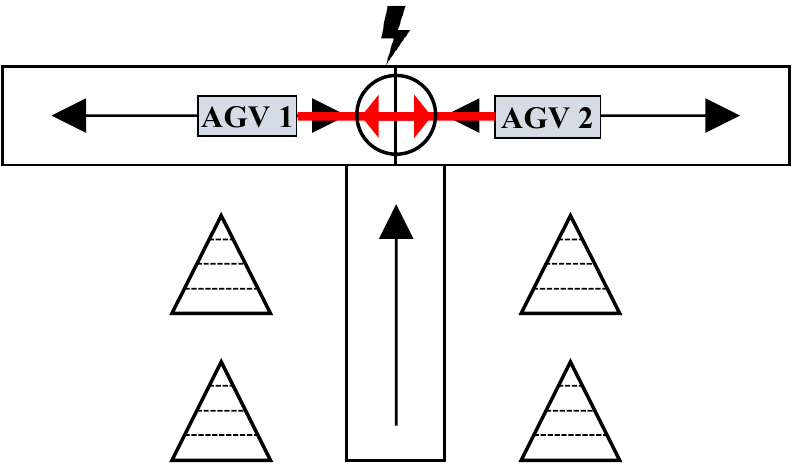} & \includegraphics[align=c, scale=0.4]{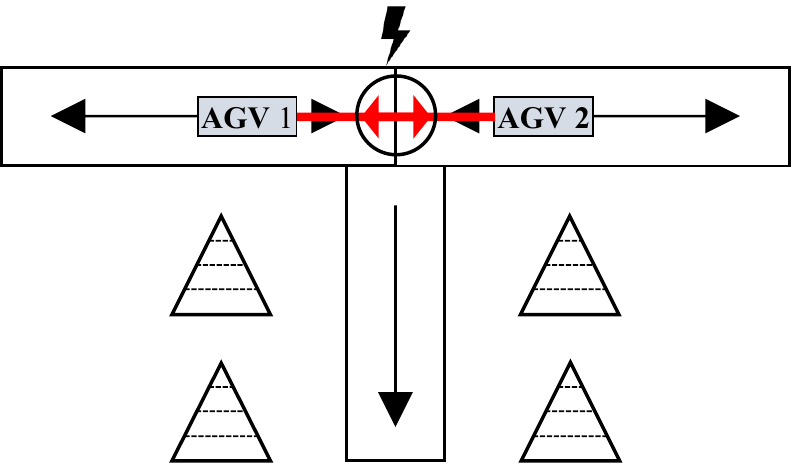} & \includegraphics[align=c, scale=0.35]{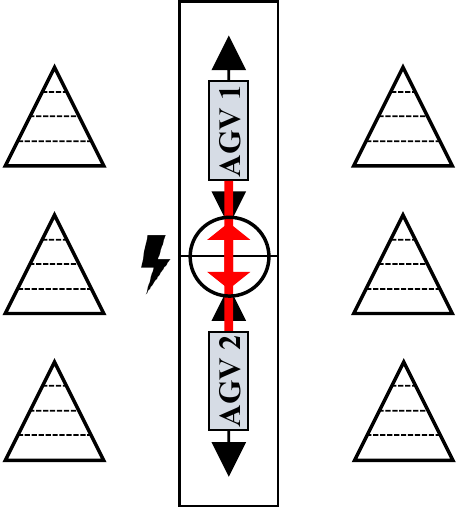} &\includegraphics[align=c, scale=0.4]{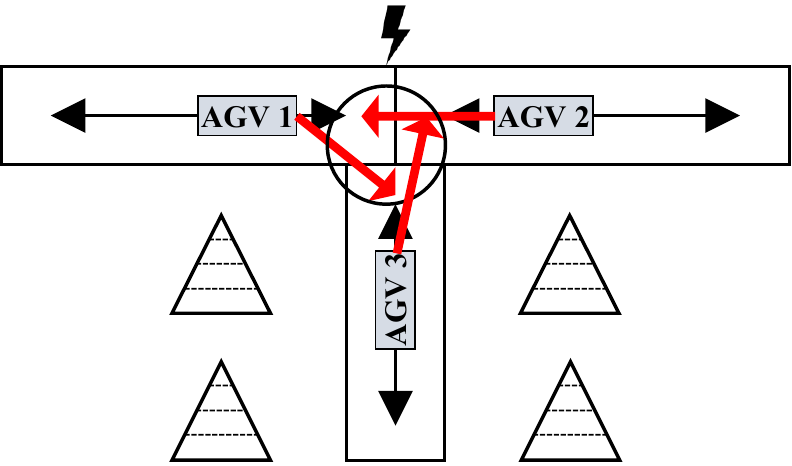} \\ 
        Possible solution & \includegraphics[align=c, scale=0.4]{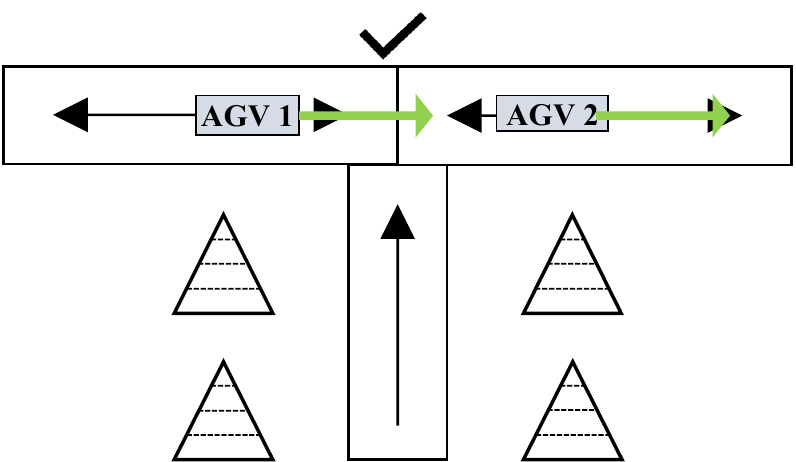} & \includegraphics[align=c, scale=0.4]{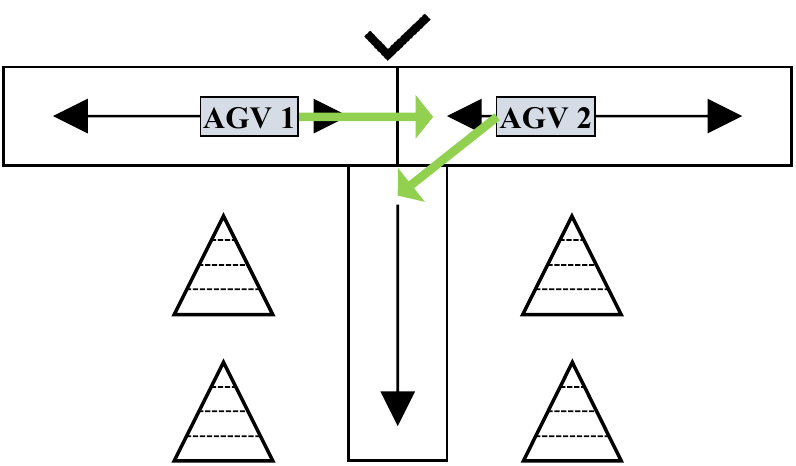} & \includegraphics[align=c, scale=0.35]{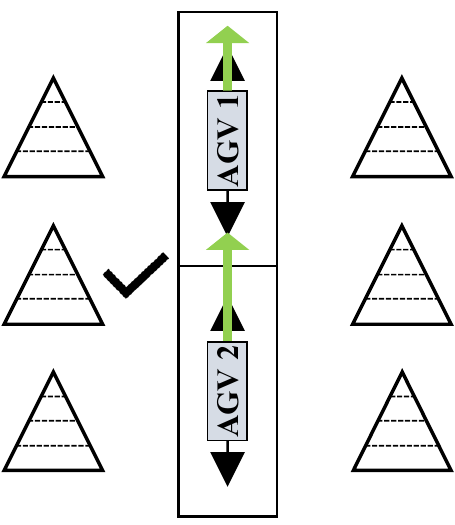} & \includegraphics[align=c, scale=0.4]{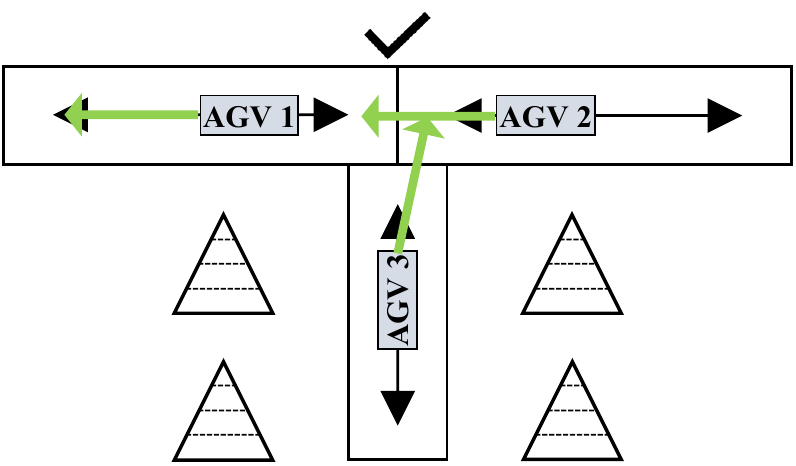}\\ 
        \bottomrule
    \end{tabular}
\end{table}

\begin{figure}[h]
    \centering
    \includegraphics[width=1\linewidth]{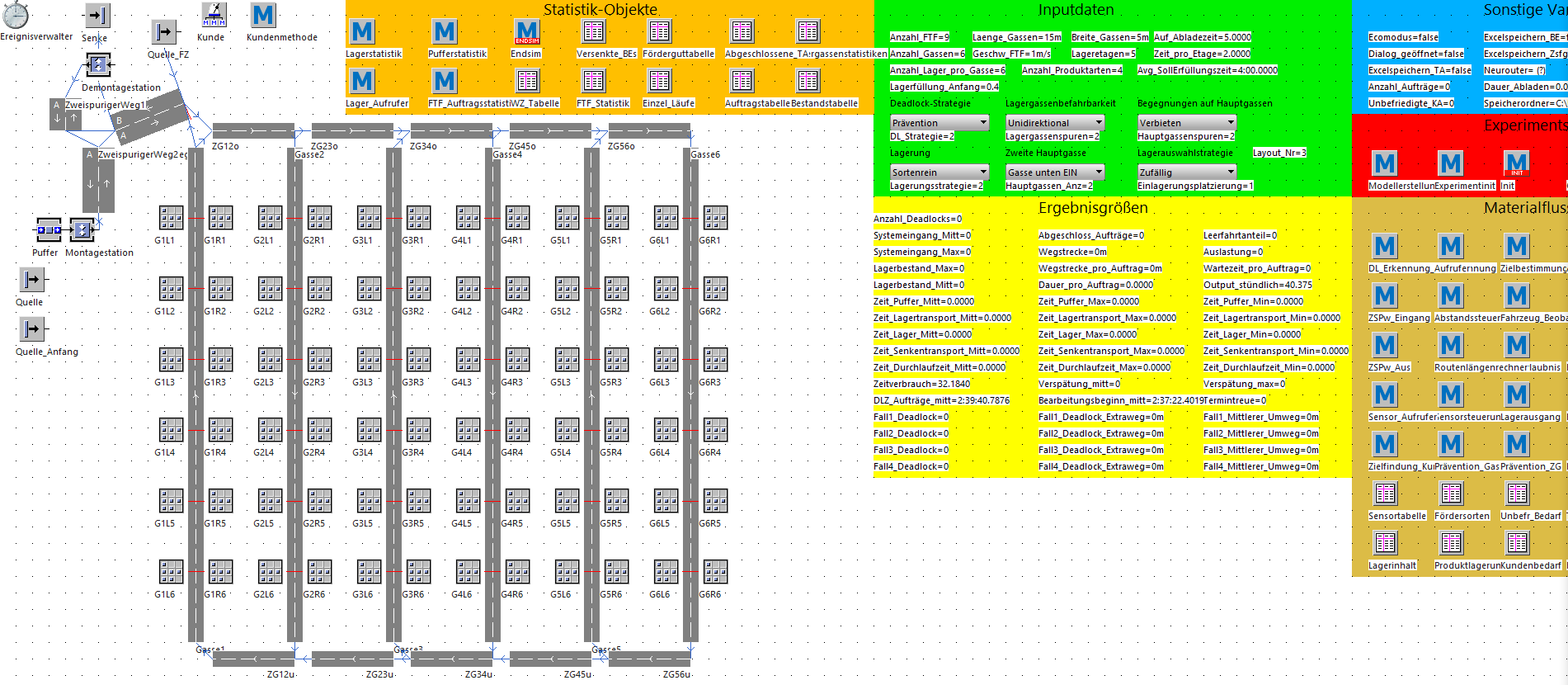}
    \caption{Simulation model of the considered warehouse based on \cite{Muller.2020}.}
    \label{fig:wsc2020-simulation-model}
\end{figure}

The experiments are conducted using the simulation software \enquote{Plant Simulation} version 15.1.1, with methods programmed in the integrated language \enquote{SimTalk 2.0}. Figure \ref{fig:wsc2020-simulation-model} provides an overview of the simulation model. The way sections and storage blocks on the left side (green area) are dynamically created at the beginning of a simulation run based on input parameters. The orange area comprises tables and methods for calculating statistics, such as completed transport orders and the accumulation of AGV waiting times. The yellow area displays all relevant results of a simulation run. The brown area contains most of the methods and is responsible for controlling the material flow, including the calculation of the shortest path, distribution of transport orders, and deadlock handling methods. The blue area is dedicated to the control of experiments and the initialization of the simulation model. The blue area also includes miscellaneous variables, primarily used for exporting data to Excel.

Table \ref{tab:wsc2020-adjust-parameters} outlines the adjustment parameters for the first series of experiments. Each experiment has a simulation period of eight hours, with ten simulation runs performed per experiment. The experiments aim to evaluate deadlock handling strategies and warehouse performance by comparing different combinations of strategies and layout designs based on varying lengths of storage aisles and the number of AGVs.

\begin{table}[h]
    \centering
    \caption{Adjustment parameters for the first series of experiments based on \cite{Muller.2020}.}
    \begin{tabular}{m{2.3cm}cccc}
    \toprule
        \textbf{Deadlock \mbox{handling}} & \textbf{Layouts} & \textbf{Number of AGVs} &  \textbf{Length of aisles} & \textbf{Combinations} \\ \hline
         Prevention & 1, 5, 6 & 3, 4, 5, 6, 7, 8, 9 & 15 m, 30 m, 45 m & \textbf{63} \\
         Avoidance & 1, 2, 3, 4, 5, 6 & 3, 4, 5, 6, 7, 8, 9 & 15 m, 30 m, 45 m & \textbf{126}\\
         Detection \mbox{and recovery} & 1, 3, 4  & 3, 4, 5, 6, 7, 8, 9 & 15 m, 30 m, 45 m & \textbf{63}\\
         &  &  & Total experiments & \textbf{252} \\
    \bottomrule
    \end{tabular}
    
    \label{tab:wsc2020-adjust-parameters}
\end{table}

In a second series of experiments, the influence of varying the number of storage aisles (4, 6, 8, 10) is investigated. The number of AGVs is fixed at seven, while all other adjustment parameters remain unchanged. This second series comprises 144 experiments, each with 10 observations.

The simulation results provide insights into the impact of deadlock handling on the overall performance of intralogistics systems. The findings show that scenarios with effective deadlock handling strategies lead to higher throughput and reduced waiting times compared to scenarios with other deadlock handling strategies. 

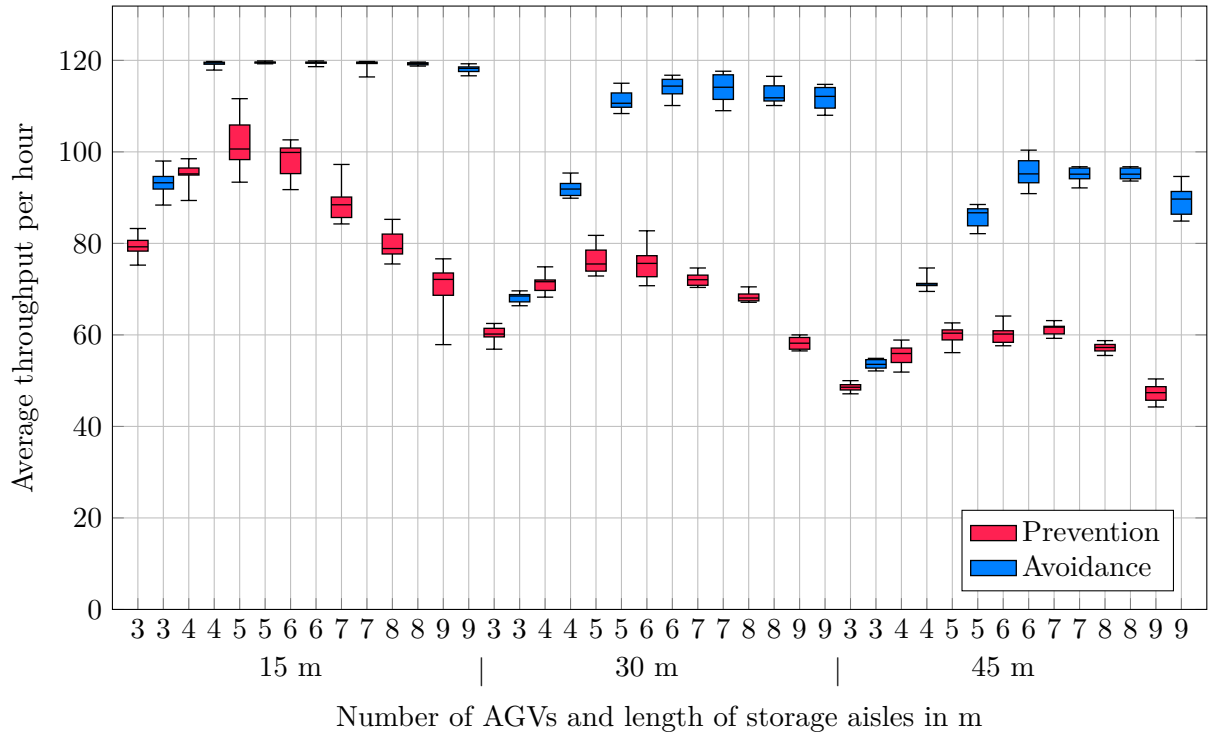
\begin{figure}[h]
    \centering
    \begin{tikzpicture}
        \begin{axis}[
            width=0.9\textwidth,
            height=8cm,
            boxplot/draw direction=y,
            ylabel={Average throughput per hour},
            ymin=0,
            xlabel={Number of AGVs and length of storage aisles in m},
            xtick={1,2,...,42},
            xticklabels={
                {3}, {3}, {4}, {4}, {5}, {5}, {6}, {6}, {7}, {7}, {8}, {8}, {9}, {9},
                {3}, {3}, {4}, {4}, {5}, {5}, {6}, {6}, {7}, {7}, {8}, {8}, {9}, {9},
                {3}, {3}, {4}, {4}, {5}, {5}, {6}, {6}, {7}, {7}, {8}, {8}, {9}, {9}
            },
            x tick label style={anchor=north},
            extra x ticks = {7, 14.5, 21, 28.5, 35},
            extra x tick labels = {15 m, $|$, 30 m, $|$, 45 m},
            extra x tick style = {tick label style={yshift=-0.5cm}, tick style= {draw=none}, grid=none},
            xmin = 0,
            xmax = 43,
            area legend,
            legend pos=south east,
            legend image post style={solid}
        ]
        
        \addplot+ [boxplot prepared={
            lower whisker=75.25, lower quartile=78.3125,
            median=79.25, upper quartile=80.65625,
            upper whisker=83.25},
        fill=redMM, draw=black, line width=0.5pt] coordinates {};
        
        \addplot+ [boxplot prepared={
            lower whisker=88.375, lower quartile=91.875,
            median=93.25, upper quartile=94.625,
            upper whisker=98},
        fill=blueMM, draw=black, line width=0.5pt] coordinates {};
        
        \addplot+ [boxplot prepared={
            lower whisker=89.375, lower quartile=94.9375,
            median=95.1875, upper quartile=96.46875,
            upper whisker=98.5},
        fill=redMM, draw=black, line width=0.5pt] coordinates {};
        
        \addplot+ [boxplot prepared={
            lower whisker=117.875, lower quartile=119.15625,
            median=119.4375, upper quartile=119.59375,
            upper whisker=119.75},
        fill=blueMM, draw=black, line width=0.5pt] coordinates {};
        
        \addplot+ [boxplot prepared={
            lower whisker=93.375, lower quartile=98.3125,
            median=100.625, upper quartile=105.875,
            upper whisker=111.625},
        fill=redMM, draw=black, line width=0.5pt] coordinates {};
        
        \addplot+ [boxplot prepared={
            lower whisker=119.25, lower quartile=119.40625,
            median=119.5625, upper quartile=119.625,
            upper whisker=119.875},
        fill=blueMM, draw=black, line width=0.5pt] coordinates {};
        
        \addplot+ [boxplot prepared={
            lower whisker=91.75, lower quartile=95.25,
            median=99.875, upper quartile=100.84375,
            upper whisker=102.625},
        fill=redMM, draw=black, line width=0.5pt] coordinates {};
        
        \addplot+ [boxplot prepared={
            lower whisker=118.625, lower quartile=119.375,
            median=119.5, upper quartile=119.625,
            upper whisker=119.875},
        fill=blueMM, draw=black, line width=0.5pt] coordinates {};
        
        \addplot+ [boxplot prepared={
            lower whisker=84.25, lower quartile=85.65625,
            median=88.4375, upper quartile=90.125,
            upper whisker=97.25},
        fill=redMM, draw=black, line width=0.5pt] coordinates {};
        
        \addplot+ [boxplot prepared={
            lower whisker=116.375, lower quartile=119.3125,
            median=119.5, upper quartile=119.625,
            upper whisker=119.75},
        fill=blueMM, draw=black, line width=0.5pt] coordinates {};
        
        \addplot+ [boxplot prepared={
            lower whisker=75.5, lower quartile=77.6875,
            median=78.875, upper quartile=82.03125,
            upper whisker=85.25},
        fill=redMM, draw=black, line width=0.5pt] coordinates {};
        
        \addplot+ [boxplot prepared={
            lower whisker=118.75, lower quartile=119.03125,
            median=119.3125, upper quartile=119.46875,
            upper whisker=119.625},
        fill=blueMM, draw=black, line width=0.5pt] coordinates {};
        
        \addplot+ [boxplot prepared={
            lower whisker=57.875, lower quartile=68.65625,
            median=72.125, upper quartile=73.53125,
            upper whisker=76.625},
        fill=redMM, draw=black, line width=0.5pt] coordinates {};
        
        \addplot+ [boxplot prepared={
            lower whisker=116.625, lower quartile=117.5625,
            median=118.25, upper quartile=118.5625,
            upper whisker=119.25},
        fill=blueMM, draw=black, line width=0.5pt] coordinates {};
        
        \addplot+ [boxplot prepared={
            lower whisker=56.875, lower quartile=59.5625,
            median=60.1875, upper quartile=61.4375,
            upper whisker=62.5},
        fill=redMM, draw=black, line width=0.5pt] coordinates {};
        
        \addplot+ [boxplot prepared={
            lower whisker=66.375, lower quartile=67.21875,
            median=68.5, upper quartile=68.8125,
            upper whisker=69.625},
        fill=blueMM, draw=black, line width=0.5pt] coordinates {};
        
        \addplot+ [boxplot prepared={
            lower whisker=68.25, lower quartile=69.71875,
            median=71.625, upper quartile=72,
            upper whisker=74.875},
        fill=redMM, draw=black, line width=0.5pt] coordinates {};
        
        \addplot+ [boxplot prepared={
            lower whisker=89.875, lower quartile=90.46875,
            median=91.875, upper quartile=93.09375,
            upper whisker=95.375},
        fill=blueMM, draw=black, line width=0.5pt] coordinates {};
        
        \addplot+ [boxplot prepared={
            lower whisker=72.875, lower quartile=73.9375,
            median=75.5, upper quartile=78.53125,
            upper whisker=81.75},
        fill=redMM, draw=black, line width=0.5pt] coordinates {};
        
        \addplot+ [boxplot prepared={
            lower whisker=108.375, lower quartile=109.75,
            median=110.625, upper quartile=112.875,
            upper whisker=115},
        fill=blueMM, draw=black, line width=0.5pt] coordinates {};
        
        \addplot+ [boxplot prepared={
            lower whisker=70.75, lower quartile=72.71875,
            median=75.625, upper quartile=77.3125,
            upper whisker=82.75},
        fill=redMM, draw=black, line width=0.5pt] coordinates {};
        
        \addplot+ [boxplot prepared={
            lower whisker=110.125, lower quartile=112.6875,
            median=114.375, upper quartile=115.84375,
            upper whisker=116.75},
        fill=blueMM, draw=black, line width=0.5pt] coordinates {};
        
        \addplot+ [boxplot prepared={
            lower whisker=70.375, lower quartile=70.84375,
            median=72.0625, upper quartile=73.0625,
            upper whisker=74.625},
        fill=redMM, draw=black, line width=0.5pt] coordinates {};
        
        \addplot+ [boxplot prepared={
            lower whisker=109, lower quartile=111.46875,
            median=114.125, upper quartile=116.84375,
            upper whisker=117.625},
        fill=blueMM, draw=black, line width=0.5pt] coordinates {};
        
        \addplot+ [boxplot prepared={
            lower whisker=67.125, lower quartile=67.4375,
            median=68.0625, upper quartile=68.9375,
            upper whisker=70.5},
        fill=redMM, draw=black, line width=0.5pt] coordinates {};
        
        \addplot+ [boxplot prepared={
            lower whisker=110.125, lower quartile=111.125,
            median=111.8125, upper quartile=114.4375,
            upper whisker=116.5},
        fill=blueMM, draw=black, line width=0.5pt] coordinates {};
        
        \addplot+ [boxplot prepared={
            lower whisker=56.5, lower quartile=56.875,
            median=58.1875, upper quartile=59.4375,
            upper whisker=60},
        fill=redMM, draw=black, line width=0.5pt] coordinates {};
        
        \addplot+ [boxplot prepared={
            lower whisker=108, lower quartile=109.5625,
            median=112.125, upper quartile=114.0625,
            upper whisker=114.75},
        fill=blueMM, draw=black, line width=0.5pt] coordinates {};
        
        \addplot+ [boxplot prepared={
            lower whisker=47.125, lower quartile=47.96875,
            median=48.5625, upper quartile=49.125,
            upper whisker=50},
        fill=redMM, draw=black, line width=0.5pt] coordinates {};
        
        \addplot+ [boxplot prepared={
            lower whisker=52.125, lower quartile=52.75,
            median=53.5625, upper quartile=54.59375,
            upper whisker=54.875},
        fill=blueMM, draw=black, line width=0.5pt] coordinates {};
        
        \addplot+ [boxplot prepared={
            lower whisker=51.875, lower quartile=53.96875,
            median=55.9375, upper quartile=57.125,
            upper whisker=58.875},
        fill=redMM, draw=black, line width=0.5pt] coordinates {};
        
        \addplot+ [boxplot prepared={
            lower whisker=69.5, lower quartile=70.78125,
            median=71.125, upper quartile=71.25,
            upper whisker=74.625},
        fill=blueMM, draw=black, line width=0.5pt] coordinates {};
        
        \addplot+ [boxplot prepared={
            lower whisker=56.125, lower quartile=58.90625,
            median=60.375, upper quartile=61.09375,
            upper whisker=62.625},
        fill=redMM, draw=black, line width=0.5pt] coordinates {};
        
        \addplot+ [boxplot prepared={
            lower whisker=82.125, lower quartile=83.84375,
            median=86.6875, upper quartile=87.5625,
            upper whisker=88.5},
        fill=blueMM, draw=black, line width=0.5pt] coordinates {};
        
        \addplot+ [boxplot prepared={
            lower whisker=57.625, lower quartile=58.375,
            median=60.1875, upper quartile=60.90625,
            upper whisker=64.125},
        fill=redMM, draw=black, line width=0.5pt] coordinates {};
        
        \addplot+ [boxplot prepared={
            lower whisker=90.875, lower quartile=93.25,
            median=95.1875, upper quartile=98.0625,
            upper whisker=100.375},
        fill=blueMM, draw=black, line width=0.5pt] coordinates {};
        
        \addplot+ [boxplot prepared={
            lower whisker=59.25, lower quartile=60.21875,
            median=61.6875, upper quartile=61.875,
            upper whisker=63.125},
        fill=redMM, draw=black, line width=0.5pt] coordinates {};
        
        \addplot+ [boxplot prepared={
            lower whisker=92.125, lower quartile=94.125,
            median=95.125, upper quartile=96.46875,
            upper whisker=96.75},
        fill=blueMM, draw=black, line width=0.5pt] coordinates {};
        
        \addplot+ [boxplot prepared={
            lower whisker=55.5, lower quartile=56.5,
            median=57.25, upper quartile=57.90625,
            upper whisker=58.75},
        fill=redMM, draw=black, line width=0.5pt] coordinates {};
        
        \addplot+ [boxplot prepared={
            lower whisker=93.625, lower quartile=94.125,
            median=95.125, upper quartile=96.46875,
            upper whisker=96.75},
        fill=blueMM, draw=black, line width=0.5pt] coordinates {};
        
        \addplot+ [boxplot prepared={
            lower whisker=44.25, lower quartile=45.71875,
            median=47.375, upper quartile=48.6875,
            upper whisker=50.375},
        fill=redMM, draw=black, line width=0.5pt] coordinates {};
        
        \addplot+ [boxplot prepared={
            lower whisker=84.875, lower quartile=86.375,
            median=89.6875, upper quartile=91.34375,
            upper whisker=94.625},
        fill=blueMM, draw=black, line width=0.5pt] coordinates {};
        
        \addlegendentry{Prevention}
        \addlegendentry{Avoidance}
        \end{axis}
    \end{tikzpicture}
    \caption{Throughput of different configurations for layout 5 according to \cite{Muller.2020}.}
    \label{fig:wsc2020-boxplot-layout5}
\end{figure}

Figure \ref{fig:wsc2020-boxplot-layout5} shows the average throughput per hour for layout 5 across different experiment configurations, varying the number of AGVs and the length of storage aisles. The configurations include both prevention and avoidance strategies for deadlock handling. The results indicate a distinct behavior in throughput as the number of AGVs increases. For the prevention strategy, throughput significantly decreases once the number of AGVs exceeds approximately 5-7. This decline is attributed to the increased average waiting time per transport order, which escalates with more AGVs. Neither the prevention nor the avoidance strategy achieves maximum throughput when storage aisle lengths are set to 30 meters or 45 meters. The performance difference between the two strategies varies depending on the specific configuration, highlighting the impact of layout design and AGV numbers on the efficiency of deadlock handling strategies. The avoidance strategy tends to maintain higher throughput levels in scenarios with longer storage aisles and more AGVs, while the prevention strategy struggles to cope with the increased traffic, leading to significant throughput reductions.

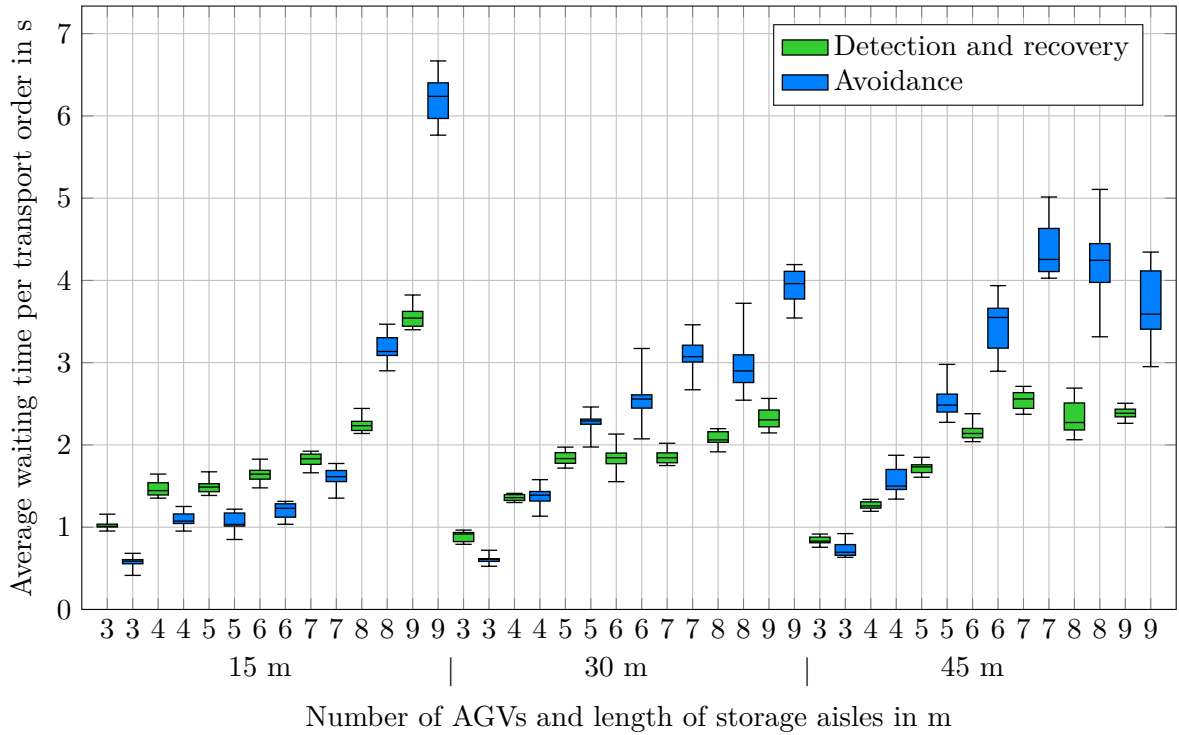
\begin{figure}[h]
    \centering
    \begin{tikzpicture}
        \begin{axis}[
            width=0.9\textwidth,
            height=8cm,
            boxplot/draw direction=y,
            ylabel={Average waiting time per transport order in s},
            ymin=0,
            xlabel={Number of AGVs and length of storage aisles in m},
            xtick={1,2,...,42},
            xticklabels={
                {3}, {3}, {4}, {4}, {5}, {5}, {6}, {6}, {7}, {7}, {8}, {8}, {9}, {9},
                {3}, {3}, {4}, {4}, {5}, {5}, {6}, {6}, {7}, {7}, {8}, {8}, {9}, {9},
                {3}, {3}, {4}, {4}, {5}, {5}, {6}, {6}, {7}, {7}, {8}, {8}, {9}, {9}
            },
            x tick label style={anchor=north},
            extra x ticks = {7, 14.5, 21, 28.5, 35},
            extra x tick labels = {15 m, $|$, 30 m, $|$, 45 m},
            extra x tick style = {tick label style={yshift=-0.5cm}, tick style= {draw=none}, grid=none},
            xmin = 0,
            xmax = 43,
            area legend,
            legend pos=north east,
            legend image post style={solid}
        ]

        \addplot+ [boxplot prepared={
            lower whisker=0.9533, lower quartile=1.0023,
            median=1.0104, upper quartile=1.0368,
            upper whisker=1.1583},
        fill=greenMM, draw=black, line width=0.5pt] coordinates {};
        
        \addplot+ [boxplot prepared={
            lower whisker=0.4141, lower quartile=0.555325,
            median=0.58665, upper quartile=0.605775,
            upper whisker=0.6824},
        fill=blueMM, draw=black, line width=0.5pt] coordinates {};
        
        \addplot+ [boxplot prepared={
            lower whisker=1.3531, lower quartile=1.391675,
            median=1.44465, upper quartile=1.54035,
            upper whisker=1.6454},
        fill=greenMM, draw=black, line width=0.5pt] coordinates {};
        
        \addplot+ [boxplot prepared={
            lower whisker=0.9527, lower quartile=1.04495,
            median=1.07415, upper quartile=1.16095,
            upper whisker=1.2517},
        fill=blueMM, draw=black, line width=0.5pt] coordinates {};
        
        \addplot+ [boxplot prepared={
            lower whisker=1.3858, lower quartile=1.4317,
            median=1.4884, upper quartile=1.528375,
            upper whisker=1.6732},
        fill=greenMM, draw=black, line width=0.5pt] coordinates {};
        
        \addplot+ [boxplot prepared={
            lower whisker=0.8505, lower quartile=1.01335,
            median=1.03355, upper quartile=1.1724,
            upper whisker=1.2191},
        fill=blueMM, draw=black, line width=0.5pt] coordinates {};
        
        \addplot+ [boxplot prepared={
            lower whisker=1.4781, lower quartile=1.5839,
            median=1.64375, upper quartile=1.691225,
            upper whisker=1.8271},
        fill=greenMM, draw=black, line width=0.5pt] coordinates {};
        
        \addplot+ [boxplot prepared={
            lower whisker=1.0353, lower quartile=1.12145,
            median=1.22965, upper quartile=1.28435,
            upper whisker=1.314},
        fill=blueMM, draw=black, line width=0.5pt] coordinates {};
        
        \addplot+ [boxplot prepared={
            lower whisker=1.6614, lower quartile=1.76405,
            median=1.83105, upper quartile=1.8882,
            upper whisker=1.9238},
        fill=greenMM, draw=black, line width=0.5pt] coordinates {};
        
        \addplot+ [boxplot prepared={
            lower whisker=1.3532, lower quartile=1.55455,
            median=1.6143, upper quartile=1.689225,
            upper whisker=1.7744},
        fill=blueMM, draw=black, line width=0.5pt] coordinates {};
        
        \addplot+ [boxplot prepared={
            lower whisker=2.1395, lower quartile=2.17575,
            median=2.2341, upper quartile=2.28665,
            upper whisker=2.4436},
        fill=greenMM, draw=black, line width=0.5pt] coordinates {};
        
        \addplot+ [boxplot prepared={
            lower whisker=2.9013, lower quartile=3.087725,
            median=3.13605, upper quartile=3.303575,
            upper whisker=3.4675},
        fill=blueMM, draw=black, line width=0.5pt] coordinates {};
        
        \addplot+ [boxplot prepared={
            lower whisker=3.4007, lower quartile=3.442975,
            median=3.5422, upper quartile=3.623725,
            upper whisker=3.8228},
        fill=greenMM, draw=black, line width=0.5pt] coordinates {};
        
        \addplot+ [boxplot prepared={
            lower whisker=5.7659, lower quartile=5.9687,
            median=6.23805, upper quartile=6.402,
            upper whisker=6.6683},
        fill=blueMM, draw=black, line width=0.5pt] coordinates {};
        
        \addplot+ [boxplot prepared={
            lower whisker=0.7921, lower quartile=0.82475,
            median=0.91745, upper quartile=0.935625,
            upper whisker=0.9647},
        fill=greenMM, draw=black, line width=0.5pt] coordinates {};
        
        \addplot+ [boxplot prepared={
            lower whisker=0.525, lower quartile=0.583825,
            median=0.6082, upper quartile=0.617675,
            upper whisker=0.7196},
        fill=blueMM, draw=black, line width=0.5pt] coordinates {};
        
        \addplot+ [boxplot prepared={
            lower whisker=1.2999, lower quartile=1.32505,
            median=1.359, upper quartile=1.3965,
            upper whisker=1.4111},
        fill=greenMM, draw=black, line width=0.5pt] coordinates {};
        
        \addplot+ [boxplot prepared={
            lower whisker=1.1337, lower quartile=1.317825,
            median=1.3891, upper quartile=1.4339,
            upper whisker=1.5778},
        fill=blueMM, draw=black, line width=0.5pt] coordinates {};
        
        \addplot+ [boxplot prepared={
            lower whisker=1.7182, lower quartile=1.77675,
            median=1.8342, upper quartile=1.906425,
            upper whisker=1.9733},
        fill=greenMM, draw=black, line width=0.5pt] coordinates {};
        
        \addplot+ [boxplot prepared={
            lower whisker=1.9747, lower quartile=2.25175,
            median=2.29125, upper quartile=2.314225,
            upper whisker=2.4614},
        fill=blueMM, draw=black, line width=0.5pt] coordinates {};
        
        \addplot+ [boxplot prepared={
            lower whisker=1.5535, lower quartile=1.77205,
            median=1.8446, upper quartile=1.9014,
            upper whisker=2.1323},
        fill=greenMM, draw=black, line width=0.5pt] coordinates {};
        
        \addplot+ [boxplot prepared={
            lower whisker=2.0738, lower quartile=2.4477,
            median=2.55785, upper quartile=2.609125,
            upper whisker=3.1725},
        fill=blueMM, draw=black, line width=0.5pt] coordinates {};
        
        \addplot+ [boxplot prepared={
            lower whisker=1.7495, lower quartile=1.783375,
            median=1.8445, upper quartile=1.9034,
            upper whisker=2.0202},
        fill=greenMM, draw=black, line width=0.5pt] coordinates {};
        
        \addplot+ [boxplot prepared={
            lower whisker=2.6706, lower quartile=3.008425,
            median=3.073, upper quartile=3.21255,
            upper whisker=3.4609},
        fill=blueMM, draw=black, line width=0.5pt] coordinates {};
        
        \addplot+ [boxplot prepared={
            lower whisker=1.9162, lower quartile=2.03125,
            median=2.0617, upper quartile=2.160775,
            upper whisker=2.1973},
        fill=greenMM, draw=black, line width=0.5pt] coordinates {};
        
        \addplot+ [boxplot prepared={
            lower whisker=2.5439, lower quartile=2.758475,
            median=2.9005, upper quartile=3.095425,
            upper whisker=3.7223},
        fill=blueMM, draw=black, line width=0.5pt] coordinates {};
        
        \addplot+ [boxplot prepared={
            lower whisker=2.146, lower quartile=2.220175,
            median=2.30475, upper quartile=2.42435,
            upper whisker=2.5655},
        fill=greenMM, draw=black, line width=0.5pt] coordinates {};
        
        \addplot+ [boxplot prepared={
            lower whisker=3.5432, lower quartile=3.773975,
            median=3.96105, upper quartile=4.110275,
            upper whisker=4.1925},
        fill=blueMM, draw=black, line width=0.5pt] coordinates {};
        
        \addplot+ [boxplot prepared={
            lower whisker=0.7559, lower quartile=0.8106,
            median=0.82995, upper quartile=0.879425,
            upper whisker=0.9167},
        fill=greenMM, draw=black, line width=0.5pt] coordinates {};
        
        \addplot+ [boxplot prepared={
            lower whisker=0.6339, lower quartile=0.6585,
            median=0.69555, upper quartile=0.788525,
            upper whisker=0.9221},
        fill=blueMM, draw=black, line width=0.5pt] coordinates {};
        
        \addplot+ [boxplot prepared={
            lower whisker=1.193, lower quartile=1.231925,
            median=1.2591, upper quartile=1.308025,
            upper whisker=1.3389},
        fill=greenMM, draw=black, line width=0.5pt] coordinates {};
        
        \addplot+ [boxplot prepared={
            lower whisker=1.3414, lower quartile=1.4582,
            median=1.50025, upper quartile=1.702,
            upper whisker=1.8742},
        fill=blueMM, draw=black, line width=0.5pt] coordinates {};
        
        \addplot+ [boxplot prepared={
            lower whisker=1.6073, lower quartile=1.663975,
            median=1.73285, upper quartile=1.758825,
            upper whisker=1.8494},
        fill=greenMM, draw=black, line width=0.5pt] coordinates {};
        
        \addplot+ [boxplot prepared={
            lower whisker=2.2747, lower quartile=2.40005,
            median=2.4841, upper quartile=2.6179,
            upper whisker=2.9792},
        fill=blueMM, draw=black, line width=0.5pt] coordinates {};
        
        \addplot+ [boxplot prepared={
            lower whisker=2.0399, lower quartile=2.0869,
            median=2.1392, upper quartile=2.200525,
            upper whisker=2.3796},
        fill=greenMM, draw=black, line width=0.5pt] coordinates {};
        
        \addplot+ [boxplot prepared={
            lower whisker=2.8957, lower quartile=3.176875,
            median=3.54975, upper quartile=3.661175,
            upper whisker=3.9365},
        fill=blueMM, draw=black, line width=0.5pt] coordinates {};
        
        \addplot+ [boxplot prepared={
            lower whisker=2.3729, lower quartile=2.444575,
            median=2.55905, upper quartile=2.635925,
            upper whisker=2.7121},
        fill=greenMM, draw=black, line width=0.5pt] coordinates {};
        
        \addplot+ [boxplot prepared={
            lower whisker=4.0271, lower quartile=4.107725,
            median=4.25555, upper quartile=4.63185,
            upper whisker=5.0142},
        fill=blueMM, draw=black, line width=0.5pt] coordinates {};
        
        \addplot+ [boxplot prepared={
            lower whisker=2.063, lower quartile=2.181825,
            median=2.2726, upper quartile=2.510475,
            upper whisker=2.6908},
        fill=greenMM, draw=black, line width=0.5pt] coordinates {};
        
        \addplot+ [boxplot prepared={
            lower whisker=3.3149, lower quartile=3.975875,
            median=4.24405, upper quartile=4.447575,
            upper whisker=5.1064},
        fill=blueMM, draw=black, line width=0.5pt] coordinates {};
        
        \addplot+ [boxplot prepared={
            lower whisker=2.2629, lower quartile=2.341225,
            median=2.38485, upper quartile=2.434875,
            upper whisker=2.5063},
        fill=greenMM, draw=black, line width=0.5pt] coordinates {};
        
        \addplot+ [boxplot prepared={
            lower whisker=2.9518, lower quartile=3.406775,
            median=3.5905, upper quartile=4.115625,
            upper whisker=4.3443},
        fill=blueMM, draw=black, line width=0.5pt] coordinates {};
        
        \addlegendentry{Detection and recovery}
        \addlegendentry{Avoidance}
        \end{axis}
    \end{tikzpicture}
    \caption{Average waiting time for different configurations of layout 3.}
    \label{fig:wsc2020-boxplot-layout3}
\end{figure}

Another example for significant deviations gives Figure \ref{fig:wsc2020-boxplot-layout3}, which shows the average waiting time for layout 3, highlighting the differences between the detection and recovery strategy and the avoidance strategy. For experiments with shorter storage aisles (15 m) and up to six AGVs, the detection and recovery strategy exhibits higher average waiting times compared to the avoidance strategy. This may be because, under the avoidance strategy, vehicles only need to wait briefly for a clear route, whereas in the detection and recovery strategy, AGVs may have to wait longer behind loading or unloading vehicles. The significance of this effect in the shortest aisle length suggests that reserved route sections are released relatively quickly. In the remaining experiments with 15 m aisles, the detection and recovery strategy results in lower waiting times. For configurations with longer storage aisles (30 m and 45 m), the detection and recovery strategy consistently outperforms the avoidance strategy, showing significantly lower waiting times for four AGVs in 30 m aisles and five AGVs in 45 m aisles.

\cite{Muller.2020} demonstrate that no single strategy dominates in deadlock handling. Instead, the effectiveness of a strategy hinges on the specific warehouse layout and operational parameters. In their experiments, layouts with unidirectional storage aisles, which easily prevent deadlocks, perform worse compared to layouts with bidirectional aisles where deadlock avoidance and detection and recovery strategies are necessary. The results show that the avoidance strategy generally achieves higher throughput than prevention, especially in more complex and dense traffic scenarios. In environments with fewer AGVs, detection and recovery strategies prove to be more efficient due to shorter waiting times. The detection and recovery of deadlocks lead to additional detours and therefore to a higher wear of the AGVs. Over an eight-hour simulation period, hundreds of deadlocks are recovered, with an average detour of approximately 20 meters for each recovered deadlock, leading to significant cumulative detours.

\cite{Muller.2021} extend these findings by investigating the impact of disruptions on deadlock handling strategies within an AGV system. Their research compares the performance of the three deadlock handling strategies under varying disruption profiles characterized by different levels of AGV availability and mean time to repair (MTTR). The simulation model employed in the article is based on \cite{Muller.2020} and adds several disruption profiles, such as frequent short disruptions and infrequent long disruptions. Table \ref{tab:asim2021-adjust-parameters} outlines the parameters for the series of experiments conducted by \cite{Muller.2021}. Each experiment involved a simulation period of eight hours, with ten simulation runs performed per experiment. The experiments evaluate deadlock handling strategies and AGV system performance under varying conditions of AGV availability and MTTR.

\begin{table}[h]
    \centering
    \caption{Adjustment parameters for the disruption experiments.}
    \label{tab:asim2021-adjust-parameters}
    \begin{tabular}{lcc}
    \toprule
        \textbf{Parameter} & \textbf{Value range} & \textbf{Experiments} \\ \hline
         Number of AGVs & 5 - 10 & 6 \\
         Deadlock handling & Prevention, avoidance, detection and recovery & 3 \\
         Availability & 91\%, 95\%, 99\% & 3 \\
         MTTR & 5, 30, 120 in min & 3 \\
    \bottomrule
    \end{tabular}
\end{table}

The avoidance strategy consistently yielded the best throughput per hour across various disruption scenarios. Specifically, under conditions of high AGV availability with 99\% and a low MTTR of 5 minutes, the avoidance strategy maximized system throughput, achieving up to 120 units per hour with seven AGVs. The prevention strategy lagged, unable to reach this throughput even with 10 AGVs. This strategy also resulted in the longest average distance traveled per transport order and the highest average waiting time. Figure \ref{fig:asim2021-diagrams} shows the simulation results of a disruption scenario with a low availability of the AGVs and presents the key figures throughput per hour, average distance per transport, and average waiting time per transport.

\begin{figure}[h]
    
    \begin{minipage}{.33\textwidth}
        \centering
        \begin{tikzpicture}
            \begin{axis}[
                height = 3 cm,
                xlabel={MTTR in min},
                ylabel={Throughput per hour},
                ymin = 0,
                xmin = 0,
                xtick={5, 30, 120},
                legend style={at={(0.3,1.2)},
                anchor=south,},
                every axis plot/.append style={thick}
            ]
            
            \addplot[color=redMM,mark=*] coordinates {(5,82.175) (30,63.9375) (120,58.625)};    
            \addplot[color=blueMM,mark=triangle*] coordinates {(5,116.9375) (30,110.925) (120,117.6875)};
            \addplot[color=greenMM,mark=square*] coordinates {(5,116.875) (30,98.075) (120,94.325)};

            \legend{Prevention, Avoidance, Detection and recovery}
            
            \end{axis}
        \end{tikzpicture}
        \caption*{10 AGVs\\91\% availability}
    \end{minipage}
    \begin{minipage}{.33\textwidth}
        \centering
        \captionsetup{justification=centering}
        \begin{tikzpicture}
            \begin{axis}[
                height = 3 cm,
                xlabel={Number of AGVs},
                ylabel={Average distance in m},
                ymin = 0,
                xtick={5, 6, 7, 8, 9, 10},
                legend style={at={(0.3,1.2)},
                anchor=south,},
                every axis plot/.append style={thick}
            ]
            
            \addplot[color=redMM,mark=*] coordinates {(5,114.85671) (6,115.38236) (7,115.66576) (8,114.56136) (9,113.18571) (10,111.45881)};
            \addplot[color=blueMM,mark=triangle*] coordinates {(5,74.67498) (6,76.25533) (7,78.52111) (8,81.03909) (9,81.84748) (10,83.36406)};
            \addplot[color=greenMM,mark=square*] coordinates {(5,73.94707) (6,75.86203) (7,80.33491) (8,82.83766) (9,84.02977) (10,84.81017)};            
            
            \legend{Prevention, Avoidance, Detection and recovery}
            
            \end{axis}
        \end{tikzpicture}
        \caption*{120 min MTTR\\91\% availability}
    \end{minipage}
    \begin{minipage}{.33\textwidth}
        \centering
        \captionsetup{justification=centering}
        \begin{tikzpicture}
            \begin{axis}[
                height = 3 cm,
                xlabel={Number of AGVs},
                ylabel={Avg. waiting time in s},
                ymin = 0,
                xtick={5, 6, 7, 8, 9, 10},
                legend style={at={(0.3,1.2)},
                anchor=south,},
                every axis plot/.append style={thick}
            ]
            
                \addplot[color=redMM,mark=*] coordinates {(5,96.50701) (6,111.83928) (7,146.5356) (8,178.66425) (9,326.42562) (10,329.38043)};
                \addplot[color=blueMM,mark=triangle*] coordinates {(5,7.14568) (6,8.87023) (7,11.87424) (8,14.82774) (9,22.4239) (10,20.63443)};
                \addplot[color=greenMM,mark=square*] coordinates {(5,20.63278) (6,32.46589) (7,26.85559) (8,40.53768) (9,41.82985) (10,55.33617)};

            \legend{Prevention, Avoidance, Detection and recovery}
            
            \end{axis}
        \end{tikzpicture}
        \caption*{120 min MTTR\\91\% availability}
    \end{minipage}
    \caption{Throughput per hour, average distance per transport, and average waiting time per transport for a low availability of AGVs according to \cite{Muller.2021}.}
    \label{fig:asim2021-diagrams}
\end{figure}

When disruptions were frequent and severe with a low availability at 91\% and high MTTR of 120 minutes, the avoidance strategy maintained its superior performance, showing a clear advantage in throughput over the detection and recovery strategy, which was only comparable in less severe disruption scenarios. Notably, the detection and recovery strategy performed well in environments with fewer and shorter disruptions but was less effective as the frequency and severity of disruptions increased. This strategy also led to significant cumulative detours, impacting the overall efficiency and wear on the AGVs.

The prevention strategy, while effective at eliminating deadlocks, performed poorly in terms of key logistical metrics such as system throughput, average waiting time, and average distance traveled per transport order. The simulations revealed that the prevention strategy's rigid approach to avoiding deadlocks compromised overall system performance, particularly in high-traffic scenarios.

\cite{Muller.2021} highlight the importance of selecting a deadlock handling strategy that aligns with the specific operational environment and disruption profile. Their findings emphasize the need for flexible and adaptive strategies in AGV systems to optimize performance and maintain efficiency under varying conditions. The avoidance strategy emerged as the most robust and effective across different disruption scenarios, demonstrating the highest throughput and lowest waiting times, especially in complex and high-density environments.

These findings show the importance of the comparison of different deadlock handling strategies in intralogistics systems. Effective handling of deadlocks can lead to substantial improvements in throughput, task completion times, and overall system resilience. Correct deadlock handling reduces operational disruptions and associated costs, demonstrating clear economic benefits. Therefore, integrating deadlock handling strategies into intralogistics planning is essential for maintaining efficient and resilient operations. The choice of the correct deadlock handling strategy is important and answers the RQ 3.1 positively.

The analysis affirms also the general importance of deadlock handling in intralogistics systems, addressing RQ 1.1 positively. Deadlocks can have a high impact on the performance of a intralogistics system. The research emphasizes the necessity of developing and implementing robust deadlock handling solutions tailored to the specific needs and configurations of intralogistics environments. These results provide a foundation for further research into advanced deadlock handling policies, such as provided by MARL.

\section{Deadlock Consideration and RL for Logistics Planning} \label{sec:methodology-logisticsplanning}
\subsection{Resilience as the fourth Logistics Objective}
To address RQ 1.2, "How to integrate RL into the planning process?", this section explores the systematic incorporation of RL techniques into logistics planning. To enhance the consideration of deadlocks in logistics planning, it is essential to recognize that the problem of deadlocks is often overlooked in the planning process. Even in the fundamental logistics objectives defined by \citet[p. 74]{Gudehus.2010}, which include performance, quality, and costs, the disruption-free and reliable nature of a system is not explicitly addressed. Deadlocks can adversely affect all three objectives. To address this oversight and encourage proactive thinking about deadlocks in the planning process, this thesis proposes adding a fourth objective: resilience.

Resilience, in the context of logistics, refers to the ability of a system to withstand, adapt to, and recover from disruptions, ensuring continuous and reliable operations. Incorporating resilience as a key objective acknowledges the importance of designing logistics systems that can handle unexpected events and maintain operational stability.

Integrating resilience into logistics planning helps mitigate the impact of deadlocks on performance, quality, and costs. Performance can suffer when deadlocks cause delays and reduce throughput. Quality may decline if deadlocks lead to missed delivery schedules or compromised delivery standards. Costs can escalate due to the inefficiencies and resource wastage associated with deadlock resolution.

By explicitly including resilience as a logistics objective, planners can develop strategies beforehand that enhance the system's ability to prevent, avoid or detect, and recover from deadlocks. This proactive approach ensures that logistics operations remain robust and efficient, even in the face of unforeseen challenges.

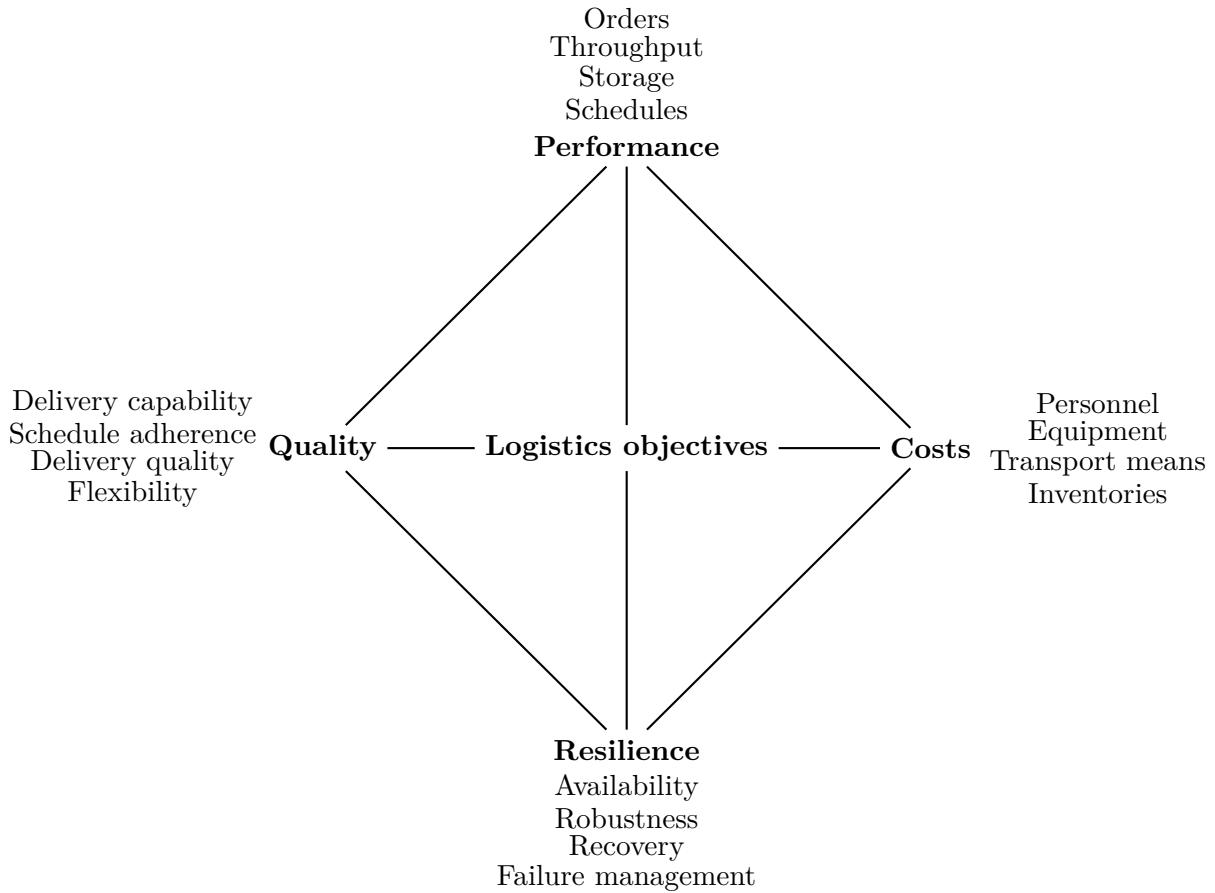
\begin{figure}[h]
    \centering
    \begin{tikzpicture}
        \node (Performance) at (0, 4) {\textbf{Performance}};
        \node (Quality) at (-4, 0) {\textbf{Quality}};
        \node (Cost) at (4, 0) {\textbf{Costs}};
        \node (Resilience) at (0, -4) {\textbf{Resilience}};

        \node at (0, 0) (Objectives) {\textbf{Logistics objectives}};
    
        \draw[thick] (Performance) -- (Quality);
        \draw[thick] (Quality) -- (Resilience);
        \draw[thick] (Resilience) -- (Cost);
        \draw[thick] (Cost) -- (Performance);
        \draw[thick] (Performance) -- (Objectives);
        \draw[thick] (Resilience) -- (Objectives);
        \draw[thick] (Quality) -- (Objectives);
        \draw[thick] (Cost) -- (Objectives);

        \node at (0, 5.7) {Orders};
        \node at (0, 5.3) {Throughput};
        \node at (0, 4.9) {Storage};
        \node at (0, 4.5) {Schedules};
    
        \node[align=right] at (-6.5, 0.6) {Delivery capability};
        \node[align=right] at (-6.5, 0.2) {Schedule adherence};
        \node[align=right] at (-6.5, -0.2) {Delivery quality};
        \node[align=right] at (-6.5, -0.6) {Flexibility};
    
        \node[align=left] at (6.2, 0.6) {Personnel};
        \node[align=left] at (6.2, 0.2) {Equipment};
        \node[align=left] at (6.2, -0.2) {Transport means};
        \node[align=left] at (6.2, -0.6) {Inventories};
    
        \node at (0, -4.5) {Availability};
        \node at (0, -4.9) {Robustness};
        \node at (0, -5.3) {Recovery};
        \node at (0, -5.7) {Failure management};
    \end{tikzpicture}
    \caption{Revised logistics objectives incorporating resilience.}
    \label{fig:logistics_objectives_revised}
\end{figure}

The revised set of logistics objectives should encompass performance, quality, costs, and resilience. This comprehensive framework ensures that logistics systems are not only efficient and cost-effective but also capable of maintaining high standards of reliability and adaptability. Figure \ref{fig:logistics_objectives_revised} expands the triangular model of logistics objectives of \citet[p. 74]{Gudehus.2010} into a quadrangle model, incorporating resilience as the fourth key objective.

The resilience objective in logistics can be broken down into several sub-levels: Availability, recovery, robustness and failure management.

Availability in logistics refers to the system's ability to maintain operational uptime and minimize downtime. This includes ensuring that the system has a high mean time between failures, meaning that the intervals between system failures are long. A high availability rate ensures that the logistics operations can continue smoothly with minimal interruptions, thereby maintaining a consistent flow of goods and services.

Robustness pertains to the system's capacity to withstand disruptions without significant performance degradation. This includes having capacity reserves and redundancies, which provide backup resources that can be utilized in case of a failure. Additionally, robustness involves ensuring that key performance indicators (KPIs) have low sensitivity to sudden changes. In this context, low sensitivity means that these KPIs do not react drastically to unexpected disruptions, allowing the system to maintain stability and reliability even under stress.

Recovery focuses on the system's ability to quickly and effectively return to normal operations following a disruption. This involves minimizing the MTTR. Recovery also includes having the requirements in place to recover a disruption.

Failure management encompasses the strategies and processes used to handle failures when they occur. This includes resource reallocation, where resources are dynamically shifted to areas most in need during a disruption, and the temporary suspension of non-critical processes to prioritize essential operations. Effective failure management ensures that the impact of disruptions is minimized, and critical functions can continue to operate, thereby maintaining overall system integrity and performance.

\subsection{Procedure Model for applying RL in Logistics Planning}
To incorporate the resilience objective effectively and to create a general procedure to apply MARL for intralogistics systems, this thesis introduces a procedure model based on established logistics planning phases. Various procedure models exist in logistics planning, but the main steps are essentially similar. The proposed procedure model is published by \cite{muller2023framework}. This thesis utilizes the planning phases of \citet[p. 70]{Gudehus.2010} for general logistics planning and the approach of \citet[p.122]{fottner2022planung} for an intralogistics-specific context as the foundation of the proposed procedure model. Table \ref{tab:comparison-procedure-models} compares these two planning methodologies and identifies necessary extensions to adequately address deadlocks using RL.

\begin{table}[h]
    \centering
    \caption{Comparison of selected procedure models for logistics system with the necessary extension for deadlock handling with RL based on \cite{muller2023framework}.}
    \label{tab:comparison-procedure-models}
    \begin{tabular}{ccc}
    \toprule
         \textbf{Logistics system} & \textbf{Intralogistic system} & \textbf{Necessary extension for}\\ 
         \cite[p. 70]{Gudehus.2010} & \cite[p. 122]{fottner2022planung} & \textbf{deadlock handling with RL}\\ 
         \hline
         Target planning &  Setting targets/ constraints & Resilience as a logistics objective \\
         &  Requirement analysis & Deadlock consideration\\
         System planning & Concepts/ planning variants & Preventive deadlock design\\
         Detailed planning & System design & Modeling \& RL integration\\
         Tender & Evaluation/ decision & \\
         System construction & Realization & Deadlock solution deployment\\
         System operation & & Monitoring \& continual learning \\
    \bottomrule
    \end{tabular}
\end{table}

In the target planning phase, traditional logistics planning involves setting targets and constraints based on performance, costs, and quality. With the introduction of resilience as a fourth logistics objective, it is essential to integrate resilience into this phase, ensuring that targets and constraints also address the system's ability to withstand and recover from disruptions. The handling of deadlocks and the resulting resilience of the system becomes a central target in the proposed procedure model. The recognition of deadlocks is not only a standalone key figure but a significant influence on other target parameters, such as throughput or lead time. This recognition means that even if the number of deadlocks are not directly considered as a target parameter by the logistics planner, their significant impact on these key figures necessitates their consideration in the planning process.

During the requirement analysis phase, intralogistics systems must explicitly consider potential deadlocks. Identifying and understanding potential deadlock scenarios early ensures that the system is designed to mitigate these issues effectively. This stage could include examining past data to identify conditions that have previously resulted in deadlocks, as well as predicting potential deadlock scenarios based on the proposed system design and operation. Additionally, this phase aims to estimate the degree of disruptions that could occur in the system, helping determine the level of effort needed for effective deadlock avoidance.

In the system planning phase, while traditional planning involves creating rough concepts and planning variants, handling deadlocks may require a preventive design approach. This approach underscores the importance of designing systems with the primary goal of reducing the risk of deadlocks as much as possible. This could be achieved by employing strategies such as optimizing the layout or routing, incorporating buffers for added resource allocation flexibility, and adhering to consistent process and flow orientation. Thus, the system planning follows standard principles while taking extra measures for potential deadlock scenarios.

The detailed planning phase must include modeling the planned system, which can be a mathematical or simulation model, and the integration of RL methods. Simulation models can represent how AMRs will interact within the system and identify potential deadlocks. The development of a simulation model that accurately represents the dynamics and interactions between agents and resources within the system can serve as a controlled environment, ideal for training the RL algorithm. Alternatively, when a system's dynamics can be accurately captured mathematically, a mathematical model can be used instead of a simulation. This alternative is particularly applicable for simpler systems or when comprehensive, high-quality data is available. RL algorithms can then be developed and tested within these simulations to find optimal strategies for deadlock avoidance or recovery.

The tender phase involves evaluation and decision-making processes, where there is no specific extension needed for deadlock handling with RL.

For the system construction phase, the focus shifts to the realization of the system where deadlock handling solutions are deployed. This ensures that the RL algorithms and trained ANN models developed during the planning and design phases are effectively implemented in the operational system.

Finally, in the system operation phase, continuous monitoring and learning are necessary. Continuous monitoring helps detect unconsidered deadlocks, while continual learning allows the RL models to adapt and improve over time \citep{abel2024definition}. Following the deployment of the deadlock handling solution, the system moves into a phase of continuous monitoring and learning. Performance is constantly tracked to identify changes or trends that might impact deadlock situations. The RL algorithm, in turn, utilizes this ongoing data to continually learn and adapt its strategy, improving its deadlock handling capabilities over time.

\begin{figure}
    \centering
    \includegraphics[width=1\linewidth]{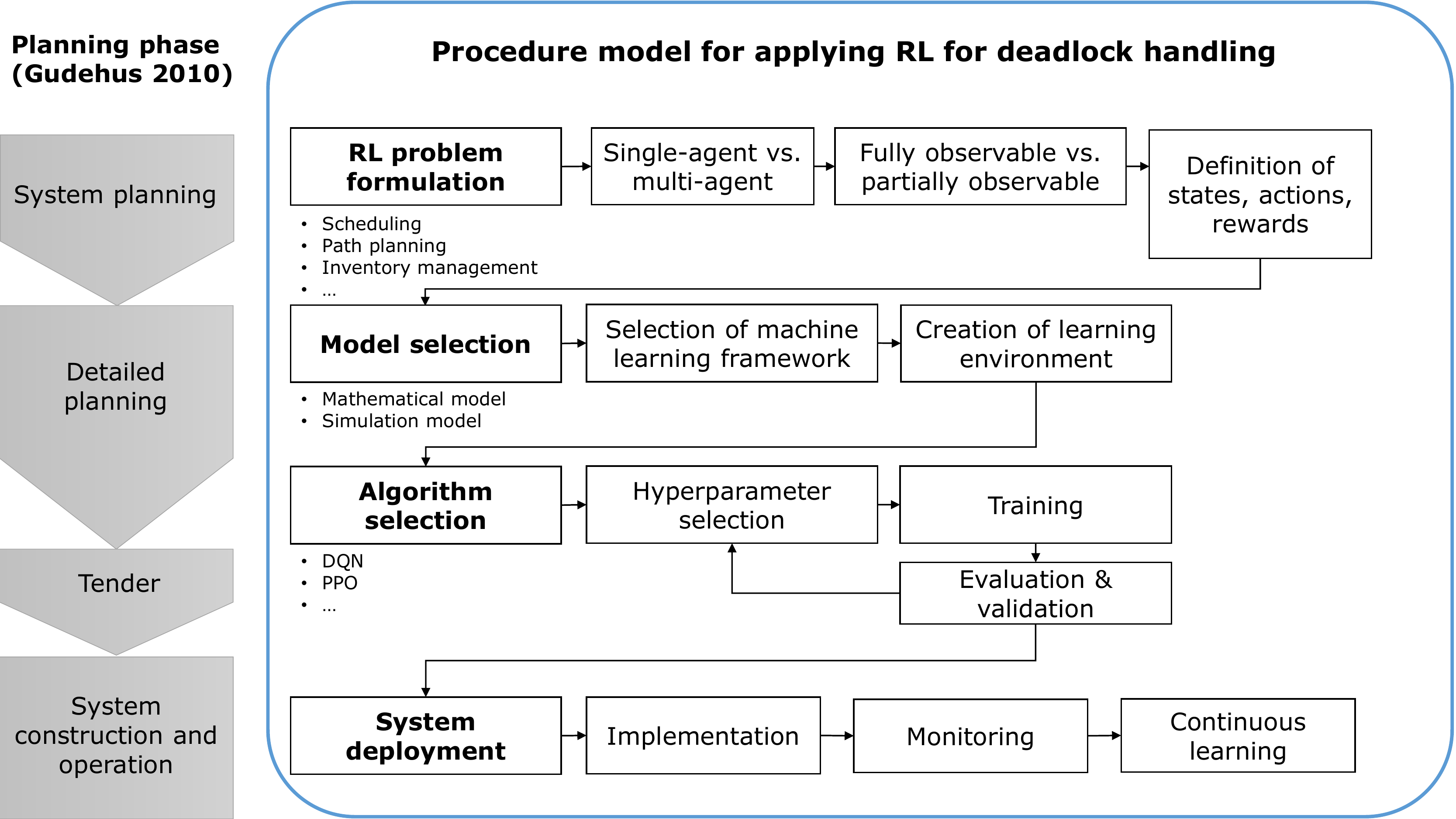}
    \caption{Procedure model for applying RL for deadlock handling in logistics based on \cite{muller2023framework}.}
    \label{fig:procedure-model}
\end{figure}

This thesis proposes a procedure model based on four main steps for applying RL for deadlock handling in logistics. Figure \ref{fig:procedure-model} shows the procedure model as it accompanies each phase of logistics systems planning according to \citet[p.70]{Gudehus.2010}. The procedure model starts from the system planning stage by preparing the use of RL and deriving conclusions from the earlier target planning.

The first stage of the framework is the \enquote{\textbf{RL problem formulation}}. This phase focuses on understanding specific challenges within the intralogistics system, such as scheduling, path planning, and inventory management.

The problem formulation process begins with a thorough understanding of the logistics system, emphasizing the nature and scope of the problem. This involves investigating the types of resources, agents, and processes involved in the system, as well as the objectives and constraints under which the system operates. A clear understanding of these elements ensures an accurate problem definition and allows for a tailored approach to find the optimal solution. This initial step includes identifying the types of resources, such as vehicles, storage units, and conveyors, along with their characteristics. It also involves determining the agents, which may include AMR and human operators, and mapping out the processes and workflows within the system. The sequence of operations and interactions between resources and agents is noted, ensuring a comprehensive view of the system. Additionally, the objectives of the system, such as minimizing lead time, maximizing throughput, or ensuring timely delivery, are clearly defined, and the constraints, such as capacity limits, operational hours, or safety regulations, are identified.

Analyzing historical data to identify patterns and trends contributing to deadlock situations is also part of this phase. Insights gained from this data help to understand the conditions under which deadlocks typically occur, enabling the early development of strategies to address these issues effectively. This step involves gathering historical data from past operations, including logs of system performance, incidents of deadlocks, and resolution attempts. By analyzing this data, patterns and common conditions that trigger deadlocks can be identified, and trends in the frequency and impact of deadlocks over time can be studied. This analysis helps to understand the effect of deadlocks on system performance and identify periods of heightened risk.

Another aspect of the RL problem formulation is determining the observations to provide to the agents. The selection process for observations should start with identifying key variables that directly impact decision-making. These variables might include positions of AMRs, inventory levels, task statuses, and the availability of resources. The goal is to ensure that the observations are both comprehensive and relevant to the tasks the agents need to perform. To determine which observations to include, one effective approach is to conduct a sensitivity analysis on potential input variables, assessing their influence on performance metrics like throughput, lead time, and system stability. Variables that significantly impact these metrics should be prioritized. An optional addition could also be the inclusion of variables that provide context for future states, such as expected task arrivals or scheduled maintenance activities, to allow the agent to plan ahead.

The form of observations, whether discrete, continuous, or a mix, should be chosen based on the nature of the variables. Discrete observations are suited for categorical or binary states, such as detecting the presence or absence of an obstacle, while continuous observations work well for variables that take on a range of values, such as the distance to the nearest object or the velocity of an AMR. In scenarios where multiple discrete categories are present for different dimensions of the state, such as tasks or agent statuses, a multi-discrete observation space can effectively represent these dimensions.

For example, in a MAPF scenario involving AMRs, typical observations may include the start and goal locations of each agent, the positions of obstacles, and the current positions of other agents. When the system is fully observable, the agents have access to the complete state, meaning that their observation space can include the entire grid, all agent positions, and their respective starts and goals. This complete information allows for optimal decision-making, as each agent has a global view of the environment and can anticipate potential conflicts or optimize paths based on the movements of others.

In logistics systems with partial observability, agents only have access to limited state information, often dictated by their sensors' field of view. In such cases, the observation space might only include a local region around the agent, and the specific field of view could vary depending on the planned sensor configuration for the AMRs. For instance, the field of view may be a 360° range around the agent if using a lidar, or it could be constrained to a forward-facing cone if relying on visual sensors. This limited view requires the agent to infer hidden parts of the state, often using memory mechanisms or RNNs to process sequences of observations and predict future states. This approach compensates for the lack of full information by allowing the agent to make decisions based on both current and historical data.

Figure \ref{fig:sensor-ranges} illustrates three different sensor range configurations for an agent in a grid-based environment with partial observability, focusing on both the spatial coverage of the sensor and the impact of obstacles on the agent's field of view. The first subfigure shows a 360° sensor range without object interference, where the agent has unobstructed perception in all directions. The blue-shaded grid cells indicate the area within the sensor's reach, and a dotted line outlines the circular boundary. Despite the presence of a vertical obstacle (shown as a gray rectangle), the sensor range is not affected in this case, as the agent can still perceive the environment uniformly.

\begin{figure}[h]
    \begin{minipage}[t]{.33\textwidth}
    \centering
        \begin{tikzpicture}[scale=0.5]
            \draw[step=1cm,gray,very thin] (0,0) grid (7,7);
            
            \fill[gray] (3,0) rectangle (4,3);
            \fill[gray] (3,4) rectangle (4,7);
            
            \node[fill=blueMM, circle] at (2.5,4.5) {};

            \fill[blue, opacity=0.2] (0,2) -- (0,7) -- (5,7) -- (5,2);

            \draw[blue, thick, dotted] (2.5,4.5) circle (2.5cm); 
        \end{tikzpicture}
        \caption*{360° sensor range}
    \end{minipage}
    \begin{minipage}[t]{.33\textwidth}
    \centering
        \begin{tikzpicture}[scale=0.5]
            \draw[step=1cm,gray,very thin] (0,0) grid (7,7);
            
            \fill[gray] (3,0) rectangle (4,3);
            \fill[gray] (3,4) rectangle (4,7);
            
            \node[fill=blueMM, circle] at (2.5,4.5) {};

            \fill[blue, opacity=0.2] (0,4) -- (0,5) -- (3,5) -- (3,4);
            \fill[blue, opacity=0.2] (0,3) -- (0,4) -- (4,4) -- (4,3);
            \fill[blue, opacity=0.2] (1,2) -- (1,3) -- (3,3) -- (3,2);

            \draw[blue, thick, dotted] (0,4.5) arc[start angle=180, end angle=360, radius=2.5cm];
            \draw[blue, thick, dotted] (0,4.5) -- (5,4.5);
        \end{tikzpicture}
        \caption*{Half-circle sensor range}
    \end{minipage}
    \begin{minipage}[t]{.33\textwidth}
    \centering
        \begin{tikzpicture}[scale=0.5]
            \draw[step=1cm,gray,very thin] (0,0) grid (7,7);
            
            \fill[gray] (3,0) rectangle (4,3);
            \fill[gray] (3,4) rectangle (4,7);
            
            \node[fill=blueMM, circle] at (2.5,4.5) {};

            \fill[blue, opacity=0.2] (2,4) -- (2,5) -- (3,5) -- (3,4);
            \fill[blue, opacity=0.2] (1,3) -- (1,4) -- (4,4) -- (4,3);
            \fill[blue, opacity=0.2] (0,2) -- (0,3) -- (3,3) -- (3,2);

            \draw[blue, thick, dotted] (2.5,4.5) -- (0.5,2.5);
            \draw[blue, thick, dotted] (2.5,4.5) -- (4.5,2.5);
            \draw[blue, thick, dotted] (0.5,2.5) arc[start angle=225, end angle=315, radius=2.828cm];

        \end{tikzpicture}
        \caption*{Cone-shaped sensor range}
    \end{minipage}
    \caption{Different sensor ranges with and without object interference.}
    \label{fig:sensor-ranges}
\end{figure}

The second subfigure presents a half-circle sensor range with object interference. The agent's field of view is directed downward, and the blue-shaded cells indicate the portion of the grid that the agent can perceive. The dotted arc outlines the half-circle sensor's boundary. In this case, the gray obstacle partially obstructs the agent's view, blocking visibility on the right side of the half-circle range. This demonstrates how an obstacle within the sensor's range can limit the agent's perception of the environment. The third subfigure displays a cone-shaped sensor range, also with object interference. The agent's perception is concentrated in a cone extending downward. The blue-shaded cells represent the grid area that falls within the agent's sensor range, while the dotted lines define the boundaries of the cone. Similar to the half-circle sensor, the obstacle interrupts the agent's field of view, resulting in an area behind the obstacle that the agent cannot perceive. This figure underscores how the shape of the sensor range, whether circular, half-circular, or conical, along with the presence of obstacles, significantly influences the agent's ability to observe and interact with its surroundings.

These examples highlight the effects of various sensor configurations and obstacles on the agent's perception. However, ignoring such restrictions regarding the field of view could help to maintain a grid-based observation model, which is highly suitable for CNNs due to their capacity to process structured grid-like data. Additionally, in logistics systems with extensive sensor networks and high levels of information exchange between agents, a fully observable grid-based approach could closely represent reality. In such systems, the aggregation of sensor data from multiple sources can provide each agent with near-complete knowledge of the environment, effectively bypassing local sensor limitations.

Defining the states, actions, and rewards for the RL problem is the last step of the RL problem formulation. The action space can be discrete, continuous, or multi-discrete. A discrete action space means that the agent can choose from a finite set of actions, while a continuous action space means that the agent can take any action within a range. A multi-discrete action space is a combination of multiple discrete action spaces, each representing a different aspect of the agent's decision-making process. The reward structure provides feedback to agents based on their actions, incentivizing desirable behaviors such as avoiding deadlocks, completing tasks efficiently, and minimizing disruptions. Rewards should be designed to align the agent's actions with the overall objectives of the logistics system.

\begin{figure}
    \centering

    \begin{tikzpicture}[node distance=2cm,
            process/.style={circle, draw, minimum size=1cm},
            resource/.style={rectangle, draw, minimum size=1cm},
            state/.style={circle,draw=black,fill=none,minimum size=180pt,inner sep=0pt},
            action/.style={circle,draw=black,fill=none,minimum size=17pt,inner sep=0pt},
            ]

            \node[state] (state_0) at (-7,-5.5){};
            \node [anchor=west, align=left] (state_0_name) at (state_0.west){\textbf{\(S_0:\)}};
            \node[state] (state_1) at (1,-2){};
            \node [anchor=west, align=left] (state_1_name) at (state_1.west){\textbf{\(S_1:\)}};
            \node[state] (state_2) at (1,-9){};
            \node [anchor=west, align=left] (state_2_name) at (state_2.west){\textbf{\(S_2:\)}};
            
            \node[action] (state_0_action_0) at (-7, -10) {$a_0$};
            \node[action] (state_0_action_1) at (-3,-4) {$a_1$};
            \node[action] (state_0_action_2) at (-4.5,-9) {$a_2$};

            \node[action] (state_1_action_2) at (2.5,-5.5) {$a_2$};

            \node[align=left](state_1_action_2_label) at (3.8,-6) {$p=1$\\$r=0$};
            
            \node[process] (p1) {\(AMR_1\)};
            \node[process, right of=p1] (p2) {\(AMR_2\)};

            \node[process] (p3) at (-8, -3.5) {\(AMR_1\)};
            \node[process, right of=p3] (p4) {\(AMR_2\)};

            \node[process] (p5) at (0,-7) {\(AMR_1\)};
            \node[process, right of=p5] (p6) {\(AMR_2\)};
        
            \node[resource, below of=p1] (r1) {\(Path_{1.1}\)};
            \node[resource, below of=p2] (r2) {\(Path_{1.2}\)};
            \node[resource, below of=r1, node distance=1.5cm] (r3) {\(Path_{2.1}\)};
            \node[resource, below of=r2, node distance=1.5cm] (r4) {\(Path_{2.2}\)};

            \node[resource, below of=p3] (r5) {\(Path_{1.1}\)};
            \node[resource, below of=p4] (r6) {\(Path_{1.2}\)};
            \node[resource, below of=r5, node distance=1.5cm] (r7) {\(Path_{2.1}\)};
            \node[resource, below of=r6, node distance=1.5cm] (r8) {\(Path_{2.2}\)};

            \node[resource, below of=p5] (r9) {\(Path_{1.1}\)};
            \node[resource, below of=p6] (r10) {\(Path_{1.2}\)};
            \node[resource, below of=r9, node distance=1.5cm] (r11) {\(Path_{2.1}\)};
            \node[resource, below of=r10, node distance=1.5cm] (r12) {\(Path_{2.2}\)};

            \begin{scope}[every path/.style={->}]
                \draw (state_0) -- (state_0_action_0);
                \draw (state_0) -- (state_0_action_1);
                \draw (state_0) -- (state_0_action_2);

                \draw (state_0_action_0.west) .. controls +(left:2cm) .. node[midway, below left, align=left] {$p=1$\\$r=0$} (state_0.south west);
                \draw (state_0_action_1) -- node[midway, above left, align=left] {$p=1$\\$r=-1$} (state_1);
                \draw (state_0_action_2) -- node[midway, above, align=left] {$p=1$\\$r=0$} (state_2.west);

                \draw(state_1) -- (state_1_action_2);
                \draw(state_1_action_2) -- (state_2);
                
                \draw (r1) -- (p1);
                \draw (r2) -- (p2);

                \draw (r5) -- (p3);
                \draw (r6) -- (p4);

                \draw (r9) -- (p5);
                \draw (r10) -- (p6);
                
                \draw (p1) -- (r2);
                \draw (p2) -- (r1);

                \draw (p4) -- (r5);

                \draw (p6) -- (r9);
                \draw (p5.west) -- (-1, -7) -- (-1, -10.5) -- (r11.west);
                
            \end{scope}
        \end{tikzpicture}
    \caption{MDP from the perspective of $AMR_1$ with the states represented as resource allocation graphs.} 
    \label{fig:ch3-rag-mdp-combined}
\end{figure}
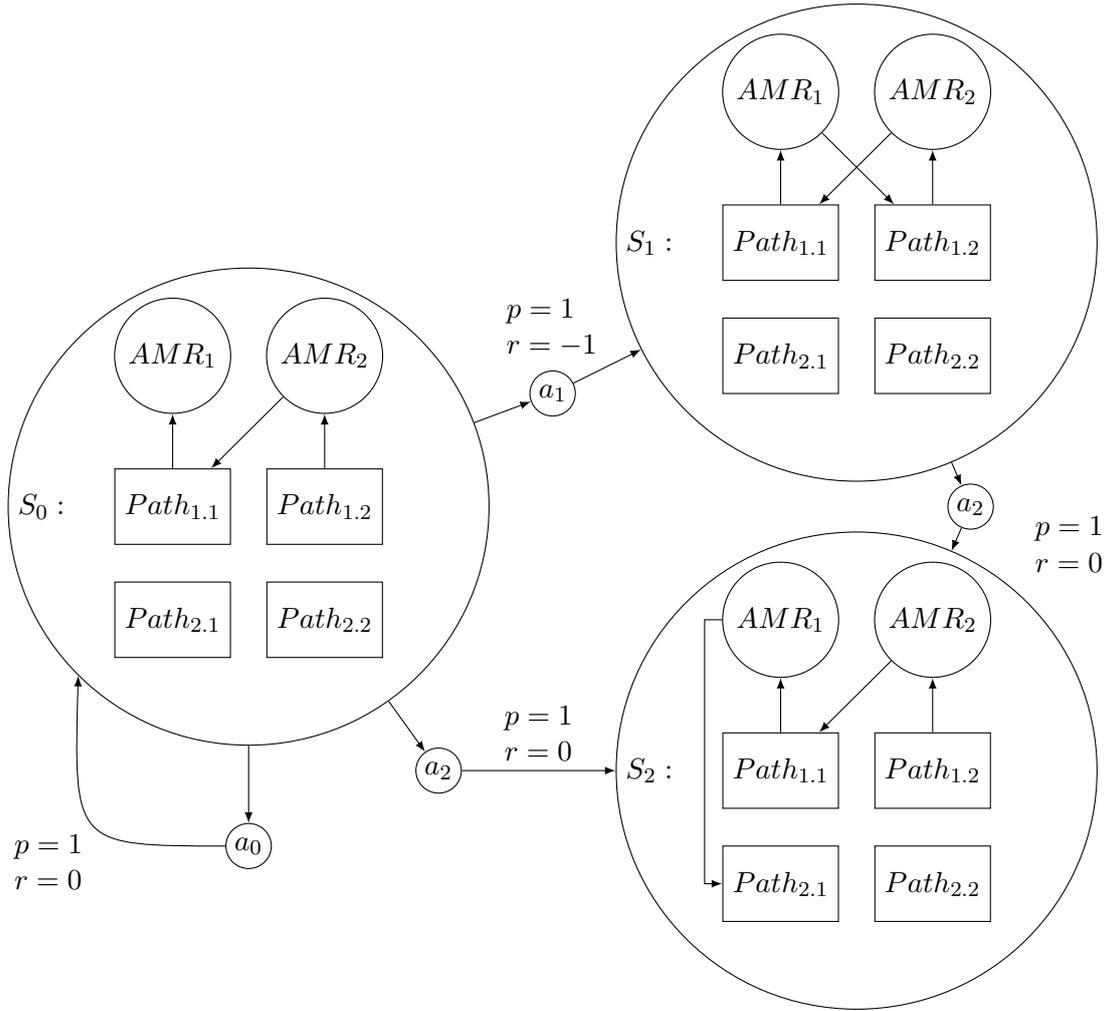

In this thesis, the state of the system is visualized using resource allocation graphs, which effectively illustrate potential deadlock scenarios. The integration of resource allocation graphs into the MDP framework, as illustrated in Figure \ref{fig:ch3-rag-mdp-combined}, provides a comprehensive visualization of the system's states and transitions. This graphical representation helps in capturing the state of the system at any given time by showing the current allocation of resources to processes/agents. In a MAS, a representation such as Figure \ref{fig:ch3-rag-mdp-combined} also only applies to a specific agent, in this case \(AMR_1\). This also means that the transitions to other states depend on how the other agents make decisions. The probabilities $p$ for \(AMR_1\) to transit to other states therefore also change with the learning process of the other agents.

In the MDP representation, three states are defined: an initial state ($S_0$), a deadlock state ($S_1$), and a non-deadlock state ($S_2$). Each state is illustrated through a distinct configuration of the resource allocation graph:

\begin{itemize}
    \item The initial state ($S_0$) shows the system's starting condition, where \(Path_{1.1}\) is allocated to $AMR_1$ and \(Path_{1.2}\) is allocated to $AMR_2$. The two AMRs are located in close proximity to each other. $AMR_2$ already requests \(Path_{1.1}\). The system is about to deadlock.
    \item The deadlock state ($S_1$) captures the situation when $AMR_1$ decides to request \(Path_{1.2}\). The resource allocation graph shows a cyclic dependency, which arises between the agents and the resources, where both AMRs are blocked because each is waiting for a path segment held by another, creating a loop that halts operations.
    \item The non-deadlock state ($S_2$) represents the decision of $AMR_1$ to take an alternative path. If the previous state was $S_1$, the state transition would be a deadlock recovery. If the transition occurs from $S_0$, it represents a deadlock avoidance.
\end{itemize}

In this model, transitioning to the non-deadlock state ($S_2$) is assigned a reward of 0, regardless of whether the transition is from $S_0$ or $S_1$. This neutral reward can be adjusted to incentivize specific strategies for handling deadlocks. For instance, by assigning a positive reward for transitions representing deadlock recovery, the policy could be encouraged to resolve deadlocks actively. However, care must be taken with such incentives, as they could inadvertently lead to the undesirable behavior of the policy deliberately entering deadlock situations to gain rewards for resolving them.

The second stage, \enquote{\textbf{Model Selection}}, necessitates the determination of an appropriate model to accurately represent the logistics system and the problem identified. Given the dynamic and stochastic characteristics of logistics systems prone to deadlocks, simulation models generally serve as the preferred choice, although mathematical models may be employed under certain circumstances. This choice depends on the complexity and nature of the system being modeled.

The initial phase of this stage involves selecting a suitable machine learning framework for RL. Options for this might range from well-established frameworks such as Ray RLlib \citep{liang2018rllib} or Stable-Baselines3 \citep{raffin2021stable-baselines3} to a custom implementation developed to meet specific requirements. The choice of the framework is a significant decision due to its impact on the scalability and development of the solution. Established frameworks provide robustness and a range of built-in functionalities, which can expedite development and testing processes. Conversely, a custom implementation, though potentially more resource-intensive to develop, can be tailored precisely to the unique characteristics and needs of the logistics system under consideration.

Following the selection of the machine learning framework, the next step is to establish the learning environment for the RL algorithm. This environment serves as the interface that bridges the RL algorithm and the simulation model. The creation process might involve integration with simulation software like Plant Simulation or AnyLogic, which are commonly used in industrial applications for their sophisticated modeling capabilities and detailed simulation options. Alternatively, a custom environment can be developed directly in a programming language such as Python, offering flexibility and control over the simulation details.

The learning environment encompasses the dynamics of the logistics system, integrating the definitions of states, actions, and rewards established during the problem formulation stage. This involves modeling the various interactions and behaviors within the system, ensuring that the environment accurately reflects the operational conditions and constraints of the logistics system. The states capture the different configurations of the system, the actions represent the decisions made by the agents, and the rewards provide feedback on the effectiveness of these decisions. The fidelity of the learning environment is important, as it determines the realism and applicability of the RL solutions developed.

An essential aspect of developing the learning environment is defining the timestep interval. The timestep refers to the duration between consecutive actions taken by the RL algorithm, during which the learning environment calculates the consequences of the last action and transitions to a new state. The interval dictates the frequency at which the RL agent can interact with the environment and update its policy.

One approach to defining timesteps is the constant timestep approach, where the environment updates at fixed time intervals. After each action taken by the RL agent, the environment simulates the system's progression over a constant time duration before allowing the RL agent to take another action. This method simplifies synchronization between the RL algorithm and the environment and ensures consistent temporal spacing between actions. However, it may not accurately capture fast or irregular dynamics, particularly if the chosen timestep is too large.

Alternatively, the environment can operate on an event-driven basis, where timesteps are defined by specific events rather than fixed intervals. In this setup, the RL agent makes an action, and the environment processes the consequences until the next significant event occurs, such as task completion, resource contention, or reaching a decision point. After the event's effects are simulated, the RL agent is prompted to act again. This approach allows for more precise handling of irregular events and can be more efficient in systems where the critical dynamics are event-triggered rather than time-driven.

The choice between these approaches depends on the nature of the logistics system and the level of detail required in the simulation. A constant timestep approach is typically easier to implement and manage, but an event-based approach can offer finer control and more accurate representation of complex systems where critical changes happen irregularly.

Additionally, it is important to consider the length of the episodes, which is defined by the number of timesteps. The duration of an episode should be as short as possible and as long as necessary. It depends on how representative the states of an episode are and whether the state spaces change over time because, for example, in the case of multi-stage processing stations, the rear processing stations are only reached by goods later in an episode if no initial stocks are defined. Longer episodes allow the RL algorithm to explore a broader range of system states and potential outcomes, fostering better learning. However, they also demand more computational resources and time per episode. The initialization of the environment at the start of each episode should also be carefully designed to represent realistic and varied operational scenarios, aiding the RL algorithm in developing robust policies.

The third stage of the framework is \enquote{\textbf{Algorithm Selection}}, which involves determining and setting up the RL algorithm best suited to address the problem defined in the learning environment. The first sub-step in this stage is the hyperparameter selection. Hyperparameters are parameters whose values are set before the learning process begins and significantly impact the training of the RL algorithm. These might include the learning rate, discount factor, number of episodes, batch size, exploration rate, and network architecture, among others. The choice of hyperparameters typically involves balancing exploration and exploitation, directly affecting the speed and effectiveness of learning. A well-chosen set of hyperparameters can facilitate efficient learning, whereas poorly chosen ones can hinder completely the algorithm's performance. 

Hyperparameter tuning often involves systematic methods such as grid search, random search, or more sophisticated techniques like Bayesian optimization. Grid search tests combinations of predefined hyperparameter values, while random search selects combinations randomly within specified ranges. Bayesian optimization iteratively refines hyperparameter values based on past evaluations to find the optimal set more efficiently.

Following hyperparameter selection, the training process is initiated. This involves the RL algorithm interacting with the learning environment over a series of episodes, learning from the reward feedback each time, and progressively improving its decision-making policy. During training, the algorithm iteratively updates its policy based on the experiences it gathers from the environment. This process is guided by the reward signals, which provide feedback on the quality of the actions taken by the algorithm.

After a set of training episodes, the performance of the algorithm is evaluated and validated. This step involves analyzing key performance metrics such as the mean episode reward, which should be maximized. Other considered metrics can be the minimization of loss and the increase of the explained variance. The evaluation and validation process helps ensure that the RL algorithm is learning effectively and can generalize well to new situations. Evaluation typically includes running the trained policy in the environment and measuring its performance against predefined benchmarks or objectives. Validation may also involve testing the algorithm in different scenarios or under varying conditions to assess its robustness and adaptability. The rendering of the environment to see the animation of the agents can also be helpful to validate the learned policy.

The hyperparameter selection, training, and evaluation \& validation steps form a cycle that is repeated until a satisfactory combination of hyperparameters is found that allows the algorithm to solve the problem effectively. Through repeated cycles of tuning and testing, the algorithm's performance is incrementally enhanced, leading to an sufficiently good set of hyperparameters and a well-trained model capable of addressing the specified intralogistics challenges.

The final stage of the procedure model is \enquote{\textbf{System deployment}}, which involves transitioning the trained RL algorithm into the actual logistics system. This stage is important for ensuring that the theoretical and simulated developments translate effectively into real-world operations.

The first sub-step in this stage is implementation. During implementation, the RL algorithm, which has been thoroughly trained and validated in a simulated environment, is integrated into the real-world logistics system. This involves configuring the algorithm to operate with the system's existing hardware and software infrastructure. Once implemented, the RL algorithm begins making decisions in real-time, utilizing the learned policy to manage operations and handle deadlock situations as they occur. The practical integration must account for real-time data streams, system latencies, and the physical constraints of the logistics environment, ensuring that the algorithm can function seamlessly within the operational workflow.

Following implementation, the next critical sub-step is monitoring. Given the dynamic and often unpredictable nature of logistics systems, continuous monitoring is essential to evaluate the performance of the deployed RL algorithm. KPIs, such as throughput, lead time, and the frequency of deadlock situations, are tracked to assess whether the system operates as intended and whether the RL algorithm effectively manages deadlocks. Monitoring also involves checking for any anomalies or unexpected behaviors that might indicate issues with the algorithm's decision-making process or its integration into the system.

The final sub-step is continual learning. As the logistics system evolves, new scenarios and challenges may emerge that the RL algorithm did not encounter during its initial training phase. Continual learning is a process that allows the RL algorithm to adapt to these new situations by learning from ongoing operations \citep{abel2024definition}. This involves periodically retraining the RL algorithm with fresh data collected from the logistics system, which may include new patterns of operation, emerging deadlock situations, or changes in the system's structure or processes. The continual learning cycle ensures that the RL algorithm refines and updates its policy over time, maintaining its effectiveness and relevance in managing the logistics system. Continual learning is not merely about retraining the algorithm but also about optimizing its parameters and possibly adjusting the learning environment to better reflect new realities. This ongoing adaptation process helps in keeping the RL solution resilient against changes in the system's operational conditions, such as shifts in demand patterns, alterations in the logistics network, or the introduction of new technologies or processes.

Section \ref{sec:methodology-logisticsplanning} outlines a comprehensive approach to integrating RL into logistics planning, focusing on the systematic alignment of RL methods with traditional logistics objectives and processes. The section details a structured framework that includes problem formulation, model selection, algorithm optimization, and continual learning. This approach ensures that RL is not only incorporated effectively into the planning stages but also remains adaptable to evolving operational conditions. In conclusion, Section \ref{sec:methodology-logisticsplanning} answers RQ 1.2 by demonstrating that integrating RL into logistics planning involves a continuous, iterative process that enhances the system's resilience and efficiency, particularly in handling deadlocks.

\section{Reference Models for Deadlock-capable MAPF problems} \label{sec:methodology-reference-models}

\subsection{Requirements and Design Criteria}
This section addresses the RQ 2.1 \enquote{What do reference scenarios for deadlock-capable MAPF problems look like?} and presents reference models specifically designed to represent scenarios with deadlock-capable MAPF problems. These models serve as standardized scenarios for evaluating and comparing various deadlock-handling strategies and algorithms, particularly in the context of AGVs and AMRs in intralogistics environments.

To effectively serve as benchmarks, the reference models must meet several criteria. Firstly, they should represent realistic scenarios, capturing typical configurations and layouts found in practical intralogistics systems. This includes factors such as bidirectional paths, shared resources, and bottlenecks. Secondly, the models must explicitly include situations where deadlocks are likely to occur, such as scenarios with limited resources, high-density traffic, and complex route networks requiring sophisticated coordination among the agents. Thirdly, the models should be scalable to accommodate different numbers of agents and varying levels of environmental complexity. This ensures their applicability to both small-scale and large-scale systems. Finally, the models should be flexible, allowing adaptation to different types of MAPF problems, whether involving ground vehicles, aerial drones, or other types of autonomous agents.

The reference models comprise several components. The environment layout includes the physical arrangement of paths, obstacles, start, and goal locations, defining constraints and potential conflict points within the system. A certain behavior and variance of the reference models should be ensured so that the algorithms do not overfit too easily and lose their generalizability. The variance can be expressed in different starting and goal locations for the agents or slight layout changes. Fundamentally randomly generated environments are not developed as part of this thesis. Agent specifications encompass the number and types of agents, their movement capabilities, and which information each agent has. The policies or control algorithms to be trained are expected to take over the path planning and execution by means of the movement capabilities of the agents for each agent from the starting point to the goal and to avoid collisions and deadlocks.

The reference models are implemented using a grid-based structure, selected for its suitability in representing discrete values and facilitating a wide range of algorithmic approaches. The grid structure is particularly advantageous in MAPF problems due to its uniform time steps and consistent distance measurements, allowing for precise control over agent movements and interactions. This approach aligns with standard practices in the field, where grid-based environments are commonly used to model and solve pathfinding problems.

The reference models are developed in Python and conform to the Gymnasium standard, ensuring compatibility with popular RL libraries such as Stable-Baselines3 and RLlib. These libraries leverage Gymnasium's standardized framework to integrate with various RL algorithms, facilitating the testing and evaluation of these algorithms in a consistent and controlled manner. We also assume that logistics planners will tend to use such RL libraries rather than developing the algorithms themselves, which also underlines the use of the Gymnasium standard. 

\subsection{Derivation Strategy for the Reference Models} \label{sec:reference-model-derivation}
The term \enquote{reference models} is used deliberately in this work to denote a structured set of standardized test environments. These models are not described as generic simulation scenarios, because their purpose is not to broadly simulate logistics operations or production processes. The developed models are also not synthetic benchmarks in the classical sense, which typically focus on algorithmic scalability or performance metrics in artificially generated, often oversimplified instances. Nor are they domain-specific layouts in the form of fully detailed case studies, as they intentionally abstract from context-specific constraints to retain general applicability and analytical control. The decision for the term \enquote{reference} reflects their intended methodological role: to serve as fixed, reusable, and interpretable baselines for the evaluation of algorithmic coordination behavior in structurally deadlock-prone MAPF settings.

The design of the reference models is grounded in a structured derivation process that proceeds from foundational principles of deadlock emergence in MAPF systems to increasingly realistic and complex logistics scenarios. The derivation follows a two-dimensional progression: from abstract to realistic and from simple to complex. At the abstract end, models represent minimal configurations that isolate specific deadlock-inducing structures such as single-lane corridors or crossings with insufficient yielding options. These minimal cases enable the controlled investigation of learning dynamics and coordination behaviors without the confounding influence of global task complexity. At the realistic end, models are informed by empirical layouts drawn from warehouse and production logistics, including bottleneck zones, overlapping routing demands, and shared spatial resources. This ensures that tested policies also engage with application-relevant scenarios.

The reference models are not intended to capture all variants of intralogistics systems, nor do they claim completeness with respect to deadlock typologies. Rather, they define a set of representative conditions under which deadlocks are structurally embedded and algorithmic strategies can be compared. 

The derivation process starts with the analysis of deadlock emergence in discretized pathfinding systems. Based on the structural analysis of agent interactions in grid-based MAPF systems, two prototypical sources of deadlock are identified: (i) mutual blocking in constrained one-dimensional passages, such as head-on encounters in single-lane corridors, and (ii) spatial contention at intersections or buffer zones where simultaneous access attempts create circular dependencies. These patterns reflect the minimal configurations in which agents cannot progress without coordinated yielding or reordering. The reference models build on this foundation by introducing controlled degrees of complexity:

\begin{itemize}
    \item Spatial structure: from open corridors to interconnected networks
    \item Agent count: from two-agent coordination to congested multi-agent episodes
    \item Task variability: via randomized start and goal sampling within predefined layout constraints
\end{itemize}

The derivation is also informed by domain analysis. Specifically, the model types reflect usage contexts of AMRs in intralogistics, including warehouse navigation, production supply chains, and closed-loop transport systems with fixed routing paths and return flows. The warehouse-type models abstract the layout of parallel aisles, pick stations, and crossing points. Reference model~3.1, in particular, is adapted from a real-world layout involving material provisioning for assembly.

In summary, the reference models are derived as a spectrum of test environments that embed known sources of deadlock, represent relevant industrial use cases, and support structured evaluation of MARL policies under controlled variation of agent density, task assignment, and layout structure.

\subsection{Model Families and Structural Variants}

The proposed reference models encompass three problem areas as use cases: 
\begin{itemize}
    \item Conflict situation (Reference models 1.1 - 1.4)
    \item Warehouse (Reference models 2.1 and 2.2)
    \item Production logistics (Reference model 3.1)
\end{itemize}

In all reference models, the grid represents the traversable environment. Cells filled in grey denote obstacles or walls, i.e., non-traversable space. White cells form the free path, which agents may occupy or move through. Each agent is represented by a filled circle in a distinct color, while its corresponding goal is shown as a diamond of the same color. These conventions are maintained throughout all reference model illustrations. Table \ref{tab:reference-models-legend} gives an overview of the used symbols.

\begin{table}[h]
    \centering
    \caption{Symbol legend for reference model illustrations.}
    \label{tab:reference-models-legend}
    \renewcommand{\arraystretch}{1.4}
    \begin{tabular}{>{\centering\arraybackslash}p{2.5cm} p{11cm}}
        \toprule
        \textbf{Symbol} & \textbf{Description} \\
        \midrule
        \begin{tikzpicture}
            \fill[gray] (0,0) rectangle (0.4,0.4);
        \end{tikzpicture}
        & Obstacle or wall (non-traversable) \\
        
        \begin{tikzpicture}
            \draw[gray, very thin] (0,0) rectangle (0.4,0.4);
        \end{tikzpicture}
        & Free path (traversable cell) \\
        
        \begin{tikzpicture}
            \node[fill=redMM, circle, scale=0.9] at (0,0.2) {};
            \node[fill=blueMM, circle, scale=0.9] at (0.4,0.2) {};
            \node[fill=greenMM, circle, scale=0.9] at (0.8,0.2) {};
            \node[fill=magenta, circle, scale=0.9] at (1.2,0.2) {};
            \node[fill=amber, circle, scale=0.9] at (1.6,0.2) {};
        \end{tikzpicture}
        & Agent start positions (circle), one color per agent \\
        
        \begin{tikzpicture}
            \node[fill=redMM, diamond, scale=0.7] at (0,0.2) {};
            \node[fill=blueMM, diamond, scale=0.7] at (0.4,0.2) {};
            \node[fill=greenMM, diamond, scale=0.7] at (0.8,0.2) {};
            \node[fill=magenta, diamond, scale=0.7] at (1.2,0.2) {};
            \node[fill=amber, diamond, scale=0.7] at (1.6,0.2) {};
        \end{tikzpicture}
        & Agent goal positions (diamond), matched by color to their start \\
    \bottomrule
    \end{tabular}
\end{table}

The reference models 1.1, 1.2, 1.3, and 1.4 represent typical conflict situations in MAPF problems, where an avoidance of a collision and a deadlock is in a few timesteps necessary. The environments of the these scenarios are very small and create the difficult situations early on in each episode. They serve as initial test cases to evaluate the baseline performance of an algorithm on collision avoidance and deadlock handling. 

Figure \ref{fig:reference-model-1-1} illustrates three variations of reference model 1.1, each designed as a two-agent scenario, with different configurations of start positions (circles) and goals (diamonds), as well as varying aisle placements. The idea behind reference model 1.1 is to create a conflict situation in which the agents learn to use the short aisle to perform an evasive maneuver.

\begin{figure}[h]
    \begin{minipage}{.33\textwidth}
    \centering
        \begin{tikzpicture}[scale=0.5]
            \draw[step=1cm,gray,very thin] (0,0) grid (10,2);
            
            \fill[gray] (0,1) rectangle (5,2);
            \fill[gray] (6,1) rectangle (10,2);
            
            \node[fill=redMM, circle] at (9.5,0.5) {};
            \node[fill=blueMM, circle] at (3.5,0.5) {};
            \node[fill=redMM, diamond, scale=0.8] at (0.5,0.5) {};
            \node[fill=blueMM, diamond, scale=0.8] at (7.5,0.5) {};
        \end{tikzpicture}
        \caption*{Basic variant}
    \end{minipage}
    \begin{minipage}{.33\textwidth}
    \centering
        \begin{tikzpicture}[scale=0.5]
            \draw[step=1cm,gray,very thin] (0,0) grid (10,2);
            
            \fill[gray] (0,1) rectangle (5,2);
            \fill[gray] (6,1) rectangle (10,2);
            
            \node[fill=redMM, circle] at (2.5,0.5) {};
            \node[fill=blueMM, circle] at (1.5,0.5) {};
            \node[fill=redMM, diamond, scale=0.8] at (0.5,0.5) {};
            \node[fill=blueMM, diamond, scale=0.8] at (4.5,0.5) {};
        \end{tikzpicture}
        \caption*{Unfavorable start/goal}
    \end{minipage}
    \begin{minipage}{.33\textwidth}
    \centering
        \begin{tikzpicture}[scale=0.5]
            \draw[step=1cm,gray,very thin] (0,0) grid (10,2);
            
            \fill[gray] (0,0) rectangle (8,1);
            \fill[gray] (9,0) rectangle (10,1);
            
            \node[fill=redMM, circle] at (9.5,1.5) {};
            \node[fill=blueMM, circle] at (3.5,1.5) {};
            \node[fill=redMM, diamond, scale=0.8] at (0.5,1.5) {};
            \node[fill=blueMM, diamond, scale=0.8] at (7.5,1.5) {};
        \end{tikzpicture}
        \caption*{Variation of aisle}
    \end{minipage}
    \caption{Reference model 1.1 with variation of start, goal and aisle positions.}
    \label{fig:reference-model-1-1}
\end{figure}

In the basic variant, the start and goal positions are arranged such that the two agents are likely to encounter each other along their paths. The blue agent must navigate from the left to the right, while the red agent moves in the opposite direction. Given this setup, the blue agent must make an evasive maneuver to avoid a collision with the red agent, who is also traversing the path towards its goal. This scenario tests the basic conflict avoidance capabilities of the agents in a straightforward setting, where the potential for deadlock arises from their simultaneous presence in a shared path. The unfavorable start/goal variant presents a scenario where the start and goal are positioned in a way that exacerbates the potential for conflict. Here, the agents' paths not only intersect but are also aligned in such a manner that each agent must make an evasive maneuver to reach their respective goals. This variant increases the complexity, as it requires both agents to take a significant detour. The variation of aisle variant illustrates a different aspect of the environment by altering the layout of the aisle. Unlike the previous  variant, this variation does not necessarily increase the difficulty. The change in aisle configuration shows how the environment can be adjusted, providing a broader range of testing conditions. 

Figure \ref{fig:reference-model-1-2} illustrates three variations of reference model 1.2. The fundamental idea behind reference model 1.2 is to evaluate how agents manage decision-making and coordination in a constrained environment where the corridor lacks aisles for evasive maneuvers. Agents must decide beforehand whether to enter the corridor or, once inside, must be prepared to potentially return to the start area if conflicts arise.

\begin{figure}[h]
    \begin{minipage}[t]{.3\textwidth}
    \centering
        \begin{tikzpicture}[scale=0.45]
            \draw[step=1cm,gray,very thin] (0,0) grid (10,3);
            
            \fill[gray] (2,0) rectangle (8,1);
            \fill[gray] (2,2) rectangle (8,3);
            
            \node[fill=redMM, circle] at (8.5,1.5) {};
            \node[fill=blueMM, circle] at (1.5,1.5) {};
            \node[fill=redMM, diamond, scale=0.8] at (0.5,1.5) {};
            \node[fill=blueMM, diamond, scale=0.8] at (9.5,1.5) {};
        \end{tikzpicture}
        \caption*{Basic variant}
    \end{minipage}
    \begin{minipage}[t]{.3\textwidth}
    \centering
        \begin{tikzpicture}[scale=0.45]
            \draw[step=1cm,gray,very thin] (0,0) grid (10,3);
            
            \fill[gray] (2,0) rectangle (8,1);
            \fill[gray] (2,2) rectangle (8,3);
            
            \node[fill=redMM, circle] at (8.5,2.5) {};
            \node[fill=blueMM, circle] at (0.5,2.5) {};
            \node[fill=greenMM, circle] at (1.5,2.5) {};
            \node[fill=redMM, diamond, scale=0.8] at (0.5,0.5) {};
            \node[fill=blueMM, diamond, scale=0.8] at (9.5,1.5) {};
            \node[fill=greenMM, diamond, scale=0.8] at (8.5,0.5) {};
        \end{tikzpicture}
        \caption*{Variation of start/goal and number of agents}
    \end{minipage}
    \begin{minipage}[t]{.4\textwidth}
    \centering
        \begin{tikzpicture}[scale=0.45]
            \draw[step=1cm,gray,very thin] (0,0) grid (13,3);
            
            \fill[gray] (2,0) rectangle (11,1);
            \fill[gray] (2,2) rectangle (11,3);
            
            \node[fill=redMM, circle, scale=0.9] at (11.5,1.5) {};
            \node[fill=blueMM, circle, scale=0.9] at (1.5,1.5) {};
            \node[fill=redMM, diamond, scale=0.7] at (0.5,1.5) {};
            \node[fill=blueMM, diamond, scale=0.7] at (12.5,1.5) {};
        \end{tikzpicture}
        \caption*{Variation of corridor length}
    \end{minipage}
    \caption{Reference model 1.2 with variation of start, goal, number of agents and corridor length.}
    \label{fig:reference-model-1-2}
\end{figure}

In the basic variant, two agents (blue and red) start from their respective positions and aim to reach their goals. The corridor is narrow, allowing no space for evasive maneuvers within the pathway itself. Agents must carefully coordinate their movements to avoid collisions, either by timing their entries or by potentially retreating to their start positions if a conflict cannot be avoided. This setup tests the basic conflict avoidance and decision-making capabilities of the agents in a straightforward, constrained environment. The variation of start/goal and number of agents increases the complexity by introducing a third agent (green) and varying the start and goal positions. This scenario requires the agents to handle multiple interactions simultaneously in a more densely populated setting. The agents need to decide not only when to enter the corridor but also how to navigate around each other or retreat if necessary, thus testing their advanced coordination and deadlock recovery strategies in a more complex environment. The variation of aisle length extends the length of the corridor while maintaining the original two-agent configuration. The start and goal positions remain the same as in the basic variant, but the longer corridor requires the agents to maintain coordination over an extended distance. This setup examines the agents' ability to plan and execute long-distance navigation without the possibility of mid-path evasive maneuvers, emphasizing sustained decision-making and conflict avoidance over longer paths.

Figure \ref{fig:reference-model-1-3} presents three variations of reference model 1.3. The primary purpose of reference model 1.3 is to evaluate how agents manage decision-making and coordination in environments with only one small passage but sufficient space before and after the passage. 

\begin{figure}[h]
    \begin{minipage}[t]{.2\textwidth}
    \centering
        \begin{tikzpicture}[scale=0.5]
            \draw[step=1cm,gray,very thin] (0,0) grid (3,5);
            
            \fill[gray] (1,0) rectangle (2,2);
            \fill[gray] (1,3) rectangle (2,5);
            
            \node[fill=redMM, circle] at (2.5,3.5) {};
            \node[fill=blueMM, circle] at (0.5,3.5) {};
            \node[fill=redMM, diamond, scale=0.8] at (0.5,1.5) {};
            \node[fill=blueMM, diamond, scale=0.8] at (2.5,1.5) {};
        \end{tikzpicture}
        \caption*{Basic variant}
    \end{minipage}
    \begin{minipage}[t]{.34\textwidth}
    \centering
        \begin{tikzpicture}[scale=0.5]
            \draw[step=1cm,gray,very thin] (0,0) grid (3,5);
            
            \fill[gray] (1,0) rectangle (2,2);
            \fill[gray] (1,3) rectangle (2,5);
            
            \node[fill=redMM, circle] at (2.5,3.5) {};
            \node[fill=blueMM, circle] at (0.5,3.5) {};
            \node[fill=greenMM, circle] at (0.5,1.5) {};
            \node[fill=redMM, diamond, scale=0.8] at (0.5,4.5) {};
            \node[fill=blueMM, diamond, scale=0.8] at (2.5,0.5) {};
            \node[fill=greenMM, diamond, scale=0.8] at (2.5,4.5) {};
        \end{tikzpicture}
        \caption*{Variation of start/goal and number of agents}
    \end{minipage}
    \begin{minipage}[t]{.45\textwidth}
    \centering
        \begin{tikzpicture}[scale=0.5]
            \draw[step=1cm,gray,very thin] (-1,-1) grid (4,6);
            
            \fill[gray] (1,0) rectangle (2,6);
            
            \node[fill=redMM, circle] at (2.5,3.5) {};
            \node[fill=blueMM, circle] at (0.5,3.5) {};
            \node[fill=redMM, diamond, scale=0.8] at (0.5,1.5) {};
            \node[fill=blueMM, diamond, scale=0.8] at (2.5,1.5) {};
        \end{tikzpicture}
        \caption*{Variation of area and passage position}
    \end{minipage}
    \caption{Reference model 1.3 with variation of start, goal, number of agents and passage position.}
    \label{fig:reference-model-1-3}
\end{figure}

In the basic variant, the two agents start from their respective positions and aim to reach their goals. The vertical wall in the middle creates a narrow corridor, which forces the agents to coordinate their movements carefully to avoid collisions. Since there is no space for evasive maneuvers within the passage itself, agents must decide on their entry timing. The variation of start/goal and number of agents increases the complexity by introducing a third agent and varying the start and goal positions. In this variant, three agents must navigate the shared passage, necessitating more sophisticated coordination to handle the increased number of interactions and potential conflicts. Agents must decide not only when to enter the passage but also how to navigate around each other or retreat if necessary, thus testing their advanced coordination and deadlock recovery strategies in a more densely populated setting. The variation of area and passage position extends the size of the environment and alters the position of the passage. This variation can maintain the original two-agent configuration or can be scaled up to a higher number of agents. The agents need to navigate a longer distance than in the basic variant and adjust to a different passage layout. 

Figure \ref{fig:reference-model-1-4} illustrates three variations of reference model 1.4, each designed to evaluate agent coordination and decision-making in an environment centered around a four-way intersection. This model is particularly focused on testing how agents manage complex interactions at a central junction where multiple paths converge.

\begin{figure}[h]
    \begin{minipage}[t]{.33\textwidth}
    \centering
        \begin{tikzpicture}[scale=0.5]
            \draw[step=1cm,gray,very thin] (0,0) grid (7,7);
            
            \fill[gray] (0,0) rectangle (3,3);
            \fill[gray] (4,0) rectangle (7,3);
            \fill[gray] (0,4) rectangle (3,7);
            \fill[gray] (4,4) rectangle (7,7);
            
            \node[fill=redMM, circle] at (5.5,3.5) {};
            \node[fill=blueMM, circle] at (1.5,3.5) {};
            \node[fill=greenMM, circle] at (3.5,5.5) {};
            \node[fill=magenta, circle] at (3.5,1.5) {};
            \node[fill=redMM, diamond, scale=0.8] at (0.5,3.5) {};
            \node[fill=blueMM, diamond, scale=0.8] at (6.5,3.5) {};
            \node[fill=greenMM, diamond, scale=0.8] at (3.5,0.5) {};
            \node[fill=magenta, diamond, scale=0.8] at (3.5,6.5) {};
        \end{tikzpicture}
        \caption*{Basic variant}
    \end{minipage}
    \begin{minipage}[t]{.33\textwidth}
    \centering
        \begin{tikzpicture}[scale=0.5]
            \draw[step=1cm,gray,very thin] (0,0) grid (7,7);
            
            \fill[gray] (0,0) rectangle (3,3);
            \fill[gray] (4,0) rectangle (7,3);
            \fill[gray] (0,4) rectangle (3,7);
            \fill[gray] (4,4) rectangle (7,7);
            
            \node[fill=redMM, circle] at (2.5,3.5) {};
            \node[fill=blueMM, circle] at (1.5,3.5) {};
            \node[fill=greenMM, circle] at (3.5,5.5) {};
            \node[fill=magenta, circle] at (3.5,0.5) {};
            \node[fill=amber, circle] at (3.5,6.5) {};
            \node[fill=redMM, diamond, scale=0.8] at (0.5,3.5) {};
            \node[fill=blueMM, diamond, scale=0.8] at (3.5,1.5) {};
            \node[fill=greenMM, diamond, scale=0.8] at (4.5,3.5) {};
            \node[fill=magenta, diamond, scale=0.8] at (5.5,3.5) {};
            \node[fill=amber, diamond, scale=0.8] at (6.5,3.5) {};
        \end{tikzpicture}
        \caption*{Variation of start/goal and number of agents}
    \end{minipage}
    \begin{minipage}[t]{.33\textwidth}
    \centering
        \begin{tikzpicture}[scale=0.5]
            \draw[step=1cm,gray,very thin] (-1,-1) grid (8,8);
            
            \fill[gray] (-1,-1) rectangle (3,3);
            \fill[gray] (4,-1) rectangle (8,3);
            \fill[gray] (-1,4) rectangle (3,8);
            \fill[gray] (4,4) rectangle (8,8);
            
            \node[fill=redMM, circle] at (6.5,3.5) {};
            \node[fill=blueMM, circle] at (0.5,3.5) {};
            \node[fill=greenMM, circle] at (3.5,6.5) {};
            \node[fill=magenta, circle] at (3.5,0.5) {};
            \node[fill=redMM, diamond, scale=0.8] at (-0.5,3.5) {};
            \node[fill=blueMM, diamond, scale=0.8] at (7.5,3.5) {};
            \node[fill=greenMM, diamond, scale=0.8] at (3.5,-0.5) {};
            \node[fill=magenta, diamond, scale=0.8] at (3.5,7.5) {};
        \end{tikzpicture}
        \caption*{Variation of path length}
    \end{minipage}
    \caption{Reference model 1.4 with variation of start, goal, number of agents and path length.}
    \label{fig:reference-model-1-4}
\end{figure}

In the basic variant, four agents (colored blue, red, pink, and green) start from different positions and must navigate through a central intersection to reach their respective goals. The layout forces all agents to pass through the intersection, creating a high potential for conflicts as they cross each other's paths. This scenario tests the agents' basic abilities in timing, coordination, and conflict avoidance within a confined but critical intersection. The variation of start/goal and number of agents increases the complexity by introducing one additional agents (yellow) and altering the start and goal positions. With more agents converging on the central intersection, the likelihood of conflicts increases. This setup tests the agents' advanced coordination strategies, requiring them to handle simultaneous interactions and make quick decisions to avoid deadlocks while ensuring all agents reach their goals. The variation of path length extends the environment by increasing the distance agents must travel to reach their goals while maintaining the original four-agent configuration. The longer paths test the agents' ability to sustain coordination over a more extended area, ensuring that their decisions at the intersection are aligned with the longer-term goal of reaching their goals efficiently. This scenario emphasizes sustained decision-making and the management of interactions over both short and long distances.

The following reference models 2.1 and 2.2 introduce more structured environments, such as warehouse layouts. The reference models 2.1 and 2.2 help to assess how algorithms handle typical MAPF problems in warehouse layouts. Traditional warehouse layouts are considered as multiple block warehouse layouts \cite[p. 368]{pohl2009traditional_warehouse_layouts} and more modern layouts such as fishbone layouts \cite[p. 390]{pohl2009fishbone_layout}. The characteristic aspect for the warehouse layout is that single-lane aisles are assumed in order to provoke conflict situations and maximize the use of space for warehouse goods.

Figure \ref{fig:reference-model-2-1} illustrates two variations of reference model 2.1, each representing a warehouse layout designed to test agent coordination and navigation in environments with varying degrees of maneuverability.

\begin{figure}[h]
    \begin{minipage}[t]{.50\textwidth}
    \centering
        \begin{tikzpicture}[scale=0.39]
            \draw[step=1cm,gray,very thin] (0,0) grid (20,10);
            
            \fill[gray] (2,1) rectangle (9,3);
            \fill[gray] (11,1) rectangle (18,3);
            \fill[gray] (2,4) rectangle (9,6);
            \fill[gray] (11,4) rectangle (18,6);
            \fill[gray] (2,7) rectangle (9,9);
            \fill[gray] (11,7) rectangle (18,9);
            
            \node[fill=redMM, circle, scale=0.9] at (5.5,3.5) {};
            \node[fill=blueMM, circle, scale=0.9] at (0.5,4.5) {};
            \node[fill=greenMM, circle, scale=0.9] at (12.5,6.5) {};
            \node[fill=magenta, circle, scale=0.9] at (14.5,3.5) {};
            \node[fill=redMM, diamond, scale=0.7] at (0.5,3.5) {};
            \node[fill=blueMM, diamond, scale=0.7] at (6.5,3.5) {};
            \node[fill=greenMM, diamond, scale=0.7] at (3.5,0.5) {};
            \node[fill=magenta, diamond, scale=0.7] at (3.5,6.5) {};
        \end{tikzpicture}
        \caption*{Basic variant}
    \end{minipage}
    \begin{minipage}[t]{.49\textwidth}
    \centering
        \begin{tikzpicture}[scale=0.39]
            \draw[step=1cm,gray,very thin] (0,0) grid (14,10);
            
            \fill[gray] (2,1) rectangle (14,3);
            \fill[gray] (2,4) rectangle (14,6);
            \fill[gray] (2,7) rectangle (14,9);
            
            \node[fill=redMM, circle, scale=0.9] at (2.5,3.5) {};
            \node[fill=blueMM, circle, scale=0.9] at (0.5,4.5) {};
            \node[fill=greenMM, circle, scale=0.9] at (10.5,9.5) {};
            \node[fill=magenta, circle, scale=0.9] at (3.5,0.5) {};
            \node[fill=amber, circle, scale=0.9] at (3.5,6.5) {};
            \node[fill=redMM, diamond, scale=0.7] at (0.5,3.5) {};
            \node[fill=blueMM, diamond, scale=0.7] at (5.5,0.5) {};
            \node[fill=greenMM, diamond, scale=0.7] at (4.5,3.5) {};
            \node[fill=magenta, diamond, scale=0.7] at (5.5,3.5) {};
            \node[fill=amber, diamond, scale=0.7] at (6.5,3.5) {};
        \end{tikzpicture}
        \caption*{Layout with dead ends}
    \end{minipage}
    \caption{Reference model 2.1 with variation of start, goal, number of agents and warehouse layout.}
    \label{fig:reference-model-2-1}
\end{figure}

In the basic variant, the layout consists of multiple blocks of warehouse goods with vertical two-lane aisles and horizontal single-lane aisles. Agents start from different positions and must navigate through the aisles to reach their respective goals. The single-lane aisles limit maneuverability, forcing agents to carefully time their movements and select paths that avoid conflicts. This setup tests the agents' basic abilities to navigate a structured environment, make decisions under spatial constraints, and manage potential conflicts in narrow passages in typical multiple blocks warehouse layouts. The layout with dead ends variation (reference model 2.1b) increases the challenge by removing vertical aisles and introducing additional agents. This configuration creates a more densely populated environment, where the risk of congestion and deadlocks is heightened. The reduced maneuverability forces agents to negotiate space more effectively and consider to wait if their path is blocked. 

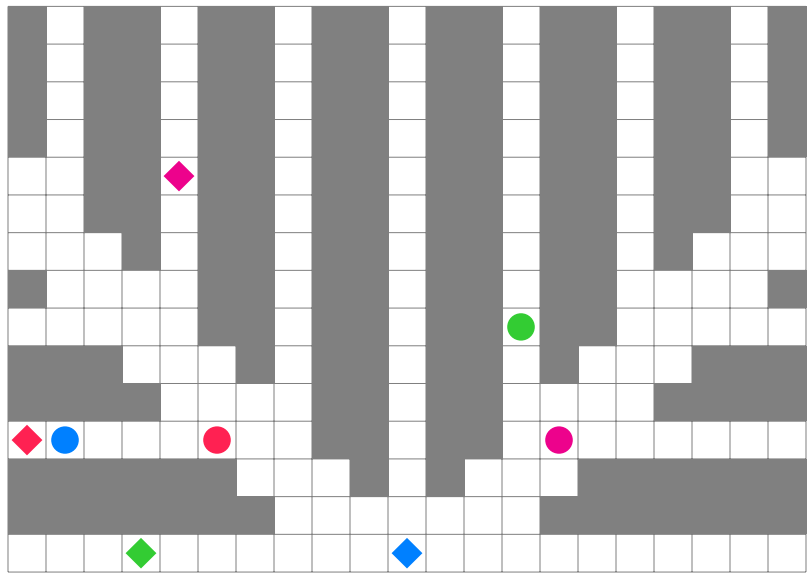
\begin{figure}[h]
    \centering
        \begin{tikzpicture}[scale=0.5]
            \draw[step=1cm,gray,very thin] (0,0) grid (21,15);
            
            \fill[gray] (0,2) rectangle (6,3);
            \fill[gray] (0,1) rectangle (7,2);
            
            \fill[gray] (0,5) rectangle (3,6);
            \fill[gray] (0,4) rectangle (4,5);
            
            \fill[gray] (0,7) rectangle (1,8);
            
            \fill[gray] (0,11) rectangle (1,15);
            
            \fill[gray] (2,9) rectangle (3,15);
            \fill[gray] (3,8) rectangle (4,15);
            
            \fill[gray] (5,6) rectangle (6,15);
            \fill[gray] (6,5) rectangle (7,15);
            
            \fill[gray] (8,3) rectangle (9,15);
            \fill[gray] (9,2) rectangle (10,15);
            
            \fill[gray] (11,2) rectangle (12,15);
            \fill[gray] (12,3) rectangle (13,15);

            \fill[gray] (14,5) rectangle (15,15);
            \fill[gray] (15,6) rectangle (16,15);
            
            \fill[gray] (17,8) rectangle (18,15);
            \fill[gray] (18,9) rectangle (19,15);

            \fill[gray] (20,11) rectangle (21,15);

            \fill[gray] (20,7) rectangle (21,8);

            \fill[gray] (18,5) rectangle (21,6);
            \fill[gray] (17,4) rectangle (21,5);

            \fill[gray] (15,2) rectangle (21,3);
            \fill[gray] (14,1) rectangle (21,2);
            
            \node[fill=redMM, circle] at (5.5,3.5) {};
            \node[fill=blueMM, circle] at (1.5,3.5) {};
            \node[fill=greenMM, circle] at (13.5,6.5) {};
            \node[fill=magenta, circle] at (14.5,3.5) {};
            \node[fill=redMM, diamond, scale=0.8] at (0.5,3.5) {};
            \node[fill=blueMM, diamond, scale=0.8] at (10.5,0.5) {};
            \node[fill=greenMM, diamond, scale=0.8] at (3.5,0.5) {};
            \node[fill=magenta, diamond, scale=0.8] at (4.5,10.5) {};
        \end{tikzpicture}
    \caption{Reference model 2.2, a warehouse with fishbone layout.}
    \label{fig:reference-model-2-2}
\end{figure}

Figure \ref{fig:reference-model-2-2} illustrates reference model 2.2, which represents a warehouse environment configured with a fishbone layout. This layout is characterized by diagonal aisles intersecting with central perpendicular aisles, creating a pattern resembling a fishbone. This design is intended to optimize space utilization and improve access to storage locations, but it also introduces unique challenges for agent navigation and coordination.

In this model, agents (colored blue, red, green, and pink) are positioned at different starting points and must navigate through the fishbone-patterned aisles to reach their designated goals, marked by diamonds. The diagonal and intersecting aisles create a complex environment where agents must carefully plan their paths to avoid collisions and efficiently reach their goals. The layout's asymmetry and varying aisle orientations increase the difficulty of pathfinding, as agents must navigate through both straight and angled passages, often requiring them to change directions multiple times. Reference model 2.2 tests the agents' ability to handle non-standard, irregular aisle layouts, emphasizing the need for dynamic path planning and real-time decision-making.

Figure \ref{fig:reference-model-3-1} presents reference model 3.1, which represents a production logistics scenario. This model is designed to reflect environments where agents must navigate within a production facility that includes fixed goal locations, such as processing units, buffers, or storage areas. The presence of these fixed locations introduces specific constraints and challenges for the agents, requiring them to carefully plan their paths to avoid conflicts and ensure timely delivery to designated locations.

\begin{figure}[h]
    \begin{minipage}[t]{.50\textwidth}
    \centering
        \begin{tikzpicture}[scale=0.25]
            \draw[step=1cm,gray,very thin] (0,0) grid (30,20);
            

            \fill[gray] (2,14) rectangle (29,18);
            \fill[gray] (2,1) rectangle (14,12);
            \fill[gray] (16,10) rectangle (30,13);
            \fill[gray] (16,1) rectangle (22,9);
            \fill[gray] (22,1) rectangle (23,2);
            \fill[gray] (23,1) rectangle (28,9);
            
            \node[fill=redMM, circle, scale=0.6] at (22.5,3.5) {};
            \node[fill=blueMM, circle, scale=0.6] at (0.5,4.5) {};
            \node[fill=greenMM, circle, scale=0.6] at (23.5,13.5) {};
            \node[fill=magenta, circle, scale=0.6] at (28.5,3.5) {};
            \node[fill=redMM, diamond, scale=0.45] at (0.5,3.5) {};
            \node[fill=blueMM, diamond, scale=0.45] at (22.5,2.5) {};
            \node[fill=greenMM, diamond, scale=0.45] at (4.5,0.5) {};
            \node[fill=magenta, diamond, scale=0.45] at (0.5,2.5) {};
        \end{tikzpicture}
        \caption*{Basic variant}
    \end{minipage}
    \begin{minipage}[t]{.49\textwidth}
    \centering
        \begin{tikzpicture}[scale=0.25]
            \draw[step=1cm,gray,very thin] (0,0) grid (30,20);
            
            \fill[gray] (2,14) rectangle (29,18);
            \fill[gray] (2,1) rectangle (14,12);
            \fill[gray] (16,10) rectangle (30,13);
            \fill[gray] (16,1) rectangle (22,9);
            \fill[gray] (22,1) rectangle (23,2);
            \fill[gray] (23,1) rectangle (28,9);

            \fill[white] (23,14) rectangle (24,15);
            \fill[white] (27,3) rectangle (28,4);
            \fill[white] (23,3) rectangle (24,4);
            \fill[white] (21,2) rectangle (22,3);
            \fill[white] (4,1) rectangle (5,2);
            
            \node[fill=redMM, circle, scale=0.6] at (23.5,3.5) {};
            \node[fill=blueMM, circle, scale=0.6] at (0.5,4.5) {};
            \node[fill=greenMM, circle, scale=0.6] at (23.5,14.5) {};
            \node[fill=magenta, circle, scale=0.6] at (27.5,3.5) {};
            \node[fill=redMM, diamond, scale=0.45] at (0.5,3.5) {};
            \node[fill=blueMM, diamond, scale=0.45] at (21.5,2.5) {};
            \node[fill=greenMM, diamond, scale=0.45] at (4.5,1.5) {};
            \node[fill=magenta, diamond, scale=0.45] at (0.5,2.5) {};
        \end{tikzpicture}
        \caption*{Layout with short goal aisles}
    \end{minipage}
    \caption{Reference model 3.1 with and without goal aisles in a production logistics scenario.}
    \label{fig:reference-model-3-1}
\end{figure}

In the basic variant, the agents are tasked with navigating through a production facility layout characterized by a combination of one- and two-lane roads, along with potential dead ends. These typical elements of production logistics environments complicate the navigation process, as agents must manage movement through both narrow and wider pathways, while also avoiding getting trapped in dead ends. The fixed goals are located within longer aisles, necessitating that agents traverse more significant distances and manage potential conflicts along these extended paths. The layout with the short goal aisles variation is designed to be less challenging by introducing short aisles that allow loading and unloading at the goals without interfering the regular transport path. 

Table \ref{tab:overview-reference-models} presents an overview of the introduced reference models.

\begin{table}[h]
    \caption{Overview of the introduced reference models.}
    \label{tab:overview-reference-models}
    \centering
    \begin{tabular}{lcccc}    
        \toprule
        \textbf{Use case} & \multicolumn{4}{c}{\textbf{Reference models}} \\ 
        \hline 
         \multirow{6}{3cm}{Conflict situation} & 1.1 & 1.2 & 1.3 & 1.4 \\ 
        & \begin{tikzpicture}[scale=0.3]
            \draw[step=1cm,gray,very thin] (0,0) grid (10,2);
            
            \fill[gray] (0,1) rectangle (5,2);
            \fill[gray] (6,1) rectangle (10,2);
            
            \node[fill=redMM, circle, scale=0.6] at (9.5,0.5) {};
            \node[fill=blueMM, circle, scale=0.6] at (3.5,0.5) {};
            \node[fill=redMM, diamond, scale=0.4] at (0.5,0.5) {};
            \node[fill=blueMM, diamond, scale=0.4] at (7.5,0.5) {};
        \end{tikzpicture}
        & \begin{tikzpicture}[scale=0.3]
            \draw[step=1cm,gray,very thin] (0,0) grid (10,3);
            
            \fill[gray] (2,0) rectangle (8,1);
            \fill[gray] (2,2) rectangle (8,3);
            
            \node[fill=redMM, circle, scale=0.6] at (8.5,1.5) {};
            \node[fill=blueMM, circle, scale=0.6] at (1.5,1.5) {};
            \node[fill=redMM, diamond, scale=0.4] at (0.5,1.5) {};
            \node[fill=blueMM, diamond, scale=0.4] at (9.5,1.5) {};
        \end{tikzpicture}
        & \begin{tikzpicture}[scale=0.3]
            \draw[step=1cm,gray,very thin] (0,0) grid (3,5);
            
            \fill[gray] (1,0) rectangle (2,2);
            \fill[gray] (1,3) rectangle (2,5);
            
            \node[fill=redMM, circle, scale=0.6] at (2.5,3.5) {};
            \node[fill=blueMM, circle, scale=0.6] at (0.5,3.5) {};
            \node[fill=redMM, diamond, scale=0.4] at (0.5,1.5) {};
            \node[fill=blueMM, diamond, scale=0.4] at (2.5,1.5) {};
        \end{tikzpicture}
        & \begin{tikzpicture}[scale=0.3]
            \draw[step=1cm,gray,very thin] (0,0) grid (7,7);
            
            \fill[gray] (0,0) rectangle (3,3);
            \fill[gray] (4,0) rectangle (7,3);
            \fill[gray] (0,4) rectangle (3,7);
            \fill[gray] (4,4) rectangle (7,7);
            
            \node[fill=redMM, circle, scale=0.6] at (5.5,3.5) {};
            \node[fill=blueMM, circle, scale=0.6] at (1.5,3.5) {};
            \node[fill=greenMM, circle, scale=0.6] at (3.5,5.5) {};
            \node[fill=magenta, circle, scale=0.6] at (3.5,1.5) {};
            \node[fill=redMM, diamond, scale=0.4] at (0.5,3.5) {};
            \node[fill=blueMM, diamond, scale=0.4] at (6.5,3.5) {};
            \node[fill=greenMM, diamond, scale=0.4] at (3.5,0.5) {};
            \node[fill=magenta, diamond, scale=0.4] at (3.5,6.5) {};
        \end{tikzpicture} \\  \\ 
        \multirow{10}{3cm}{Warehouse} & \multicolumn{2}{c}{2.1} & \multicolumn{2}{c}{2.2} \\ 
        & \multicolumn{2}{c}{\begin{tikzpicture}[scale=0.3]
            \draw[step=1cm,gray,very thin] (0,0) grid (20,10);
            
            \fill[gray] (2,1) rectangle (9,3);
            \fill[gray] (11,1) rectangle (18,3);
            \fill[gray] (2,4) rectangle (9,6);
            \fill[gray] (11,4) rectangle (18,6);
            \fill[gray] (2,7) rectangle (9,9);
            \fill[gray] (11,7) rectangle (18,9);
            
            \node[fill=redMM, circle, scale=0.6] at (5.5,3.5) {};
            \node[fill=blueMM, circle, scale=0.6] at (0.5,4.5) {};
            \node[fill=greenMM, circle, scale=0.6] at (12.5,6.5) {};
            \node[fill=magenta, circle, scale=0.6] at (14.5,3.5) {};
            \node[fill=redMM, diamond, scale=0.4] at (0.5,3.5) {};
            \node[fill=blueMM, diamond, scale=0.4] at (6.5,3.5) {};
            \node[fill=greenMM, diamond, scale=0.4] at (3.5,0.5) {};
            \node[fill=magenta, diamond, scale=0.4] at (3.5,6.5) {};
        \end{tikzpicture}}
        & \multicolumn{2}{c}{\begin{tikzpicture}[scale=0.25]
            \draw[step=1cm,gray,very thin] (0,0) grid (21,15);
            
            \fill[gray] (0,2) rectangle (6,3);
            \fill[gray] (0,1) rectangle (7,2);
            
            \fill[gray] (0,5) rectangle (3,6);
            \fill[gray] (0,4) rectangle (4,5);
            
            \fill[gray] (0,7) rectangle (1,8);
            
            \fill[gray] (0,11) rectangle (1,15);
            
            \fill[gray] (2,9) rectangle (3,15);
            \fill[gray] (3,8) rectangle (4,15);
            
            \fill[gray] (5,6) rectangle (6,15);
            \fill[gray] (6,5) rectangle (7,15);
            
            \fill[gray] (8,3) rectangle (9,15);
            \fill[gray] (9,2) rectangle (10,15);
            
            \fill[gray] (11,2) rectangle (12,15);
            \fill[gray] (12,3) rectangle (13,15);

            \fill[gray] (14,5) rectangle (15,15);
            \fill[gray] (15,6) rectangle (16,15);
            
            \fill[gray] (17,8) rectangle (18,15);
            \fill[gray] (18,9) rectangle (19,15);

            \fill[gray] (20,11) rectangle (21,15);

            \fill[gray] (20,7) rectangle (21,8);

            \fill[gray] (18,5) rectangle (21,6);
            \fill[gray] (17,4) rectangle (21,5);

            \fill[gray] (15,2) rectangle (21,3);
            \fill[gray] (14,1) rectangle (21,2);
            
            \node[fill=redMM, circle, scale=0.6] at (5.5,3.5) {};
            \node[fill=blueMM, circle, scale=0.6] at (1.5,3.5) {};
            \node[fill=greenMM, circle, scale=0.6] at (13.5,6.5) {};
            \node[fill=magenta, circle, scale=0.6] at (14.5,3.5) {};
            \node[fill=redMM, diamond, scale=0.4] at (0.5,3.5) {};
            \node[fill=blueMM, diamond, scale=0.4] at (10.5,0.5) {};
            \node[fill=greenMM, diamond, scale=0.4] at (3.5,0.5) {};
            \node[fill=magenta, diamond, scale=0.4] at (4.5,10.5) {};
        \end{tikzpicture}} \\ \\ 
         \multirow{10}{3cm}{Production logistics} & \multicolumn{4}{c}{3.1} \\ 
         & \multicolumn{4}{c}{\begin{tikzpicture}[scale=0.25]
            \draw[step=1cm,gray,very thin] (0,0) grid (30,20);
            

            \fill[gray] (2,14) rectangle (29,18);
            \fill[gray] (2,1) rectangle (14,12);
            \fill[gray] (16,10) rectangle (30,13);
            \fill[gray] (16,1) rectangle (22,9);
            \fill[gray] (22,1) rectangle (23,2);
            \fill[gray] (23,1) rectangle (28,9);
            
            \node[fill=redMM, circle, scale=0.6] at (22.5,3.5) {};
            \node[fill=blueMM, circle, scale=0.6] at (0.5,4.5) {};
            \node[fill=greenMM, circle, scale=0.6] at (23.5,13.5) {};
            \node[fill=magenta, circle, scale=0.6] at (28.5,3.5) {};
            \node[fill=redMM, diamond, scale=0.45] at (0.5,3.5) {};
            \node[fill=blueMM, diamond, scale=0.45] at (22.5,2.5) {};
            \node[fill=greenMM, diamond, scale=0.45] at (4.5,0.5) {};
            \node[fill=magenta, diamond, scale=0.45] at (0.5,2.5) {};
        \end{tikzpicture}} \\
        \bottomrule
    \end{tabular}
\end{table}

\section{Selection of MAPF Algorithms and Their Implementations for Comparison}
\label{sec:methodology-mapf-algorithms}

This section introduces a selection of MAPF algorithms for the comparison with the RL approaches and details their implementations. The chosen MAPF algorithms provide a structured baseline for evaluating RL-based approaches in deadlock handling. The selection focuses on two widely studied approaches: a multi-agent A* variant (MA-A*) and CBS, each representing a distinct planning paradigm.

The MA-A* follows a decentralized execution paradigm. Agents begin moving immediately and adjust their paths dynamically upon encountering others. Unlike standard A*, which precomputes full paths before execution, this variant allows agents to react to real-time interactions. This aligns with RL-based approaches, where agents make decisions locally without relying on precomputed global plans. In contrast, CBS operates under a centralized planning paradigm. It assumes full knowledge of all planned paths and resolves conflicts by introducing constraints and recomputing affected paths. This approach is included to highlight a key limitation of centralized MAPF methods: they require global coordination, meaning every agent must be aware of all planned trajectories. While this assumption simplifies conflict resolution, it does not align with decentralized multi-agent systems, where agents typically lack access to the full global plan. Comparing CBS to RL-based approaches provides insights into the impact of such global coordination assumptions on deadlock resolution.

These two algorithms are chosen because they represent opposite ends of the planning spectrum: reactive, decentralized decision-making (MA-A*) versus preplanned, centralized conflict resolution (CBS). This contrast allows for an evaluation of how deadlocks emerge and are handled in different MAPF solution paradigms. The pseudocode for MA-A* is shown in Algorithm \ref{alg:a-star-multi-agents}.

\begin{algorithm}
\caption{Multi-agent A* variant (MA-A*)}
\label{alg:a-star-multi-agents}
\begin{algorithmic}[1]
    \State \textbf{Input:} Grid $grid$, Start positions $starts = [start_1, \dots, start_n]$, Goal positions $goals = [goal_1, \dots, goal_n]$
    \State \textbf{Output:} Paths for all agents or \texttt{None}
    
    \State Initialize priority queue $open\_set \gets \{(0, starts, [starts])\}$
    \State Initialize visited set $visited \gets \{starts\}$
    
    \While{$open\_set$ is not empty}
        \State Pop $(f\_cost, current\_positions, paths)$ from $open\_set$
        \If{all $current\_positions = goals$}
            \State \Return $paths$
        \EndIf
        \For{each $moves = [move_1, \dots, move_n]$}
            \State Compute $next\_positions$ based on $moves$
            \If{$next\_positions$ are valid and no collisions}
                \State Compute $f\_cost$ and update $open\_set$ with new state
            \EndIf
        \EndFor
    \EndWhile
    \State \Return \texttt{None}
\end{algorithmic}
\end{algorithm}

The multi-agent variant of the A* algorithm addresses routing decisions for multiple agents simultaneously and computes collision-free paths from their start positions to their respective goals in a grid environment. A priority queue ($open\_set$) is used to store states along with their associated costs, while previously processed states are tracked in a visited set to avoid redundant computations. At each step, the algorithm evaluates all possible movements for all agents, assuming simultaneous movement. It ensures that no collisions occur, whether due to overlapping positions (node collisions) or agents swapping positions (edge collisions). The cost function integrates the path length ($g\_cost$) and the heuristic estimate to the goals ($f\_cost$). The algorithm terminates successfully when all agents reach their goals and return to their paths; otherwise, it continues until no valid states remain. The pseudo

\begin{algorithm}
\caption{Conflict-based search (High-level CBS)}
\label{alg:cbs}
\begin{algorithmic}[1]
    \State \textbf{Input:} $grid$, $starts$, $goals$
    \State \textbf{Output:} Paths for all agents or \texttt{None}
    
    \State Initialize $root\_constraints \gets \emptyset$
    \State Compute initial paths for all agents using \textsc{Low-level A*}$(grid, start_i, goal_i, root\_constraints, i)$
    \State Initialize $root\_node$ with $root\_constraints$, $paths$, and $cost$
    \State Add $root\_node$ to $open\_list$
    
    \While{$open\_list$ is not empty}
        \State Pop $current\_node$ with the lowest cost from $open\_list$
        \State Detect conflicts in $current\_node.paths$
        \If{no conflicts exist}
            \State \Return $current\_node.paths$
        \Else
            \State Identify the conflict $(agents, conflict\_type, details, t)$
            \For{each $agent$ in $agents$ involved in the conflict}
                \State Create a child node with a new constraint to resolve the conflict
                \State Recompute the path for $agent$ using \textsc{Low-Level A*}$(grid, start, goal, constraints, agent)$
                \If{path is valid}
                    \State Update the child node with the new path and cost
                    \State Add the child node to $open\_list$
                \EndIf
            \EndFor
        \EndIf
    \EndWhile
    
    \State \Return \texttt{None}
\end{algorithmic}
\end{algorithm}

\begin{algorithm}
\caption{Low-level A* search with constraints}
\label{alg:a-star}
\begin{algorithmic}[1]
    \State \textbf{Input:} $grid$, $start$, $goal$, $constraints$, $agent\_id$
    \State \textbf{Output:} Path for the agent or \texttt{None}
    
    \State Extract vertex and edge constraints for $agent\_id$
    \State Initialize $open\_list$, $came\_from$, and $g\_score$
    
    \While{$open\_list$ is not empty}
        \State Pop $(current\_cell, current\_time)$ with the lowest $f\_score$
        \If{$current\_cell = goal$}
            \State Reconstruct and return the path
        \EndIf
        \For{each neighbor of $current\_cell$}
            \State Check if neighbor satisfies vertex and edge constraints
            \If{neighbor is valid}
                \State Compute tentative $g\_score$ and update if better
                \State Push neighbor into $open\_list$
            \EndIf
        \EndFor
    \EndWhile
    
    \State \Return \texttt{None}
\end{algorithmic}
\end{algorithm}
The CBS algorithm operates on two levels: a high-level search and a low-level A* search. The high-level CBS coordinates paths for multiple agents, identifying and resolving conflicts between them. It begins by initializing a root node with no constraints and computing initial paths for all agents using the low-level A* algorithm. These initial paths are then checked for conflicts, which can involve one or more agents. Conflicts are categorized as vertex conflicts, where multiple agents occupy the same cell at the same time, or edge conflicts, where two agents swap positions during the same time step.

When a conflict is detected, CBS resolves it by creating child nodes. Each child node introduces constraints to avoid the conflict. Constraints can prevent an agent from occupying a specific cell at a given time (node constraint) or disallow an edge transition between two cells (edge constraint). Depending on the conflict, paths are recomputed for one or more agents using the low-level A* algorithm, while paths for unaffected agents remain unchanged. The child nodes are then added to a priority queue, which is sorted by the total cost of the paths. The high-level CBS continues by exploring the node with the lowest cost, ensuring that it prioritizes solutions with minimal path costs. If a node is found with no conflicts, the algorithm terminates successfully and returns the set of paths for all agents. If no valid solutions exist, CBS terminates and reports that no solution could be found.

The low-level A* algorithm is responsible for computing paths for individual agents while respecting the constraints imposed by the high-level CBS. It uses a priority queue to explore the grid, maintaining a record of visited nodes and their associated costs. At each step, the algorithm evaluates neighboring cells and checks for conflicts against vertex and edge constraints. Only valid nodes, which comply with the constraints, are considered for further exploration. The cost of a node is calculated as the sum of the path length so far ($g\_score$) and a heuristic estimate to the goal, typically using Manhattan distance. If a path to the goal is found, A* reconstructs the path and returns it. If no valid path exists under the given constraints, it reports failure.

In cases of conflicts involving more than two agents, the high-level CBS generates constraints for all agents involved in the conflict, and their paths are recomputed as necessary. This iterative process ensures collision-free paths for all agents while maintaining optimality in terms of the total path cost. By addressing conflicts iteratively and resolving them hierarchically, CBS efficiently finds a feasible and optimal solution for multi-agent pathfinding in grid-based environments.

\section{Conclusion}

This chapter presents the methodological framework for addressing deadlock handling between AMRs in intralogistics. The findings emphasize the significance of deadlocks in logistics systems, demonstrating their impact on operational efficiency and system resilience. The analysis in Section \ref{sec:methodology-significance} highlights that deadlocks are not merely an isolated technical issue but a critical factor influencing key logistics performance indicators. The choice of deadlock handling strategy, whether prevention, avoidance, or detection and recovery, significantly affects throughput, efficiency, and system robustness. Selecting the most effective strategy is challenging, as small variations in logistics system parameters can shift the advantage toward different handling approaches.

To address these challenges, Section \ref{sec:methodology-logisticsplanning} introduces resilience as a formal logistics objective and integrates deadlock handling into logistics planning. This perspective extends beyond traditional performance-driven optimization and underscores the necessity of strategically considering deadlocks during system design. A procedural model is proposed to systematically integrate RL-based solutions into the established logistics planning process. 

In support of RL-based deadlock handling, Section~\ref{sec:methodology-reference-models} develops reference models for deadlock-capable MAPF problems. These models serve as standardized benchmarks for evaluating RL and traditional MAPF algorithms in controlled but representative logistics scenarios. The reference models capture the complexity of multi-agent interactions, ensuring that the developed MARL solutions address real-world constraints such as communication limitations, and varying levels of freedom of movement of the agents due to restricting layouts.

Furthermore, this chapter introduces selected MAPF algorithms to provide a structured baseline for evaluating RL-based approaches. The chosen algorithms, including MA-A* and CBS, represent distinct paradigms in MAPF. MA-A* operates in a decentralized manner, enabling real-time adaptations, while CBS employs centralized conflict resolution with complete trajectory awareness. This contrast highlights fundamental trade-offs between centralized and decentralized planning, informing the subsequent comparison with MARL-based deadlock handling strategies.

By establishing a structured methodology, defining reference models, and selecting comparable MAPF baselines, this chapter lays the groundwork for evaluating RL-based approaches in the subsequent chapter. The following chapter applies this framework in controlled experiments, assessing the effectiveness of MARL in mitigating deadlocks under various logistics conditions.

%

\chapter{Application of MARL in Grid-Based and Continuous-Space AMR Environments} \label{ch:evaluation}
Chapter \ref{ch:evaluation} applies the methods of Chapter \ref{ch:methodology}. Two use cases are presented, followed by an explanation of their requirements and the technical implementation. In both cases, a simulation model serves as the learning environment for the agents.

\begin{itemize} 
    \item Section \ref{sec:evaluation-planning} present the experimental setup for evaluating MARL's effectiveness in MAPF with possible deadlock situations. The section focuses on the experiment design for the use cases and the selection of the considered RL algorithms. 
    \item Section \ref{sec:application-reference-models} introduces grid-based learning environments. The observation space, action space, and reward function are formalized to reflect the agents' performance. This section also details the experimental setup, hyperparameter search, and comparison of MARL with traditional MAPF algorithms. 
    \item Section \ref{sec:evaluation-external-simulation-software} extends the evaluation to an external simulation software, where MARL is tested in a continuous-space environment. The conceptual and simulation models are outlined, and observations focus on hyperparameter influence, adaptation to dynamic conditions, and overall algorithm performance. 
    \item Section \ref{sec:evaluation-conclusion} summarizes the experimental findings, highlighting key insights and their implications for MARL-based deadlock handling. 
\end{itemize}

\section{Experimental Setup} \label{sec:evaluation-planning}
In this chapter, we address the evaluation of MARL's effectiveness in MAPF with possible deadlock situations, focusing on answering research questions RQ 2.2 \enquote{When and in which configurations is MARL effective for MAPF with deadlocks?} and RQ 3.2 \enquote{Is CTDE the best approach in deadlock-capable MAPF problems?}. To achieve this, it is necessary to compare various RL algorithms and training approaches, ensuring that they are configured as optimally as possible through a careful hyperparameter search. Although the hyperparameter search does not yield fully optimal parameters, each variant is allocated a limited computational budget, and parameters found for basic reference models are applied across similar models to balance computational efficiency and performance comparability.

The experimental setup for the use cases is illustrated in Figure~\ref{fig:experiment-design} and follows the simulation-based optimization framework of \cite{marz2010simulation}. The input parameters are the grid layout, the initial positions, and the goals of each agent. The experimental parameters are the algorithmic choice, namely which specific algorithm is used, the training and execution modes for the MARL methods, and the number of agents. These parameters influence the output variables computational time, success rate, and timesteps. We omit the classical path‑length metric because the timestep count already covers every movement and waiting action. Computational time is logged, yet the analysis focuses on success rate and timesteps because success rate is meaningful only within the given computation budget. Timesteps denote the number of steps until all agents reach their respective goals. The experiments use the MARL algorithms PPO and IMPALA and the MAPF algorithms MA‑A* and CBS. The reference models exclude collisions by applying action masks in the RL algorithms, and MA‑A* and CBS avoid collisions by design.

\begin{figure}[h]
\centering
\begin{tikzpicture}[node distance=1cm, auto,>=latex']
    \node [rectangle, draw, minimum height=5cm, minimum width=6.5cm] (central_box) {};
    
    \node [rectangle, draw, align=center, left = of central_box] (input) {
    Grid layout\\
    Starts for each agent\\
    Goals for each agent
    };
    \node [above = 0.1cm of input.north] (input_name) {Input parameters};

    \node [rectangle, draw, align=center, above=of central_box] (experimental_parameters) {
    MARL/MAPF algorithm\\
    Training and execution mode\\
    Number of agents
    };
    \node [above = 0.1cm of experimental_parameters.north] (experimental_parameters_name) {Experimental parameters};

    \node [rectangle, draw, right = of central_box, align=center, minimum height=2em] (output) {
    Computational time\\
    Success rate\\
    Timesteps
    };
    \node [above = 0.1cm of output.north] (output_name) {Output variables};
    
    \node [rectangle, draw, align=center, below = 0.3cm of central_box.north] (optimization) {
    \textbf{Optimization}\\
    MARL algorithms: PPO, IMPALA\\
    MAPF algorithms: MA-A*, CBS
    };

    \node [rectangle, draw, align=center, below = of optimization] (simulation) {
    \textbf{Simulation}\\
    Mapping of transport processes\\
    Mapping of peripheral processes
    };

    \node [rectangle, draw, fill=gray!20, above = 0.1cm of optimization.north east] (python) {Python};
    \node [rectangle, draw, fill=gray!20, align=center, below = 0.1cm of simulation.south east] (python_plant) {Python\\
    Tecnomatix Plant Simulation};

    \draw [->] (experimental_parameters.south) -- (central_box.north);
    \draw [->] (input.east) -- (central_box.west);
    \draw [->] (central_box.east) -- (output.west);
    \draw [->] (optimization.south) -- (simulation.north);

\end{tikzpicture}
\caption{Experiment design for the use cases based on \citet[p. xii]{marz2010simulation}.}
\label{fig:experiment-design}
\end{figure}

The interaction between optimization and simulation is conducted using Python. While the MARL and MAPF rely completely on Python, the learning environment/simulation is modeled either in Python (Section~\ref{sec:application-reference-models}) or with Tecnomatix Plant Simulation as an example for the usage of more complex material flow modeling software (Section~\ref{sec:evaluation-external-simulation-software}). 

For the RL we employ RLlib from Ray version 2.35.0 for this thesis due to its comprehensive support for MARL and its scalability, which are essential for the complexity of MAPF with deadlocks \citep{liang2018rllib}. RLlib offers a built-in multi-agent framework, simplifying the implementation of cooperative and competitive agent settings, which is particularly beneficial for experiments involving MARL. This stands in contrast to other libraries, such as Stable-Baselines3, which provide limited native support for multi-agent environments and focus more on single-agent algorithms. RLlib's flexible environment customization further supports the implementation of advanced methods, such as CTDE, used in this thesis. RLlib provides ready-to-use implementations of a range of RL algorithms. This is particularly advantageous for this research, as it enables the immediate application of these algorithms to complex scenarios like MAPF with deadlocks without the need for extensive customization. The availability of these algorithms, maintained and regularly updated by an active community, ensures that they are both robust and up-to-date, meeting the evolving needs of real-world logistics applications. RLlib's continued maintenance and support ensure compatibility with modern hardware and software environments, providing a stable and reliable platform suitable for long-term projects. The availability of pre-implemented algorithms, along with strong community support, makes RLlib a robust choice for both research and practical applications in logistics planning systems. Its design allows for seamless integration with minimal configuration effort, facilitating efficient deployment in diverse experimental and operational contexts.

In terms of environment interaction, we adhere to the Gymnasium standard version 0.28.1, which provides a stable and widely adopted API for defining RL environments. Gymnasium ensures compatibility with RLlib and other RL libraries by offering a consistent interface across a variety of environments. This standard facilitates reproducibility and comparability of experimental results. 

We review four MAPF libraries that appear to offer similar environments for the considered use case of deadlock-capable AMR systems but each omits at least one property essential for the goals of this thesis.

\cite{stern2019MAPF} released a grid-based MAPF benchmark composed of 24 maps, each accompanied by 25 scenario files. Each scenario lists up to \(1\,000\) start–goal pairs selected to guarantee that both endpoints lie within the same connected component, ensuring single-agent reachability. The benchmark is purely static: it provides maps and coordinate lists but no step-wise API, reward definition, or termination condition. Consequently, using it in a RL pipeline (for example, with Gymnasium) requires a custom wrapper that loads the map, samples a subset of \(k\) start–goal pairs, advances the simulation in discrete steps, assigns rewards, and defines episode termination rules. The paper reports baseline results for an improved CBS variant but does not include precomputed joint paths. As a result, learning agents might still experience collisions and resource–holding loops depending on the environment dynamics. The corpus is designed for evaluating offline search algorithms in MAPF. Although it can serve as a test-bed for on-line learning after substantial engineering, it is not an RL benchmark by itself.

Flatland targets railway traffic management rather than AMRs \citep{mohanty2020flatland}. On every call to \texttt{reset()}, a \texttt{RailEnv} samples a new track graph and timetable through its rail- and line-generator hooks, yielding a fresh problem instance but complicating ablation studies that require a \emph{fixed} infrastructure and schedule \citep{mohanty2020flatland}. The simulator hard-codes several train-specific assumptions: tracks are \emph{directed} (agents may only advance in their current heading), motion follows a two-mode speed profile (stop or cruise, without gradual acceleration), and each grid cell is reserved exclusively by a single agent, effectively locking the full length of the train's path segment ahead of time \citep{mohanty2020flatland}. While these choices realistically capture heavy rail dispatching, they do not match the behavior of AMRs that operate in short, bi-directional aisles with fine-grained speed control and cooperative sharing of tiles. Adapting Flatland to AMRs would therefore require deep changes to the low-level finite-state motion model and the collision checker, which currently assume one-way, full-reservation traffic.

POGEMA~\citep{skrynnik2024pogema} provides a Gymnasium-compatible environment for partially observable MAPF and, by default, procedurally regenerates obstacles, agent starts, and goals on each call to \texttt{reset()} unless the user specifies a \texttt{seed} or explicit \texttt{GridConfig}. Each agent executes one of five basic actions (up, down, left, right, wait) per timestep, which matches the primitive action set in our environment. More importantly, while POGEMA integrates with PettingZoo, PyMARL, and SampleFactory, it does not implement RLlib's \texttt{MultiAgentEnv} interface. This gap requires a custom adapter layer to map list-based observations and rewards into RLlib's expected per-agent dictionaries and to manage RLlib's policy-mapping callback. Because the objective of this thesis is to evaluate deadlock-handling strategies in logistics layouts using RLlib's built-in policy tools and distributed roll-out infrastructure, RLlib compliance is essential. While POGEMA provides a solid foundation for generic MAPF studies, adapting it for persistent deadlocks and seamless RLlib integration would involve extensive modification or forking. The custom environment, built on top of Gymnasium's \texttt{MultiAgentEnv} interface, thus offers a more direct, maintainable, and reproducible platform for our experiments.

RWARE simulates goods-to-person warehouse operations, where robots lift shelves and carry them to workstations \citep{christianos2020rware}. As soon as an agent issues the load/unload action, the shelf disappears from the aisle, removing the long-lived static obstacles that induce resource-holding loops. RWARE follows the Gymnasium API, returning tuples of observations and rewards, but it does not emit the per-agent dictionaries that RLlib's \texttt{MultiAgentEnv} requires. Adapting RWARE would therefore involve both semantic changes (static shelves as obstacles) and interface adjustments.

None of these four candidates delivers the triad of features essential for this study: an AMR movement behavior, walls as static obstacles inside the layout which can lead to deadlock situations, and native RLlib compatibility. Implementing these properties directly in a Gymnasium 0.28.1 \texttt{MultiAgentEnv} proved simpler than forking and maintaining any of the existing alternatives. The mathematical foundations of the resulting environment are described in Section \ref{sec:application-reference-models}.

RLlib from Ray version 2.35.0 offers a variety of RL algorithms, as presented in Table \ref{tab:rllib-algorithms} \citep{liang2018rllib}. The table details the algorithms selected for this thesis, alongside the reasons for excluding others. The algorithms PPO and IMPALA are chosen as they support both discrete action spaces and multi-agent environments, aligning with the requirements of the MAPF problems with deadlocks addressed in this research. Several algorithms are excluded due to limitations in their applicability to the problem domain. DQN, in the form of the implementation from RLlib, does not support multi-discrete action spaces, making it unsuitable for a CTE approach. SAC is also excluded due to its restriction to continuous action spaces, which conflicts with the discrete nature of the environment. Algorithms such as DreamerV3 \citep{hafner2024dreamerv3}, BC \citep{torabi2018behavioral}, monotonic advantage re-weighted imitation learning \citep{wang2018exponentially}, and curiosity-driven exploration by self-supervised prediction \citep{pathak2017curiosity} are not selected because RLlib does not provide support for the mentioned algorithms in multi-agent environments, a fundamental aspect of the experiments conducted in this thesis.

\begin{table}[h]
    \centering
    \caption{Chosen algorithms from RLlib 2.35.0 and reasons for exclusion.}
    \label{tab:rllib-algorithms}
    \begin{tabular}{p{5.5cm}cc}
    \toprule
        \textbf{Algorithm} & \textbf{Chosen} & \textbf{Reason for exclusion} \\ \hline
         PPO & Yes & -  \\
         DQN & No & Does not support multi-discrete action space. \\
         SAC & No & Does not support discrete action space. \\
         IMPALA & Yes  & - \\
         APPO & No & Already covered by PPO. \\
         DreamerV3 & No & Does not support multi-agent environments. \\
         BC & No & Does not support multi-agent environments. \\
         Monotonic advantage re-weighted imitation learning & No & Does not support multi-agent environments. \\
         Curiosity-driven exploration by self-supervised prediction & No & Does not support multi-agent environments. \\
    \bottomrule
    \end{tabular}
\end{table}

\section{Reference Models as Grid-based Learning Environments}\label{sec:application-reference-models}
\subsection{Observation Space, Action Space, and Reward Function}

The experiments in this Section~\ref{sec:application-reference-models} are based on the predefined reference models of Section~\ref{sec:methodology-reference-models} to systematically test the effectiveness of RL and pathfinding algorithms for MAPF problems involving deadlocks. The reward structure in these models is designed to reflect the agents' performance, where the maximum reward is achieved when all agents successfully reach their goals. While this indicates a successful pathfinding solution, an optimal solution would also minimize the number of timesteps. The structure incentivizes agents to maximize their reward by reaching their goals as quickly as possible, thus implicitly encouraging more efficient solutions. The reward structure is designed to incentivize agents to reach the goal for the first time and to coordinate such that all agents reach the goal. The episode terminates when all agents are at the goal at the same timestep. To describe the observation spaces, action spaces, and  reward function formally, we define the following sets and parameters:

\begin{itemize}
    \item $I = \{1, 2, \dots, n\}$: the set of $n$ agents.
    \item $T_{\max} = 100$: the maximum number of discrete timesteps.
    \item $T = \{0, 1, \dots, T_{\max}\}$: the set of timesteps.
    \item $T'=T \setminus \{T_{max}\}: T' \subset T.$
    \item $s_i(t)$: the state of each agent $i \in I$ at time $t \in T$.    
    \item $M\times N$, where $M$, $N \in \mathbb{N}_0$: the grid dimensions.
    \item $Pos = \{(x,y) | x \in \{0, 1, \dots, M \}, y \in \{0, 1, \dots, N\}\}$: the set of $M \times N$ positions.
    \item $pos_i(t) \in Pos$: the position of agent $i \in I$ at time $t \in T$, represented by its grid coordinates $(x_i(t), y_i(t))$.
    \item $G_i=(x_i,y_i) \in Pos$: where $G_i$ is the goal position for each agent $i \in I$.
    \item $A_i = \{0,\ 1,\ 2,\ 3,\ 4\}$: represents the action set for each agent $i \in I$, where each action corresponds to:
        \begin{equation}
            A_i=
            \begin{cases}
                0, & \text{stay (no-op), $(x, y) \rightarrow (x, y)$}\\
                1, & \text{move up, $(x, y) \rightarrow (x-1, y)$}\\
                2, & \text{move right, $(x, y) \rightarrow (x, y+1)$}\\
                3, & \text{move down, $(x, y) \rightarrow (x+1, y)$}\\
                4, & \text{move left, $(x, y) \rightarrow (x, y-1)$}\\
            \end{cases}
            \quad \forall i \in I.
        \end{equation}
    \item $F_i(t)$: a flag indicating whether agent $i \in I$ has reached the goal before $T_{\max}$.
\end{itemize}

\subsubsection*{Observation and Action Space for CTE}
In the CTE paradigm, a single agent (the centralized controller) has access to the complete state of the environment and determines the actions for all individual agents simultaneously. At each discrete timestep \( t \in T \), the observation available to the centralized controller is the full state of the environment:

\begin{equation}
    O(t) = s(t) \in S \quad \forall t \in T,
\end{equation}

where \( s(t) \) represents the complete grid configuration at time \( t \), and \( S \) denotes the set of all possible grid states. The action selected by the centralized controller is a joint action vector comprising the actions of all agents:

\begin{equation}
    A(t) = \left( a_1(t),\ a_2(t),\ \dots,\ a_n(t) \right) \in \mathcal{A}  \quad \forall t \in T,
\end{equation}

where $a_i(t) \in A_i$ is the action chosen for each agent $i$ at time $t$, \( n \) is the number of agents, and \( \mathcal{A} \) is the joint action space defined as the Cartesian product of individual action spaces:

\begin{equation}
    \mathcal{A} = \prod_{i=1}^{n} A_i = A_1 \times A_2 \times \dots \times A_n.
\end{equation}

\subsubsection*{Observation Space for CTDE and DTE}

In both the CTDE and DTE paradigms, each agent operates based on its own local observations without access to the full state of the environment. The policies are decentralized, and each agent makes decisions independently. At each timestep \( t \in T \), each agent \( i \) receives an observation \( o_i(t) \) composed of several components:

\begin{equation}
    o_i(t) = \left( o_i^{\text{grid}}(t),\ o_i^{\text{pos}}(t),\ o_i^{\text{goal}}(t) \right)  \quad \forall i\in I,  t \in T.
\end{equation}

The elements of \( o_i(t) \) are defined as follows:

The \textbf{local grid observation \( o_i^{\text{grid}}(t) \) }provides a partial view of the environment. At each timestep \(t \in T\), each agent $i \in I$ observes a local grid centered around its current position $pos_i(t)$. This local grid represents the environment within the agent's sensor range $R$ and is defined as: 

\begin{equation}
    o_i^{\text{grid}}(t) \in \{0,\ 1,\ 2,\ 3,\ 4\}^{(2R + 1) \times (2R + 1)}\quad \forall i\in I,  t \in T.
\end{equation}

Each cell in $o_i^{\text{grid}}(t)$ contains a value from the set $\{0,\ 1,\ 2,\ 3,\ 4\}$, representing different features within the environment. A value of $c = 0$ signifies that the cell is an empty space, meaning it is unoccupied and contains neither a goal nor an obstacle. When a cell contains a wall or obstacle, or if the cell is outside the grid environment, it is represented by $c = 1$, indicating either a static barrier within the environment or an out-of-bounds location. If the cell is occupied by another agent $j \neq i$, it is denoted by $c = 2$. Formally, this condition is expressed as $c = 2$ if there exists an agent $j \in I$, with $j \neq i$, such that $pos_j(t)=(x_j,y_j)$, where $(x_j,y_j)$ are the coordinates of the cell in question. The representation of goal locations depends on whether the goal belongs to the observing agent or another agent. The goal of the observing agent $i$ is indicated by $c = 3$; mathematically, $c = 3$ if $G_i=(x_i,y_i)$. If the cell contains the goal of another agent $j \neq i$, it is represented by $c = 4$. This is formally defined as $c = 4$ if there exists an agent $j \in I$, with $j \neq i$, such that $G_j = (x_j,y_j)$.

The \textbf{agent's own position \( o_i^{\text{pos}}(t) \)} on the grid is given by:

\begin{equation}
    o_i^{\text{pos}}(t) \in \mathbb{N}_0 \quad \forall i \in I, t \in T, pos \in Pos,
\end{equation}

where \( \mathbb{N}_0 \) denotes the set of natural numbers (including zero), representing the grid coordinates.

The known \textbf{position of the goal \( o_i^{\text{goal}}(t) \)} is:

\begin{equation}
    o_i^{\text{goal}}(t) = G_i \quad \forall i\in I,  t \in T.
\end{equation}

\subsubsection*{Policy Definitions}

In the CTE paradigm, the centralized policy \( \pi \) is a mapping from the observation space to the joint action space:

\begin{equation}
    \pi: S \rightarrow \mathcal{A}.
\end{equation}

The policy determines the joint action \( A(t) \) based on the full environment state \( s(t) \).

In the CTDE and DTE paradigms, each agent \( i \) employs an individual policy \( \pi_i \) that maps its local observation to an action:

\begin{equation}
    \pi_i: O_i \rightarrow A_i \quad \forall i\in I,
\end{equation}

where \( O_i \) represents the observation space of each agent \( i \). Each agent makes decisions autonomously based on its perception of the environment. The environment for all training and execution modes transitions from state \( s(t) \) to \( s(t+1) \) according to the transition function \( f \):

\begin{equation}
    s(t+1) = f\left( s(t),\ A(t) \right) \quad \forall t \in T',
\end{equation}

where \(T'=\{0,1, \dots, T_{\max}-1\}\) is the set of timesteps without the final timestep $T_{max}$ and \( A(t) \) is the joint action taken at time \( t \). The transition function encapsulates the environment's dynamics, including agent movements and interactions with obstacles.

\subsubsection*{Reward Function}
Initially, no agent has reached the goal:

\begin{equation}
    F_i(0) = 0, \quad \forall i \in I.
\end{equation}

At each timestep $t$, the flag $F_i(t)$ is updated as:

\begin{equation}
    F_i(t) = 
    \begin{cases}
    1, & \text{if } F_i(t-1) = 1 \text{ or } o_i^{\text{pos}}(t) = G_i, \\
    0, & \text{otherwise},
    \end{cases}
    \quad \forall i \in I, t \in T.
\end{equation}

This ensures that once an agent reaches the goal, $F_i(t)$ remains $1$ for all subsequent time steps.

We define the following indicator functions:

\begin{itemize}
    \item $L_i(t)$: indicates if agent $i$ is at the goal at time $t$,
    \begin{equation}
        L_i(t) = 
        \begin{cases}
        1, & \text{if } s_i(t) = G_i, \\
        0, & \text{otherwise},
        \end{cases}
        \quad \forall i \in I, t \in T.
    \end{equation}
    \item $J_i(t)$: indicates if agent $i$ reaches the goal for the first time at time $t$,
    \begin{equation}
        J_i(t) = 
        \begin{cases}
        1, & \text{if } F_i(t-1) = 0 \text{ and } s_i(t) = G_i, \\
        0, & \text{otherwise},
        \end{cases}
        \quad \forall i \in I, t \in T.
    \end{equation}
    \item $K(t)$: indicates if all agents are at the goal at time $t$,
    \begin{equation}
        K(t) = 
        \begin{cases}
        1, & \text{if } L_i(t) = 1 \quad \forall i \in I, \\
        0, & \text{otherwise},
        \end{cases}
        \quad \forall t \in T.
    \end{equation}
    \item $C_i(t)$: indicates if an agent is involved in a collision at time $t$,
    \begin{equation}
        C_i(t) = 
        \begin{cases}
        1, & \text{if agent $i$ is involved in a collision at time $t$,} \\
        0, & \text{otherwise},
        \end{cases}
        \quad \forall i \in I, t \in T.
    \end{equation}
\end{itemize}

An agent \( i \) is considered to be involved in a collision at time \( t \) if either agent \( i \) attempts to move into a cell that is currently occupied by another agent \( j \neq i \) or if agent \( i \) and another agent \( j \neq i \) both attempt to move into the same cell at the same time. Formally, the first case occurs when
\begin{equation}
    pos_i(t+1) = pos_j(t), \quad \forall i, j \in I,\, j \neq i, t \in T',
\end{equation}
and the second case occurs when
\begin{equation}
    pos_i(t+1) = pos_j(t+1), \quad \forall i, j \in I,\, j \neq i, t \in T'.
\end{equation}
Note that scenarios in which two agents attempt to swap positions, where agent \( i \) moves to \( pos_j(t) \) and agent \( j \) moves to \( pos_i(t) \), are included in the first condition, as each agent is attempting to enter a cell currently occupied by the other.

 Let $\alpha$ represent the weight for the first-time goal reach reward, $\beta$ the weight for the all-agents-at-goal reward, and $\gamma$ the penalty weight for collisions. The reward $r_i(t)$ for each agent $i$ at time $t$ is defined as:

\begin{equation} 
    r_i(t) = \alpha \cdot J_i(t) + \beta \cdot K(t) + \gamma \cdot C_i(t), \quad \forall i \in I, \ t \in T. 
\end{equation}

We assign the following values to the constants:

\begin{equation} \alpha = 0.5, \quad \beta = 1, \quad \gamma = -1. \end{equation}

The reward $r_i(t)$ for each agent $i$ at time $t$ can be broken down into three components:

\begin{enumerate}
    \item First-time goal reach reward ($\alpha \cdot J_i(t)$): An agent receives a reward of $+0.5$ only when it reaches the goal for the first time.

    \item All agents at goal reward ($\beta \cdot K(t)$): If all agents are at the goal at the same time $t$, each agent receives an additional reward of $+1$, and the episode terminates.

    \item Collision penalty (\(\gamma \cdot C_i(t)\)): An agent $i$ incurs a penalty of \(-1\) if it is involved in a collision at time \( t \).
\end{enumerate}

The episode ends at time $t$ if either condition is met:

\begin{enumerate}
    \item All agents are simultaneously at the goal state:
    \begin{equation}
        \text{If } K(t) = 1, \quad \text{then the episode terminates at time } t.
    \end{equation}
    \item The maximum time limit is reached:
    \begin{equation}
        \text{If } t = T_{\max} = 100, \quad \text{then the episode terminates at time } t.
    \end{equation}
\end{enumerate}

The total reward for each agent $i$ over the episode is:

\begin{equation}
R_i = \sum_{t=1}^{T_{\text{end}}} r_i(t) \quad \forall i \in I,
\end{equation}

where $T_{\text{end}}$ is the time step at which the episode ends. This reward structure incentivizes agents to reach the goal for the first time and to coordinate with other agents to reach the goal simultaneously, maximizing their cumulative reward.

Table \ref{table:reference-models-spaces} provides an overview of the observation space, action space, and reward structure for the reference model environments. In terms of the observation space, for CTE, agents receive a complete view of the grid, including information about empty cells, obstacles, agent positions, and goals. For CTDE and DTE, the observation is limited to a partial view based on the agent's sensor range, along with the agent's current position and its goal.

\begin{table}[h]
    \centering
    \caption{Observation space, action space, and rewards for the reference model environments.}
    \begin{tabular}{ll}
        \toprule
            \multicolumn{2}{c}{\textbf{Observation space}} \\ \hline
            CTE & Complete grid with empty cells, obstacles, agent positions, and goals\\
            CTDE \& DTE & Partial view of the grid, limited by the sensor range\\
            & Current position of the considered agent \\
            & Goal of the considered agent \\
            \multicolumn{2}{c}{\textbf{Action space}} \\ 
            CTE & MultiDiscrete([5] * $n$) \\
            & (0: No-op, 1: move up, 2: Move right, 3: move down, 4: move left) \\
            CTDE \& DTE & Discrete (5) \\
            & (0: No-op, 1: move up, 2: Move right, 3: move down, 4: move left) \\
            \multicolumn{2}{c}{\textbf{Rewards}} \\ 
            First-time goal reach & $+0.5$ for reaching the goal the first time in an episode\\ 
            All agents at goal & $+1$ for every agent if all agents are on their goal position
             \\ 
            Collision & $-1$ for each collision (for each agent involved) \\ 
        \bottomrule
    \end{tabular}
    \label{table:reference-models-spaces}
\end{table}

\subsection{Technical Implementation of the Learning Environment}
\begin{figure}[h]
    \centering
    \includegraphics[width=0.7\linewidth]{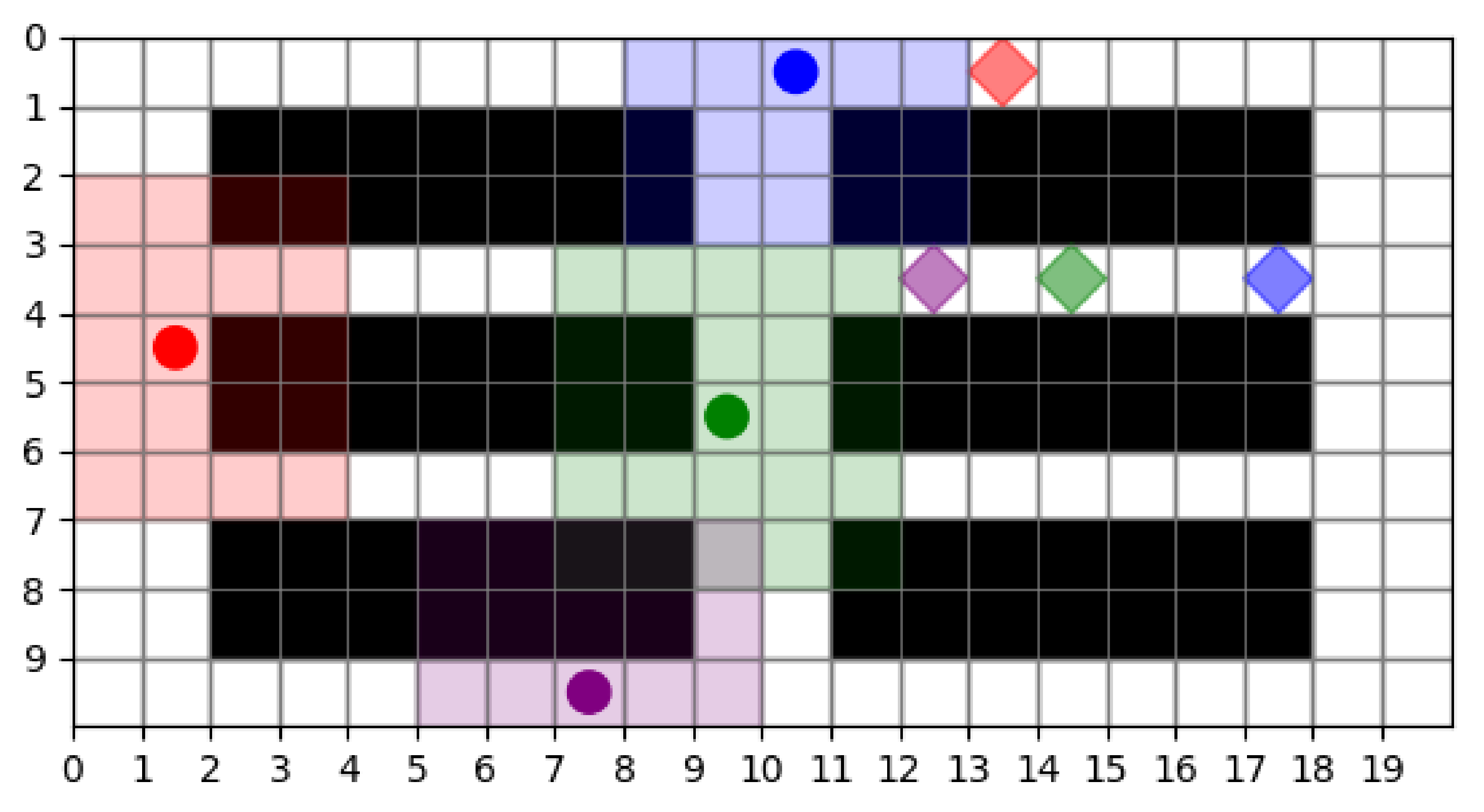}
    \caption{Rendered environment of reference model 2.1 with four agents.}
    \label{fig:reference-models-rendered}
\end{figure}

The technical implementation of the learning environment is based on a 2D grid, where each cell is defined as empty, an obstacle, an agent's position, or a goal. Figure \ref{fig:reference-models-rendered} illustrates the rendered structure, with agents shown as colored circles and their corresponding goals as matching diamonds. The grid state is represented as a 2D NumPy array, allowing efficient manipulation and evaluation during simulation.

The environment is defined in the class \texttt{ReferenceModel}. For multi-agent scenarios, it inherits from \texttt{MultiAgentEnv} in Ray RLlib, supporting multiple agents interacting within the same grid. For single-agent setups, the class inherits from \texttt{gym.Env} in Gymnasium, designed for standard RL use cases. This separation ensures compatibility with the respective frameworks. Starting and goal positions for agents are either preconfigured for deterministic scenarios or randomly assigned to ensure variability in training. Each episode begins with a reset function that sets the initial states, including agent positions, goal locations, and internal counters. Rendering is implemented with Matplotlib, providing real-time visualization of agent movements, obstacles, and goal locations. The environment updates the grid state at each timestep, adjusting the visual output accordingly. Obstacles are displayed as black cells, while agents and goals are color-coded for easy identification. This visualization aids in debugging and ensures alignment between theoretical expectations and practical outcomes. The implementation is designed to be modular and scalable, supporting both single-agent and multi-agent configurations. It allows systematic evaluation of RL algorithms under varying levels of complexity and interaction dynamics. New grid layouts can easily be added with a NumPy array.

The source code for the learning environment, as well as for training and evaluation of the agents used in this use case, is available in a public repository maintained by the author: \url{https://github.com/Nerozud/dl_reference_models}.

\subsection{Four-Phase Evaluation of MARL}

The experimental setup for the reference models follows a structured four-phase process to ensure a comprehensive evaluation of the RL algorithms and their respective training and execution modes. This consecutive approach is designed to assess their performance under varying conditions, resulting in reliable and reproducible outcomes.

\textbf{Phase 1: Hyperparameter configuration and exploration.}\\
In the first phase, random samples are generated for each hyperparameter configuration, which consists of a RL algorithm paired with a specific training and execution mode, within a defined computational budget. This limited timeframe is used to conduct a hyperparameter search, exploring the hyperparameters batch size, learning rate, entropy coefficient, number of epochs, clipping parameter, value function loss coefficient and the dimensions of the hidden layers of the ANN. This phase is essential for identifying effective configurations efficiently, as a full-scale hyperparameter optimization would be computationally prohibitive.

\textbf{Phase 2: Validation of selected configurations.}\\
The second phase entails further testing of the best configurations identified during the hyperparameter search. Each selected configuration is evaluated using 10 independent samples to ensure comparability between different algorithms and training modes. This phase ensures that the results reflect consistent performance across multiple runs, rather than being influenced by random variability. All experiments are conducted under walltime constraints, ensuring fair comparison across RL algorithms by providing equal computational resources.

\textbf{Phase 3: Comparative analysis of RL Policies and classical MAPF algorithms.}\\
In the third phase, starting from reference model 2.1, the RL policies are compared with classical MAPF algorithms MA-A* and CBS. This comparison focuses on success rates and the number of timesteps required to complete tasks. Additionally, policy behavior is visualized and analyzed using heat maps to better understand agent decision-making.

\textbf{Phase 4: Assessment of generalization and scalability.}\\
The final phase assesses the generalization and scalability of the trained RL policies. This is achieved by varying the number of agents and evaluating policy performance on previously unseen environments, thus testing the adaptability of the learned behaviors beyond the training setup.

All RL experiments are conducted on a system equipped with an Intel Core i7-10700K central processing unit (CPU) with 3.80 GHz, utilizing 11 CPUs. Of these, 10 CPUs are allocated to the learning environment, each managing 2 environments, while one CPU is reserved for the execution of the algorithm. A single NVIDIA GeForce RTX 3080 GPU is available for accelerating computations where applicable. The system provides 16 GB of RAM.

\subsection{Hyperparameter Configuration and Exploration}
To ensure comparability across different algorithms and training or execution modes, each variant is given an equal opportunity to identify suitable parameters for the respective configuration. Each variant is allocated one hour of computational budget for reference model 1.1 and four hours for reference models 2.1 and 3.1, representing a reasonable amount of time on conventional computing resources to perform a hyperparameter search in logistics planning. To reduce computational effort during hyperparameter search, only the basic variants of reference models 1.1, 2.1, and 3.1 are used to represent their respective categories. For example, the parameters chosen for reference model 1.1 are applied to reference model 1.3. This approach results in nine hyperparameter search experiments for each algorithm in the first experiment series.

The hyperparameters are selected at random from a range of typical values for the respective hyperparameter. Recommended hyperparameters ranges are oriented on conclusions of \cite{yu2022surprisingPPO} and \cite{papoudakis2021benchmarking}. Table \ref{tab:hyperparameter-search-PPO} shows the fixed and varied hyperparameters for all PPO configurations for the reference models. 

\begin{table}[h]
    \centering
    \caption{Hyperparameter search for PPO in the reference models 1.1, 2.1, and 3.1.}
    \label{tab:hyperparameter-search-PPO}
    \begin{tabular}{lc}
    \toprule
        \textbf{Hyperparameter} & \textbf{Value} \\ \hline
         Training batch size & 4,000 \\
         Minibatch size & 4,000 \\
         Discount factor $\gamma$ & 0.99  \\
         Parameter $\lambda$ for GAE & 0.95 \\
         Initial coefficient for KL divergence & 0.2 \\
         Target value for KL divergence & 0.01 \\
         Number of epochs & 5 - 14 \\
         Clipping parameter $\epsilon$ & 0.05, 0.1, 0.2, 0.3 \\
         Learning rate $\alpha$ & 0.0001, 0.0003, 0.0005, 0.001 \\
         Entropy coefficient & 0, 0.001, 0.01 \\
         Dimension of hidden ANN layers & [32, 32], [64, 64], [256, 256] \\
    \bottomrule
    \end{tabular}
\end{table}

For the PPO algorithm, hyperparameters such as the training batch size, discount factor, and $\lambda$ for GAE are fixed based on established values known to provide stable performance in RL tasks. For example, the discount factor is fixed at $\gamma = 0.99$ to prioritize long-term rewards, while $\lambda = 0.95$ for GAE helps strike a balance between bias and variance. The number of epochs is sampled between 5 and 14, allowing flexibility in the number of passes through the data, with a higher number of epochs potentially leading to better convergence, though at the cost of longer training times. The clipping parameter, learning rate, and entropy coefficient are sampled from predefined categorical values rather than a continuous range. Specifically, the clipping parameter $\epsilon$ is sampled from the values 0.05, 0.1, 0.2, and 0.3 to test varying degrees of constraint on policy updates. The learning rate is selected from the categorical values 0.0001, 0.0003, 0.0005, and 0.001, representing typical values used in RL to balance stability and convergence speed. Similarly, the entropy coefficient is chosen from 0, 0.001, and 0.01 to control the exploration-exploitation trade-off, with higher values promoting more exploration.

Table \ref{tab:hyperparameter-search-IMPALA} outlines the hyperparameter search space for the IMPALA algorithm in the same reference models. IMPALA also includes the value function loss coefficient, as we follow the standard RLlib configuration of IMPALA, where the hidden layers of the value function are shared with the policy network.

\begin{table}[h]
    \centering
    \caption{Hyperparameter search for IMPALA in the reference models 1.1, 2.1, and 3.1.}
    \label{tab:hyperparameter-search-IMPALA}
    \begin{tabular}{lc}
    \toprule
        \textbf{Hyperparameter} & \textbf{Value} \\ \hline
         Training batch size & 500, 1,000, 4,000 \\
         Discount factor $\gamma$ & 0.99  \\
         Rollout fragment length & 50 \\
         Value function loss coefficient & 0, 0.1, 0.3, 0.5, 0.9, 1 \\
         Learning rate $\alpha$ & 0.0001, 0.0003, 0.0005, 0.001 \\
         Entropy coefficient & 0, 0.001, 0.01 \\
         Dimension of hidden ANN layers & [32, 32], [64, 64], [256, 256] \\
    \bottomrule
    \end{tabular}
\end{table}

For IMPALA, the training batch size is varied more widely, from 500 to 4,000, to test how different batch sizes affect learning stability and sample efficiency. Smaller batch sizes provide faster updates but may introduce more variability in gradients, whereas larger batches help smooth gradient updates. The rollout fragment length is fixed at 50 steps to ensure that updates occur frequently enough, while the value function loss coefficient is varied to test its influence on the balance between policy optimization and value function learning. As with PPO, the learning rate is sampled from the categorical values 0.0001, 0.0003, 0.0005, and 0.001 to evaluate its effect on training speed and stability. Similarly, the entropy coefficient and hidden layer dimensions are varied to test their impact on the exploration-exploitation balance and the network's capacity to represent more complex functions. The hidden layers are tested with sizes of [32, 32], [64, 64], and [256, 256] to explore the trade-off between model complexity and computational cost.

Each episode has a maximum of 100 timesteps unless it ends earlier when all agents reach their goals. A trial is stopped after 200,000 timesteps for PPO and 500,000 for IMPALA for reference model 1.1 and after 2,000,000 timesteps for reference model 2.1 and 3.1 for both algorithms. CTE is modeled to encompass the complete grid, while the training and execution modes of CTDE and DTE represent the learning environments as POMDPs, using an observation grid of 5x5 cells. The reference model 1.1 uses the fixed starting and goals of the basic variant as in Figure \ref{fig:reference-model-1-1} to learn directly the proposed conflict situation. For reference model 2.1 and 3.1 we use random start and goals to learn a policy for general MAPF on the respective layout with a fixed number of four agents. The best hyperparameter configurations are selected based on the mean episode reward across 100 episodes. If the mean reward is identical, the configuration with the shortest average episode length over the last 100 episodes is chosen. The episode length reflects how quickly the agents reach their goals, prioritizing configurations that require fewer timesteps.

Table \ref{tab:hyperparameter-found-PPO-1} presents the best hyperparameters found for PPO in reference model 1.1, across CTE, CTDE, and DTE. The table compares the number of trials, timesteps per trial, mean episode reward, mean episode length, and the found best hyperparameters. All training and execution modes are able to find the best solution regarding the mean reward in one hour. There are only minor differences regarding the mean episode length, which is close to the optimum of nine timesteps for reference model 1.1.

\begin{table}[h]
    \centering
    \caption{Results of the hyperparameter search for PPO in reference model 1.1.}
    \label{tab:hyperparameter-found-PPO-1}
    \begin{tabular}{lccc}
    \toprule
        \textbf{Parameter} & \textbf{CTE} & \textbf{CTDE} & \textbf{DTE} \\ \hline
         Number of trials & 46 & 32 & 24 \\
         Timesteps per trial & 200,000 & 200,000 & 200,000 \\
         Mean episode reward & 3 & 3 & 3 \\
         Mean episode length & 11.01  & 12.1 & 12.03 \\
         Number of epochs & 10 & 6 & 12 \\
         Clipping parameter $\epsilon$ & 0.3 & 0.3 & 0.2 \\
         Learning rate $\alpha$ & 0.0005 & 0.001 & 0.0005 \\
         Entropy coefficient & 0.001 & 0 & 0.001 \\
         Dimension of hidden ANN layers & [256, 256] & [64, 64] & [32, 32] \\
    \bottomrule
    \end{tabular}
\end{table}

Table \ref{tab:hyperparameter-found-PPO-2} displays the hyperparameter search results for PPO in reference model 2.1. Here, CTDE emerges as the most successful approach, achieving the highest mean episode reward of 5.4, compared to 2.28 in DTE and a negative mean episode reward of -3.57 in CTE. The differences in mean episode length are also notable, with CTDE showing a considerably shorter episode length of 34.84 compared to 75.89 in DTE and 100 in CTE, indicating higher success rates for reaching the goal in an episode. Hyperparameter choices also vary significantly, particularly in terms of the clipping parameter and hidden ANN layer dimensions, which reflect the tuning required for each training and execution mode's performance in this more complex scenario.

\begin{table}[h]
    \centering
    \caption{Results of the hyperparameter search for PPO in reference model 2.1.}
    \label{tab:hyperparameter-found-PPO-2}
    \begin{tabular}{lccc}
    \toprule
        \textbf{Parameter} & \textbf{CTE} & \textbf{CTDE} & \textbf{DTE} \\ \hline
         Number of trials & 24 & 11 & 6 \\
         Timesteps per trial & 2,000,000 & 2,000,000 & 2,000,000 \\
         Mean episode reward & -3.57 & 5.4 & 2.28 \\
         Mean episode length & 100  & 34.84 & 75.89 \\
         Number of epochs & 6 & 12 & 6 \\
         Clipping parameter $\epsilon$ & 0.2 & 0.05 & 0.1 \\
         Learning rate $\alpha$ & 0.0005 & 0.001 & 0.0005 \\
         Entropy coefficient & 0.001 & 0.001 & 0.01 \\
         Dimension of hidden ANN layers & [64, 64] & [32, 32] & [256, 256] \\
    \bottomrule
    \end{tabular}
\end{table}

Table \ref{tab:hyperparameter-found-PPO-3} illustrates the hyperparameter search results for PPO in reference model 3.1. In this case, none of the training and execution modes are able to find a successful policy in the given number of timesteps, as indicated by the negative mean episode rewards across all approaches. The mean episode length remains close to the maximum possible value of 100, suggesting that the episodes often truncate and not all agents reach the goals within the episode length. However, as expected, CTDE achieved the best result due to the better utilization of timesteps via the shared policy, which enables more efficient learning compared to DTE. With more timesteps we expect that CTDE and DTE find good policies with the given hyperparameter configuration.

\begin{table}[h]
    \centering
    \caption{Results of the hyperparameter search for PPO in reference model 3.1.}
    \label{tab:hyperparameter-found-PPO-3}
    \begin{tabular}{lccc}
    \toprule
        \textbf{Parameter} & \textbf{CTE} & \textbf{CTDE} & \textbf{DTE} \\ \hline
         Number of trials & 24 & 10 & 7 \\
         Timesteps per trial & 2,000,000 & 2,000,000 & 2,000,000 \\
         Mean episode reward & -3.73 & -0.45 & -2.59 \\
         Mean episode length & 100  & 94.45 & 99.86 \\
         Number of epochs & 10 & 7 & 8 \\
         Clipping parameter $\epsilon$ & 0.2 & 0.1 & 0.05 \\
         Learning rate $\alpha$ & 0.0001 & 0.001 & 0.0005 \\
         Entropy coefficient & 0.01 & 0.01 & 0 \\
         Dimension of hidden ANN layers & [32, 32] & [256, 256] & [64, 64] \\
    \bottomrule
    \end{tabular}
\end{table}

Table \ref{tab:hyperparameter-found-IMPALA-1} presents the hyperparameter search results for IMPALA in reference model 1.1. Similar to the PPO results for the same model, all training and execution modes reach the optimal mean episode reward of 3, with only minor variations in mean episode length across CTE, CTDE, and DTE. Despite notable differences in certain hyperparameters, such as the training batch size and the value function loss coefficient, their impact on performance appears minimal in this case. The performance across modes remains robust, with the best configurations leading to similar results, suggesting that for this specific scenario, the hyperparameters did not have a high visible influence on the outcome.

\begin{table}[h]
    \centering
    \caption{Results of the hyperparameter search for IMPALA in reference model 1.1.}
    \label{tab:hyperparameter-found-IMPALA-1}
    \begin{tabular}{lccc}
    \toprule
        \textbf{Parameter} & \textbf{CTE} & \textbf{CTDE} & \textbf{DTE} \\ \hline
         Number of trials & 27 & 18 & 13 \\
         Timesteps per trial & 500,000 & 500,000 & 500,000 \\
         Mean episode reward & 3 & 3 & 3 \\
         Mean episode length & 11  & 12.23 & 12.02 \\
         Training batch size & 500 & 4,000 & 500 \\
         Value function loss coefficient & 0 & 0.3 & 0.9 \\
         Learning rate $\alpha$ & 0.0005 & 0.001 & 0.0003 \\
         Entropy coefficient & 0.01 & 0.001 & 0.001 \\
         Dimension of hidden ANN layers & [64, 64] & [32, 32] & [256, 256] \\
    \bottomrule
    \end{tabular}
\end{table}

Table \ref{tab:hyperparameter-found-IMPALA-2} presents the hyperparameter search results for IMPALA in reference model 2.1. Compared to PPO, IMPALA converges more slowly per timestep, which is evident from the lower mean episode rewards across all modes. While CTDE and DTE show some potential, with rewards of -1.27 and -3.14 respectively, they do not reach the higher rewards seen in PPO for the same model. The timesteps restriction during the trials allows for enough sampling to identify suitable configurations, which is necessary for CTDE and DTE. Due to IMPALA's slower convergence, IMPALA has not fully learned better rewards within the allocated time. Despite this, the best hyperparameters found for CTDE and DTE indicate that further training could lead to improved results, reflecting effective tuning within the given constraints.

\begin{table}[h]
    \centering
    \caption{Results of the hyperparameter search for IMPALA in reference model 2.1.}
    \label{tab:hyperparameter-found-IMPALA-2}
    \begin{tabular}{lccc}
    \toprule
        \textbf{Parameter} & \textbf{CTE} & \textbf{CTDE} & \textbf{DTE} \\ \hline
         Number of trials & 27 & 12 & 7 \\
         Timesteps per trial & 2,000,000 & 2,000,000 & 2,000,000 \\
         Mean episode reward & -3.59 & -1.27 & -3.14 \\
         Mean episode length & 100  & 96.93 & 100 \\
         Training batch size & 1,000 & 500 & 1,000 \\
         Value function loss coefficient & 0 & 1 & 0.1 \\
         Learning rate $\alpha$ & 0.0003 & 0.0001 & 0.0003 \\
         Entropy coefficient & 0.01 & 0.001 & 0 \\
         Dimension of hidden ANN layers & [32, 32] & [256, 256] & [256, 256] \\
    \bottomrule
    \end{tabular}
\end{table}

Table \ref{tab:hyperparameter-found-IMPALA-3} shows the hyperparameter search results for IMPALA in reference model 3.1. Similar to the previous results, none of the training and execution modes are able to achieve positive rewards, with all modes showing negative mean episode rewards close to -3.5. The episode lengths remain fixed at 100, indicating that the policies failed to effectively learn within the allocated time. Despite these poor results, CTDE performs slightly better than CTE and DTE, with a mean episode reward of -3.16, likely due to the more efficient usage of the timesteps via the shared policy. This advantage of CTDE, while subtle, reflects its ability to utilize the experience of multiple agents more effectively. The overall performance suggests that IMPALA has not had sufficient time to converge to better rewards in this complex scenario. 

\begin{table}[h]
    \centering
    \caption{Results of the hyperparameter search for IMPALA in reference model 3.1.}
    \label{tab:hyperparameter-found-IMPALA-3}
    \begin{tabular}{lccc}
    \toprule
        \textbf{Parameter} & \textbf{CTE} & \textbf{CTDE} & \textbf{DTE} \\ \hline
         Number of trials & 27 & 12 & 7 \\
         Timesteps per trial & 2,000,000 & 2,000,000 & 2,000,000 \\
         Mean episode reward & -3.76 & -3.16 & -3.5 \\
         Mean episode length & 100  & 100 & 100 \\
         Training batch size & 4,000 & 4,000 & 500 \\
         Value function loss coefficient & 0.3 & 0.1 & 0.5 \\
         Learning rate $\alpha$ & 0.001 & 0.0005 & 0.0001 \\
         Entropy coefficient & 0.001 & 0.001 & 0.001 \\
         Dimension of hidden ANN layers & [256, 256] & [32, 32] & [32, 32] \\
    \bottomrule
    \end{tabular}
\end{table}

The hyperparameter search is restricted to four hours to ensure a fair comparison across algorithms. Given these constraints, the focus is on completing at least some trials within the allotted time. Both CTDE and DTE show a relatively low number of trials under these restrictions. Although the time limit prevents further training or hyperparameter tuning, this constraint is necessary to maintain consistency across the experiments.

\subsection{Validation of Selected Configurations}
This subsection provides a comparison of RL algorithms and training modes across the reference models. The configuration used for PPO in reference model 1.1 is directly applied to reference models 1.2, 1.3, and 1.4, maintaining consistency in the evaluation. In these models, the stopping criterion for a trial is set to either achieving an episode mean reward equal to 1.5 times the number of agents or reaching a maximum wall time of ten minutes. For reference models 2.1, 2.2, and 3.1, the configuration from model 2.1 is reused for model 2.2, ensuring that a uniform setup is applied across these more complex scenarios. In these models, the stopping criterion is also based on reaching an episode mean reward of 1.5 times the number of agents, but with an extended wall time of 30 minutes due to the higher complexity of the tasks.

\begin{figure}[h]
    \centering
    \includegraphics[width=1\linewidth]{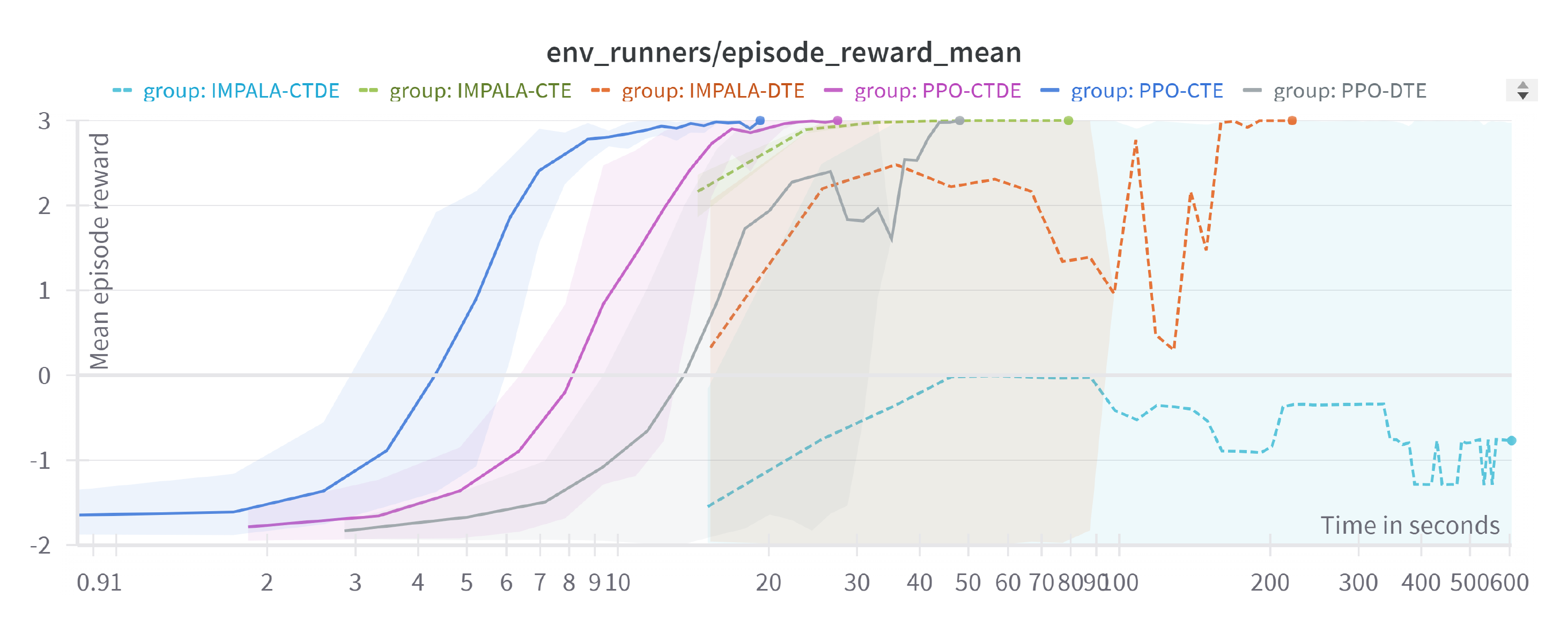}
    \caption{Mean episode rewards for PPO and IMPALA in reference model 1.1 with different training and execution modes.}
    \label{fig:learning-curves-reference-model-1-1}
\end{figure}

Figure \ref{fig:learning-curves-reference-model-1-1} presents the mean episode rewards over time for PPO and IMPALA under different training and execution modes in reference model 1.1. The solid lines represent the average mean episode reward of the ten samples, while the shaded areas indicate the range between the minimum and maximum mean episode rewards across all samples at each time step. The logarithmic scale for the reference models 1.1 - 1.4 is chosen to provide better visibility of early learning dynamics, where significant changes in mean episode rewards occur. In a linear scale, these early changes are compressed, making it difficult to observe differences in learning speed and stability between configurations. By using a logarithmic scale, early time steps are expanded, offering clearer insights into how quickly the algorithms converge. The logarithmic scale allows for a balanced view of both early learning phases and long-term trends in a single plot.

For PPO, all configurations (CTE, CTDE, and DTE) lead to stable solutions, with PPO-CTE showing the fastest convergence. PPO-CTDE and PPO-DTE follow, both eventually reaching similar performance. The narrow shaded areas indicate low variability, demonstrating consistent results across all PPO modes. For IMPALA, the CTE configuration leads to stable and consistent learning, quickly achieving high rewards. IMPALA-DTE also converges, but only after an initial period of instability, where fluctuations are more pronounced. IMPALA-CTDE, on the other hand, displays considerable instability, failing to reach a reliable solution and showing large variability in its performance. It is important to note an approximately 15-second initialization time for IMPALA across all modes, during which no learning occurs. This delay reflects some additional computational overhead, particularly in the distributed setting of IMPALA.

\begin{figure}[h]
    \centering
    \includegraphics[width=1\linewidth]{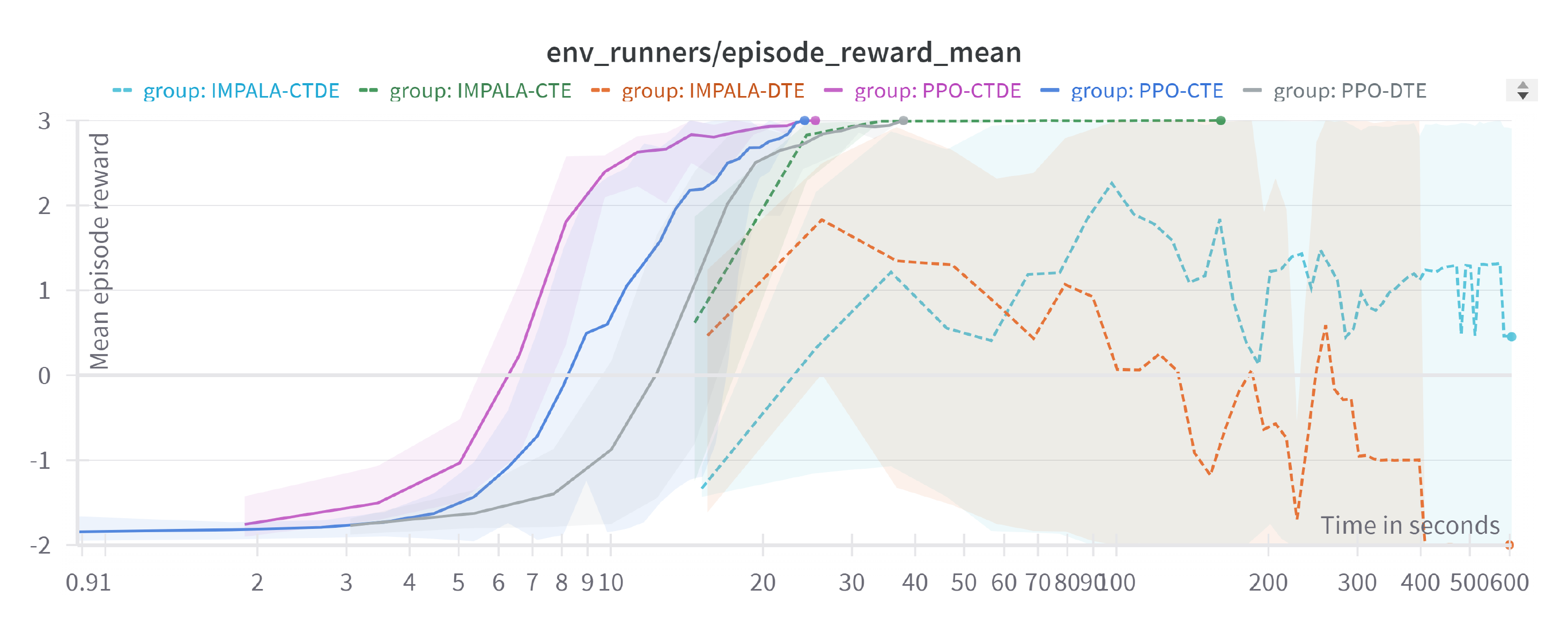}
    \caption{Mean episode rewards for PPO and IMPALA in reference model 1.2 with different training and execution modes.}
    \label{fig:learning-curves-reference-model-1-2}
\end{figure}

Figure \ref{fig:learning-curves-reference-model-1-2} illustrates the mean episode rewards over time for PPO and IMPALA across different training and execution configurations. For PPO, all variants (CTE, CTDE, and DTE) manage to solve the environment stably, with PPO-CTDE demonstrating the fastest convergence. This indicates that PPO is effective and stable regardless of the specific training and execution mode for reference model 1.2, with CTDE offering a slight advantage in speed. In contrast, for IMPALA, only CTE exhibits stable and reliable learning, consistently reaching high rewards. IMPALA-CTDE and IMPALA-DTE display significant variability in performance, with both modes sometimes finding good solutions but failing to maintain stability. IMPALA-DTE shows particularly erratic behavior, highlighting the challenges of decentralized training and execution in this setup. Figure \ref{fig:learning-curves-reference-model-1-2} suggests that PPO's robustness extends across different modes, while IMPALA struggles with decentralized configurations, especially in DTE, where stable learning is rarely achieved.

\begin{figure}[h]
    \centering
    \includegraphics[width=1\linewidth]{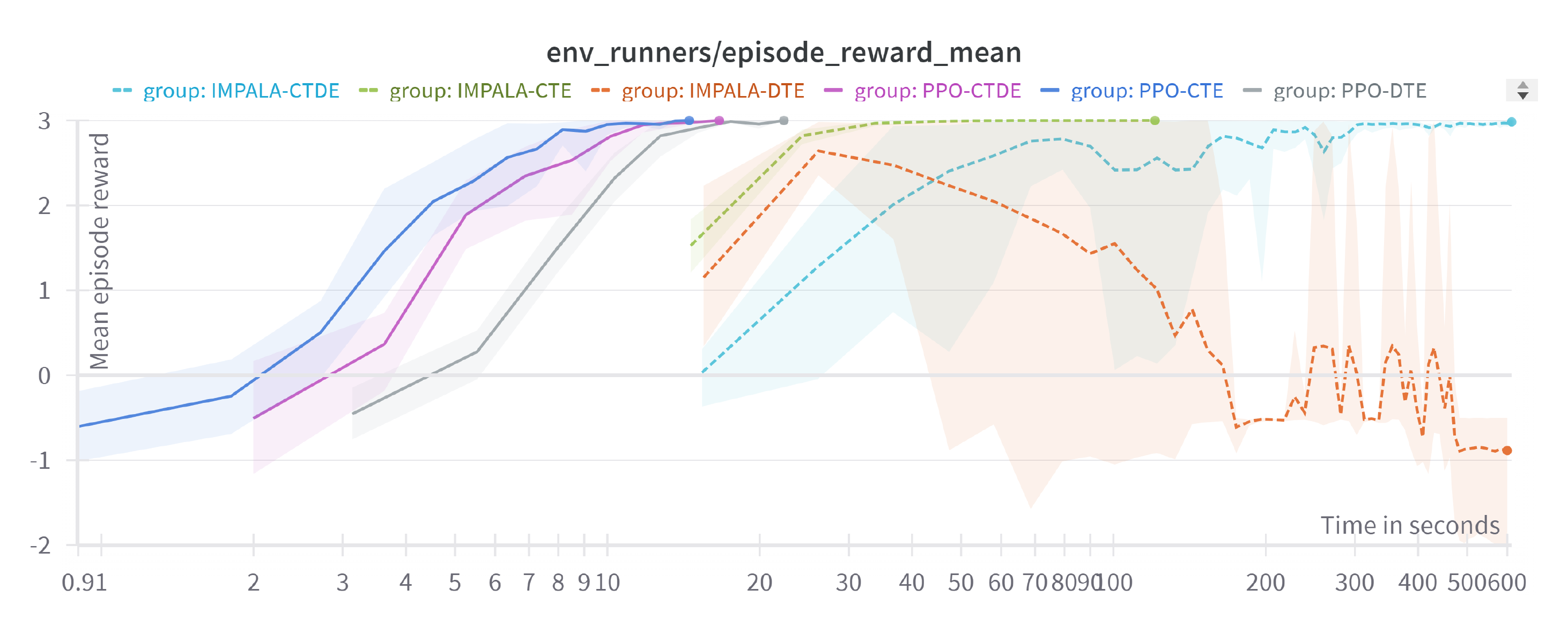}
    \caption{Mean episode rewards for PPO and IMPALA in reference model 1.3 with different training and execution modes.}
    \label{fig:learning-curves-reference-model-1-3}
\end{figure}

Figure \ref{fig:learning-curves-reference-model-1-3} shows the mean episode rewards over time for PPO and IMPALA across different training and execution modes for reference model 1.3. For PPO, all training and execution modes converge to stable solutions, with PPO-CTE showing the fastest convergence, followed by PPO-CTDE and PPO-DTE. The small variability in rewards indicates stable performance across trials for all PPO configurations. For IMPALA, CTE again delivers stable learning, while CTDE also reaches a stable solution after some initial instability. However, IMPALA-DTE struggles, often failing to converge and showing high variability in rewards.

\begin{figure}[h]
    \centering
    \includegraphics[width=1\linewidth]{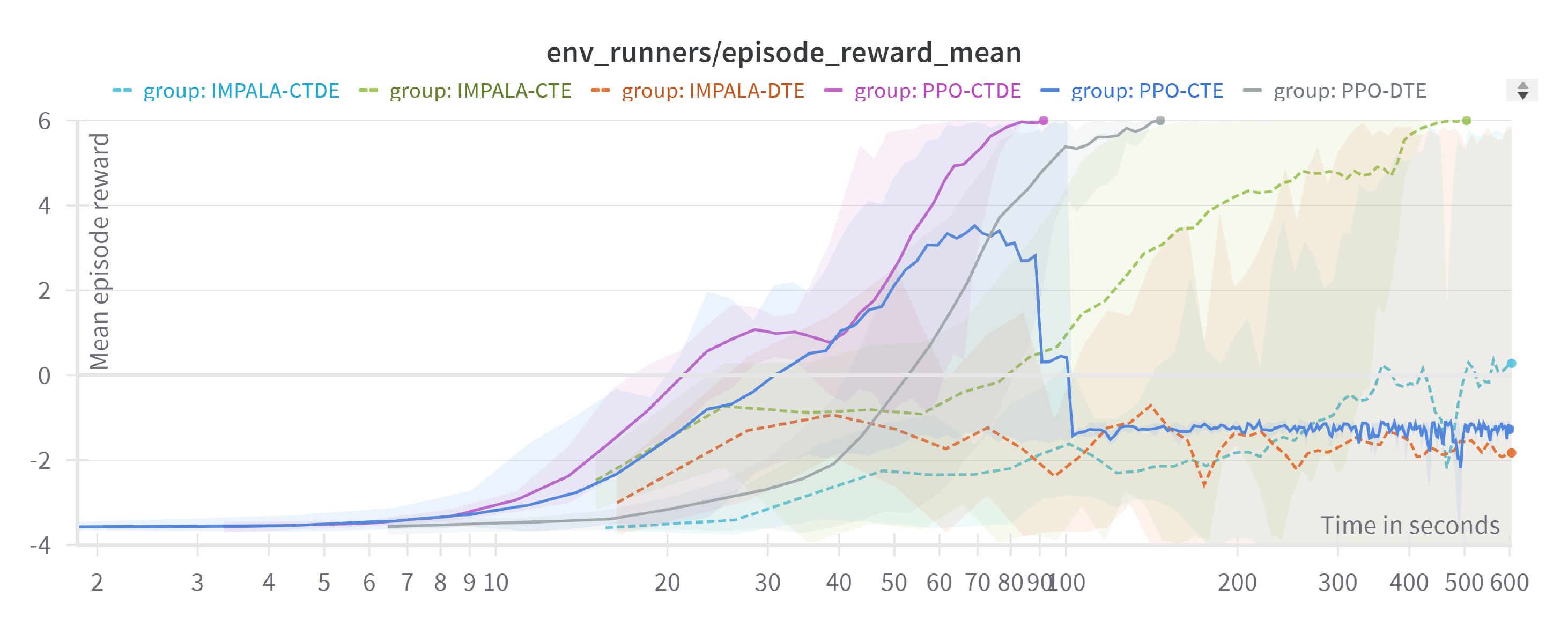}
    \caption{Mean episode rewards for PPO and IMPALA in reference model 1.4 with different training and execution modes.}
    \label{fig:learning-curves-reference-model-1-4}
\end{figure}

Figure \ref{fig:learning-curves-reference-model-1-4} presents the mean episode rewards for PPO and IMPALA across different training and execution configurations in reference model 1.4. PPO-CTDE reaches the highest rewards most quickly, showing stable convergence, followed by PPO-DTE, which converges slightly slower but also stabilizes at a high reward level. PPO-CTE often converges as fast as PPO-CTDE but failed in three of the ten samples to find a good policy. IMPALA displays more variability across configurations. IMPALA-CTE shows consistent learning, achieving stable high rewards over time, though at a slower pace than PPO. IMPALA-CTDE, while taking longer to learn than PPO-CTDE, also struggles with few samples not learning a suitable policy. IMPALA-DTE shows high fluctuations and fails to reach the same performance levels as the other configurations.

The learning curves for there reference models 1.1 - 1.4 highlight the continued robustness of PPO across training and execution modes, with CTE and CTDE offering the fastest convergence. IMPALA also shows stable learning under CTE, but its performance under CTDE and  DTE remains unreliable, as reflected by the large variability in rewards and slower convergence in these modes. 

\begin{figure}[h]
    \centering
    \includegraphics[width=1\linewidth]{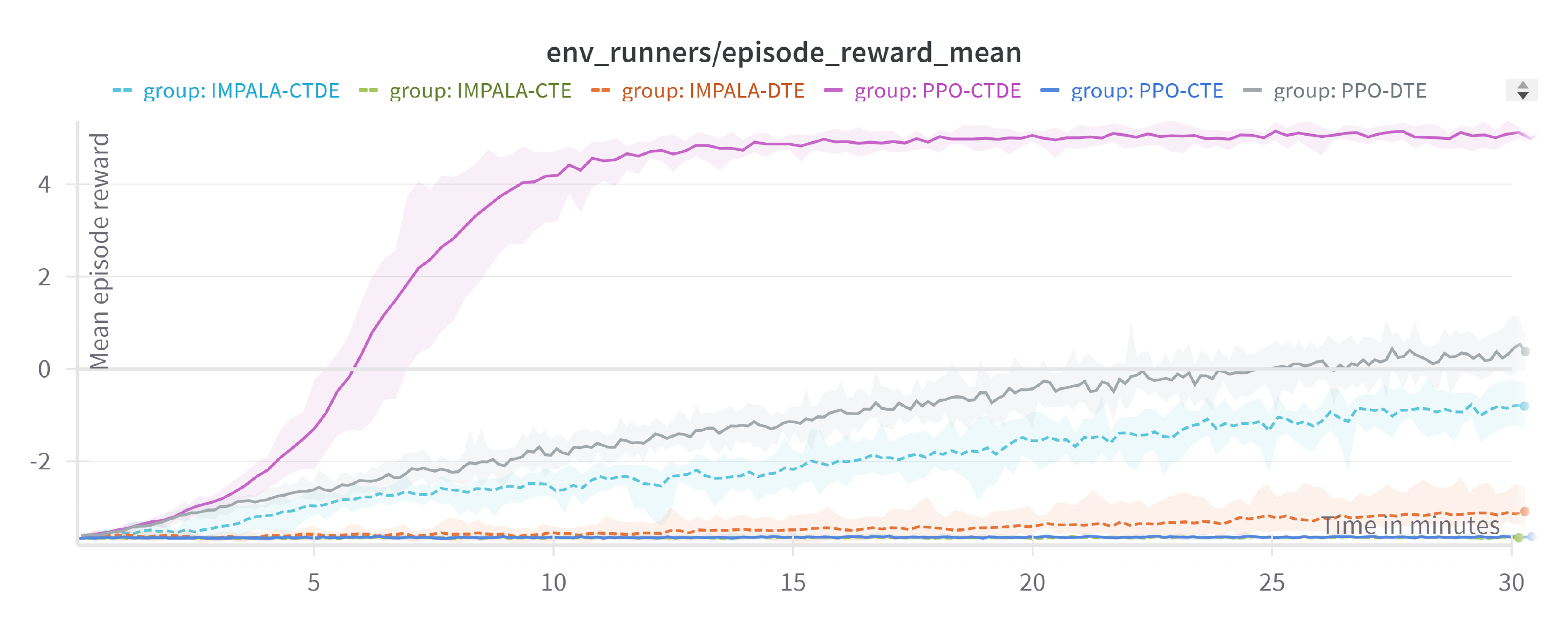}
    \caption{Mean episode rewards for PPO and IMPALA in reference model 2.1 with different training and execution modes.}
    \label{fig:learning-curves-reference-model-2-1}
\end{figure}

Figure \ref{fig:learning-curves-reference-model-2-1} illustrates the mean episode rewards over time for PPO and IMPALA across different training and execution configurations in reference model 2.1. For PPO, the CTDE configuration shows the fastest and most stable convergence, reaching the highest rewards. PPO-DTE also converges well, though at a slower rate, while PPO-CTE fails to learn a suitable policy. For IMPALA, CTDE shows a faster convergence similar to PPO-DTE. IMPALA-DTE struggles, showing minimal improvement in rewards over time. IMPALA-CTE fails like PPO-CTE to find a good policy.

\begin{figure}[h]
    \centering
    \includegraphics[width=1\linewidth]{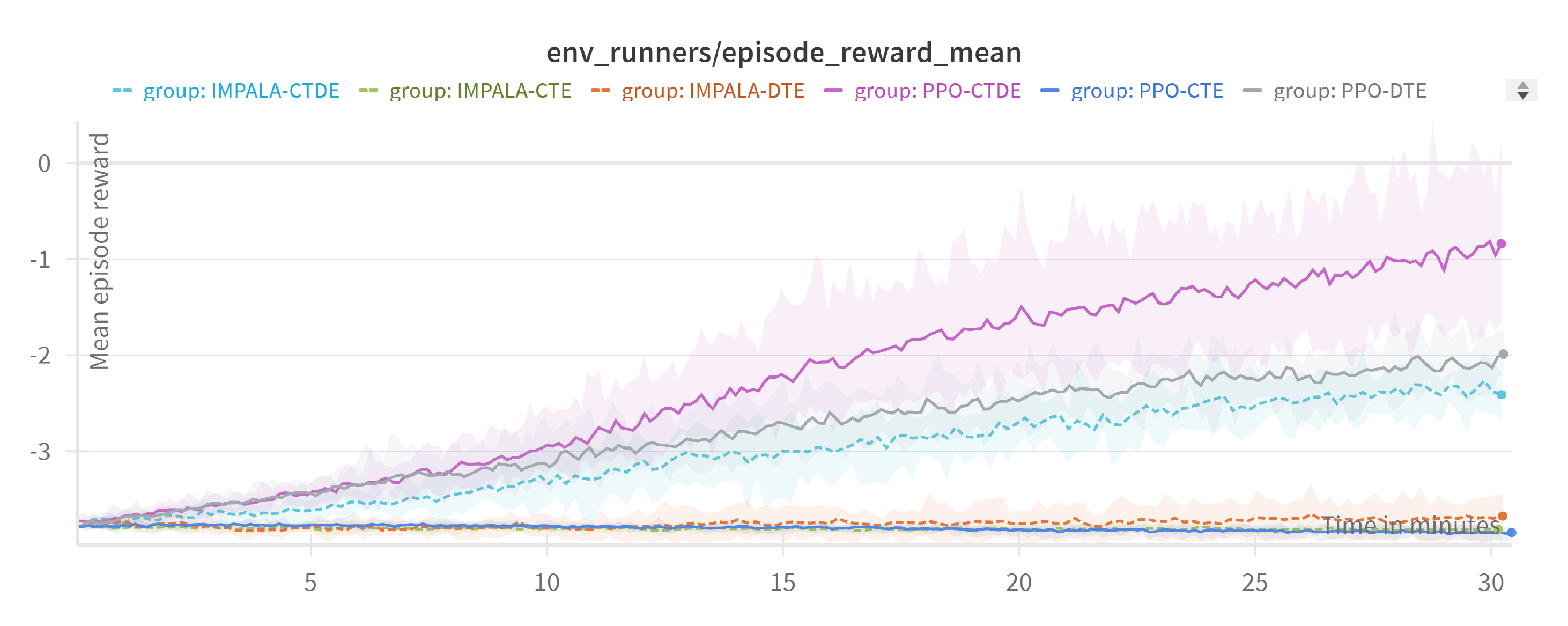}
    \caption{Mean episode rewards for PPO and IMPALA in reference model 2.2 with different training and execution modes.}
    \label{fig:learning-curves-reference-model-2-2}
\end{figure}

Figure \ref{fig:learning-curves-reference-model-2-2} illustrates the mean episode rewards over time for PPO and IMPALA across different training and execution configurations in reference model 2.2. This model appears to be more challenging for learning, as no configuration achieves a significant plateau in the learning curve over the given time period. PPO-CTDE shows the best learning progress, though its performance is more variable compared to reference model 2.1, with larger fluctuations in the rewards. PPO-DTE and IMPALA-CTDE also show some improvement, but their learning remains slower, indicating that these configurations struggle with the complexity of this environment. PPO-CTE, IMPALA-DTE, and IMPALA-CTE exhibit minimal learning progress, remaining close to the initial negative reward levels throughout the experiment, suggesting that these configurations are not effective in solving this particular environment.

The warehouse layouts of reference model 2.1 and 2.2 present significant difficulties, particularly for IMPALA, with PPO-CTDE showing the most potential but with varying degrees of stability across the different models. This suggests that while PPO generally performs robustly across diverse layouts, increased layout complexity, as seen in model 2.2, introduces greater variability and challenges for all configurations.

\begin{figure}[h]
    \centering
    \includegraphics[width=1\linewidth]{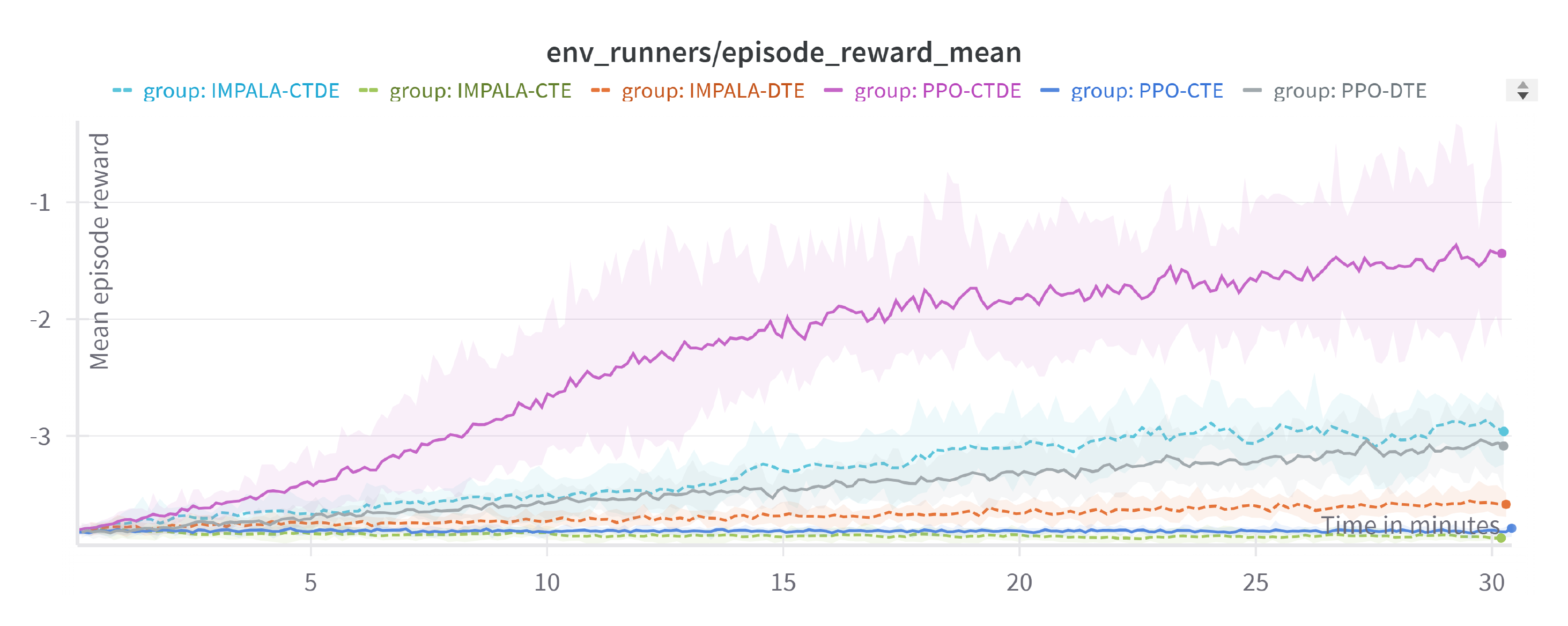}
    \caption{Mean episode rewards for PPO and IMPALA in reference model 3.1 with different training and execution modes.}
    \label{fig:learning-curves-reference-model-3-1}
\end{figure}

Figure \ref{fig:learning-curves-reference-model-3-1} presents the mean episode rewards over time for PPO and IMPALA across different training and execution modes in reference model 3.1. This model appears to challenge the learning algorithms further, as none of the configurations reach positive reward levels within the given timeframe. PPO-CTDE shows the most significant learning progress, achieving the highest mean episode reward among the configurations, though it still remains below zero. Notably, PPO-CTDE exhibits considerable variability, as indicated by the broad shaded region around its learning curve. This suggests that while PPO-CTDE shows potential, its performance is inconsistent across trials for this model. PPO-DTE follows a similar trend but at a lower reward level and with slightly less variability, indicating a more stable, albeit less effective, learning outcome. PPO-CTE shows minimal improvement, remaining close to its initial values throughout the experiment. For IMPALA, the learning results are more limited. IMPALA-CTDE and IMPALA-DTE exhibit slow improvement, with IMPALA-CTDE performing slightly better. IMPALA-CTE shows no significant progress, remaining stagnant at low reward levels, which reflects its continued struggle with a completely centralized approach in more complex environments.

\subsection{Comparative Analysis of RL Policies and Classical MAPF Algorithms}
In this section, we evaluate the performance of MARL algorithms relative to traditional MAPF algorithms. This comparison aims to assess how effectively MARL-based methods can handle the complex pathfinding requirements of warehouse and logistics environments compared to classical, heuristic-based approaches commonly used in MAPF. To ensure a fair and reliable comparison, we use the same configuration of start and goal positions across both MARL and MAPF experiments. This consistency is important for comparability, as it eliminates variations due to random placement of agents and goals, focusing the comparison on algorithmic performance rather than differences in initial conditions. A deterministic seeding process (seed = 42) generates unique start and goal positions for each agent, ensuring reproducibility. By using a controlled environment configuration, we are able to fairly assess the effectiveness of both MARL and MAPF approaches on identical task setups.

The evaluation metrics for this comparison include the number of timesteps required to solve the MAPF problem, computational time, and success rate. Timesteps measure the total steps needed for all agents to reach their respective goals. Computational time reflects the duration taken by the selected algorithm to compute paths for all agents. Success rate represents the proportion of successfully computed paths within a one-minute time limit.

This comparative analysis focuses on reference models 2.1, 2.2, and 3.1, each with four agents. Reference models 2.1 and 2.2 mimic typical warehouse layouts. For reference model 2.1 both the basic variant (block layout) and the layout with dead ends (2.1 b) are considered. Reference model 3.1 presents a production logistics setting, allowing us to assess the robustness of each approach under various environmental conditions.

In the experiments in this section, we make slight deviations from the optimal parameters identified in the earlier hyperparameter search. Informal preliminary testing revealed that certain modifications, such as the use of an LSTM layer in PPO, yield improved performance in the specific MAPF tasks addressed here. These adjustments, while not strictly based on the formal hyperparameter tuning process, are found to yield better results in practical application, thus justifying their use in the current comparison setup.

Both RL algorithms, PPO and IMPALA, use a network with two fully connected layers with 64 neurons each. An LSTM layer with a cell size of 64 is added after the fully connected layers. The PPO model is trained with a batch size of 4,000 and a minibatch size of 4,000, iterated over 12 epochs per update. The algorithm learning rate is set at 0.001, with a gamma value of 0.99 to discount future rewards and an entropy coefficient of 0.001. A clipping parameter of 0.05 is applied to stabilize policy updates. The IMPALA configuration includes a training batch size of 8,000, with a learning rate of 0.001 and a gamma value of 0.99. A value function loss coefficient of 0.5 is applied to balance policy optimization with accurate value estimation, and an entropy coefficient of 0.0 is used. Both models incorporate no custom action-masking mechanism. Both RL algorithms use CTDE and leverage again parallel processing with 10 environment runners, each handling two environments, to improve sample efficiency and stability in large-scale, multi-agent settings.

\begin{figure}[h]
    \centering
    \begin{tikzpicture}

        \begin{axis}[
            width=0.9\textwidth,
            height=0.35\textwidth,
            ybar,
            ymin=0,
            ymax=110,
            ylabel={Success rate in \%},
            xtick = {2.25, 6.25, 10.75},
            xticklabels = {Block layout, Dead ends, Fishbone layout},
            tick pos=left,
            nodes near coords,
            nodes near coords align={vertical},
            grid=none,
            legend style={at={(0.5, 1.05)}, anchor=south, legend columns=-1},
            ]

            \addplot[fill=redMM] coordinates {(1.5, 100) (5.5,23) (10, 44)};
            \addlegendentry{MA-A*}
            \addplot[fill=magenta] coordinates {(2, 83) (6,33) (10.5, 80)};
            \addlegendentry{CBS}
            \addplot[fill=blueMM] coordinates {(2.5,96) (6.5, 100) (11,94)};
            \addlegendentry{PPO}
            \addplot[fill=greenMM] coordinates {(3,90) (7,88) (11.5,98)};
            \addlegendentry{IMPALA}
        \end{axis}
    \end{tikzpicture}
    \caption{Success rates of reference model 2.1 and 2.2.}
    \label{fig:succes-rate-reference-model-2}
\end{figure}
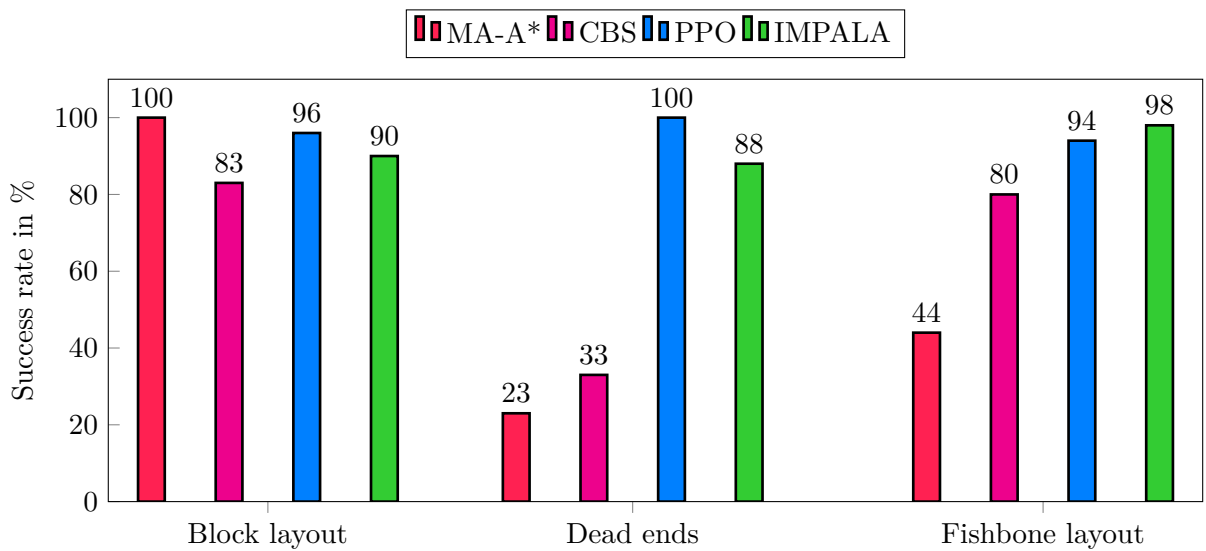

Figure \ref{fig:succes-rate-reference-model-2} illustrates the success rates of the MA-A*, CBS, PPO, and IMPALA algorithms across the block layout and layout with dead ends of reference model 2.1, as well as the fishbone layout of reference model 2.2 for 100 random start and goal configurations. Random agents are tested as a baseline and consistently achieve a success rate of 0 \% across all layouts. In the block layout, all algorithms display relatively high performance. MA-A* achieves a success rate of 100 \%, with PPO and IMPALA reaching 96 \% and 90 \%, respectively. CBS, with a success rate of 83 \%, demonstrates moderate performance compared to the others. In the dead ends layout, distinct differences in performance emerge. MA-A* achieves only 23 \%, highlighting its difficulty in handling constrained environments. CBS improves on this, achieving a success rate of 33 \%, though it remains far behind PPO and IMPALA, which achieve 100 \% and 88 \%, respectively. A similar trend is observed in the fishbone layout. MA-A* achieves a success rate of 44 \%, while CBS improves to 80 \%, and PPO and IMPALA maintain high performance with success rates of 94 \% and 98 \%. These results underscore the limitations of MA-A* and, to a lesser extent, CBS in complex or restricted layouts. PPO and IMPALA demonstrate robust performance across all scenarios, reflecting their adaptability to stochastic and constrained environments.

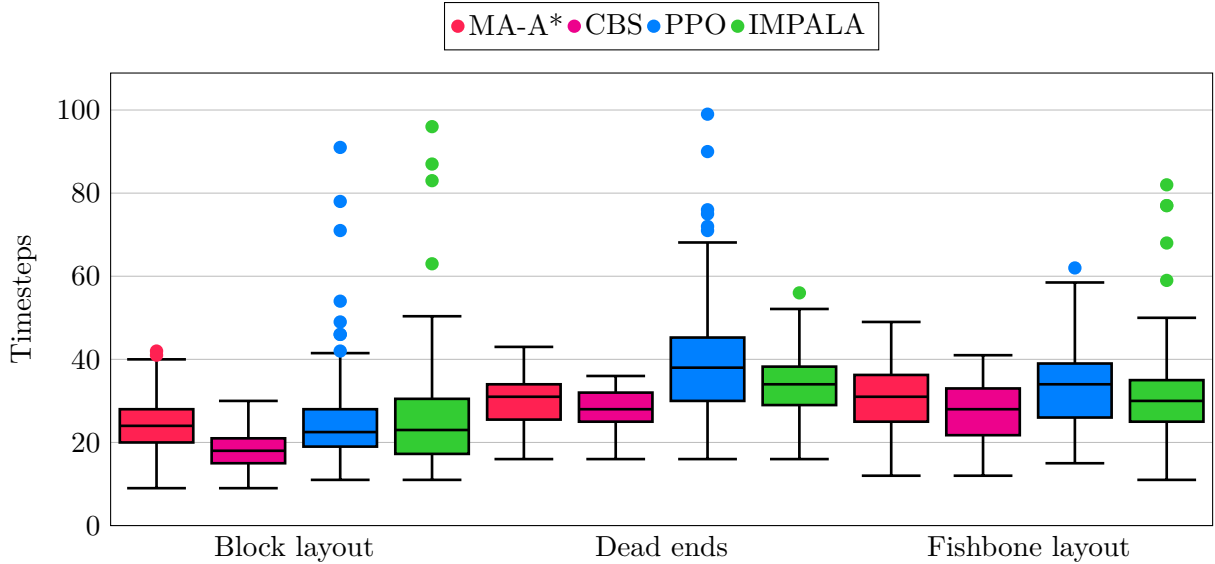
\begin{figure}[h]
    \centering
    \begin{tikzpicture}
        \begin{axis}[,
            width=14.5cm,       
            height=6cm,
            boxplot/draw direction=y,
            ylabel={Timesteps},
            xtick = {2.5, 6.5, 10.5},
            xticklabels = {Block layout, Dead ends, Fishbone layout},
            xmin=0.5,
            xmax=12.5,
            ymin=0,
            legend style={at={(0.5, 1.05)}, anchor=south, legend columns=-1},
            ymajorgrids=true,
            xmajorgrids=false,
            tick style={draw=none} 
        ]
            \addlegendimage{only marks, mark=*, color=redMM}
            \addlegendentry{MA-A*}
            \addlegendimage{only marks, mark=*, color=magenta}
            \addlegendentry{CBS}
            \addlegendimage{only marks, mark=*, color=blueMM}
            \addlegendentry{PPO}
            \addlegendimage{only marks, mark=*, color=greenMM}
            \addlegendentry{IMPALA}
            
            \addplot[
                boxplot prepared={
                    median=24,
                    upper quartile=28,
                    lower quartile=20,
                    upper whisker=40,
                    lower whisker=9,
                },
                fill=redMM,
                draw=black,
                mark options={color=redMM},
                ] coordinates {
                    (1,41) (1,42)
                };
            
            \addplot[
            boxplot prepared={
                median=18.0,
                upper quartile=21.0,
                lower quartile=15.0,
                upper whisker=30.0,
                lower whisker=9.0,
            },
            fill=magenta,
            draw=black,
            mark options={color=magenta},
            ] coordinates { };

            \addplot[
                boxplot prepared={
                    median=22.5,
                    upper quartile=28,
                    lower quartile=19,
                    upper whisker=41.5,
                    lower whisker=11,
                },
            fill=blueMM,
            draw=black,
            mark options={color=blueMM},
            ] coordinates {
             (2,46) (2,91) (2,78) (2,54) (2,42) (2,46) (2,49) (2,71)
            };

            \addplot[
                boxplot prepared={
                    median=23,
                    upper quartile=30.5,
                    lower quartile=17.25,
                    upper whisker=50.375,
                    lower whisker=11,
                },
            fill=greenMM,
            draw=black,
            mark options={color=greenMM},
            ]coordinates {
            (3,63) (3,83) (3,87) (3,96)
            };

            \addplot[
            boxplot prepared={
                    median=31,
                    upper quartile=34,
                    lower quartile=25.5,
                    upper whisker=43,
                    lower whisker=16,
                },
            fill=redMM,
            draw=black,
            mark options={color=redMM},
            ]coordinates {};

            \addplot[
            boxplot prepared={
                median=28.0,
                upper quartile=32.0,
                lower quartile=25.0,
                upper whisker=36.0,
                lower whisker=16.0,
            },
            fill=magenta,
            draw=black,
            mark options={color=magenta},
            ] coordinates { };

            \addplot[
            boxplot prepared={
                median=38,
                upper quartile=45.25,
                lower quartile=30,
                upper whisker=68.125,
                lower whisker=16,
            },
            fill=blueMM,
            draw=black,
            mark options={color=blueMM},
            ] coordinates {
                (5,71.0) (5,75.0) (5,76.0) (5,72.0) (5,90.0) (5,99.0)
            };

            \addplot[
                boxplot prepared={
                    median=34,
                    upper quartile=38.25,
                    lower quartile=29.0, 
                    upper whisker=52.125,
                    lower whisker=16,
                },
            fill=greenMM,
            draw=black,
            mark options={color=greenMM},
            ]coordinates {
            (6,56) 
            };  
            \addplot[
            boxplot prepared={
                median=31,
                upper quartile=36.25,
                lower quartile=25,
                upper whisker=49,
                lower whisker=12,
            },
            fill=redMM,
            draw=black,
            mark options={color=redMM},
            ] coordinates { };

            \addplot[
            boxplot prepared={
                median=28.0,
                upper quartile=33.0,
                lower quartile=21.75,        
                upper whisker=41.0,
                lower whisker=12.0,
            },
            fill=magenta,
            draw=black,
            mark options={color=magenta},
            ] coordinates { };

            \addplot[
            boxplot prepared={
                median=34,        
                upper quartile=39,
                lower quartile=26,
                upper whisker=58.5, 
                lower whisker=15, 
            },
            fill=blueMM,
            draw=black,
            mark options={color=blueMM},
            ] coordinates { (8,62) };
            
            \addplot[
                boxplot prepared={
                    median=30.0,
                    upper quartile=35.0,
                    lower quartile=25.0,
                    upper whisker=50.0,
                    lower whisker=11.0,
                },
            fill=greenMM,
            draw=black,
            mark options={color=greenMM},
            ]coordinates {
            (9,77) (9,82) (9,68) (9,59) (9,77)
            }; 
        \end{axis}
    \end{tikzpicture}
    \caption{Timesteps of successful pathfinding for all agents of reference model 2.1 and 2.2.}
    \label{fig:timesteps-reference-model-2}
\end{figure}

Figure \ref{fig:timesteps-reference-model-2} shows the timesteps required for successful pathfinding of all agents using MA-A*, CBS, PPO, and IMPALA across the block layout, the layout with dead ends of reference model 2.1, and the fishbone layout of reference model 2.2 for 100 random start and goal configurations. The boxplots depict the distribution of timesteps, including the median, whiskers, and outliers. The whiskers represent the range of typical values, determined based on the interquartile range. The lower whisker is set to the larger of either the smallest value in the data or 1.5 times the interquartile range below the first quartile. Similarly, the upper whisker is set to the smaller of either the largest value in the data or 1.5 times the interquartile range above the third quartile. Any data points outside these whiskers are considered outliers.

In the block layout, CBS demonstrates the lowest median timesteps at 18, followed by PPO (22.5 timesteps), IMPALA (23 timesteps), and MA-A* (24 timesteps). Despite the slight differences in medians, CBS, MA-A*, and PPO exhibit relatively narrow interquartile ranges, indicating consistent performance across different start-goal configurations. PPO and IMPALA show higher variability, with several outliers exceeding 60 timesteps. This increased variability in PPO and IMPALA is likely due to their premature training termination, suggesting that extended training could reduce both timesteps and deviations. CBS demonstrates both efficiency and low variability, outperforming all other algorithms in terms of median timesteps. In the layout with dead ends, CBS achieves a median of 28 timesteps, which is lower than MA-A* (31 timesteps), PPO (38 timesteps), and IMPALA (34 timesteps). However, CBS's lower median timesteps must also be interpreted in light of its relatively low success rate, similar to MA-A*. Although MA-A* and CBS exhibit lower median timesteps, this performance is less meaningful given their inability to consistently find successful paths in this layout. PPO and IMPALA, which achieve significantly higher success rates, display higher median values, likely due to the increased complexity posed by dead ends. In the fishbone layout, CBS maintains competitive performance with a median of 28 timesteps, slightly lower than PPO (34 timesteps), IMPALA (30 timesteps), and MA-A* (31 timesteps). These results suggest that CBS offers a viable trade-off between pathfinding efficiency and robustness in layouts without significant constraints, such as the block and fishbone layouts. However, its lower success rate in the dead ends layout limits its applicability in more restricted environments. While MA-A* maintains consistently low median timesteps, its performance is even more constrained by its inability to handle restricted layouts effectively. In contrast, PPO and IMPALA demonstrate robustness and adaptability across all layouts, albeit with increased variability due to incomplete optimization.

\begin{figure}[h]
    \begin{minipage}[t]{.49\textwidth}
        \centering
        \caption*{MA-A*}
        \includegraphics[width=1\textwidth]{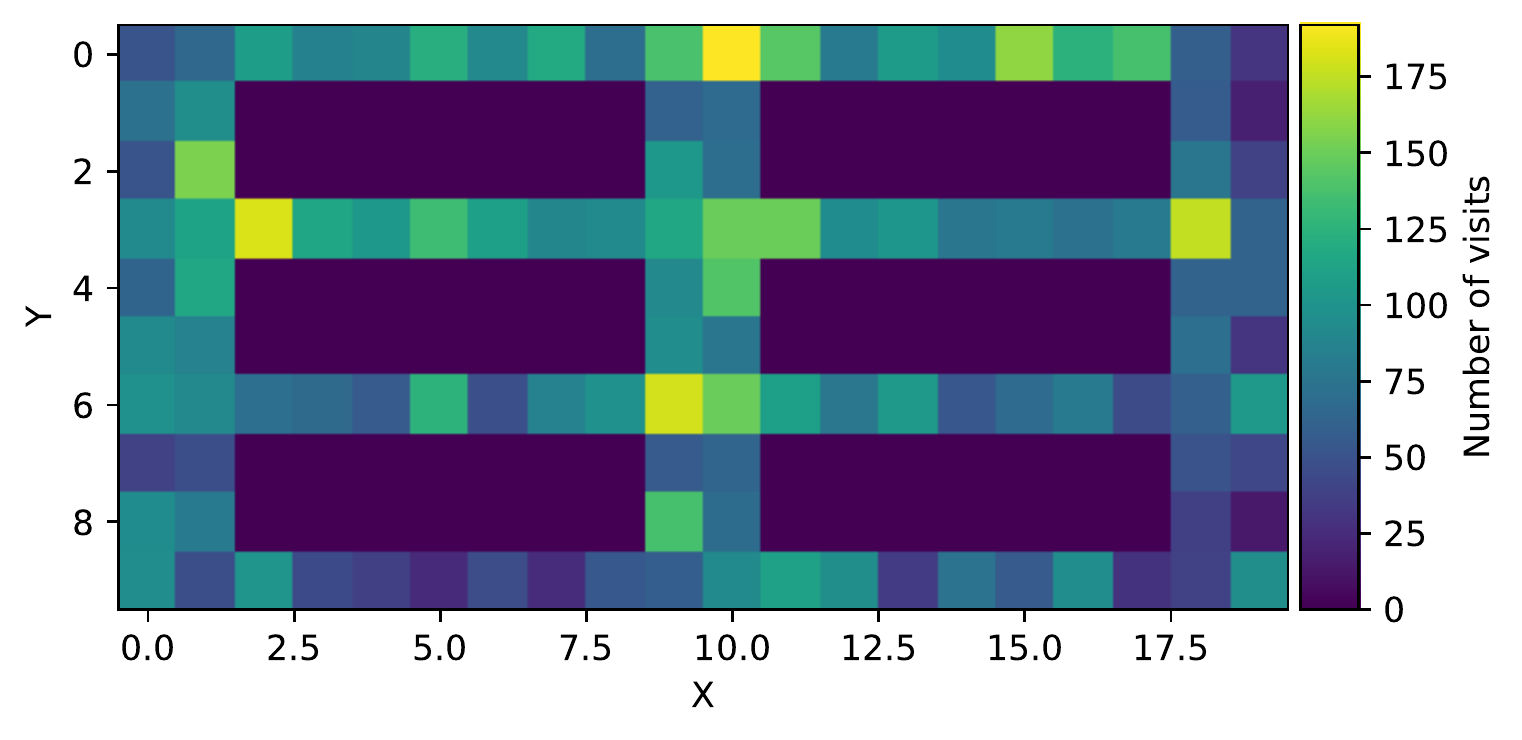}
    \end{minipage}
    \begin{minipage}[t]{.49\textwidth}
        \centering
        \caption*{CBS}
        \includegraphics[width=1\textwidth]{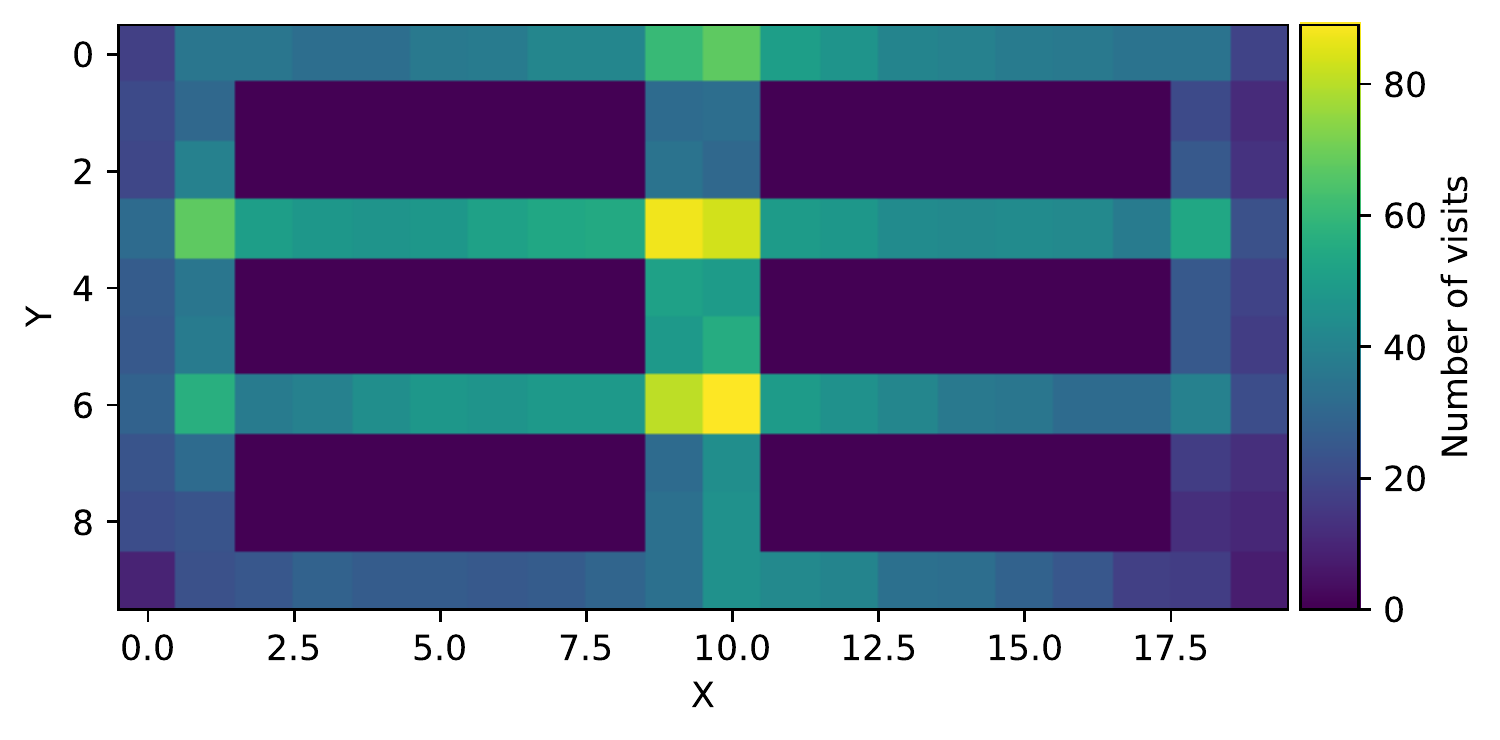}
    \end{minipage}
    
    \begin{minipage}[t]{.49\textwidth}
        \centering
        \caption*{PPO}
        \includegraphics[width=1\textwidth]{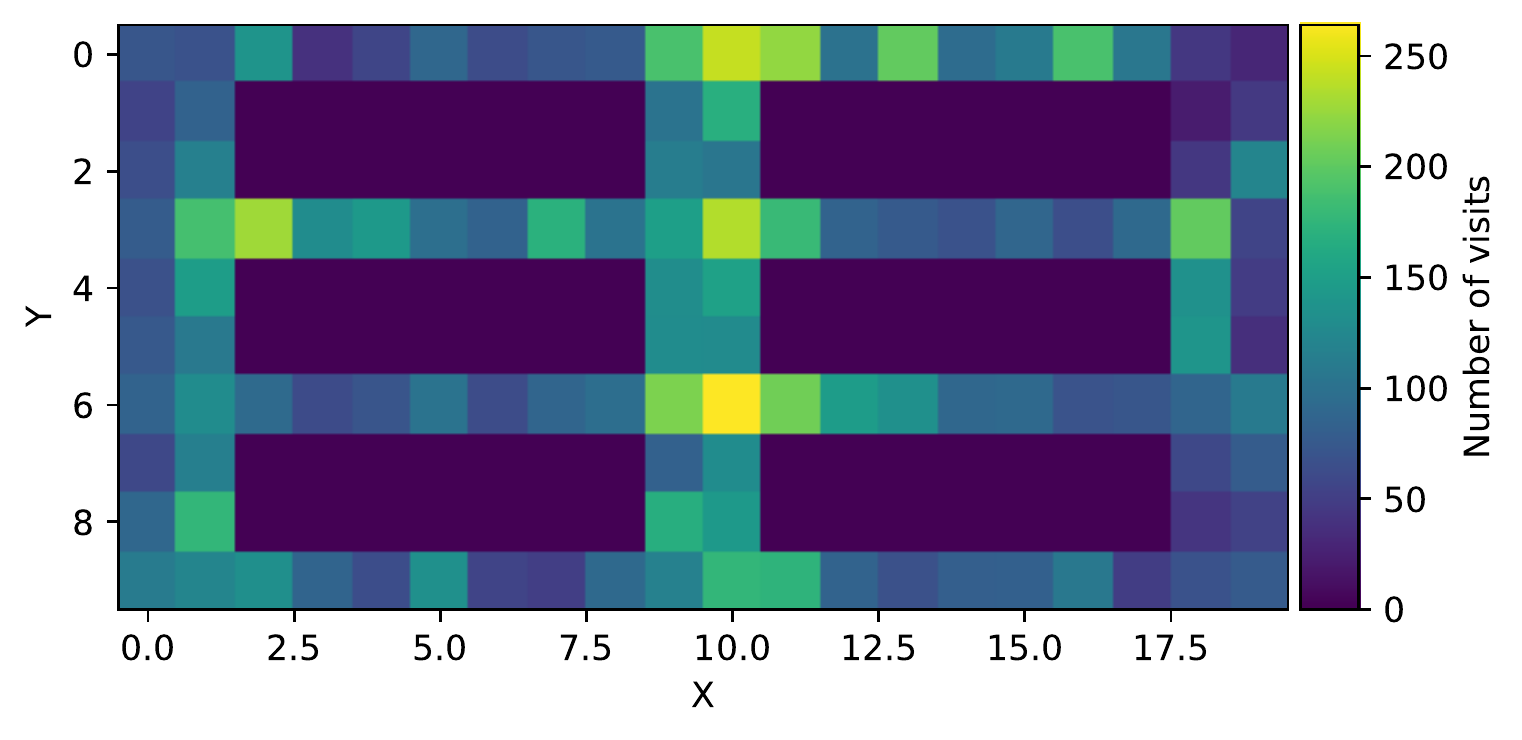}        
    \end{minipage}    
    \begin{minipage}[t]{.49\textwidth}
        \centering
        \caption*{IMPALA}
        \includegraphics[width=1\textwidth]{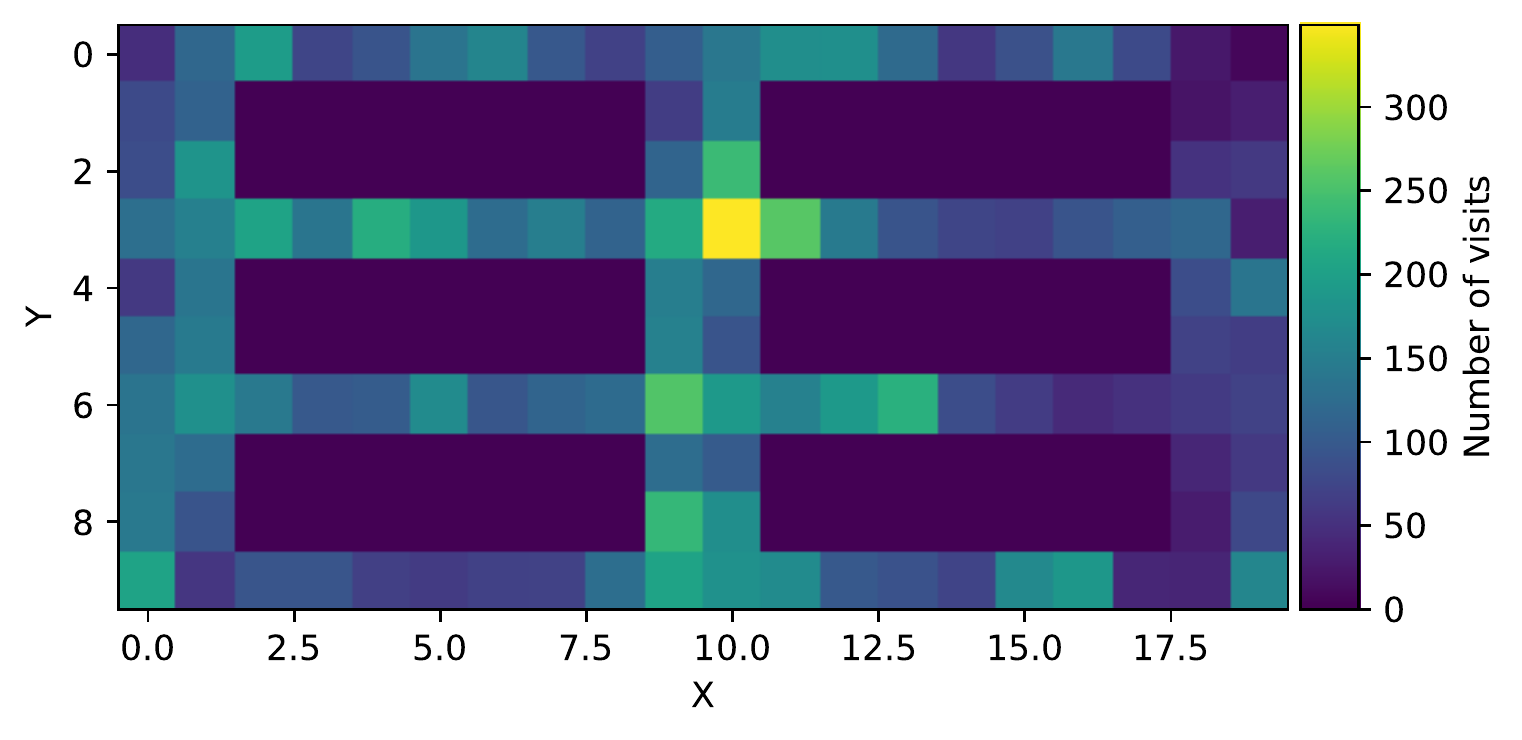}
    \end{minipage}

    \begin{minipage}[t]{.49\textwidth}
        \centering
        \caption*{Random}
        \includegraphics[width=1\textwidth]{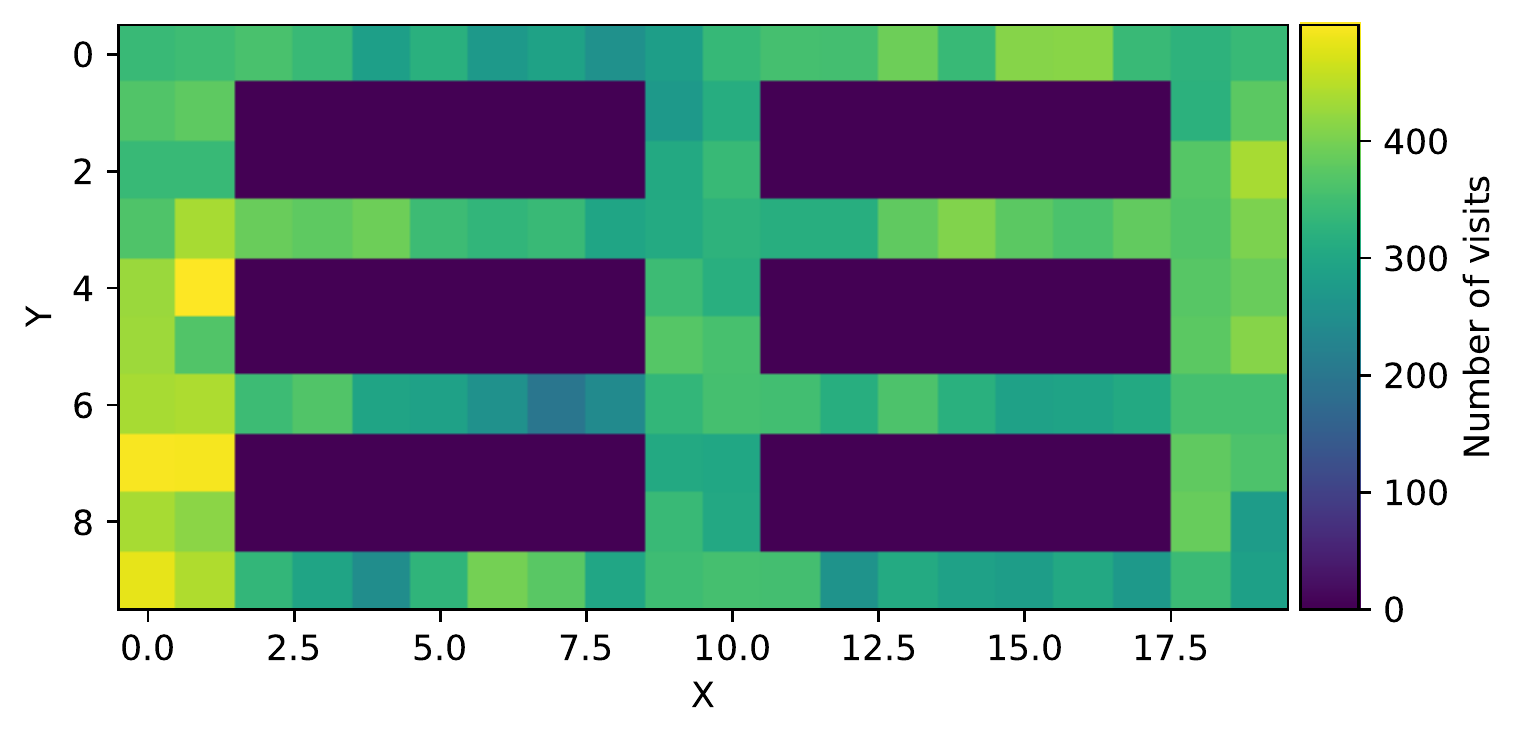}
    \end{minipage}
    \caption{Heat maps of the number of visits for reference model 2.1.}
    \label{fig:heatmaps-2-1}
\end{figure}

Figure \ref{fig:heatmaps-2-1} shows the number of visits for each grid cell during successful pathfinding attempts by agents using MA-A*, CBS, PPO, IMPALA, and the random agent in the block layout of reference model 2.1. The heat maps primarily reflect the influence of the layout structure, with high visit counts concentrated around choke points and intersections where paths converge.

CBS exhibits a distinct and more structured pattern of visits compared to MA-A*, PPO, and IMPALA. The clear and consistent pathways observed in the CBS heat map are likely due to the nature of its underlying approach. Unlike the other algorithms, CBS pre-computes collision-free paths for all agents before execution, resulting in fewer unnecessary detours and a more deterministic traversal pattern. This contrasts sharply with MA-A*, PPO, and IMPALA, which rely on local decisions made dynamically during execution. Consequently, these algorithms exhibit more exploratory behavior and a higher degree of revisits, leading to increased clutter in their heat maps. MA-A* demonstrates a somewhat more distributed pattern compared to PPO and IMPALA, with lower maximum visit counts per cell despite achieving a 100 \% success rate. This reflects its reliance on real-time heuristic search, which can result in slightly more scattered paths but avoids excessive concentration at key points. PPO and IMPALA, on the other hand, show higher maximum visit counts at critical crossings, indicative of more frequent detours and path corrections due to their policy-based learning approach, which adapts to the observed environment but lacks full pre-planned coordination. The random agent exhibits a markedly different behavior, with a nearly uniform distribution of visits across the entire grid. This behavior results from its lack of a directed strategy, leading to a high number of visits throughout the layout. The higher visit counts, particularly in less critical regions, highlight the inefficiency of the random agent compared to the deliberate strategies of MA-A*, CBS, PPO, and IMPALA. The random agent's behavior serves as a baseline, emphasizing the more goal-oriented navigation strategies of the other algorithms.

\begin{figure}[h]
    \begin{minipage}[t]{.49\textwidth}
        \centering
        \caption*{MA-A*}
        \includegraphics[width=1\textwidth]{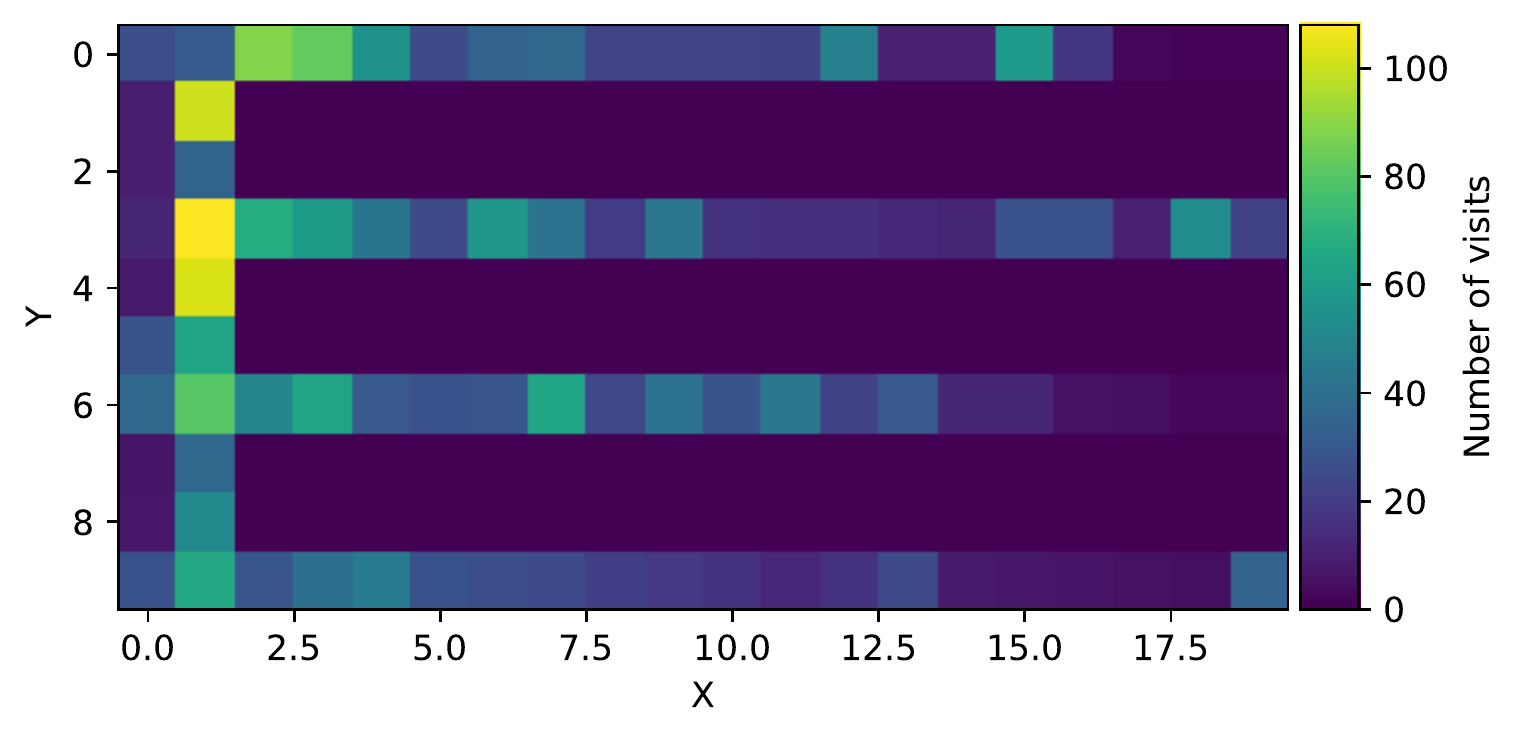}
    \end{minipage}
    \begin{minipage}[t]{.49\textwidth}
        \centering
        \caption*{CBS}
        \includegraphics[width=1\textwidth]{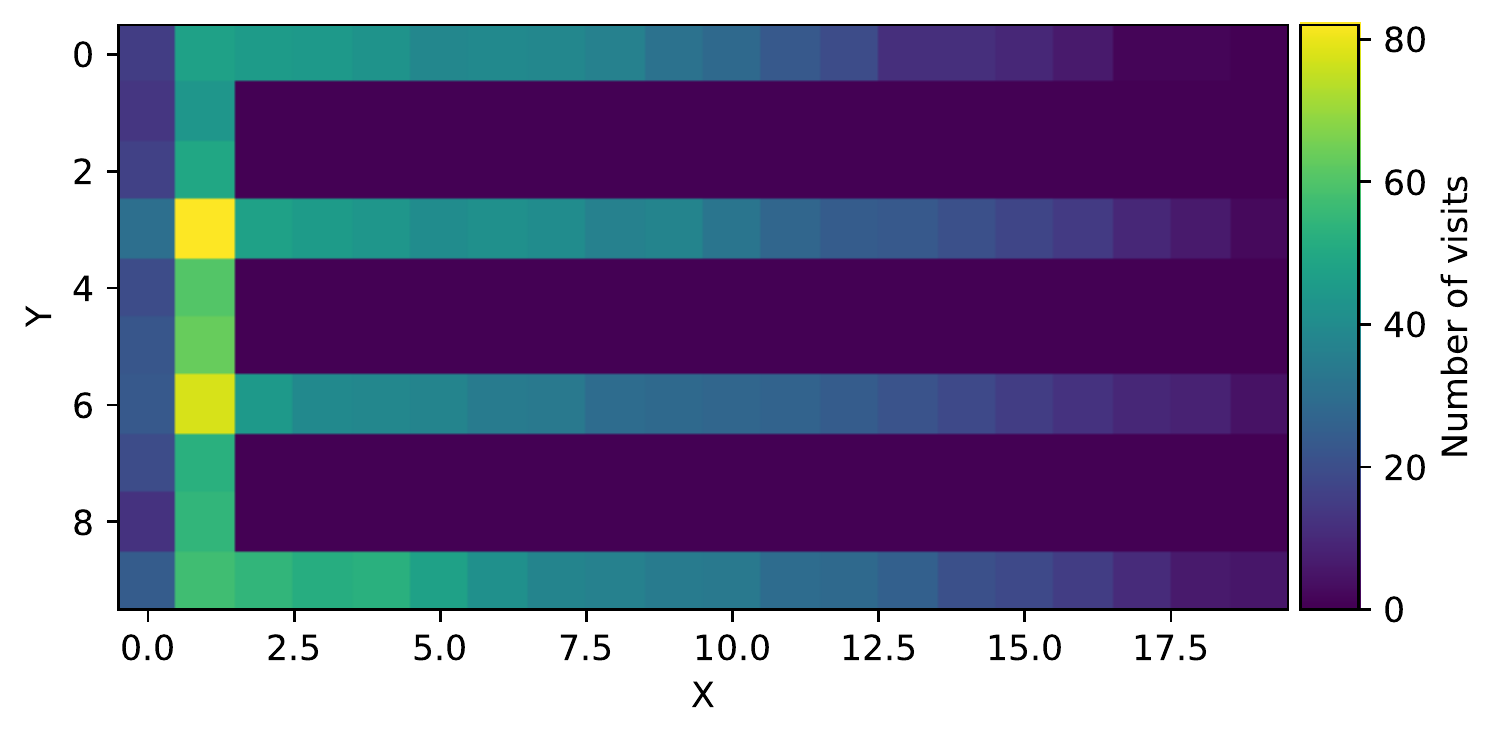}
    \end{minipage}

    \begin{minipage}[t]{.49\textwidth}
        \centering
        \caption*{PPO}
        \includegraphics[width=1\textwidth]{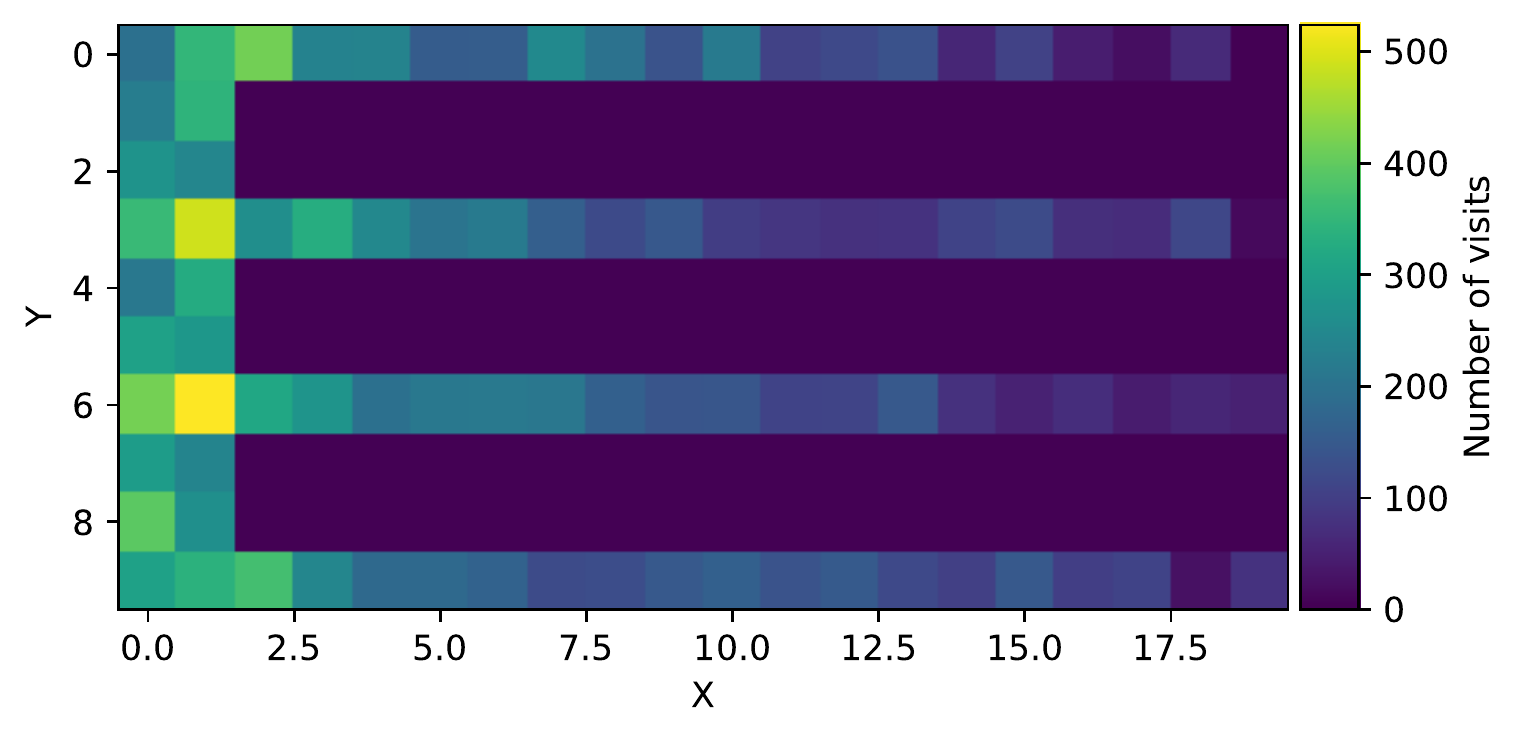}
    \end{minipage}
    \begin{minipage}[t]{.49\textwidth}
        \centering
        \caption*{IMPALA}
        \includegraphics[width=1\textwidth]{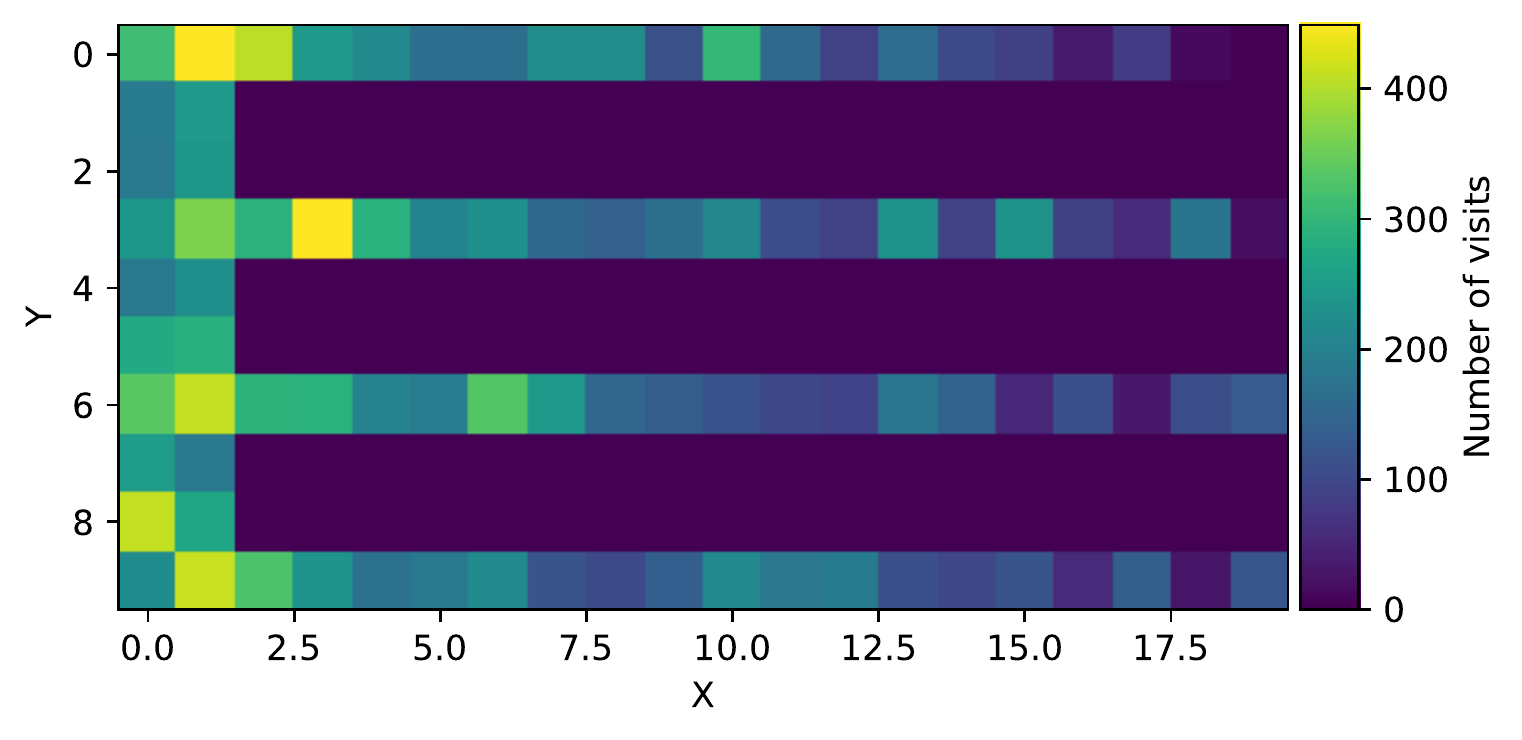}
    \end{minipage}

    \begin{minipage}[t]{.49\textwidth}
        \centering
        \caption*{Random}
        \includegraphics[width=1\textwidth]{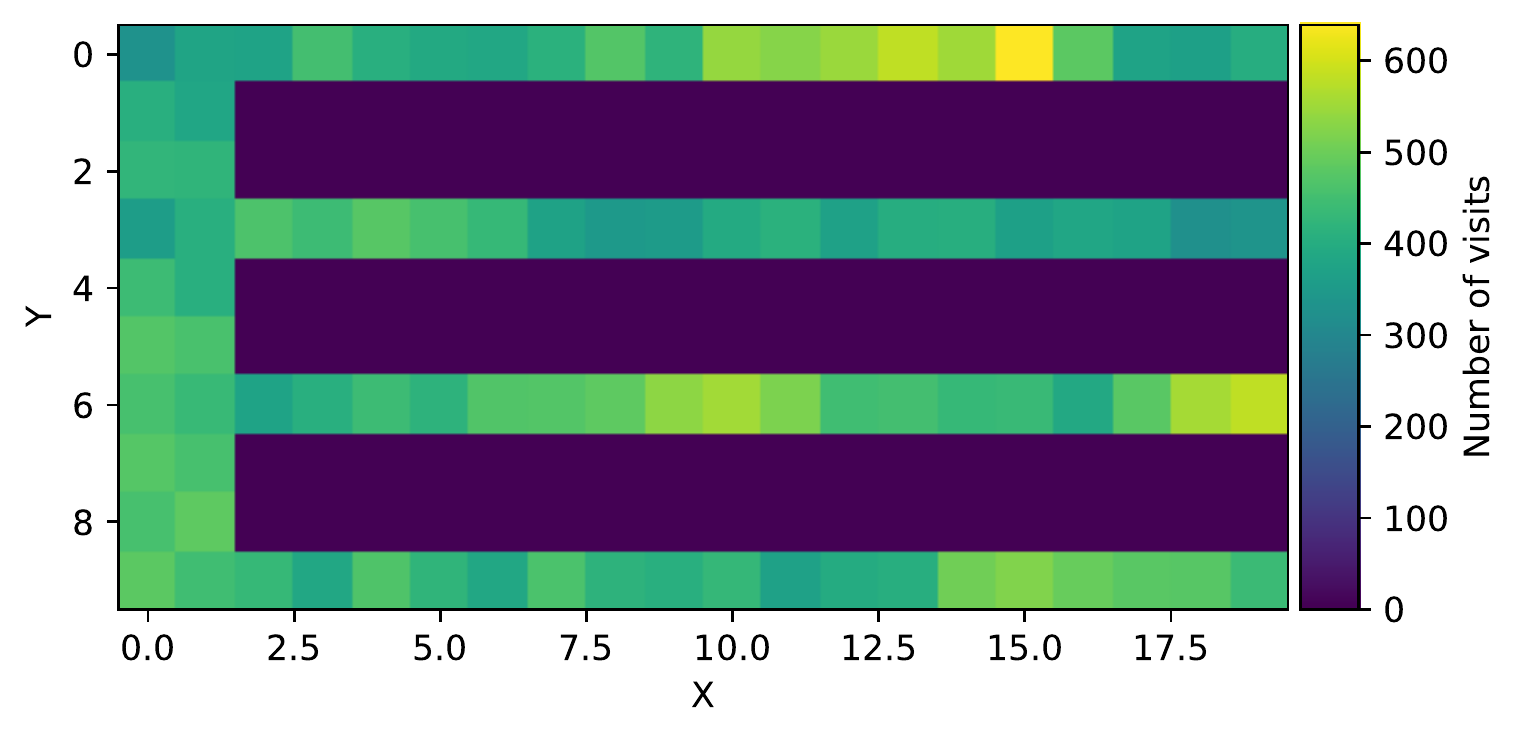}
    \end{minipage}
    \caption{Heat maps of the number of visits for reference model 2.1 with dead ends.}
    \label{fig:heatmaps-2-1b}
\end{figure}

Figure \ref{fig:heatmaps-2-1b} illustrates the number of visits for each grid cell during successful pathfinding attempts by agents using MA-A*, CBS, PPO, and IMPALA in the layout with dead ends of reference model 2.1. The highest visit counts are concentrated at choke points and intersections where paths converge, reflecting the influence of the layout's structure.

PPO and IMPALA show a generally higher number of visits compared to MA-A* and CBS, which can be attributed to their higher success rates in this layout. Their scattered patterns, combined with increased visit counts, indicate more exploratory behavior during navigation. In contrast, CBS once again exhibits a more structured and deterministic traversal pattern with fewer revisits, consistent with its centralized planning approach discussed earlier. MA-A* shows a broader but less efficient exploration pattern, in line with its heuristic-based search strategy. Overall, while CBS demonstrates clear and efficient paths, PPO and IMPALA's higher success rates lead to greater overall visit counts. The patterns confirm that all algorithms are influenced heavily by the layout's choke points and dead ends, which limit agent mobility.

\begin{figure}[h]
    \begin{minipage}[t]{.49\textwidth}
        \centering
        \caption*{MA-A*}
        \includegraphics[width=1\textwidth]{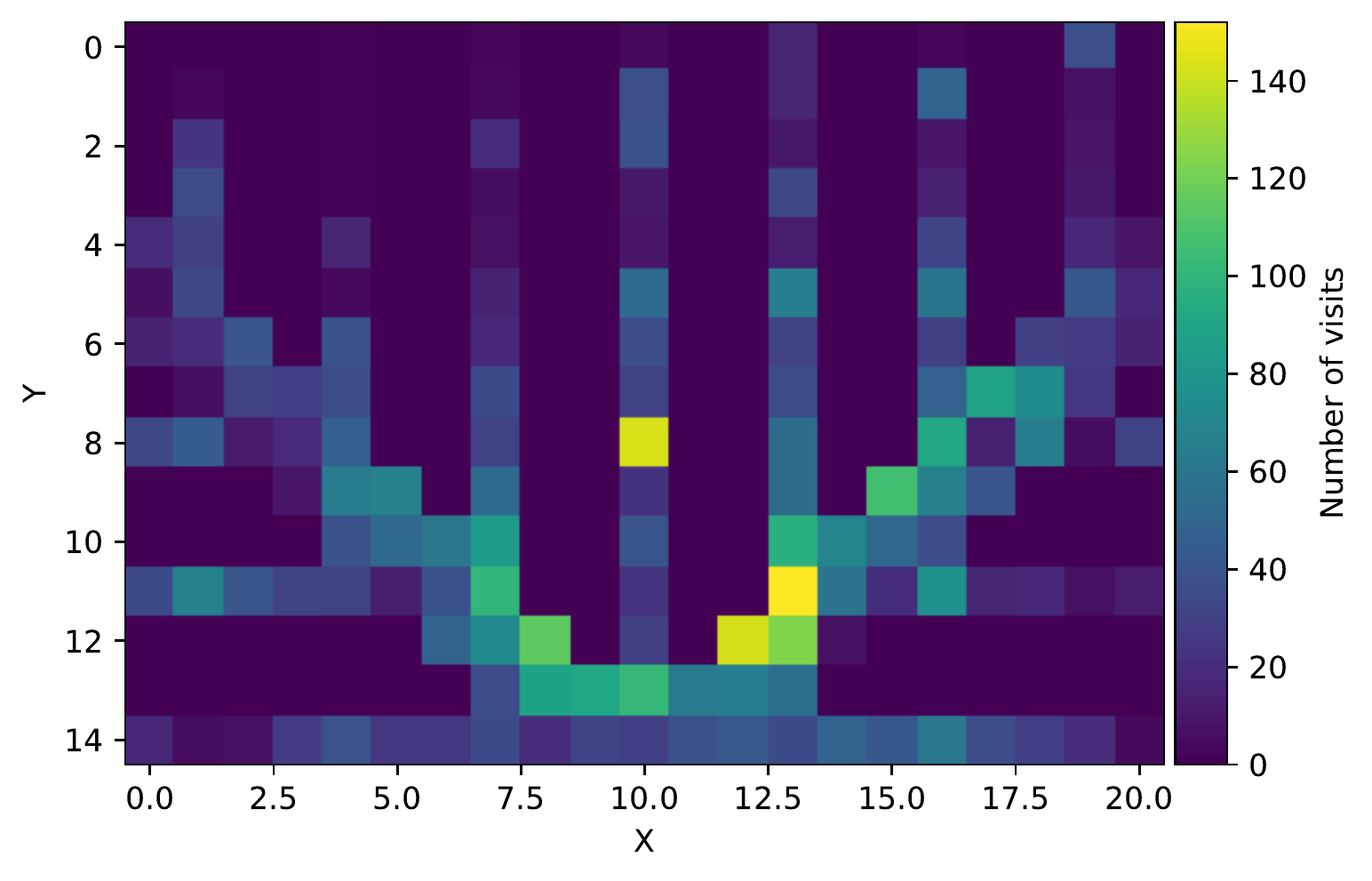}
    \end{minipage}
    \begin{minipage}[t]{.49\textwidth}
        \centering
        \caption*{CBS}
        \includegraphics[width=1\textwidth]{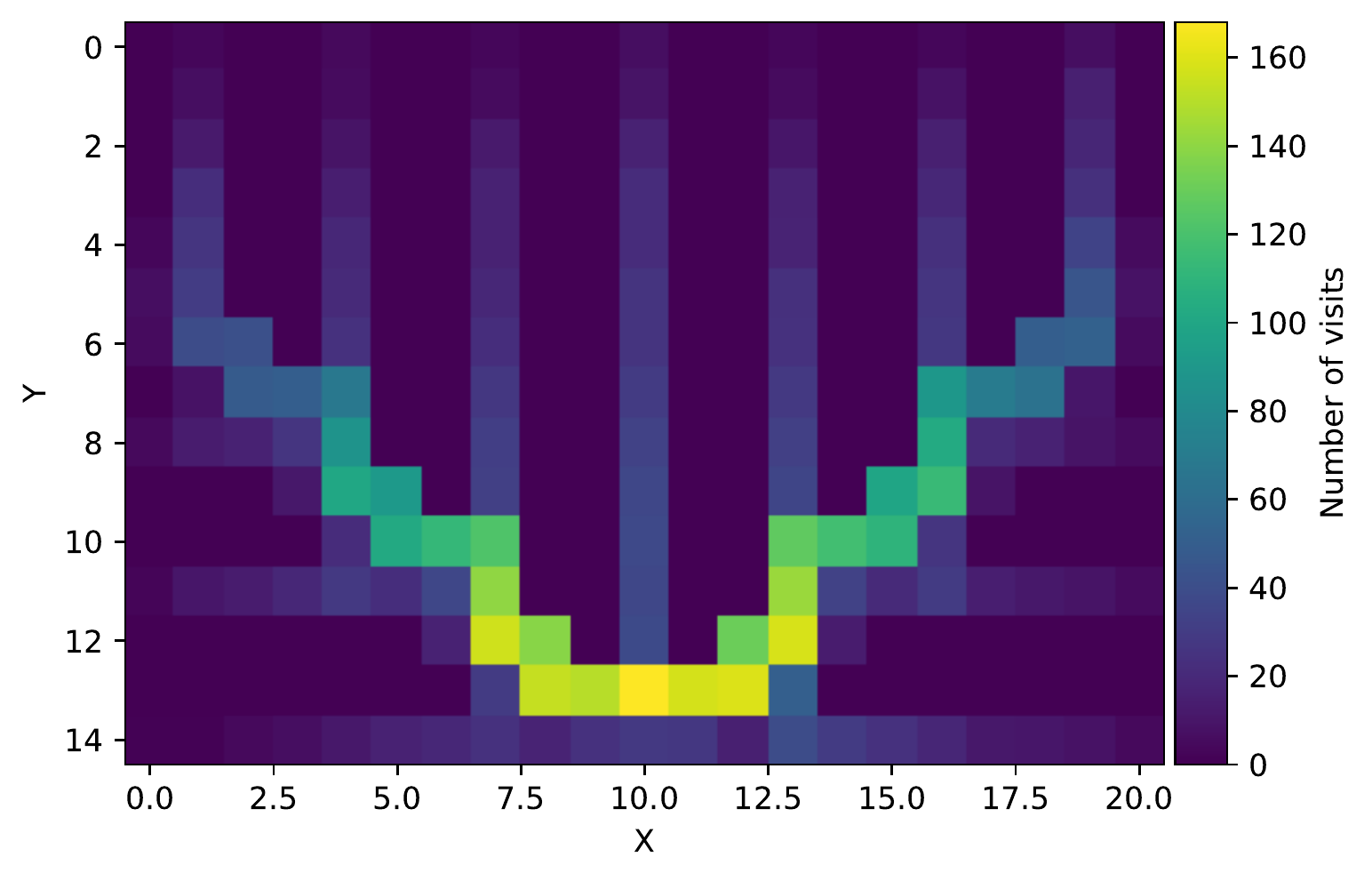}
    \end{minipage}
    
    \begin{minipage}[t]{.49\textwidth}
        \centering
        \caption*{PPO}
        \includegraphics[width=1\textwidth]{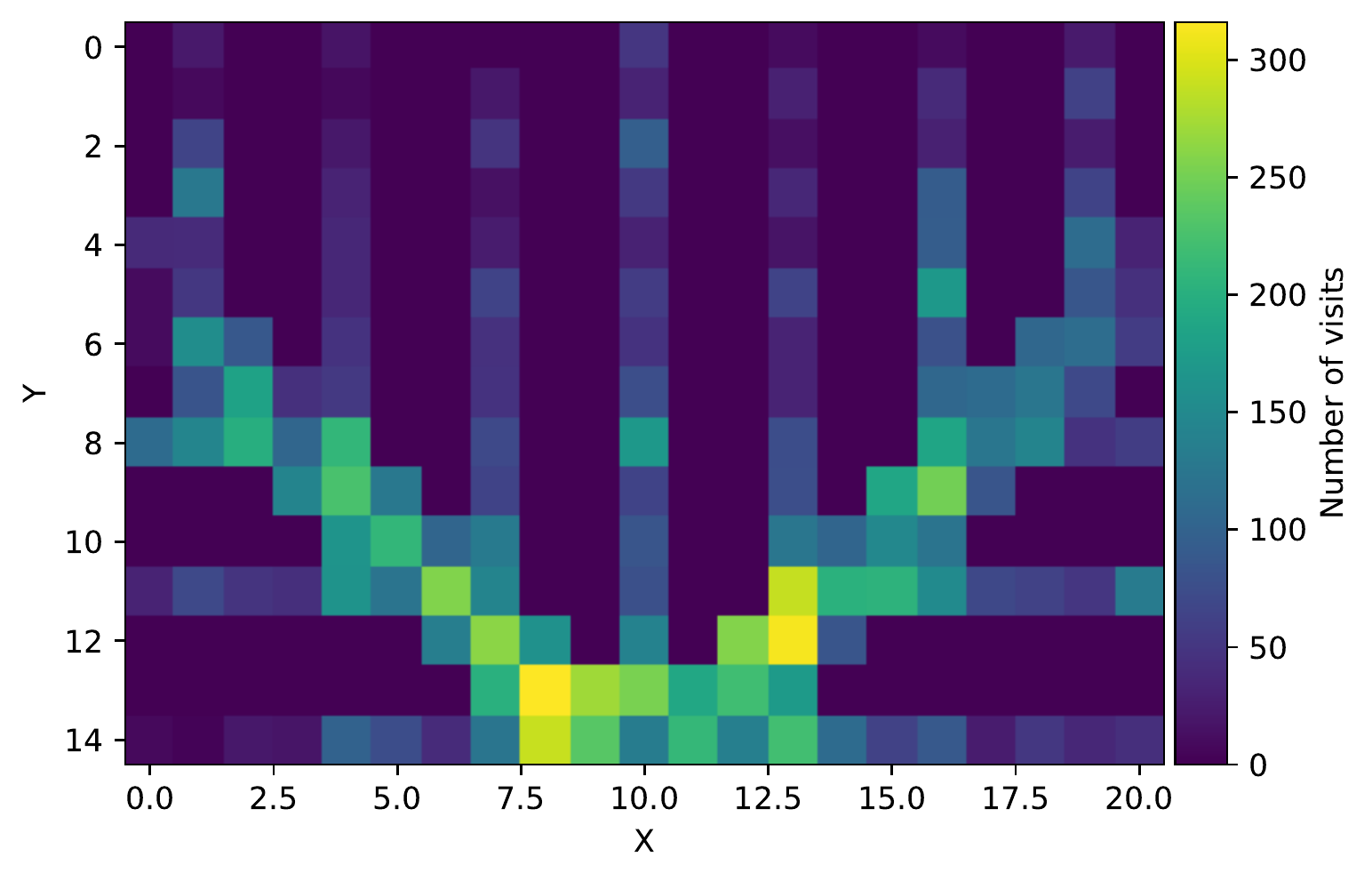}
    \end{minipage}
    \begin{minipage}[t]{.49\textwidth}
        \centering
        \caption*{IMPALA}
        \includegraphics[width=1\textwidth]{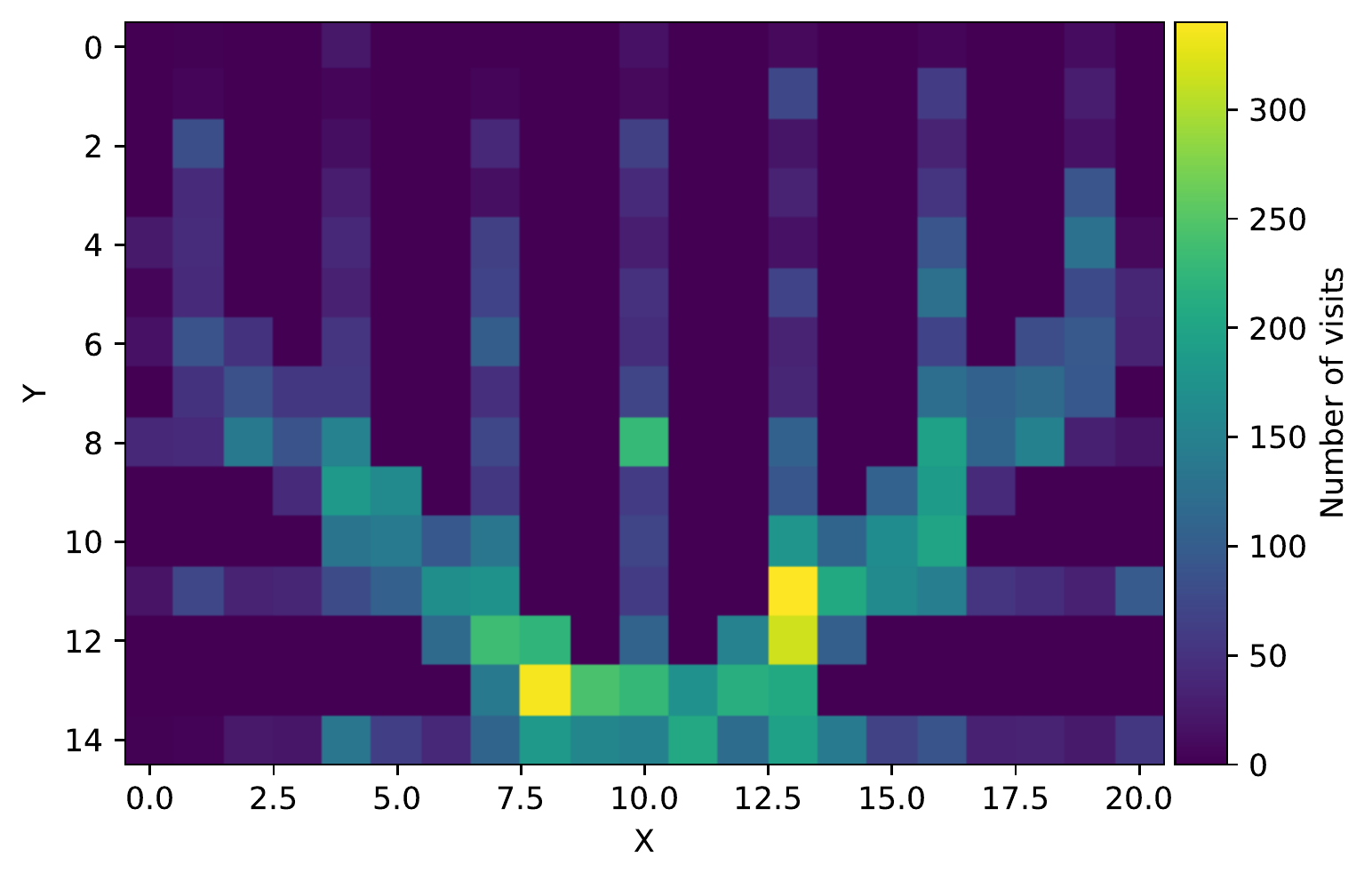}
    \end{minipage}

    \begin{minipage}[t]{.49\textwidth}
        \centering
        \caption*{Random}
        \includegraphics[width=1\textwidth]{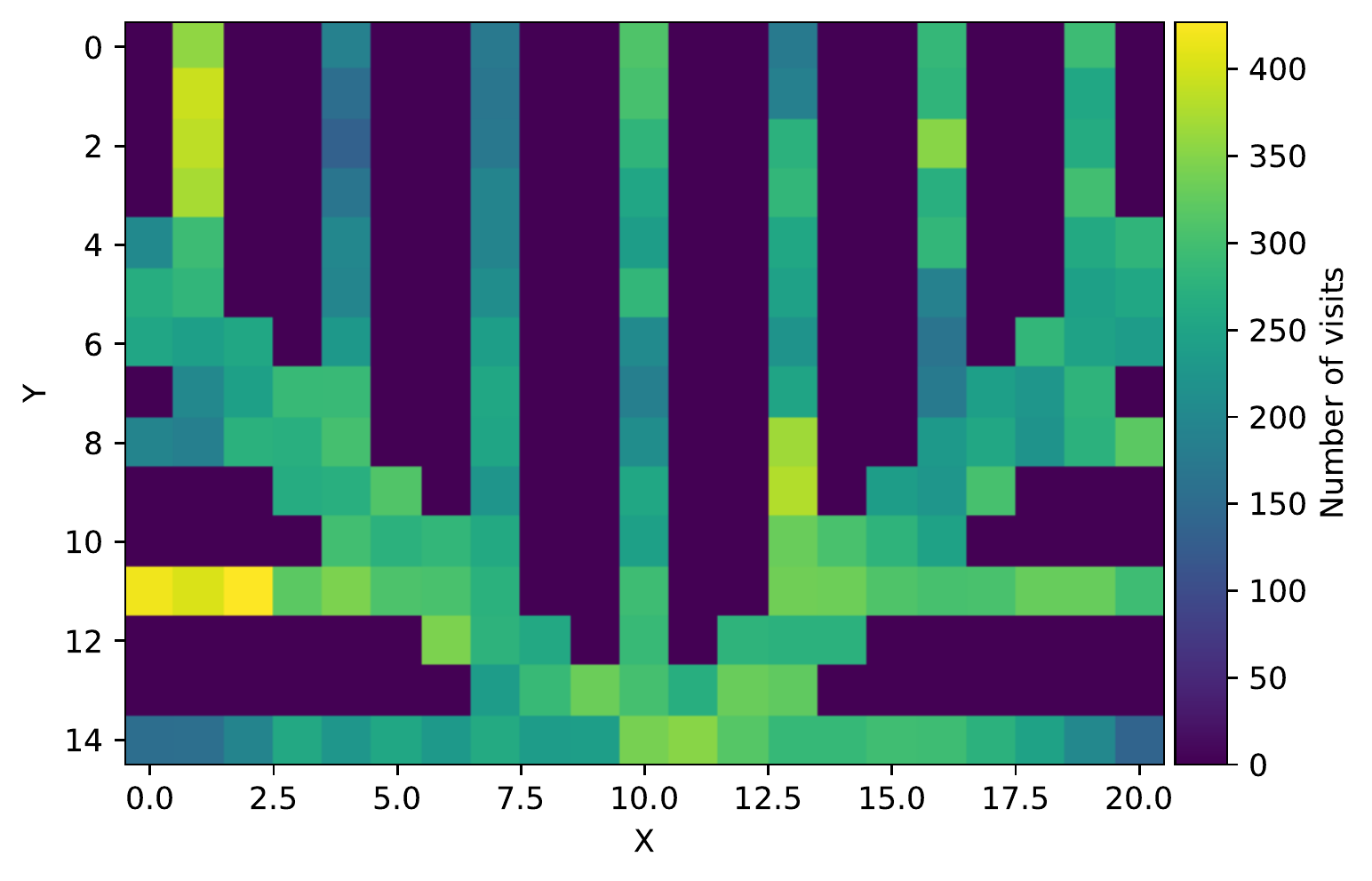}
    \end{minipage}
    \caption{Heat maps of the number of visits for reference model 2.2.}
    \label{fig:heatmaps-2-2}
\end{figure}

In the fishbone layout, the number of visits for each grid cell during successful pathfinding attempts is shown in Figure \ref{fig:heatmaps-2-2} for the MA-A*, CBS, PPO, and IMPALA algorithms. The heat maps highlight the strong influence of the layout structure on agent behavior, with all algorithms showing similar visit patterns concentrated at key intersections and along the primary paths of the layout.

MA-A*, PPO, and IMPALA exhibit comparable distributions, with concentrated visit counts at key junctions where paths converge. This similarity suggests that despite their different underlying strategies, these algorithms navigate the layout in a comparable manner. The higher visit counts for PPO and IMPALA can be attributed to their greater number of successful episodes, resulting in more data points being represented in the heat map. In contrast, CBS once again demonstrates a more structured and deterministic pattern of visits. The paths taken by CBS are clearly defined, with minimal revisits and fewer detours compared to the other algorithms. This behavior reflects its centralized planning mechanism, which computes collision-free paths for agents in advance, ensuring consistent and efficient navigation. While CBS achieves fewer successful episodes overall, its traversal remains more predictable and orderly, as indicated by its heat map.

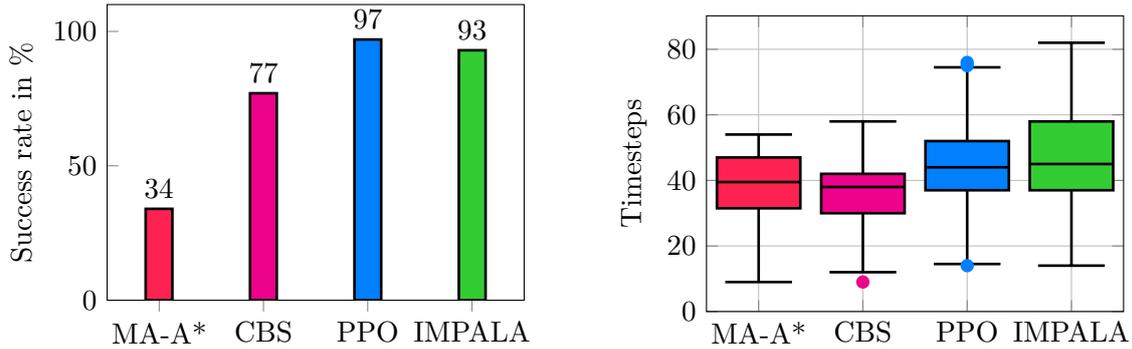
\begin{figure}[h]
    \begin{minipage}[t]{.49\textwidth}
        \centering
        \begin{tikzpicture}
            \begin{axis}[
                width=0.7\textwidth,
                height=0.5\textwidth,
                ybar,
                ymin=0,
                ymax=110,
                xmin=0.5,
                xmax=4.5,
                ylabel={Success rate in \%},
                xtick={1, 2, 3, 4},
                xticklabels={MA-A*, CBS, PPO, IMPALA},
                tick pos=left,
                nodes near coords,
                nodes near coords align={vertical},
                grid=none
                ]
    
                \addplot[fill=redMM, xshift=0.63cm] coordinates {(1, 34)};
                \addplot[fill=magenta, xshift=0.21cm] coordinates {(2, 77)};
                \addplot[fill=blueMM, xshift=-0.21cm] coordinates {(3,97)};
                \addplot[fill=greenMM, xshift=-0.63cm] coordinates {(4,93)};
    
            \end{axis}
        \end{tikzpicture}
    \end{minipage}
    \begin{minipage}[t]{.49\textwidth}
        \centering
        \begin{tikzpicture}
        \begin{axis}[
            width=0.7\textwidth,       
            height=0.5\textwidth,
            boxplot/draw direction=y,
            ylabel={Timesteps},
            xtick={1, 2, 3, 4},
            xticklabels={MA-A*, CBS, PPO, IMPALA},
            xmin=0.5,
            xmax=4.5,
            ymin=0,
        ]
            \addplot[
                boxplot prepared={
                    median=39.5, 
                    upper quartile=47.0,
                    lower quartile=31.5,
                    upper whisker=54.0,
                    lower whisker=9.0,
                },
                fill=redMM,
                draw=black,
                mark options={color=redMM},
                ] coordinates { };

            \addplot[
                boxplot prepared={
                    median=38.0,
                    upper quartile=42.0,
                    lower quartile=30.0,
                    upper whisker=58.0,
                    lower whisker=12.0,
                },
                fill=magenta,
                draw=black,
                mark options={color=magenta},
                ] coordinates {(1,9.0) };
                
            \addplot[
                boxplot prepared={
                    median=44.0,
                    upper quartile=52.0,
                    lower quartile=37.0,
                    upper whisker=74.5,
                    lower whisker=14.5,
                },
            fill=blueMM,
            draw=black,
            mark options={color=blueMM},
            ] coordinates {
             (2,75.0) (2,76.0) (2,14.0)
            };

            \addplot[
                boxplot prepared={
                    median=45.0,        
                    upper quartile=58.0,
                    lower quartile=37.0,
                    upper whisker=82.0, 
                    lower whisker=14.0, 
                },
            fill=greenMM,
            draw=black,
            mark options={color=greenMM},
            ]coordinates {};
        \end{axis}
    \end{tikzpicture}     
    \end{minipage}
    
    \caption{Success rate and timesteps for reference model 3.1.}
    \label{fig:success-rate-timesteps-reference-model-3-1}
\end{figure}

Under reference model 3.1, Figure \ref{fig:success-rate-timesteps-reference-model-3-1} shows the success rates and timesteps for MA-A*, CBS, PPO, and IMPALA. MA-A* achieves a success rate of 34 \%, which is significantly lower compared to 77 \% for CBS, 97 \% for PPO, and 93 \% for IMPALA. While MA-A* demonstrates a relatively low success rate, it has a slightly lower median number of timesteps (39.5) compared to PPO (44) and IMPALA (45). This lower median does not imply better performance overall, as MA-A* struggles to consistently find successful paths.

CBS, with a success rate of 77 \%, performs significantly better than MA-A* in terms of pathfinding success, though it still falls behind PPO and IMPALA. Its median timesteps (38) are slightly lower than those of MA-A* and significantly lower than those of PPO and IMPALA. The interquartile range for CBS (30 to 42 timesteps) indicates relatively low variability in its pathfinding duration, which suggests consistent performance across different scenarios. CBS's centralized planning approach likely accounts for both its improved success rate compared to MA-A* and its lower variability in timesteps. However, its success rate remains limited in this environment, possibly due to the inherent challenges posed by dynamic interactions and complex agent coordination in reference model 3.1. PPO and IMPALA achieve notably higher success rates, with 97 \% and 93 \% respectively. The median timesteps of PPO and IMPALA are higher than those of CBS and MA-A*, and their greater interquartile ranges (37 to 52 for PPO and 37 to 58 for IMPALA) suggest increased variability in navigation times. This variability reflects the nature of their learning-based strategies, where policies adapt dynamically to encountered situations but may result in occasional inefficiencies or longer paths. PPO's slight edge in success rate over IMPALA likely results from differences in the underlying training process, though both algorithms display comparable overall performance.

Overall, CBS shows a clear improvement over MA-A* in both success rate and consistency, with lower median timesteps and reduced variability. While PPO and IMPALA achieve higher success rates, CBS maintains competitive efficiency in terms of timesteps. The results highlight the distinct strengths and weaknesses of each algorithm: CBS excels in consistency and efficiency due to its pre-computed paths, while PPO and IMPALA demonstrate robustness in handling dynamic and stochastic environments at the cost of increased variability in pathfinding duration.

\begin{figure}[h]
    \begin{minipage}[t]{.49\textwidth}
        \centering
        \caption*{MA-A*}
        \includegraphics[width=1\textwidth]{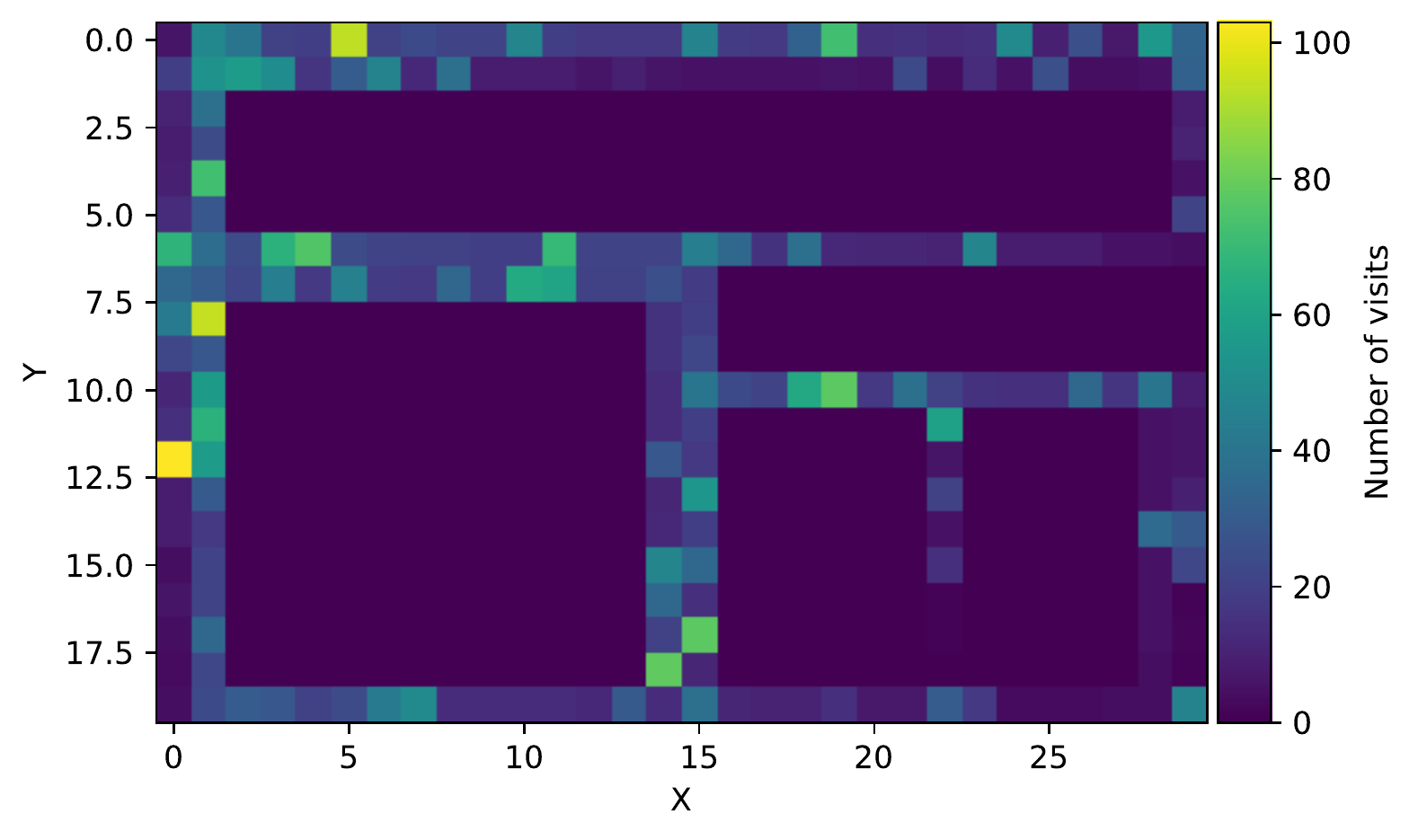}
    \end{minipage}
    \begin{minipage}[t]{.49\textwidth}
        \centering
        \caption*{CBS}
        \includegraphics[width=1\textwidth]{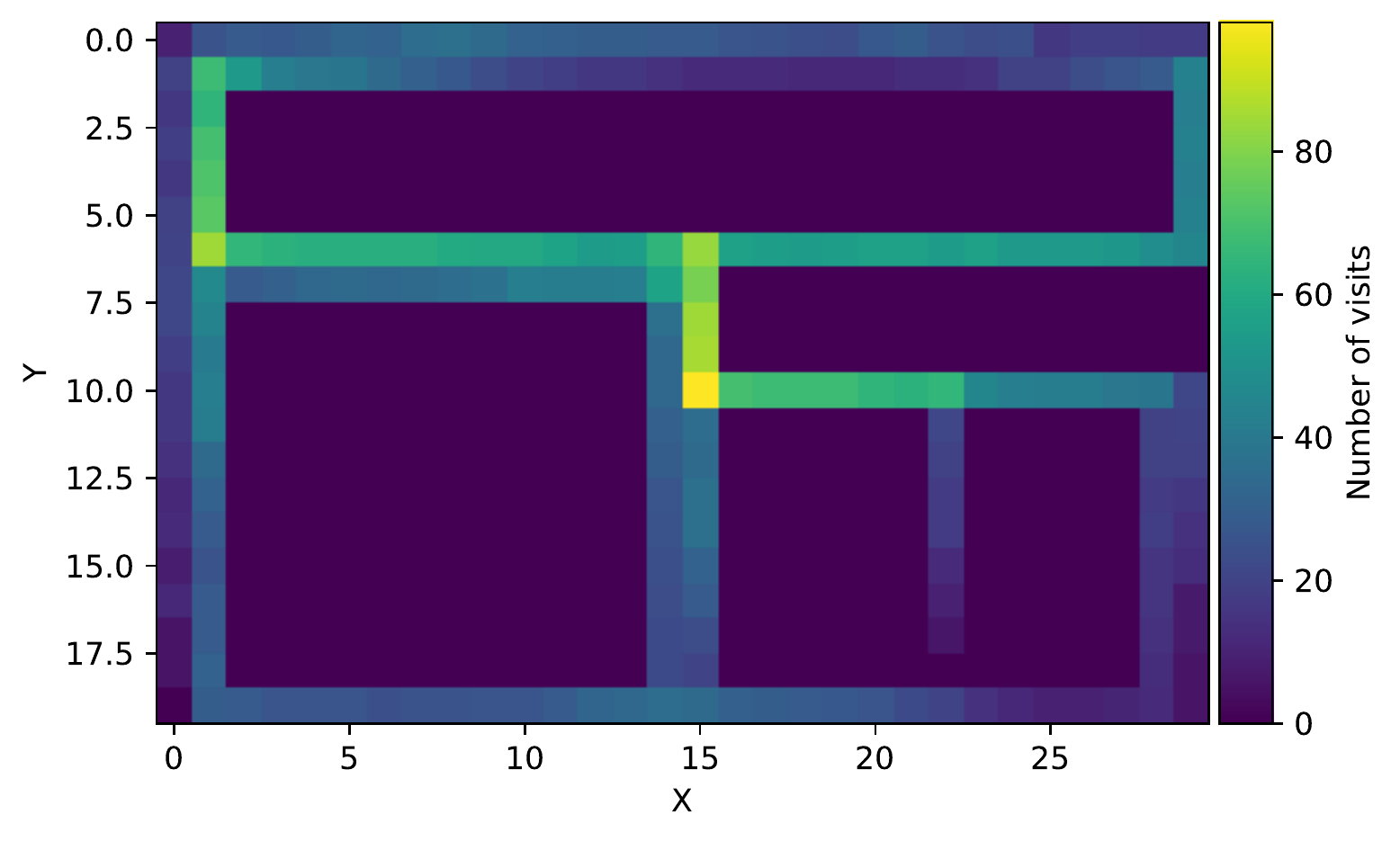}
    \end{minipage}
    
    \begin{minipage}[t]{.49\textwidth}
        \centering
        \caption*{PPO}
        \includegraphics[width=1\textwidth]{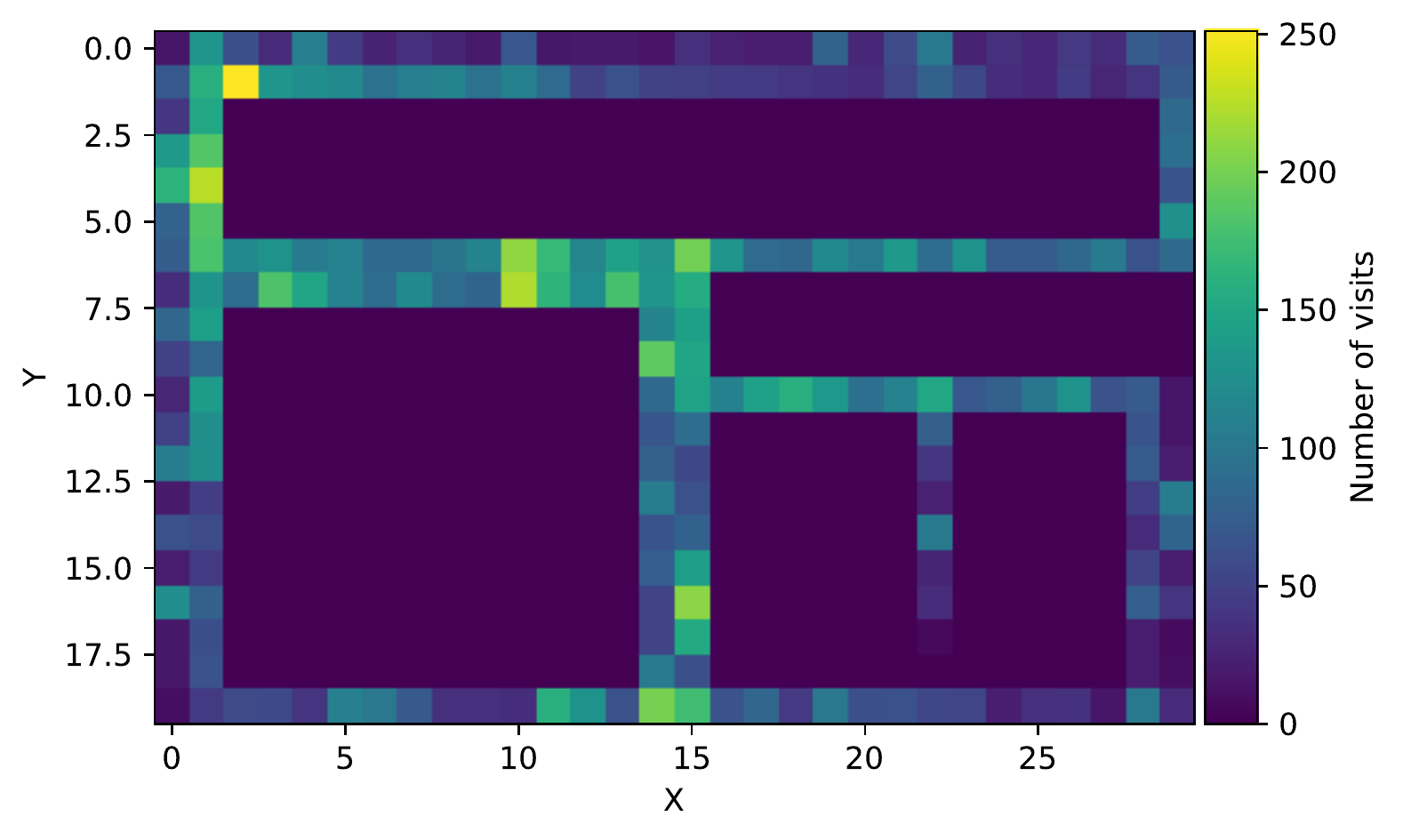}
    \end{minipage}
    \begin{minipage}[t]{.49\textwidth}
        \centering
        \caption*{IMPALA}
        \includegraphics[width=1\textwidth]{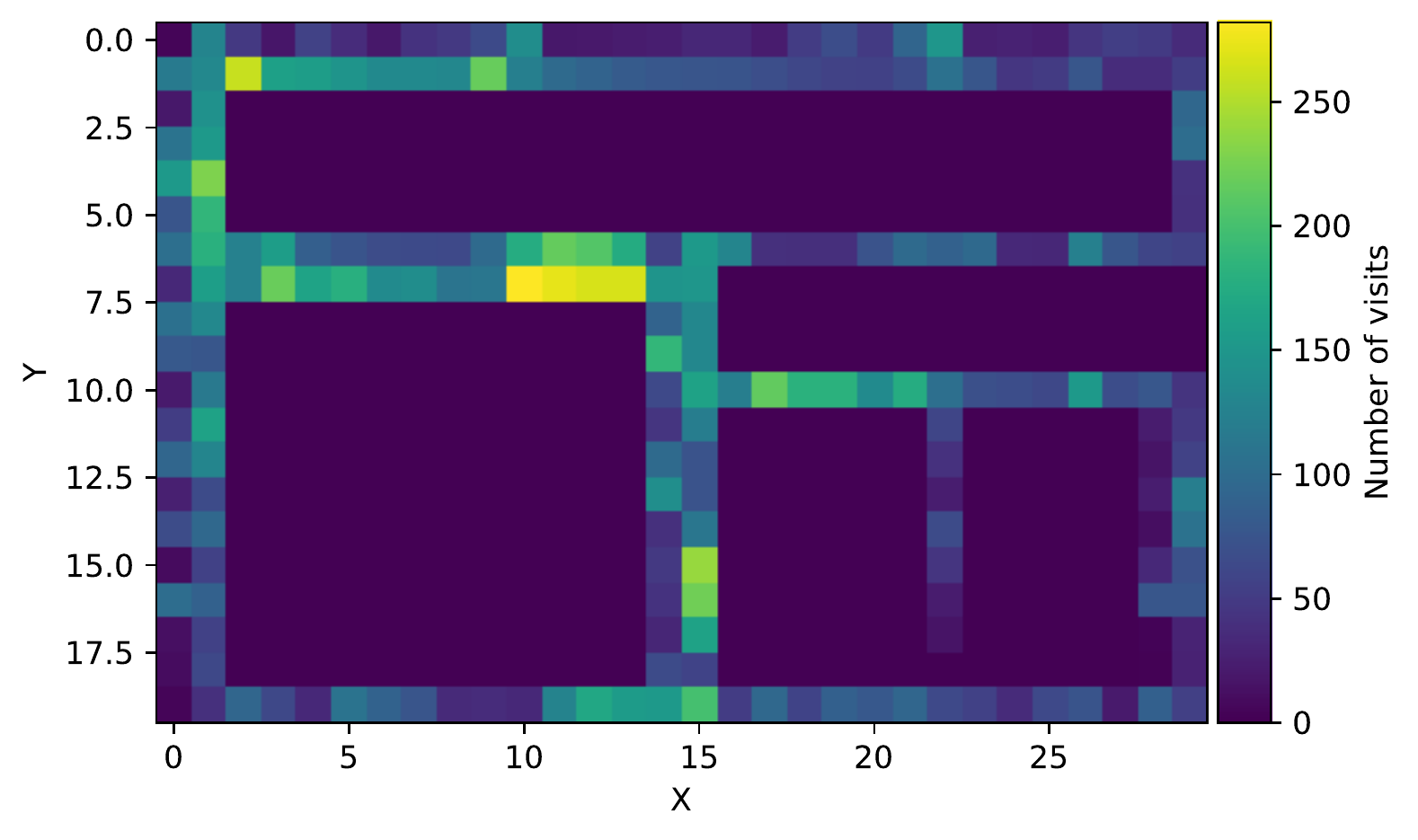}
    \end{minipage}

    \begin{minipage}[t]{.49\textwidth}
        \centering
        \caption*{Random}
        \includegraphics[width=1\textwidth]{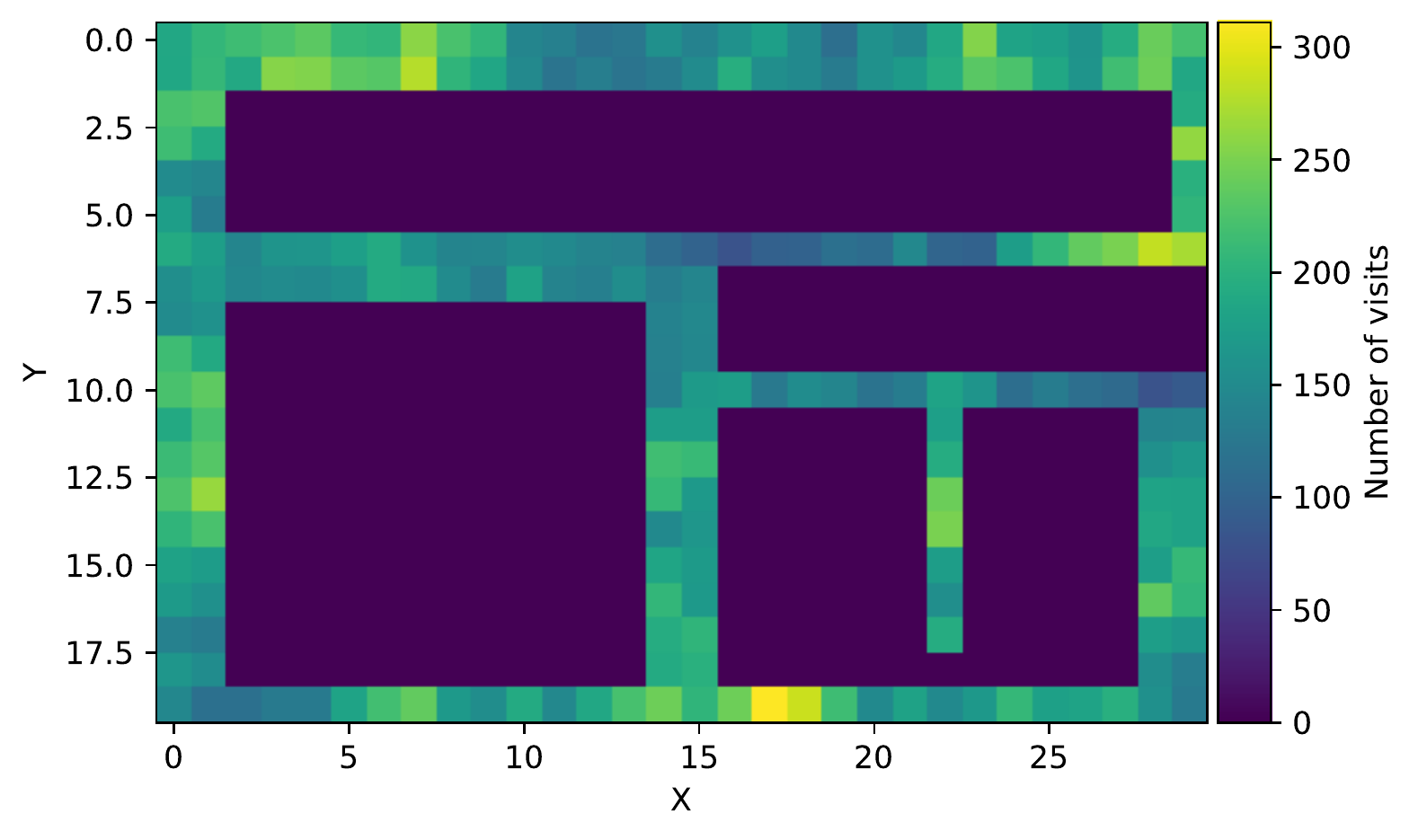}
    \end{minipage}
    \caption{Heat maps of the number of visits for reference model 3.1.}
    \label{fig:heatmaps-3-1}
\end{figure}

Under reference model 3.1, the heat maps in Figure \ref{fig:heatmaps-3-1} illustrate the spatial distribution of visits for the random policy, MA-A*, CBS, PPO, and IMPALA. The random policy exhibits a relatively uniform distribution of visits across the environment, with visit densities evenly spread throughout most regions. This behavior reflects the lack of goal-directed movement, leading to widespread and relatively balanced exploration.

MA-A*, PPO, and IMPALA exhibit similar patterns, with high visit densities concentrated at key intersections and along frequently used corridors. The visit distributions of these algorithms show that while they follow different approaches the resulting paths tend to converge on critical areas of the layout. The broad but consistent spread of visits highlights their behavior in avoiding collisions with detours. CBS, in contrast, displays a distinctly structured and orderly traversal pattern. The visit distribution for CBS is tightly focused along well-defined paths, with minimal spread beyond these routes. This behavior results from CBS's centralized planning approach, where paths are pre-computed for agents while ensuring collision-free movement. Consequently, CBS shows less exploratory behavior and more direct traversal compared to the other algorithms. The high-density regions observed in the middle crossing for all algorithms, except the random policy, highlight the importance of this central junction in navigating the layout of reference model 3.1. While CBS exhibits efficient and deterministic paths, MA-A*, PPO, and IMPALA rely on more dynamic strategies, resulting in broader but less structured coverage.

\subsection{Assessment of Scalability and Generalization}

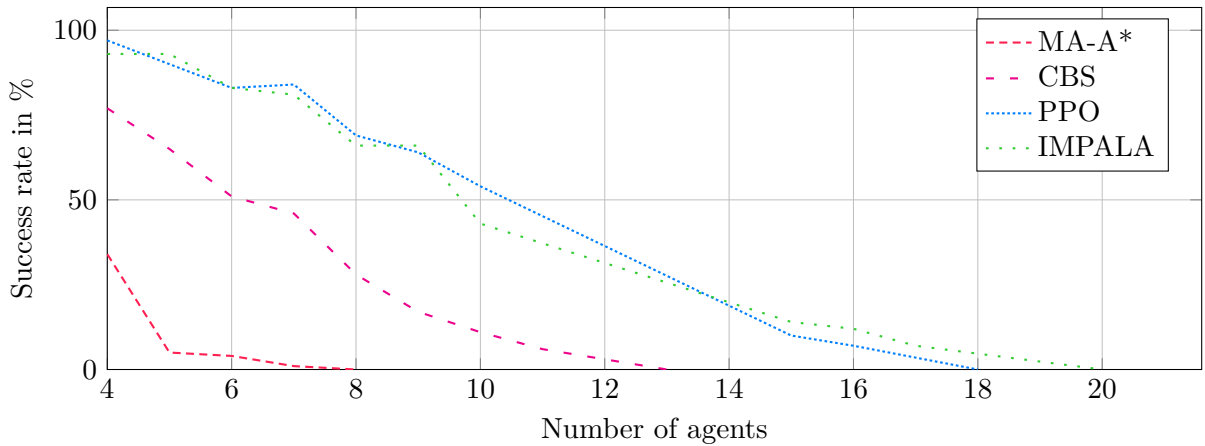
\begin{figure}[h]
    \centering
    \begin{tikzpicture}
        \begin{axis}[
            width=0.9\textwidth,
            height=0.3\textwidth,
            xlabel={Number of agents},
            ylabel={Success rate in \%},
            ymin=0, 
            xmin=4, 
            legend pos=north east,
            grid=both,
            major grid style={line width=0.2pt, draw=gray!50},
            minor grid style={line width=0.1pt, draw=gray!20}
        ]
            \addplot[densely dashed, redMM, thick] coordinates {
                (4, 34) (5, 5) (6, 4) (7, 1) (8,0)
            };
            \addlegendentry{MA-A*}

            \addplot[loosely dashed, magenta, thick] coordinates {
                (4, 77) (5,65) (6, 51) (7, 46) (8, 28) (9, 17) (10, 11) (11, 6) (12,3) (13,0)
            };
            \addlegendentry{CBS}

            \addplot[densely dotted, blueMM, thick] coordinates {
                (4, 97) (5, 90) (6, 83) (7, 84) (8, 69) (9, 64) (10, 54) (15, 10) (16, 7) (18, 0)
            };
            \addlegendentry{PPO}

            \addplot[loosely dotted, greenMM, thick] coordinates {
                (4, 93) (5, 93) (6, 83) (7,81) (8, 66) (9, 66) (10, 43) (15, 14) (16, 12) (17, 7) (20, 0)
            };
            \addlegendentry{IMPALA}
        \end{axis}
    \end{tikzpicture}
    \caption{Success rate versus number of agents for reference model 3.1.}
    \label{fig:number-of-agents-3-1}
\end{figure}

The MARL policy, trained with four agents, is evaluated in environments with increasing agent counts to assess its scalability. Figures \ref{fig:number-of-agents-2-1} and \ref{fig:number-of-agents-3-1} show the success rate as a function of the number of agents for reference models 2.1 and 3.1. The results indicate that classical MAPF approaches, such as MA-A* and CBS, struggle to scale effectively. In both models, MA-A* exhibits a rapid decline in success rate. In reference model 2.1, its success rate drops below 50 \% at six agents and fails entirely beyond eight. The performance is even worse in reference model 3.1, where MA-A* is already highly unreliable at five agents and fails completely at eight.

 \begin{figure}[h]
    \centering
    \begin{tikzpicture}
        \begin{axis}[
            width=0.9\textwidth,
            height=0.3\textwidth,
            xlabel={Number of agents},
            ylabel={Success rate in \%},
            ymin=0, 
            xmin=4, 
            legend pos=north east,
            grid=both,
            major grid style={line width=0.2pt, draw=gray!50},
            minor grid style={line width=0.1pt, draw=gray!20}
        ]
            \addplot[densely dashed, redMM, thick] coordinates {
                (4, 100) (5, 89) (6, 46) (7, 20) (8, 4) (9,0)
            };
            \addlegendentry{MA-A*}

            \addplot[loosely dashed, magenta, thick] coordinates {
                (4, 83) (5, 72) (6, 51) (7, 49) (8,37) (9, 26) (10, 11) (11, 8) (12, 1) (13,0)
            };
            \addlegendentry{CBS}

            \addplot[densely dotted, blueMM, thick] coordinates {
                (4, 96) (5,89) (6, 84) (7, 82) (8, 76) (9, 52) (10, 31) (15, 4) (16, 1) (17,0)
            };
            \addlegendentry{PPO}

            \addplot[loosely dotted, greenMM, thick] coordinates {
                (4, 90) (5,92) (6, 77) (7,78) (8,66) (9,54) (10,41) (15,4) (16,0)
            };
            \addlegendentry{IMPALA}
        \end{axis}
    \end{tikzpicture}
    \caption{Success rate versus number of agents for reference model 2.1.}
    \label{fig:number-of-agents-2-1}
\end{figure}

CBS performs better than MA-A* but also struggles with increasing agent counts. In model 2.1, CBS maintains moderate success rates up to nine agents, after which performance deteriorates, failing entirely beyond thirteen. In model 3.1, CBS starts with a lower initial success rate but declines at a similar rate, failing entirely at thirteen agents.

MARL-based policies, PPO and IMPALA, exhibit significantly better scalability. In model 2.1, PPO maintains high success rates up to eight agents before gradually declining, failing completely beyond seventeen. IMPALA follows a similar pattern but with slightly lower success rates, failing at sixteen agents. Model 3.1 presents a more challenging setting, reflected in the lower success rates across all methods. PPO maintains high success rates up to eight agents but declines more steeply beyond ten agents, reaching failure at eighteen. IMPALA performs similarly but retains marginal effectiveness up to seventeen agents, where it eventually fails.

Despite these advantages, MARL policies still experience a decline in performance as the number of agents increases. This decline is expected, as the state and action space grows combinatorially, increasing coordination complexity. Unlike traditional MAPF algorithms, which rely on predefined heuristics, MARL agents dynamically adapt their strategies, resulting in more robust behavior under moderate scaling. The differences in performance across reference models suggest that scalability depends on environment-specific factors, underscoring the need for further analysis. Future experiments will extend this investigation by examining the generalization capabilities of MARL policies in environments with varying agent distributions and task structures.

The following experiment evaluates how well trained policies generalize to alternative layouts under identical sampling budgets. Each policy is tested on 1,000 episodes per layout (seed = 42), using randomly sampled start-goal pairs to ensure statistical stability. Evaluation includes four reference models: 2.1, 2.1b, 2.2, and 3.1. Depending on the training setup, some of these environments are seen during training while others are not. This allows for a controlled comparison of how well different training configurations generalize beyond their original layout distribution.

\begin{figure}[h]
  \centering
  \begin{minipage}[t]{0.45\linewidth}
    \includegraphics[width=\linewidth]{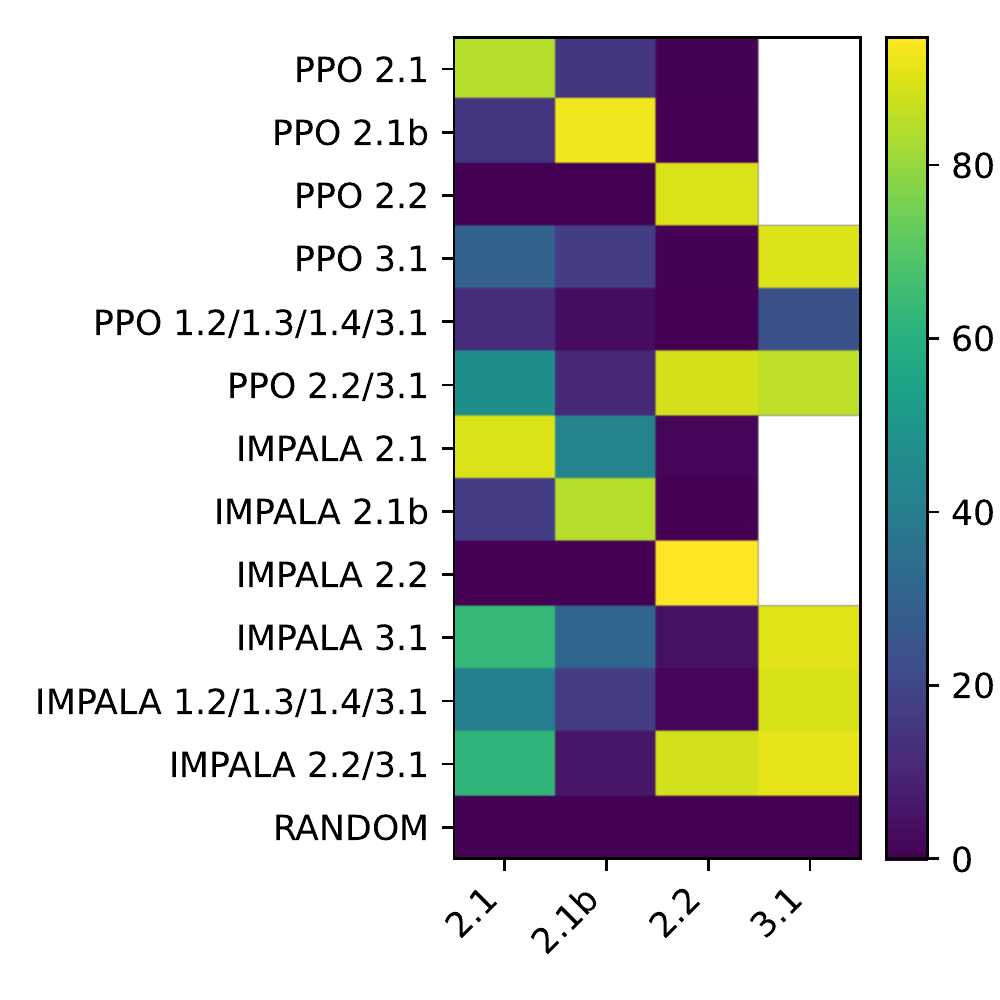}
    \caption*{Success rate on evaluation environments}
  \end{minipage}
  \begin{minipage}[t]{0.54\linewidth}
    \centering
    \begin{tikzpicture}
      \begin{axis}[
          width=\linewidth*0.8,
          height=5.8cm,
          xmin=0,
          xlabel={Environment steps sampled in millions},
          ylabel={Mean success rate in \%},
          grid=both,
          grid style={dashed,gray!30},
          legend style={at={(0.98,0.02)},anchor=south east},
      ]
      \addplot[only marks, mark=o, color=blueMM]
        coordinates {
          (1.944 , 33.2)
          (12.816, 35.7)
          (3.824 , 29.9)
          (14.36 , 34.3)
          (14.36 ,  9.7)
          (14.36 , 57.9)
        };
      \addlegendentry{PPO}
      \addplot[only marks, mark=square*, color=greenMM]
        coordinates {
          (40.09 , 44.4)
          (64.861, 33.8)
          (56.194, 31.6)
          (90.598, 47.1)
          (90.612, 37.1)
          (90.602, 61.9)
        };
      \addlegendentry{IMPALA}

        \node[anchor=west, font=\scriptsize] at (axis cs:1.944 , 33.2) {2.1};
        \node[anchor=south, font=\scriptsize] at (axis cs:12.816, 35.7) {2.1b};
        \node[anchor=west, font=\scriptsize] at (axis cs:3.824 , 29.9) {2.2};
        \node[anchor=west, font=\scriptsize] at (axis cs:14.36 , 34.3) {3.1};
        \node[anchor=west, font=\scriptsize] at (axis cs:14.36 , 9.7) {1.2–1.4/3.1};
        \node[anchor=west, font=\scriptsize] at (axis cs:14.36 , 57.9) {2.2+3.1};
        
        \node[anchor=west, font=\scriptsize] at (axis cs:40.09 , 44.4) {2.1};
        \node[anchor=west, font=\scriptsize] at (axis cs:64.861, 33.8) {2.1b};
        \node[anchor=east, font=\scriptsize] at (axis cs:56.194, 31.6) {2.2};
        \node[anchor=east, font=\scriptsize] at (axis cs:90.598, 47.1) {3.1};
        \node[anchor=east, font=\scriptsize] at (axis cs:90.612, 37.1) {1.2–1.4/3.1};
        \node[anchor=east, font=\scriptsize] at (axis cs:90.602, 61.9) {2.2+3.1};
      \end{axis}
    \end{tikzpicture}
    \caption*{Success rate vs.\ training budget}
  \end{minipage}
\caption{Generalization of PPO and IMPALA on different layouts for 1,000 evaluation episodes.}
\label{fig:generalization-results}
\end{figure}

Figure \ref{fig:generalization-results} (left) presents the average success rate of each policy on these four layouts. We extend the set of trained policies to include additional combinations, while always matching the training budget to that of the reference model 3.1 runs, the most sample-intensive configuration. This normalization ensures that differences in performance stem from the diversity of training layouts, not from unequal access to environment interactions. Figure \ref{fig:generalization-results} (right) plots the same generalization performance against the total number of sampled steps, making it possible to assess how success scales with training effort.

Two training strategies are of particular interest. One uses only reference models 1.2–1.4 and 3.1, which contain the conflict scenarios with frequent deadlock situations. The other extends reference model 3.1 training with reference model 2.2, adding more geometric diversity without increasing the total budget. Despite similar step counts, the resulting generalization differs sharply. Policies trained only on 1.2–1.4 and 3.1 show poor transfer performance. In contrast, training on 2.2 alongside 3.1 substantially improves generalization. The relative gain is more pronounced for PPO, where success increases from 34.3 \% to 57.9 \%. For IMPALA, the increase is from 47.1 \% to 61.9 \%. While IMPALA reaches the highest absolute success rate, the stronger relative improvement is observed in PPO.

The scatter plot shows that PPO reaches moderate generalization performance with relatively low sample budgets. When trained on layout 3.1 alone, PPO saturates around 14 million environment steps. Adding layout 2.2 substantially improves success from 34.3 \% to 57.9 \%, despite identical sampling budgets. This indicates that PPO benefits strongly from structural diversity and achieves good transfer with fewer samples. In contrast, PPO trained solely on conflict-heavy layouts (1.2–1.4) and 3.1 performs poorly (9.7 \%), underlining that layout complexity alone does not ensure generalizable policies. IMPALA operates with significantly higher budgets (up to 90 million steps). While training on 3.1 yields decent success (47.1 \%), only the addition of reference model 2.2 raises performance to 61.9 \%. The relative improvement is smaller than for PPO, but the trend is consistent: layout diversity improves generalization. Compared to IMPALA, PPO appears more sample-efficient under equivalent layout exposure.

The results support the view that generalization in multi-layout environments depends less on absolute training volume and more on the structural variation in the training curriculum. However, all policies are trained with a single seed. While the evaluation uses 1,000 episodes per policy and layout to ensure statistical stability, the trends observed should be understood as empirical indications rather than conclusive evidence. Broader conclusions would require multi-seed training runs to assess variability across initializations.

\section{Evaluation with an external Simulation Software} \label{sec:evaluation-external-simulation-software}
This section evaluates the performance of MARL algorithms in the external simulation software Plant Simulation. Unlike the Python-based reference models, Plant Simulation represents operational challenges in real production and warehouse systems more accurately. The software models and simulates complex logistics systems with a level of detail that surpasses the simplified representations of the developed reference models. Plant Simulation is a standard in industrial applications and provides a benchmark to compare MARL algorithms against established simulation modeling practices. The experiments in this section build upon the findings of \cite{muller2023mappo} and only use a CTDE approach.

\subsection{Conceptual Model}
The conceptual model represents a production logistics environment where AGVs navigate single-lane tracks with a high risk of deadlocks, providing a structured basis for evaluating the effectiveness of MARL in complex operational settings. This setup is designed to replicate typical industrial constraints, including confined pathways and intersecting routes, where AGVs must coordinate their movements to avoid congestion and deadlock situations. The environment has been further modified to intensify deadlock-prone scenarios, requiring AGVs to handle complex maneuvers such as waiting or reversing to clear paths for others.

\begin{figure}[h]
    \centering
    \includegraphics[width=1\linewidth]{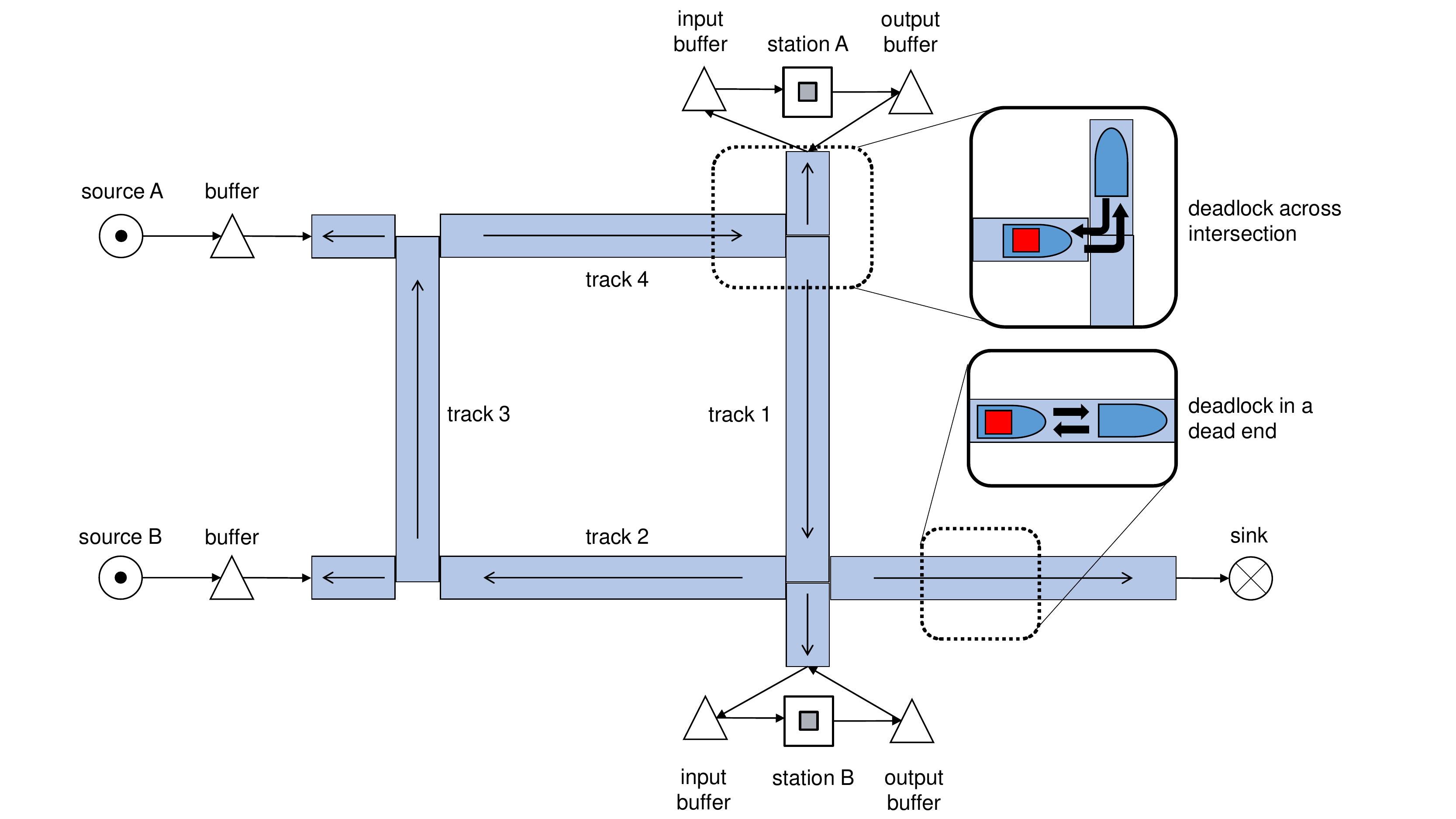}
    \caption{Conceptual model of the considered production logistics system.}
    \label{fig:wsc2023-conceptual-model}
\end{figure}

The considered production logistics system comprises two sources of products, labeled source A and source B, each responsible for producing a distinct product type. Products from these sources are transported by AGVs to dedicated processing units—one for each product type. After processing, the products are directed to a shared sink, which serves as the final destination in the system. The components of this system are interconnected via single-lane, bidirectional tracks, simulating the restrictive pathways typically encountered in warehouse and production facilities. Buffers are strategically positioned after the sources, as well as before and after each processing unit, to manage flow and provide temporary holding areas for products. The AGVs, represented as agents in this model, range in number from one to four. Their forward movement is indicated by directional arrows on the tracks, as shown in Figure \ref{fig:wsc2023-conceptual-model}. These AGVs are capable of reversing but cannot perform turning maneuvers, reflecting realistic limitations in physical AGV systems where direction changes are constrained. To navigate effectively, AGVs must coordinate their movements to avoid both collisions and deadlocks, particularly in narrow segments of the track network where they encounter oncoming traffic.

The model's routing algorithm prioritizes output buffers of processing units over input buffers at the sources, ensuring that processed products reach the sink without unnecessary delays. This prioritization rule helps in managing congestion and maintaining an efficient flow within the system, as AGVs dynamically adjust their routes based on buffer occupancy. In addition, dead-end tracks and intersections create scenarios requiring complex coordination, where AGVs may need to reverse or wait in designated buffer areas to clear pathways for others.

Through this environment, we aim to evaluate MARL's potential for solving real-world logistics challenges by rigorously testing its capacity to manage pathfinding, collision avoidance, and deadlock handling. By comparing MARL's performance against traditional MAPF algorithms under identical conditions, this learning environment provides a comprehensive framework for assessing the adaptability of MARL in production logistics systems with an established simulation software.

\subsection{Observation Space, Action Space, and Reward Function}
We define the following sets and parameters:
\begin{itemize} 
    \item $I = \{1, 2, \dots, n\}$: the set of $n$ AGVs (agents). 
    \item $T_{\max}=1,000$: the maximum number of discrete timesteps. 
    \item $T = \{0, 1, \dots, T_{\max}\}$: the set of timesteps. 
\end{itemize}

\subsubsection*{State Variables}

For each agent $i \in I$ at time $t \in T$:

\begin{itemize} 
    \item $pos_i(t) = (x_i(t), y_i(t)) \in \mathbb{R}$: the position of agent $i$. 
    \item $v_i(t) \in \mathbb{R}$: the current speed of agent $i$. 
    \item $G_i(t) \in \mathcal{G}$: the goal of agent $i$, where $\mathcal{G} = \{0, 1, 2, 3, 4, 5\}$ corresponds to specific locations:
    \begin{equation}
    G = \begin{cases} 
        0, & \text{no goal,} \\
        1, & \text{station A,} \\
        2, & \text{station B,} \\
        3, & \text{sink,} \\
        4, & \text{buffer 1,} \\
        5, & \text{buffer 2.}
        \end{cases} 
    \end{equation}
    \item $C_i(t) \in {0, 1}$: the carrying status of agent $i$: 
    \begin{equation} 
        C_i(t) = \begin{cases} 
            1, & \text{if agent } i \text{ is carrying a product}, \\ 
            0, & \text{if agent } i \text{ is not carrying a product}. 
            \end{cases} 
    \end{equation} 
    \item $L_i(t) \in \mathbb{R}_+$: the remaining route length to the goal of agent $i$. 
\end{itemize}

For the stations:

\begin{itemize} 
    \item $S = \{\text{A}, \text{B}\}$: the set of stations. 
    \item $status_s(t) \in \{0, 1\}$: the operational status of station $s \in S$ at time $t \in T$: 
    \begin{equation} status_s(t) = 
        \begin{cases} 
        1, & \text{if station } s \text{ is working}, \\ 
        0, & \text{if station } s \text{ is not working}. 
        \end{cases} 
    \end{equation} 
    \item $buffer_s(t) \in \mathbb{N}_0$: the number of items in the buffer of source $s \in S$ at time $t \in T$. 
\end{itemize}

\subsubsection*{Observation Space}

At each timestep $t \in T$, each agent $i \in I$ receives an observation $o_i(t)$ comprising:

\begin{equation} 
o_i(t) = \left( i,\ status_{\text{A}}(t),\ status_{\text{B}}(t),\ buffer_{\text{A}}(t),\ buffer_{\text{B}}(t),\ pos_i(t),\ v_i(t),\ G_i(t),\ C_i(t),\ L_i(t) \right). 
\end{equation}

Here:

\begin{itemize} 
    \item $i$: the ID of the considered agent. 
    \item $status_{\text{A}}(t),\ status_{\text{B}}(t)$: statuses of stations A and B. 
    \item $buffer_{\text{A}}(t),\ buffer_{\text{B}}(t)$: number of products in the buffers of sources A and B. 
    \item $pos_i(t)$: the $x$ and $y$ positions of agent $i$. 
    \item $v_i(t)$: the current speed of agent $i$. 
    \item $G_i(t)$: the goal of agent $i$. 
    \item $C_i(t)$: the carrying status of agent $i$. 
    \item $L_i(t)$: the remaining route length to the goal of agent $i$. 
\end{itemize}

\subsubsection*{Action Space}

For each agent $i \in I$, the action space is defined as:

\begin{equation} A_i = \{0,\ 1,\ 2\}, \end{equation}

where:

\begin{equation}
    A_i=
    \begin{cases}
        0, & \text{hold (no movement),} \\
        1, & \text{move forwards,} \\
        2, & \text{move backwards.}
    \end{cases}
\end{equation}

\subsubsection*{Policy Definitions}

In the CTDE paradigm, each agent $i$ employs an individual policy $\pi_i$ that maps its local observation to an action:

\begin{equation} \pi_i: O_i \rightarrow A_i, \quad \forall i \in I, \end{equation}

where $O_i$ is the observation space of agent $i$.

\subsubsection*{Transition Function}

The environment transitions from state $s(t)$ to $s(t+1)$ according to the transition function $f$:

\begin{equation} s(t+1) = f\left( s(t),\ A(t) \right), \quad \forall t \in T', \end{equation}

where:

\begin{itemize} 
    \item $T' = T \setminus \{T_{\max}\} = \{0, 1, \dots, T_{\max} - 1\}$. 
    \item $A(t) = \left( a_1(t),\ a_2(t),\ \dots,\ a_n(t) \right) \in \mathcal{A}$ is the joint action vector. 
    \item $\mathcal{A} = \prod_{i=1}^{n} A_i = A_1 \times A_2 \times \dots \times A_n \quad i \in I$. 
\end{itemize}

The function $f$ encapsulates the dynamics of the agents and the environment, including movement, interactions with stations, and collision handling.

\subsubsection*{Reward Function}

We define indicator functions for specific events:

\begin{itemize} 
    \item Delivering a finished product to the sink: 
    \begin{equation} D_i(t) = 
        \begin{cases} 
        1, & \text{if agent } i \text{ delivers a finished product to the sink at time } t, \\ 
        0, & \text{otherwise}, 
        \end{cases}
        \quad \forall i \in I.
    \end{equation} 
    \item Picking up an item from the source: 
    \begin{equation} 
        P_i^{\text{pickup}}(t) = 
        \begin{cases} 
        1, & \text{if agent } i \text{ picks up a product from a source at time } t, \\ 
        0, & \text{otherwise},
        \end{cases}
        \quad \forall i \in I.
    \end{equation} 
    \item Placing an item in the correct station: 
    \begin{equation} P_i^{\text{place}}(t) = 
        \begin{cases} 
        1, & \text{if agent } i \text{ places a product in the correct station at time } t, \\ 
        0, & \text{otherwise},
        \end{cases}
        \quad \forall i \in I.
    \end{equation} 
    \item Stations not working: 
    \begin{equation} W_s(t) = 
        \begin{cases} 
        1, & \text{if } status_s(t) = 0, \\ 
        0, & \text{if } status_s(t) = 1, 
        \end{cases} 
        \quad \forall s \in S. 
    \end{equation} 
    \item Collision involvement:
    \begin{equation} 
        Col_i(t) = 
        \begin{cases} 
        1, & \text{if agent } i \text{ is involved in a collision at time } t, \\ 
        0, & \text{otherwise},
        \end{cases}
        \quad \forall i \in I.
    \end{equation} 
\end{itemize}

Let $\alpha, \beta, \gamma, \delta,$ and $\epsilon$ represent the weights for different components of the reward function:
\begin{itemize}
    \item $\alpha = 1$: weight for delivering a finished product to the sink.
    \item $\beta = 0.1$: weight for picking up an item from the source.
    \item $\gamma = 0.1$: weight for placing an item in the correct processing unit.
    \item $\delta = - 0.01$: penalty weight per station not working.
    \item $\epsilon = -1$: penalty weight for collisions.
\end{itemize}

The reward $r_i(t)$ for each agent $i \in I$ at time $t \in T$ is calculated as:

\begin{equation} 
r_i(t) = \alpha \cdot D_i(t) + \beta \cdot P_i^{\text{pickup}}(t) + \gamma \cdot P_i^{\text{place}}(t) + \delta \cdot \sum_{s \in S} W_s(t) + \epsilon \cdot Col_i(t). 
\end{equation}

Explanation of the reward components:

\begin{enumerate} 
    \item Delivering product ($\alpha \cdot D_i(t)$): agents receive a reward of $+1$ for delivering a finished product to the sink. 
    \item Picking up item ($\beta \cdot P_i^{\text{pickup}}(t)$): agents receive a reward of $+0.1$ for picking up an item from a source. 
    \item Placing in unit ($\gamma \cdot P_i^{\text{place}}(t)$): agents receive a reward of $+0.1$ for placing an item in the correct processing unit. 
    \item Stations not working penalty ($\delta \cdot \sum_{s \in S} W_s(t)$): agents incur a penalty of $-0.01$ for each station that is not working at time $t \in T$. 
    \item Collision penalty ($\epsilon \cdot Col_i(t)$): agents incur a penalty of $-1$ if they are involved in a collision at time $t \in T$. The penalty applies to the oncoming AGV. 
\end{enumerate}

The total accumulated reward for agent $i$ over the episode is:

\begin{equation} 
    R_i = \sum_{t=0}^{T_{\text{max}}} r_i(t) \quad \forall i. 
\end{equation}

The episode ends when the maximum timestep $T_{\max}$ is reached.

Table \ref{table:wsc2023-summary} shows the summary of the defined observation space, action space and rewards for the considered use case with the external simulation software.
 
\begin{table}[h]
\centering
\caption{Summary of observation space, action space, and rewards.}
\begin{tabular}{ll}
\toprule
    \multicolumn{2}{c}{\textbf{Observation space}} \\ \hline
    General & ID of the considered AGV/agent \\
    & status of stations A and B \\
    & number of items in buffers of sources A and B \\ 
    For each agent $i$ & $x$ and $y$ positions of agent $i$ \\
    & current speed of agent $i$  \\
    & goal of agent $i$ \\
    & emptiness status of agent $i$\\
    & remaining route length to goal of agent $i$ \\ 
    \multicolumn{2}{c}{\textbf{Action space}} \\ 
    Driving behavior & $0$ (hold), $1$ (forwards), $2$ (backwards) \\ 
    \multicolumn{2}{c}{\textbf{Rewards}} \\ 
    Delivering product & $+1$ for delivering a finished product to the sink \\ 
    Picking up item & $+0.1$ for picking up an item from the source 
     \\ 
    Placing in unit & $+0.1$ for placing it in the correct processing unit \\ 
    Stations not working & $-0.01$ for every agent per second/timestep \\
    & for every station not working \\ 
    Collision & $-1$ for each collision (for the oncoming AGV) \\ 
\bottomrule
\end{tabular}
\label{table:wsc2023-summary}
\end{table}

\subsection{Simulation Model}
\begin{figure}[h]
    \centering
    \includegraphics[width=1\linewidth]{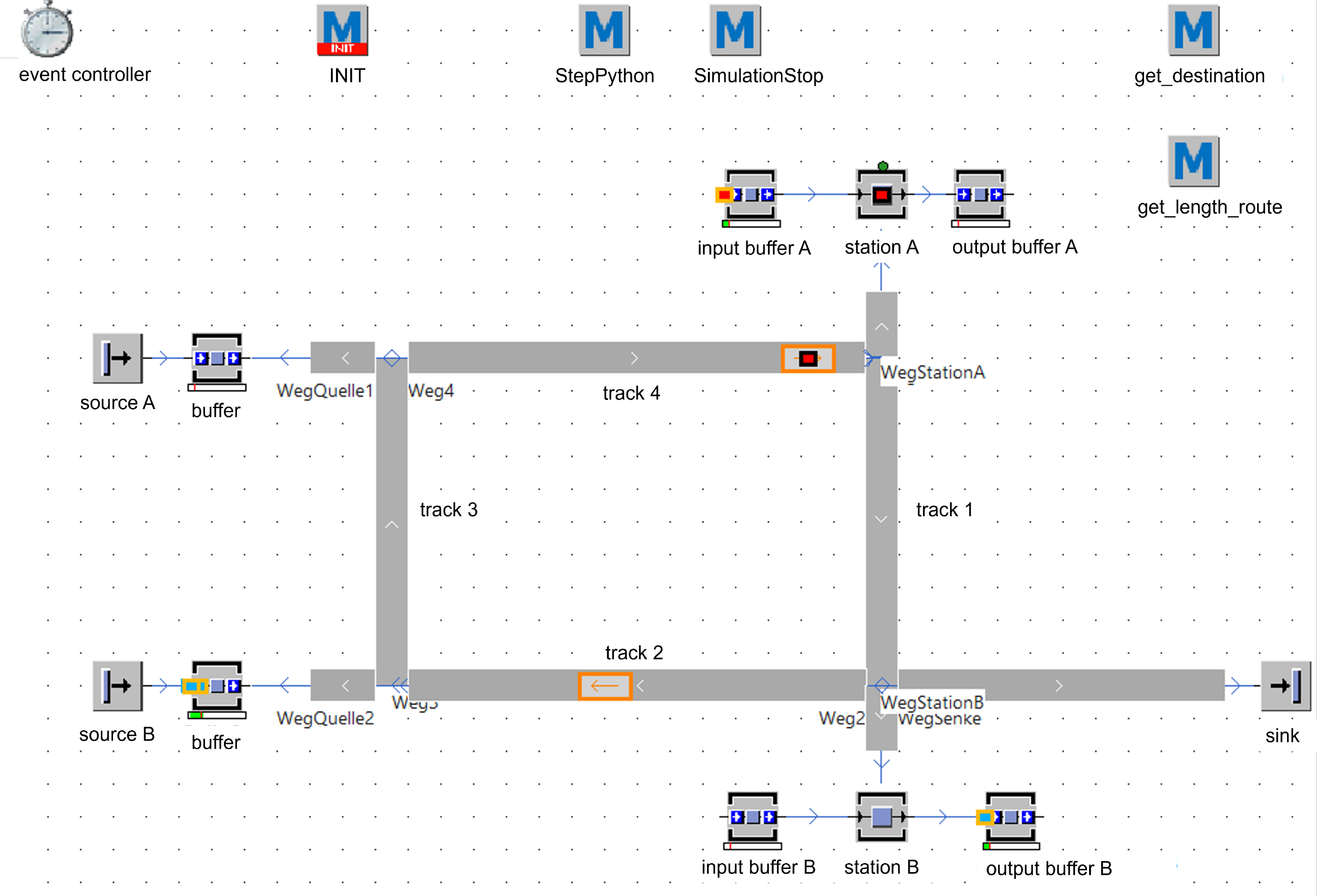}
    \caption{Simulation model during the simulation of two agents.}
    \label{fig:wsc2023-simulation-model}
\end{figure}

The simulation model uses Tecnomatix Plant Simulation version 2201. Figure \ref{fig:wsc2023-simulation-model} shows the simulation model during the simulation of two agents. At the start of the simulation, agents are generated at random positions along tracks corresponding to their IDs. For instance, agent 2 always starts on track 2. This initialization ensures reproducibility while introducing controlled variability into the environment.

SimTalk 2.0, the scripting language of Tecnomatix Plant Simulation, facilitates the implementation of the simulation logic and interaction between the used components. The  function, \enquote{StepPython}, connects to a Python script via a component object model (COM) interface. This function halts the simulation after one second of simulation time by invoking \enquote{SimulationStop}. The Python script monitors the event controller, resuming the simulation after each halt. This process defines a single timestep within the RL loop, ensuring precise synchronization between the learning algorithm and the simulation environment. The Python script integrates with Ray 2.35.0 and RLlib to manage the MARL experiments. A custom environment adhering to Gymnasium standards translates the simulation into a format suitable for RL. The environment defines the required reset and step methods. The reset method initializes the simulation by resetting the step counter, the simulation model, and the random seed, ensuring reproducibility across runs. The step method applies agent actions, updates the simulation, retrieves observations, and computes rewards based on the defined reward structure. The simulation model and source code used for this use case are available in a public repository maintained by the author: \url{https://github.com/Nerozud/FTS_simpel}.

\subsection{Hyperparameter Search and Evaluation}

The ANN uses a fully connected feedforward architecture to model a shared policy of agents operating in the considered multi-agent environment. The ANN consists of two hidden layers, each containing 256 neurons, with the hyperbolic tangent (tanh) activation function applied to each layer. This architecture is designed to support simultaneous operation of one to four agents in the environment. The training process is distributed across two worker nodes, each utilizing a single CPU. Agent performance is evaluated in parallel with training by conducting separate evaluation episodes. Following the implementation of the ANN and the training process, we perform a comprehensive search for the optimal hyperparameters of the PPO algorithm. We employ Bayesian optimization to explore the hyperparameter space. Inspired by the hyperparameter ranges outlined by \cite{AurelianTactics.2018}, we select a broad initial range for the PPO hyperparameters, encompassing the learning rate, clipping parameter, and regularization terms. This approach is motivated by the lack of prior knowledge regarding PPO's learning behavior in this specific environment. To explore the hyperparameter space, we use Bayesian optimization with the upper confidence bound (UCB) acquisition function, parameterized by the exploration-exploitation trade-off parameter $\kappa$ and the minimum improvement threshold $\xi$. For the hyperparameter search, we set the number of agents to two. This configuration ensures that the learning environment includes scenarios with collisions and deadlocks, while maintaining sufficient freedom of movement and avoidance strategies compared to environments with three or four agents. Table \ref{tab:wsc2023_bayes_optimization_runs} provides a detailed overview of the hyperparameter search space for both Bayesian optimization runs, including the initial point selected for the second run and the configuration of the BayesOptSearch algorithm. The initial point for the second run is determined by selecting the hyperparameter set that achieves the highest mean reward over 100 evaluation episodes during the first run.

\begin{table}[h]
    \centering
    \caption{Hyperparameters for Bayesian optimization runs.}
    \begin{tabular}{lccc}
    \toprule
        \textbf{Hyperparameter} & \textbf{Run 1} & \textbf{Run 2} & \textbf{Initialization for run 2} \\ \hline
        Number of samples & 225 & 43 & - \\ 
        Random search samples & 50 & 0 & - \\ 
        Minibatch size & 512 & 512 & - \\ 
        Epoch range & 20 & 20 & - \\ 
        Clipping range & [0.1, 0.3] & [0.1, 0.3] & 0.1749 \\ 
        Learning rate $\alpha$ & [5e-6, 0.003] & [5e-6, 0.001] & 0.000179 \\
        KL initialization range & [0.3, 1] & [0.3, 1] & 0.7191 \\
        KL target range & [0.003, 0.03] & [0.003, 0.03] & 0.007212 \\ 
        Discount factor $\gamma$ & [0.8, 0.9997] & [0.8, 0.9997] & 0.9462 \\ 
        GAE parameter $\lambda$ & [0.9, 1] & [0.9, 1] & 0.9156 \\ 
        Value function coefficient & [0.5, 1] & [0.5, 1] & 0.9331 \\ 
        Entropy coefficient & [0, 0.01] & [0.001, 0.01] & 0.009507 \\ 
        \multicolumn{4}{c}{\textbf{BayesOptSearch parameters}} \\ 
        Acquisition function & UCB & UCB & - \\ 
        Exploration parameter $\kappa$ & 2.5 & 0.5 & - \\ 
        Minimum improvement threshold $\xi$ & 0.0 & 0.0 & - \\ 
    \bottomrule
    \end{tabular}
    \label{tab:wsc2023_bayes_optimization_runs}
\end{table}

The first run reaches a high-performing configuration during the initial random sampling phase. Additional samples generated through Bayesian optimization do not yield further improvements. As a result, the second run narrows the search and applies early stopping. Specifically, it uses an iteration stopper after 500 trials and a trial plateau stopper that halts trials when the standard deviation of the episode mean reward remains below 0.2 over 100 iterations.

\begin{figure}[h]
    \centering
    \includegraphics[width=1\linewidth]{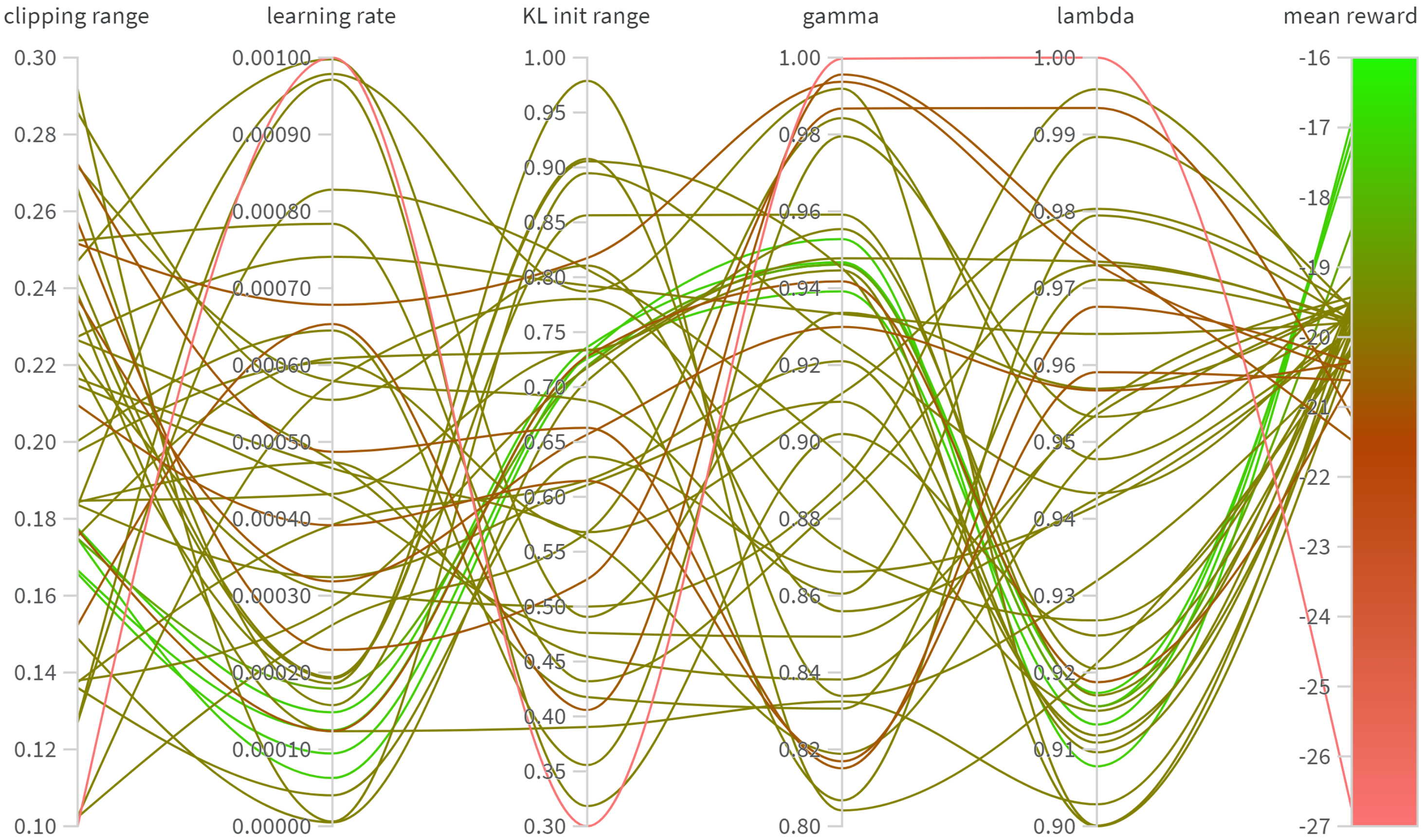}
    \caption{Hyperparameter ranges and impact on the mean reward for Bayesian optimization run 2.}
    \label{fig:wsc2023_run2_hyperparameter_ranges}
\end{figure}

Figure~\ref{fig:wsc2023_run2_hyperparameter_ranges} shows the distribution of hyperparameters in the second run. The best configuration achieves a mean reward of -16.9; the worst reaches -26.7. Many configurations lead to policies where agents remain stationary to avoid collisions, which triggers penalties for inactivity. This behavior results in mean rewards around -20. Better performance correlates with lower learning rates and reduced $\lambda$. A clipping parameter near 0.17, KL divergence initialization above 0.7, and a discount factor $\gamma$ close to 0.95 are consistent among the best-performing configurations.

\begin{figure}[h]
    \centering
    \includegraphics[width=1\linewidth]{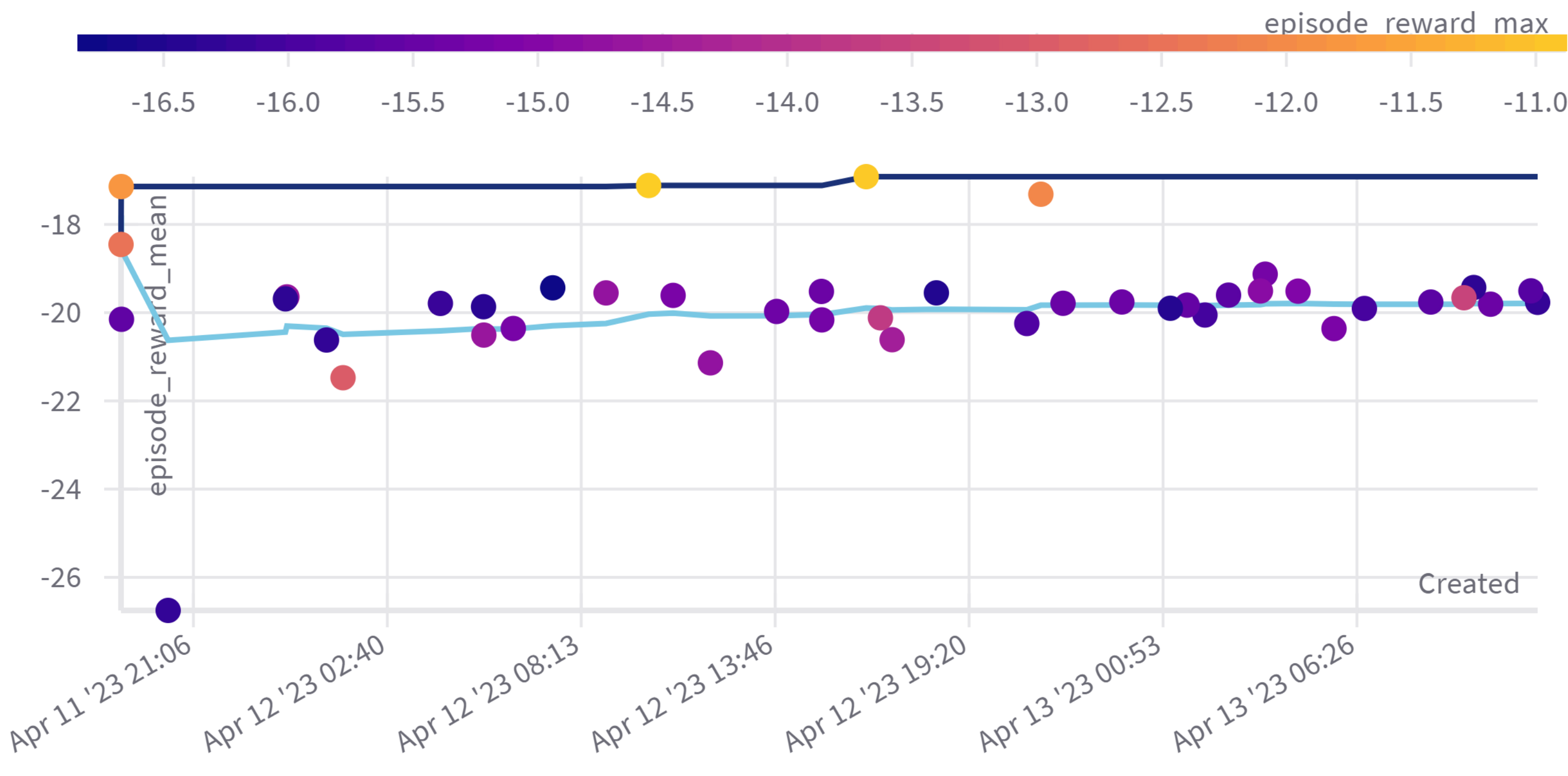}
    \caption{Scatter plot of mean episode reward and maximum reward of the samples in Bayesian optimization run 2.}
    \label{fig:wsc2023_mean_reward_over_time}
\end{figure}

Figure~\ref{fig:wsc2023_mean_reward_over_time} shows the progression of the mean episode reward during the second Bayesian optimization run. Each point represents a sampled configuration, colored by the maximum reward achieved. Over time, small variations in the hyperparameter settings lead to slight but consistent improvements over the best solution from the first run. Although no abrupt jumps in performance occur, the rolling average of the mean reward exhibits an upward trend, indicating gradual optimization progress.

\begin{figure}[h]
    \centering
    \includegraphics[width=0.9\linewidth]{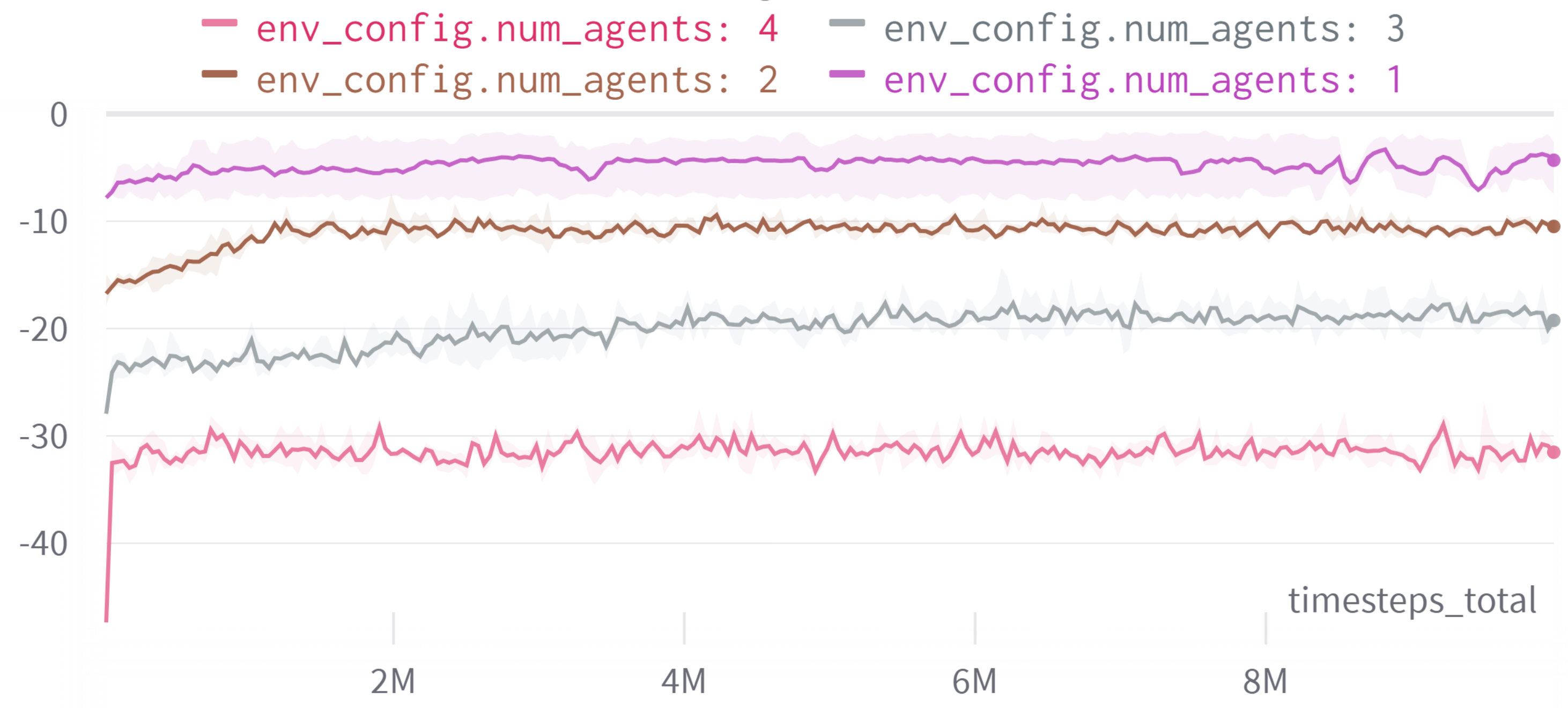}
    \caption{Mean episode reward during training process grouped by number of agents.}
    \label{fig:wsc2023_comparison_agents.pdf}
\end{figure}

To evaluate the robustness of the optimized hyperparameters, further experiments extend the training horizon to 2,500 iterations and vary the number of agents from one to four. For each configuration, three independent runs are conducted to assess the variance in learning outcomes. Figure~\ref{fig:wsc2023_comparison_agents.pdf} presents the resulting learning curves.

Performance decreases with the number of agents, reflecting the higher coordination demands and increased likelihood of agents remaining inactive due to blocked paths. This behavior leads to penalties from non-operational stations and lower mean rewards. The standard deviation of the mean reward increases with the number of agents, consistent with the growing complexity of the task. An exception is the single-agent setting, which shows the highest variance. In this case, sparse rewards and limited exploration capabilities hinder stable policy learning, resulting in diverging outcomes across runs.

For single-agent scenarios, the absence of collisions simplifies the learning task, reducing the challenge of escaping stationary behavior. With two agents, the steepest learning curve is observed, as agents learn to balance collision avoidance with productive throughput and the hyperparameter are tuned on this configuration. For three agents, a learning effect is still discernible, but it diminishes as the system becomes more congested. In the four-agent case, the mean reward stabilizes with limited improvement, and individual episodes achieve rewards as low as -26 due to the increased likelihood of collisions. This highlights the challenges of achieving consistent performance in highly congested environments.



\section{Conclusion} \label{sec:evaluation-conclusion}

Chapter \ref{ch:evaluation} investigates MARL's capability to resolve deadlocks in AMR systems by addressing the MAPF problem in dynamic, multi-agent environments. The chapter employs a grid-based simulation framework to systematically evaluate MARL strategies against traditional MAPF algorithms, such as MA-A* and CBS, under varying conditions of agent density, task complexity, and environmental constraints. These simulations provide critical insights into the interaction dynamics, solution efficiency, and computational trade-offs associated with MARL.

The results reveal MARL's adaptability and efficiency in managing deadlock scenarios in small to moderately complex systems. Its decentralized learning paradigm allows agents to independently adapt their behavior based on local observations, enabling flexible and dynamic conflict resolution. The analysis further emphasizes the importance of reward structures and hyperparameter tuning, which significantly affect the speed of convergence and the quality of the learned policies. By adjusting parameters such as learning rate, entropy coefficients, and discount factor, MARL can be optimized for specific operational goals, such as minimizing collision risks or prioritizing throughput.

Scalability remains limited. When the number of agents rises from two to four, the mean reward falls by 18 \% and its standard deviation doubles, indicating that current methods struggle to coordinate in dense settings. Future work should measure how training time and resource usage grow with agent count and how to handle larger AMR fleets with MARL.

The chapter therefore shows both the strengths and present limits of MARL for deadlock-capable MAPF. The method lifts performance in small systems yet loses efficiency in larger ones, motivating research on hierarchical control and communication mechanisms. Chapter \ref{ch:discussion} situates these findings within the thesis research questions and outlines the next steps toward scalable MARL for industrial AMR fleets.

%

\chapter{Discussion} \label{ch:discussion}
Chapter \ref{ch:discussion} critically examines and evaluates the findings from the use cases presented in Chapter \ref{ch:evaluation}. By interpreting these results, conclusions are drawn, and where applicable, inductive reasoning is applied. The chapter is structured as follows:
\begin{itemize}
    \item Section \ref{sec:discussion-research-questions} revisits the RQs introduced in Section~\ref{sec:methodology-researchgap}. These questions are addressed based on the findings from Chapter \ref{ch:evaluation}.
    \item Section \ref{sec:discussion-requirements} shifts the focus back to the end-user, specifically the logistics planner. It explores whether and to what extent the proposed methodology benefits the logistics planning process.
\end{itemize}
\section{Answering the Research Questions} \label{sec:discussion-research-questions}



This section systematically addresses the RQs outlined in Section~\ref{sec:methodology-researchgap}, using the findings presented in Chapter \ref{ch:methodology} and \ref{ch:evaluation}. Each question is revisited to draw conclusions from the results and contextualize them within the broader academic and practical framework of deadlock-capable intralogistics systems.

\subsection*{RQ 1.1: Is deadlock handling significant?}

The results indicate that logistics planners should consider deadlock handling as a critical aspect of intralogistics system performance. Deadlocks can lead to severe disruptions, including system-wide halts, reduced throughput, and increased operational costs. The analysis presented in Section~\ref{sec:methodology-significance} shows that even a small number of agents can lead to a significant number of deadlocks and can have a compounding negative effect on overall system efficiency \citep{Muller.2020}. In certain scenarios where the system operates well below its capacity or where simple layouts minimize the risk of conflicts, deadlock handling may be less critical. In such cases, logistics planners might opt for simpler strategies or even ignore deadlock handling without a significant loss in performance. This distinction highlights that while deadlock handling is generally important, its relevance depends on specific operational conditions, which needs to be considered during the planning of logistic systems. The decision to incorporate deadlock handling mechanisms should therefore be guided by the planner's primary objectives. If the focus of the planner is on performance or resilience as logistics objective, as presented in Figure \ref{fig:logistics_objectives_revised} before, ignoring deadlocks is not advisable. In less demanding environments with sufficient performance buffers, planners may prioritize other concerns over deadlock handling. These findings suggest that while deadlocks are not universally critical, they should not be disregarded without careful consideration of the operational context.

While the presented analysis in Section~\ref{sec:methodology-significance} provides qualitative guidance on when deadlocks must be considered, an exact characterization of the boundary conditions under which deadlocks can be ignored remains an open question. Future work could focus on deriving characteristic curves that specify the critical thresholds for system parameters, such as agent density, system capacity, and layout complexity, at which deadlock handling becomes necessary or optional. Such characteristic curves would enable logistics planners to make more precise, data-driven decisions about whether to incorporate deadlock handling mechanisms during the design and planning phases of logistics systems.

\subsection*{RQ 1.2: How to integrate RL into the planning process?}  
Integrating RL into the logistics planning process involves more than developing an RL solution. It requires embedding RL into the established stages of logistics system design and operation. The procedure model presented in Section~\ref{sec:methodology-logisticsplanning} provides a structured framework for creating and deploying MARL solutions, but its true value lies in how it supports and enhances key phases of the planning process.

In traditional logistics planning, decisions are often made based on static models, historical data, and predefined rules. This approach works well under predictable conditions but struggles in dynamic environments where agent interactions and system states evolve over time. MARL addresses this limitation by learning policies directly from modeled learning environments with stochastic behavior, enabling adaptive decision-making. The first phase of the procedure model, RL problem formulation, integrates with the initial stages of logistics planning, such as demand analysis and system modeling. By defining states, actions, and rewards based on actual logistics objectives, for example throughput maximization or resilience improvement, MARL solutions can be aligned with the specific goals of the planner. Once the RL problem is defined, the model selection phase complements the design and simulation phases commonly used in logistics planning. Simulation models serve as virtual testbeds where planners can evaluate various scenarios, including potential deadlock situations, without disrupting real operations. This integration allows planners to explore different strategies and assess their impact on performance before implementation. The algorithm selection phase corresponds to the optimization stage in traditional planning. Unlike static optimization methods, MARL algorithms can learn and adapt continuously, offering planners a flexible tool for handling dynamic logistics environments. The study demonstrates that using algorithms such as PPO and IMPALA allows for the effective management of complex MAS, where interactions between agents can lead to emergent behaviors that static methods fail to capture. The deployment phase integrates with the execution and monitoring stages of logistics operations. Real-time monitoring and continual learning mechanisms ensure that MARL solutions remain effective as system conditions evolve. Unlike static approaches that require manual reconfiguration when conditions change, MARL solutions can adapt automatically, providing ongoing decision support to logistics planners.

The procedure model not only guides the development of MARL solutions but also ensures their seamless integration into the logistics planning process. It aligns with traditional stages such as target planning, system design, system operation, and evaluation. By embedding MARL into the planning process, logistics planners can improve performance, enhance resilience, and reduce operational disruptions, particularly in dynamic and complex environments, as the results in Chapter \ref{ch:evaluation} suggest.

Although continual learning mechanisms are presented as a key benefit of MARL deployment, further studies could investigate how well these mechanisms perform in long-term operational settings. Specifically, future research could focus on quantifying the impact of continual learning on system stability, policy robustness, and computational overhead. Understanding these aspects would provide valuable insights into whether continual learning reliably enhances logistics operations over extended periods or whether additional safeguards are necessary to prevent performance degradation.

\subsection*{RQ 2.1: What do reference scenarios for deadlock-capable MAPF problems look like?}  
Reference scenarios for deadlock-capable MAPF problems are designed to emulate the key characteristics of environments where deadlocks are likely to occur, providing a basis for evaluating and benchmarking algorithms. These scenarios are constructed to represent realistic and diverse operational settings, ensuring their relevance for both research and practical applications. The developed reference scenarios include several critical features. First, they explicitly incorporate elements that are known to exacerbate deadlocks, such as shared resources, high agent density, and narrow passages or bottlenecks. These features reflect the operational challenges faced in typical intralogistics systems, such as warehouses or production facilities. Second, the scenarios are built with scalability in mind, allowing for variations in the number of agents and the complexity of the environment. This scalability ensures that algorithms can be tested across a range of problem sizes, from small-scale setups to large industrial systems. The scenarios are constructed around three primary use cases: conflict situations, warehouse layouts, and production logistics. Conflict scenarios focus on small-scale setups where deadlocks can arise immediately due to tightly constrained resources or paths. These serve as baseline tests for initial algorithm evaluation. Warehouse layouts are larger, structured environments that introduce challenges such as single-lane aisles, shared access points, and densely packed storage configurations. These layouts are designed to simulate the operational constraints of real-world intralogistics systems. Finally, production logistics scenarios include dynamic resource demands and fixed goals, requiring agents to coordinate effectively under complex routing requirements. To facilitate benchmarking, the scenarios use a standardized grid-based implementation, which allows precise control over agent movements and interactions. This approach ensures that the scenarios are compatible with common MAPF frameworks and RL libraries, enabling reproducible evaluations. Each scenario is parameterized to introduce variability, ensuring that algorithms are tested under diverse conditions and remain generalizable rather than overfitting to specific cases.

While these reference scenarios provide a robust foundation for testing and benchmarking, some limitations should be noted. One potential weakness is their reliance on grid-based environments, which, while widely used in research, may oversimplify real-world systems. For instance, continuous-space environments with more complex movement dynamics might not be fully captured within the constraints of a grid. Another challenge is balancing realism with computational efficiency. Highly detailed scenarios might offer greater fidelity but could also impose significant computational costs, limiting their practical applicability in iterative testing processes. Future work in this area could explore the development of hybrid reference models that combine grid-based designs with more realistic continuous-space features. Additionally, expanding the scenarios to include stochastic elements, such as unexpected failures or dynamic obstacles, would better reflect real-world uncertainties. Further studies could also investigate how well these scenarios generalize across different logistics domains, such as drone-based delivery systems or autonomous maritime logistics, where the characteristics of deadlocks may differ significantly.

In summary, reference scenarios for deadlock-capable MAPF problems are characterized by their flexibility, scalability, and focus on deadlock-prone features. These scenarios provide a systematic way to evaluate and compare algorithms, ensuring their robustness and adaptability in handling deadlocks across a range of operational settings. The design of the reference models could benefit from future advancements that increase fidelity and extend applicability to more diverse and uncertain environments.

\subsection*{RQ 2.2: When and in which configurations is MARL effective for MAPF with deadlocks?}  

The effectiveness of MARL for solving MAPF with deadlocks depends on both the algorithmic configuration and the characteristics of the environment. The results in Chapter \ref{ch:evaluation} demonstrate that MARL approaches are particularly suited for scenarios characterized by dynamic agent interactions and restricted layouts. These conditions exacerbate deadlock risks and challenge traditional heuristic-based MAPF methods. Experiments conducted with PPO and IMPALA in various training and execution modes reveal that CTDE achieves the most consistent performance across different reference scenarios. This approach leverages centralized learning to optimize policies while maintaining decentralized execution for scalability. CTDE is especially effective in environments when agent population have the same type of transport orders, as it allows agents to learn in parallel the same task. MARL is particularly effective in dynamic and complex environments where traditional MAPF algorithms, such as MA-A* or CBS, struggle with adaptability. Scenarios involving high agent density and restricted freedom of movement pose significant challenges for deterministic methods but are well-suited for MARL solutions. For instance, the experiments show that PPO and IMPALA, when configured with CTDE, demonstrate superior performance in environments with low levels of freedom of movement of the agents. PPO-CTDE consistently achieves higher success rates and faster convergence compared to other configurations. MARL's effectiveness diminishes in simpler configurations where deterministic algorithms can perform efficiently with lower computational overhead. In structured environments with minimal overlap of agent paths and predictable paths, classical algorithms such as CBS or MA-A* achieve comparable or better results without the need for extensive training. These findings suggest that MARL is most beneficial in situations where adaptability and learning from stochastic behavior are critical. Hyperparameter optimization also plays an important role in MARL's effectiveness. For example, PPO outperforms IMPALA across most scenarios, particularly when paired with CTDE. PPO's stability and capacity for handling high-dimensional state spaces contribute to its robustness. IMPALA shows higher variability and slower convergence, limiting its reliability in complex environments.

Despite its strengths, MARL has limitations. Training MARL models can be computationally expensive, particularly in large-scale systems. Additionally, MARL policies may require fine-tuning to generalize effectively to new environments. While CTDE offers a balance between centralized coordination and decentralized execution, its reliance on centralized training poses challenges for real-time applications where communication delays or failures can occur. Future work could address these limitations by exploring hybrid approaches that combine the adaptability of MARL with the efficiency of classical MAPF algorithms. Additionally, further research could investigate methods to reduce MARL's computational overhead, such as transfer learning or model simplifications, and assess the long-term robustness of MARL policies in dynamic and stochastic settings.

In summary, MARL is most effective for MAPF with deadlocks in configurations characterized by high complexity and dynamic interactions. Its adaptability and learning capabilities make it a valuable tool in such scenarios, but its computational demands and generalization challenges warrant careful consideration when selecting it for specific use cases.

\subsection*{RQ 3.1: Does the deadlock handling strategy matter?}

The results in Section~\ref{sec:methodology-significance} confirm that the choice of deadlock handling strategy significantly impacts the performance of intralogistics systems. For systems employing MARL, the question becomes not only whether deadlock handling matters but also how the responsibility for deadlock handling is distributed between system-level prevention and MARL-based policies. The findings provide evidence that MARL can learn and apply typical deadlock handling strategies, such as avoidance or recovery, depending on the system's characteristics and the logistics planner's KPIs. If deadlocks are effectively prevented during the planning phase, MARL policies can focus entirely on optimizing other system objectives, such as throughput, efficiency, and cost reduction. Prevention strategies are particularly well-suited to systems with predictable dynamics and sufficient performance headroom. They offer resilience by eliminating the possibility of deadlocks, though often at the cost of flexibility and scalability in dynamic or high-performance environments. When MARL policies are tasked with handling deadlocks directly, the strategy adopted tends to align with classical approaches: avoidance or detection and recovery. In the current configurations, MARL learns avoidance strategies in environments where agents have sufficient information and deterministic conditions, especially in CTE, leveraging centralized decision-making to preemptively allocate resources and paths for every agent. In scenarios with limited information or higher uncertainty, MARL tend to adopt detection and recovery strategies, resolving deadlocks reactively as they occur. While these observations suggest MARL's adaptability, the exact conditions under which one strategy is preferred over the other remain unclear. This uncertainty does not diminish the primary finding that MARL effectively learns to apply classical deadlock handling strategies within the operational context provided.

The choice of strategy also strongly correlates with the logistics planner's objectives. Resilience often favors prevention strategies, which eliminate deadlocks entirely at the cost of flexibility. Performance-driven or cost-sensitive systems benefit from delegating deadlock handling to MARL, as this allows for dynamic, situation-dependent strategies such as avoidance or detection and recovery. These findings validate that deadlock handling strategies remain critical, even in MARL-based systems, and should be aligned with the overall system goals. Although the results in Chapter \ref{ch:evaluation} support these conclusions, they also highlight an open question: how critical is it to precisely determine the conditions under which MARL policies adopt one strategy over another? While such an investigation might provide additional theoretical insights, it is not essential for addressing the primary RQ. The key takeaway is that the choice of deadlock handling strategy continues to matter in MARL systems and must be carefully considered in the context of system constraints, agent information, and desired outcomes.

In conclusion, the results clearly demonstrate that deadlock handling strategies are vital, even in MARL-based systems. MARL effectively learns to implement avoidance or recovery strategies, depending on the configuration, reinforcing the importance of aligning the strategy with specific operational goals. While further research could refine our understanding of strategy preferences, the findings underscore the continued relevance of deadlock handling for system performance and resilience.

\subsection*{RQ 3.2: Is CTDE the best approach in deadlock-capable MAPF problems?}  

The results demonstrate that CTDE is a highly effective approach for addressing deadlock-capable MAPF problems. CTDE combines the advantages of centralized information during training to improve policy learning with the flexibility of decentralized decision-making during execution. This hybrid approach enables agents to learn efficiently using a shared policy, avoiding the exponential growth of the action space as the number of agents increases. Experiments in Chapter \ref{ch:evaluation} highlight the strengths of CTDE across various reference scenarios. For example, PPO-CTDE consistently achieves the highest mean episode rewards and demonstrates the most significant learning progress compared to other training and execution modes, particularly in scenarios where agents share similar tasks and are required to solve diverse MAPF configurations. However, CTDE's performance is not without limitations. Comparisons with DTE and CTE, as presented in Section~\ref{sec:application-reference-models}, reveal important insights. DTE, while requiring considerably more time to converge, allows for learning agent-specific policies, which can be advantageous in asymmetric scenarios. This flexibility comes at the cost of lower coordination efficiency in high-density environments. On the other hand, CTE performs consistently well in smaller settings with a limited number of agents but fails to scale effectively in larger and more stochastic environments. These comparisons highlight that CTDE provides a strong balance of coordination and scalability, particularly when managing a high number of agents with overlapping objectives. These findings support the conclusion that CTDE is among the most effective approaches for deadlock-capable MAPF problems, especially in environments that require both high coordination and the ability to handle numerous agents. Nevertheless, the question of whether CTDE is universally the \enquote{best} approach remains open and context-dependent, as the effectiveness of any training and execution mode depends on the specific characteristics of the problem and the operational requirements.

Future research could focus on optimizing CTDE for larger-scale systems to address its computational demands during training. Additionally, further studies could investigate CTDE's applicability in highly dynamic environments characterized by limited communication or other forms of partial observability beyond the configurations explored in this thesis. Another promising research direction involves applying mixed policy populations for heterogeneous agent types or tasks. For instance, it would be valuable to explore how MARL systems can develop specialized policies for specific groups of transport orders, enabling more efficient coordination and resource allocation.

In conclusion, CTDE proves to be a highly effective approach for deadlock-capable MAPF problems in complex environments. It offers a compelling balance of centralized training benefits and decentralized execution flexibility, effectively managing computational complexity and maintaining smaller observation and action spaces. While it demonstrates clear advantages across many scenarios, its applicability should always be evaluated against the specific requirements and constraints of the system under consideration.

\section{Meeting the Requirements of Logistics Planners} \label{sec:discussion-requirements}

Logistics planners operate under the constraint of maintaining high system availability, efficiency, and cost-effectiveness in increasingly dynamic and complex environments. While performance, cost, and quality have traditionally guided system design, the concept of resilience—defined as the ability of a system to maintain or recover its function in the face of disturbances—has received less attention in the early phases of logistics planning. This thesis addresses this shortcoming by explicitly incorporating deadlock handling into planning frameworks and proposing RL as a tool for dynamic and adaptive decision-making under structural constraints.

Deadlocks are a critical failure mode in autonomous transport systems. Their occurrence disrupts throughput, increases delay variability, and can render large subsystems temporarily inoperable. These effects directly impact KPIs such as order fulfillment, buffer levels, and transport cycle times. Despite this, industrial practice continues to rely predominantly on deadlock prevention through rigid planning constraints, such as conservative path allocation, exclusive-use zones, or temporal separation. These measures are effective in eliminating deadlocks but often lead to underutilization of resources and overly static system behavior.

This thesis proposes a shift from exclusive reliance on prevention to an integrated approach that includes adaptive deadlock handling as part of operational strategy. It introduces a structured planning model that extends the classical phases of logistics system design~\citep{Gudehus.2010} to explicitly account for deadlock-prone situations. In the target planning phase, resilience is defined as a primary planning objective alongside throughput and cost. This reframes system availability not merely as a matter of component reliability or redundancy, but as the capacity to dynamically resolve coordination failures.

In the requirement specification phase, the framework incorporates structural deadlock risk into early layout and route design decisions. Using simulation-based models and historical pattern analysis, planners can anticipate high-risk configurations and quantify their impact under variable load conditions. The resulting planning documents do not assume deadlock-free operation but define margins and control mechanisms that permit dynamic avoidance or recovery.

In system and detailed planning, the introduction of RL-based agents represents a paradigm shift. Unlike static rule systems, RL agents can continuously adapt to local observations and evolving traffic conditions, learning strategies for yielding, rerouting, or delaying based on empirical outcomes. These behaviors are trained and evaluated within reference models that simulate structurally critical scenarios under controlled variation. For system planners and providers, this enables the creation of digital twins that serve not only for layout evaluation but also for policy training and performance stress testing.

From a practical standpoint, this approach changes the nature of trade-offs available to planners. Instead of dimensioning systems to avoid all conflicts, which leads to oversized buffers and exclusive paths, planners can accept certain contention scenarios if RL policies can resolve them efficiently. This creates new options for layout design, path configuration, and fleet sizing. For system providers, it opens a path toward offering self-adaptive coordination logic as part of AMR or AGV control stacks, potentially reducing the need for centralized dispatching systems.

The application of RL in logistics planning is not without constraints. Deployment requires technical expertise, robust simulation environments, and computational resources. Furthermore, while this thesis demonstrates effectiveness in controlled environments, transfer to real-world operations, where partial observability, delayed actuation, and stochastic disturbances are present, remains an open challenge.

Table~\ref{tab:planning-comparison} contrasts conventional logistics planning with the RL-based resilient planning developed in this thesis. Classical approaches typically address deadlock handling through static design measures, such as conservative layouts or deadlock-free path planning with algorithms like MA-A* or CBS. These methods assume ideal conditions like a comprehensive communication between agents or prevent deadlocks by excluding conflict situations at design time. In contrast, the proposed approach uses adaptive policies that learn to anticipate and resolve deadlocks during execution, allowing for both deadlock avoidance and deadlock recovery.

\begin{table}[h]
    \centering
    \caption{Comparison of traditional logistics planning with the RL-based resilient planning approach proposed in this thesis.}
    \label{tab:planning-comparison}
    \begin{tabular}{p{3.5cm} p{5cm} p{6cm}}
        \toprule
        \textbf{Aspect} & \textbf{Traditional planning approach} & \textbf{RL-based resilient planning (this work)} \\
        \hline
        Deadlock handling & Deadlock prevention through conservative layout or deadlock avoidance with MAPF algorithms & Adaptive deadlock avoidance or deadlock recovery using learned RL policies \\
        Planning objective & Performance, quality, and costs & Performance, quality, costs, and resilience \\
        Design philosophy & Eliminate conflict potential via structural separation & Accept conflicts as manageable; focus on dynamic resolution \\
        Flexibility & Low; routes and behavior predefined & High; agents adapt based on local state and experience; continual learning possible \\
        Scalability &  Limited by preventive layouts and the poor scaling of MAPF algorithms with agent count & Better scaling through decentralized execution and shared policies; experience collection scales linearly with agents, though overall limits remain subject to further investigation\\ 
        Use of simulation & For evaluation of different planned variants & For training, testing, and evaluation of RL policies \\
        \bottomrule
    \end{tabular}
\end{table}

Regarding the planning objective, traditional methods optimize for performance, quality, and cost under the assumption of deterministic behavior. This work extends those objectives by explicitly incorporating resilience, defined as the ability to maintain functionality under dynamic disturbances such as congestion, agent delays, or layout perturbations.

The underlying design philosophy also diverges. Whereas conventional systems aim to eliminate the potential for conflict through structural separation and strict control logic, the RL-based approach assumes that conflicts are unavoidable in complex, dynamic environments. It therefore focuses on runtime resolution rather than avoidance by design.

In terms of flexibility, traditional systems exhibit low adaptability: routes and agent behavior are predefined, and any deviation requires replanning. RL agents in this work, by contrast, adapt their behavior based on local state and past experience, and their policies can in principle be updated continually as new data becomes available.

Scalability is another major point of divergence. Classical approaches suffer from two sources of scaling limitations: the structural constraints imposed by preventive layouts and the exponential runtime growth of MAPF algorithms with increasing agent count. While RL methods are not immune to scaling challenges, the use of shared policies and decentralized execution allows for more favorable scaling. Experience collection grows linearly with the number of agents, and decentralized coordination mitigates bottlenecks in centralized decision-making. Nevertheless, further research is needed to determine the upper limits of scalability in large-scale settings.

Finally, the use of simulation differs fundamentally. In traditional planning, simulation serves mainly to evaluate a finite set of predefined layout or routing variants. In this thesis, simulation is integrated into the learning process itself: it is used not only for evaluation but also for training, testing and evaluating RL policies under a wide range of conditions.



%

\chapter{Conclusion and Future Work}
\label{ch:conclusion}

This thesis advances the application of MARL to solve deadlock-capable MAPF problems in intralogistics systems. By integrating MARL into logistics planning, it provides a novel approach to improving system performance, resilience, and adaptability while addressing deadlocks, a critical and often overlooked challenge in logistics.

The work explicitly incorporates resilience into logistics objectives, alongside performance and cost-efficiency, to ensure that systems can adapt to and recover from disruptions. This inclusion addresses a significant gap in current methodologies, which focus primarily on static and deterministic planning approaches. The proposed procedure model enables logistics planners to integrate MARL into their processes, aligning it with traditional planning phases while extending these phases to handle deadlocks proactively. By leveraging MARL, systems dynamically adapt to operational changes, optimizing throughput and resource allocation in real time.

The results demonstrate that MARL policies, especially those trained using CTDE, excel in addressing dynamic and complex environments. CTDE policies avoid the exponential growth of action spaces by using shared training while maintaining decentralized decision-making during execution. 

Despite these contributions, the thesis identifies challenges. MARL solutions require computational resources and technical expertise. There are also further studies and improvements necessary to make MARL work in more complex scenarios with a higher number of agents.

\begin{figure}[h]
\centering
\begin{tikzpicture}[
    node distance=1.5cm
]

\node (start) [] {\textbf{Future work}};
\node(generalization_heading) [below of=start] {\textbf{Generalization}};
\node(improvement_heading) [left of=generalization_heading, xshift=-3.5cm,] {\textbf{Improvement}};
\node(dissemination_heading) [right of=generalization_heading, xshift=3.5cm] {\textbf{Dissemination}};

\node (improvement) [below of=improvement_heading, align=left, anchor=north, yshift=1cm] {
Environment sampling\\
throughput/efficiency\\ 
- JAX/C-based environments\\
\\
More realistic environments\\
- Continuous speed/turning \\ 
- Line-of-sight obstacles \\ 
- Dynamic obstacles\\
- Unexpected failures\\
\\
Combination with other\\
decision-making problems\\
- Prioritizing transport orders\\
- Fleet management \\
- Production scheduling\\
- Setup AMRs\\
\\
Better reward shaping\\
- Inverse RL from\\
optimal paths
};
\node (generalization) [below of=generalization_heading, align=center, anchor=north, yshift=1cm] {
Broader deadlock scenarios\\
- Foundation models for\\
deadlock handling\\
\\
Variation of\\
observation space\\
\\
Safe RL\\
- Requirements for\\
continual learning\\
- Prevention of\\
performance degradation\\
\\
Mixed policy populations\\
Large scale agent populations
};
\node (dissemination) [below of=dissemination_heading, align=right, anchor=north, yshift=1cm] {
German industry\\ 
- Practical guidance \\
- Blog/video content \\
in German\\
\\
Research RL community\\
- Open-source environments\\
- Published benchmarks \\ 

};

\draw (start) -- (generalization_heading);
\draw (start) -- (improvement_heading);
\draw (start) -- (dissemination_heading);

\end{tikzpicture}
\caption{Visualization of the future work after this thesis.}
\label{fig:future_work}
\end{figure}
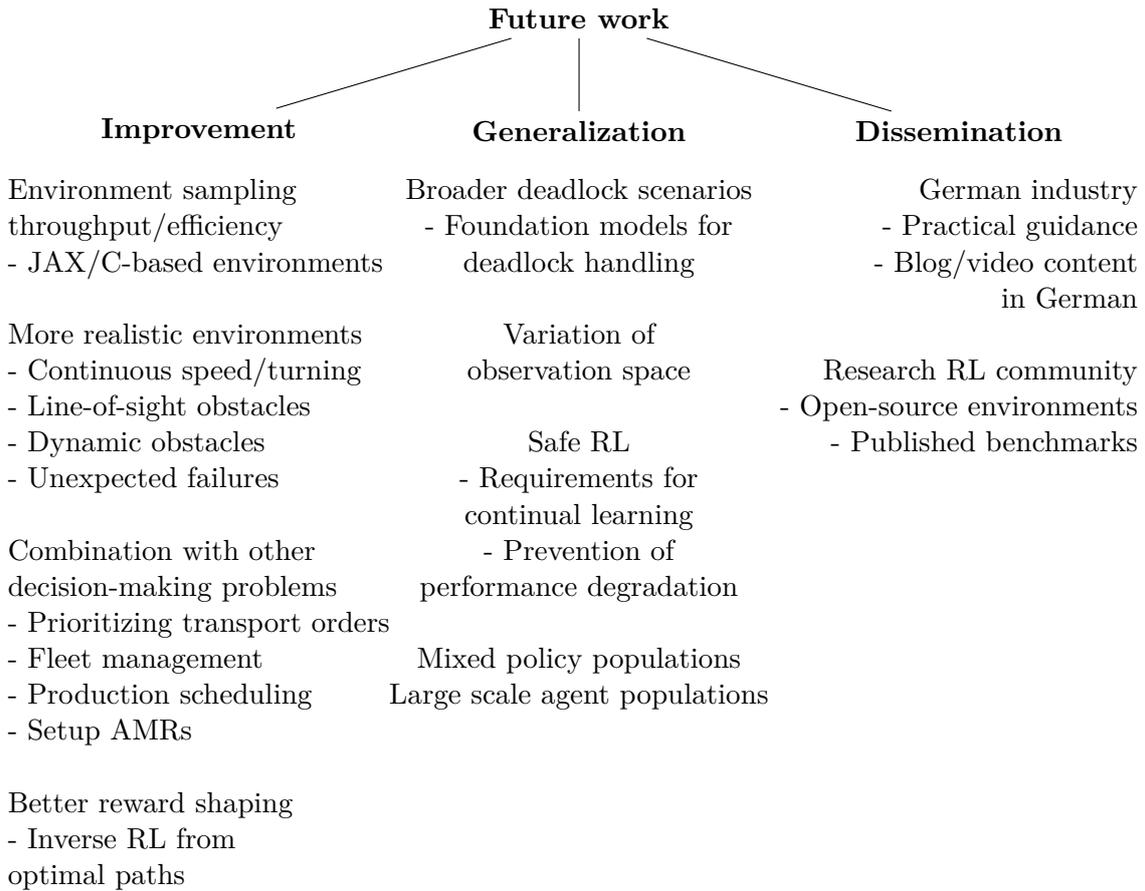

Future research should refine MARL-based solutions to better reflect real-world scenarios. For example, improving simulation environments by using JAX or C-based frameworks can enhance computational efficiency and throughput. Adapting action and observation spaces to align with robotic driving behavior, such as continuous speed and turning, and incorporating obstacles that obstruct the line of sight would make the solutions more realistic. Exploring dynamic obstacles and adding unexpected failures of the AMRs would provide a robust test of MARL policies under more challenging conditions. Combining MAPF with other decision-making problems—such as prioritizing transport orders, fleet management, production scheduling, and recharging of the AMRs, can further extend the practical relevance of the reference models. Figure \ref{fig:future_work} illustrates the directions for future work, grouped into three main areas: improvement, generalization, and dissemination.

The scope of this work can be extended to address broader deadlock scenarios in logistics, including UAV coordination, maritime traffic, and conveyor-based transport systems. Generalizing RL algorithms into foundational models for resolving deadlocks across diverse contexts would require a systematic approach. Training agents across a variety of logistics environments, while leveraging techniques such as transfer learning or meta-learning, could facilitate the reuse of policies. This approach would enable faster adaptation to new scenarios without extensive retraining. Embedding shared foundational capabilities into these models would enhance their robustness and versatility, significantly reducing development time and computational overhead when applied to new logistics settings. The primary advantage lies in scalability and efficiency. Foundational models could streamline deployment in previously unseen or evolving scenarios, ensuring smooth operations and improved throughput while minimizing manual intervention.

Safe RL represents another critical avenue for further exploration. Safe RL refers to the subset of RL methodologies that prioritize adherence to predefined safety constraints during both training and deployment phases. Safety constraints are typically formalized as restrictions on agent behavior to prevent adverse outcomes, such as physical harm, financial losses, or violations of regulatory requirements. These constraints can be probabilistic or deterministic, depending on the application domain. To ensure robustness in dynamic environments, it is necessary to establish clear requirements for continual learning while simultaneously addressing performance degradation. For instance, safe RL frameworks could be designed to identify and mitigate risks during training and execution, safeguarding against unintended behaviors. Safe RL frameworks could employ reward shaping, constraint satisfaction techniques, or dynamic policy adjustments to maintain safety standards. Mixed policy populations, where agents adopt heterogeneous strategies, could also provide insights into managing complexity. These populations may mirror real-world scenarios, such as multi-national fleets or hybrid logistics systems consisting of different types of AMRs, and studying their interactions could uncover strategies for achieving greater coordination and resilience.

Effectively communicating the findings of this thesis is essential for maximizing its impact. Publishing benchmarks and simulations as open-source tools would enable the research community to validate, reproduce, and extend these methods. Creating dissemination strategies tailored to industry practitioners, such as technical blog posts, video demonstrations, or practical guides in German, would help logistics planners and engineers in Germany adopt these solutions. By offering detailed case studies or step-by-step implementation guides, these resources could bridge the gap between academic research and real-world application. Additionally, publishing in academic journals and conferences would ensure that the work is recognized and built upon, contributing to the advancement of RL in logistics and related fields.

This thesis examines decision-making in logistics systems, where agents, like the thirsty man in al-Ghaz\={a}l\={\i}'s parable, risk paralysis when caught between equal choices. By shaping behavior through learned rewards, agents begin to act with purpose, turning stasis into movement. The results point toward logistics systems that do not depend on fixed rules but adapt through experience. Systems that move not because they must but because they know how.


\bibliographystyle{apa}
\bibliography{literatur}

\appendix

\pagestyle{empty}



\end{document}